\documentclass[reprint,
onecolumn,
amsmath,amssymb,
aps,
]{revtex4-2}

\usepackage{graphicx}
\usepackage{dcolumn}
\usepackage{bm}

\usepackage[T1]{fontenc}
\usepackage[utf8]{inputenc}
\usepackage[margin=0.8in]{geometry}

\usepackage{color}

\usepackage[table]{xcolor}
\usepackage{amsmath}
\graphicspath{ {./Images/} }
\usepackage{subfig}
\usepackage{float}
\usepackage[maxfloats=100]{morefloats}
\usepackage{multirow}
\usepackage{caption}
\usepackage{array}

\begin{document}

\preprint{APS/123-QED}

\title{Collision Dynamics and Deformation Behaviors of Multi-Core Compound Droplet Pairs in Microchannel Flow}

\author{S M Abdullah Al Mamun}
\author{Samaneh Farokhirad}%
 \email{samaneh.farokhirad@njit.edu}
\affiliation{Department of Mechanical and Industrial Engineering, New Jersey Institute of Technology, Newark, NJ 07114
}





\begin{abstract}

We numerically investigate the collision dynamics and deformation behaviors of double-core compound droplet pairs within confined shear flows using free-energy-based lattice Boltzmann method. While significant research has advanced our understanding of simple droplet pair interactions, the collision behaviors of core-shell compound droplets, where each shell contains one or more core droplets—remain largely unexplored. Even the pair-wise interaction of single-core compound droplets has not been extensively studied. We address how the interplay between physical parameters (i.e., density and viscosity ratios of immiscible fluids and Capillary number) and geometric parameters (i.e., initial offset distance between the shell droplets) affects the interaction time and collision outcomes of compound droplets. Our findings reveal that the presence of inner droplets significantly influences the deformation and stability of shell droplets, as well as the collision outcomes of both shell and core droplets. We identify several distinct collision outcomes, including (i) shell droplets’ coalescence, with core droplets remaining separate and rotating in a planetary-like motion, (ii) shell droplets’ pass-over, where core droplets maintain separation and exhibit both rotational and translational motion, (iii) core droplets’ coalescence, with shell droplets passing each other, and (iv) shell droplets’ pass-over, with the coalescence of core droplets. We demonstrate that the transition between these collision outcomes is governed by varying the Capillary number and initial offset. Additionally, we observed that increasing the density and viscosity ratios from unity to larger values always results in the pass over of shell droplets, with the core droplets remain separated and experience rotational motion.

\end{abstract}

\maketitle


\section{\label{sec:level1}Introduction}

In recent years, compound droplets—characterized by an inner droplet(s) encapsulated within an outer fluid shell—have emerged as a unique class of soft materials due to their structural complexity and wide-ranging applications~\cite{lissant1974emulsions, datta201425th, yaqoob2006multiple}. The versatility of compound droplets is underscored by their prevalence in nature and their utility in various engineering domains, including but not limited to food processing~\cite{serdarouglu2013multiple, maan2011spontaneous, cho2003evaluation}, microfluidics~\cite{xu1999gas, zhao2001co, zhao2011microfluidic, halliday2006improved}, cosmetics~\cite{bernardo2005integrated, patravale2008novel, kouhi2020edible}, emulsification~\cite{nilsson2006adsorption, lobo2003coalescence, jafari2008re}, materials synthesis~\cite{lee2008double, zhao2012bio}, and biological systems ~\cite{abkarian2007swinging, fedosov2010multiscale, pivkin2008accurate}. The ability to control these complex structures has sparked considerable interest, particularly in microfluidic applications, where the encapsulation and controlled release of active ingredients are essential for processes such as targeted drug delivery~\cite{patra2018nano, zhao2013multiphase, singh2017nanoemulsion, sarker2005engineering}, precision-controlled lab-on-a-chip devices~\cite{jayamohan2017advances, gupta2016lab, yilmaz2018lab, gupta2016lab}, and enhanced oil recovery~\cite{li2020advances, son2014enhanced, jia2021potential}. Advances in microfluidics have enabled precise control over droplet generation, making it possible to create double and multiple emulsions with multiple inner droplets encapsulated within a single shell. This development has led to compound droplet systems with increasing degrees of internal complexity, such as double-core and multi-core configurations.

Within the realm of droplet collision research, considerable progress has been made in understanding the behaviors of simple droplets without any core under shear flow conditions, but much less is known about the pairwise interactions and their deformation dynamics when these droplets contain one, two or even multiple cores. Although there is some research on single compound droplets with multiple cores, investigations into compound droplet pair collisions remain limited. The behavior of compound droplets with inner cores under shear flow has demonstrated that core-shell interaction dynamics can drastically alter their deformation and interface structures compared to simple, single-phase droplets~\cite{tiribocchi2020novel}. These phenomena are especially significant in systems where precise control over droplet morphology and stability is required, as the interaction forces and internal configurations of compound droplets impact emulsion stability, active ingredient release, and overall efficiency of droplet-based processes.

Studies on simple droplet pairs colliding under confined shear flow have shown that collision outcomes are influenced by factors such as droplet size, confinement ratio, initial offset, and flow characteristics. Allan and Mason~\cite{allan1962particle} first established the foundational concepts of droplet collisions, highlighting the behaviors of simple, single-phase droplets subjected to shear flow, which can exhibit pass-over, coalescence, or reverse-back motions depending on the interaction forces and droplet positions. Later studies by Guido and Simeone~\cite{guido1998binary}, along with numerous experimental investigations~\cite{de2013effect, de2014effects, chen2009effect}, expanded on these findings, revealing that confinement enhances the likelihood of coalescence due to increased interface interactions. These studies introduced key parameters like Capillary ($Ca$) number, viscosity ratios, and confinement effects as critical factors determining collision outcomes. Recent computational work has extended these insights by exploring how interfacial diffusivity and hydrodynamic forces, such as shear, inertia, and viscous forces, interact with geometric constraints to influence droplet behaviors~\cite{sarkar2013spatial, vananroye2006effect, grizzuti1997effects, shardt2013simulations, chen2014hydrodynamic, bayareh2011binary, loewenberg1997collision}. Our recent study further examined these interactions, demonstrating that specific combinations of initial offsets, confinement, viscosity and density ratios, as well as $Ca$ effects produce distinct collision regions, highlighting the complex interplay between geometry and flow parameters~\cite{al2022probing, al2024ml}.

While these studies have provided significant insights into single-phase droplet collision dynamics, the more complex behaviors of droplets, in the context of core-shell compound droplets, remain underexplored, particularly in two-core or multi-core configurations where additional interfaces and internal forces are introduced. Unlike simple droplets, compound droplets possess distinct internal morphologies, with encapsulated droplets that can deform independently, creating more intricate interactions within shear flows. Existing research on single-core compound droplets has shown that the presence of the inner droplets alters collision dynamics by adding layers of structural complexity that affect both the droplet's morphology and the overall hydrodynamic response. Studies by Liu et al.~\cite{liu2021deformation} and Luo et al.~\cite{luo2015deformation} have shown that factors like core-to-shell size ratios, interfacial tension, and relative viscosities contribute to diverse deformation behaviors and even breakup under certain flow conditions. Chen et al.~\cite{chen2015deformation} combined experimental results with numerical simulations to study compound droplet deformation under shear flow, identifying two distinct deformation patterns linked to different physical mechanisms. Vu et al.~\cite{vu2018numerical} found that compound droplets deform and can break into smaller droplets depending on flow conditions. Chen et al.~\cite{chen2013hydrodynamics} numerically examined droplet breakup and deformation in double emulsions under shear flow. Hua et al.~\cite{hua2014dynamics} investigated the effects of parameters like radius, interfacial tension, and inner droplet position on compound droplet deformation. 

However, these studies focused on the deformation and dynamic behavior of individual compound droplets and studies on the interactions between compound droplet pairs with multiple (and even one) core droplets remain to be explored. Recent advances in microfluidics have enabled the use of multi-core emulsions ~\cite{utada2005monodisperse, vladisavljevic2017microfluidic, ding2019double, chu2007controllable, abate2009high, wang2011controllable, clegg2016one}, making the study of hydrodynamic interactions and collision outcomes between droplets pairs essential. Advancing the dynamic behavior of single compound droplets under confined flows, some recent studies have explored the hydrodynamics of core-shell compound droplet pair interactions and their dynamics upon collision. Specifically, Vu et al.~\cite{vu2019parametric} and Liu et al.~\cite{liu2018passing} demonstrated that core-shell compound droplets display distinct collision behaviors when subjected to shear flow, with the presence of inner droplets influencing collision modes such as pass-over and reverse-back. Nguyen et al.~\cite{nguyen2020collision} have also highlighted that the position of the inner droplets can influence collision behavior within a certain range of $Ca$ numbers, revealing the sensitivity of core-shell droplets to flow conditions. In our recent work~\cite{al2024hydrodynamic}, we examined how the core-shell size ratio, initial offset, and wall confinement collectively govern the transitions between collision modes, as well as the morphology, trajectory, and final state of single-core compound droplet pairs. These initial explorations suggest that increasing the number of core droplets potentially introduces additional degrees of freedom in the motion and deformation of multi-core compound droplet pairs during collisions, particularly under confined conditions that impose spatial restrictions on droplet behaviors.

Building upon this foundational research, our current study aims to investigate the dynamic interactions of double-core compound droplet pairs in confined shear flow. This configuration, where each shell droplet encapsulates two core droplets, represents a further evolution in compound droplet complexity, providing an opportunity to examine how multiple inner droplets interact with each other and with the outer shell droplets during collision events. By extending our previous research on single-core compound droplet pairs~\cite{al2024hydrodynamic}, we delve into a novel and largely uncharted domain of multi-core droplet interactions that are relevant to the design and optimization of emulsion-based microfluidic devices, where collision control and precise manipulation of inner droplet distributions are critical. The significance of studying double-core compound droplets lies in the intricate interplay of forces—capillary, shear, and viscous—that occur as the internal droplets deform, reposition, and interact within the encapsulating shell under varying flow conditions.


To capture these complex interactions, we employ the lattice Boltzmann method (LBM), a mesoscopic approach that has proven effective in simulating multiphase flows with large topological interface changes~\cite{lee2010lattice, lee2009effects, shan1993lattice, qian1997finite, chen1998lattice, xu2006morphologies, aidun2010lattice, wen2014galilean, succi2015lattice, huang2015multiphase, li2016lattice,mshad2017}. Based on the Cahn-Hilliard diffuse interface theory for binary fluids~\cite{lee2010lattice, lee2009effects}, our LBM framework provides smooth transitions of physical quantities across interfaces, making it well-suited for handling the intricate interactions inherent to double-core droplets in confined channels. This approach eliminates the need for dynamic interface tracking and reconstruction~\cite{lafaurie1994modelling, tryggvason2001front, luo2015conservative, harvie2006analysis, ryu2012comparative, zahedi2012spurious}, which is particularly advantageous in the context of compound droplets with evolving morphologies and multiple interacting interfaces. While LBM can be prone to numerical instabilities, the formulation used here has been validated in our prior studies~\cite{farokhirad2023spreading, farokhirad2015coalescence, farokhirad2017coalescence, farokhirad2013effects}, including the pair-wise collision of single-core compound droplets, demonstrating its robustness in mitigating spurious currents at equilibrium.

Through this study, we aim to contribute a systematic analysis of collision behaviors for double-core compound droplet pairs in shear flow, focusing on how parameters such as, $Ca$ number, viscosity and density ratio, initial offset, and confinement jointly affect collision modes, deformation, and morphological evolution. The results presented here offer new insights into the complex hydrodynamics of multi-core compound droplets, establishing a foundation for future advancements in the design and control of emulsion-based microfluidic devices and enabling a deeper understanding of multi-phase fluid dynamics in confined geometries. The organization of this paper is as follows. Section II outlines the computational model developed to simulate the dynamics and deformation behaviors of compound droplet pairs during collisions in confined shear flow. In Section III, we present the primary numerical results along with a discussion on the implications of hydrodynamic interactions, deformation, and stability in double-core compound droplet collisions. Finally, Section IV summarizes the key findings and major insights derived from this study.

\section{\label{sec:level2}Method}

Using a lattice Boltzmann method (LBM)~\cite{lee2010lattice}, which was developed based on the Cahn-Hilliard diffuse interface theory for a binary system of incompressible and immiscible fluids, we study the hydrodynamics and interfacial deformation of compound droplet pairs during collision in confined shear flows. The method solves the continuity and the Navier–Stokes equations by solving the evolution of two distinct density distribution functions, $h_{\alpha}$ and $g_{\alpha}$, respectively. A brief overview of the method is provided here, with readers directed to previous works~\cite{mohamad2011lattice,bhatnagar1954model,he1998discrete,lee2009effects,lee2014physical} for more details on the theory.


The motion of the fluid interface is tracked by the Cahn-Hilliard equation~\cite{lee2010lattice}, which describes the evolution of a scalar order parameter $C$ (i.e., composition) as it advances toward its equilibrium state:
\begin{equation}
    \frac{\partial C}{\partial t}+\mathbf{u} \cdot \nabla C =M {\nabla}^2 \mu_c
    \label{CH-2}
\end{equation}
Here, $\mu_c$ is the chemical potential and $M$ represents the mobility that accounts for the interface diffusion. 

In the mesoscopic LBM simulation, the advection-diffusion equations are solved through a two-step process: (i) a streaming step, wherein the density distribution functions at each lattice node propagate to neighboring nodes along a set of discrete velocities $\mathbf{e}_{\alpha}$, where $\alpha$ identifies the discrete direction; and (ii) a collision step, which accounts for the effects of collisions between molecules and relaxes the density
distribution function at each lattice node towards an equilibrium distribution. The type of lattice arrangement used depends on the number of dimensions in the simulation and the number of directions used for the information propagation. In this study, we have adopted the D2Q9 model of lattice, where D2 denotes a two-dimensional domain and Q9 refers to the speed model with 9 linkages or directions. 

Out of the two distribution functions, $g_{\alpha}$ and $h_{\alpha}$, the former is used to compute the pressure and momentum of the two-component mixture, while the latter serves as a phase-field function for the transport of the composition. These two distribution functions evolve according to the following equations~\cite{lee2010lattice}:

\begin{equation}
\begin{aligned}
\frac{\partial g_{\alpha}}{\partial t} + \mathbf{e}_{\alpha} \cdot \nabla g_{\alpha} = - \frac{1}{\lambda}(g_{\alpha} - g_{\alpha}^{eq}) 
+ (\mathbf{e}_{\alpha} - \mathbf{u}) \cdot [\nabla \rho c_s^2 (\Gamma_{\alpha}-\Gamma_{\alpha}(0))+ \mu_c \nabla C \Gamma_{\alpha} ], \\
\frac{\partial h_{\alpha}}{\partial t} + \mathbf{e}_{\alpha} \cdot \nabla h_{\alpha} =  -\frac{1}{\lambda} (h_{\alpha}-h_{\alpha}^{eq}) + M\nabla^2 \mu_c \Gamma_{\alpha} + (\mathbf{e}_{\alpha}-\mathbf{u}) \cdot [\nabla C - \frac{C}{\rho c_s^2} (\nabla p - \mu_c \nabla C)]\Gamma_{\alpha}.
\end{aligned}
\end{equation}


Here, $g_{\alpha}=f_{\alpha}c_s^2+(p-\rho c_s^2)\Gamma_{\alpha}(0)$ and $h_{\alpha}=(C/\rho)f_{\alpha}$ are the new distribution functions corresponding to the discretized microscopic velocity $\mathbf{e}_{\alpha}$, $\mathbf{u}$ is the volume-averaged velocity, $c_s$ is the lattice speed of sound, $\rho$ is the mixture density, $p$ is the hydrodynamic pressure, and $\lambda$ is the relaxation time. $f_{\alpha}$ is the traditional distribution function and $\Gamma_{\alpha}$ is defined as $\Gamma_{\alpha}=  \Gamma_{\alpha}(\textbf{u})=f_{\alpha}^{eq}/C$. The equilibrium distribution functions are then defined as:

\begin{equation}
\begin{aligned}
f_{\alpha}^{eq} = t_{\alpha} \rho \left [ 1+ \frac{\mathbf{e}_{\alpha}\cdot \mathbf{u}}{c_s^2} + \frac{(\mathbf{e}_{\alpha}\cdot \mathbf{u})^2}{2c_s^4} - 
\frac{\mathbf{u}\cdot \mathbf{u}}{2c_s^2} 
\right ], \\
g_{\alpha}^{eq} = t_{\alpha} \left [ p + \rho c_s^2 \left ( \frac{\mathbf{e}_{\alpha}\cdot \mathbf{u}}{c_s^2} + \frac{(\mathbf{e}_{\alpha}\cdot \mathbf{u})^2}{2c_s^4} - 
\frac{\mathbf{u}\cdot \mathbf{u}}{2c_s^2} \right )
\right ], \\
h_{\alpha}^{eq} = t_{\alpha} C \left [ 1+ \frac{\mathbf{e}_{\alpha}\cdot \mathbf{u}}{c_s^2} + \frac{(\mathbf{e}_{\alpha}\cdot \mathbf{u})^2}{2c_s^4} - 
\frac{\mathbf{u}\cdot \mathbf{u}}{2c_s^2} 
\right ].
\end{aligned}
\end{equation}

In the above equations $t_{\alpha}$ is the weighting factor corresponding to the $\mathbf{e}_{\alpha}$ direction for a particular quadrature. The dimensionless relaxation time $\tau=\lambda/\delta t$ is related to the kinematic viscosity by $\nu=\tau c_s^2 \delta t$.
The density can be related to the composition using the linear function~\cite{lee2010lattice} $\rho = C \rho_1 + (1-C)\rho_2$, with $\rho_1$ and $\rho_2$ being the bulk densities of two fluids. The chemical potential, $\mu_c$ is defined as the derivative of the free energy with respect to the order parameter, with the free energy defined as~\cite{lee2014physical}:
\begin{equation}
    E(C) = E_0(C) + \frac{\kappa}{2} \left | \nabla C \right |^2,
\end{equation}
where $E(C)$ is the total free energy, $E_0(C) = \beta C^2(C-1)^2$~\cite{lee2010lattice} is the bulk free energy with $\beta$ being a constant, $\nabla C$ is the composition gradient, and $\kappa$ is the gradient parameter related to the surface tension of the interface between two phases. The plane interfacial profile at equilibrium is then evaluated as $C(z) = \frac{1}{2}+\frac{1}{2}\text{tanh}(\frac{2z}{\xi})$~\cite{lee2010lattice}, where $z$ denotes the coordinate normal to the interface plane and $\xi$ is the interface thickness. Given $\xi$ and $\beta$, the gradient parameter $\kappa=\beta\xi^2/8$ and the fluid-fluid surface tension $\sigma=\sqrt(2\kappa\beta)/6$ are prescribed.
Finally, the macroscopic variables such as composition, momentum, and hydrodynamic pressure can be obtained by taking the moments of $h_{\alpha}$ and $g_{\alpha}$:

\begin{equation}
\begin{aligned}
    C = \sum_{\alpha}h_\alpha, \\
    \rho \mathbf{u} = \frac{1}{c_s^2}\sum_{\alpha} \mathbf{e}_{\alpha}g_{\alpha}, \\
    p = \sum_{\alpha} g_{\alpha}.
\end{aligned}
\end{equation}

For detailed discretization of
Eqs.(2) readers are referred to Ref.~\cite{lee2010lattice}.


\section{\label{sec:level3}Results and Discussion}

\subsection{Simulation Framework}

\begin{figure}[H]
	\centering
		\includegraphics[scale = 0.33]{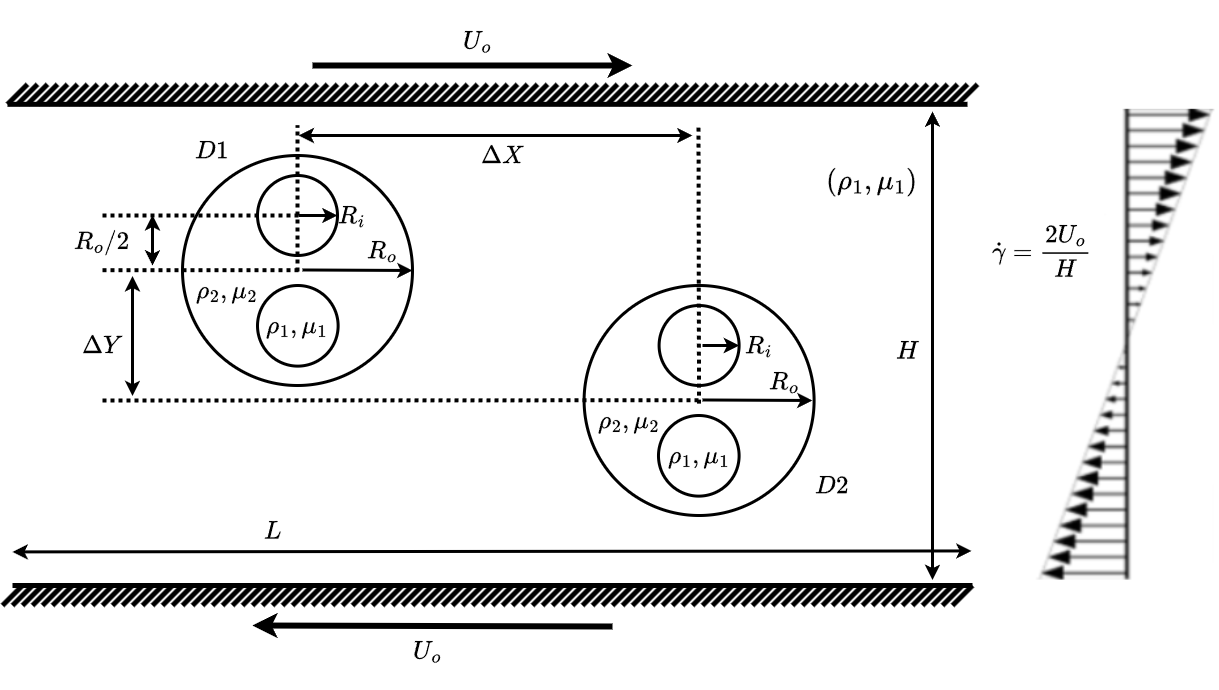} \\
		
	\caption{Schematic representation of a compound droplet pair (each shell droplet having two core droplets within) in a shear flow,  confined by two parallel plates separated by a distance $H$. The droplet initially at left is labeled as $D1$, and the one at right as $D2$. The core and shell droplets have initial radii of $R_i$ and $R_o$, respectively, and the core droplets are positioned halfway from the center of shell droplet with vertical symmetry. $\Delta X$ and $\Delta Y$ represent the horizontal and vertical distances, respectively, between the centers of the droplets at the onset.}
	\label{fig:schematic-3}
\end{figure}

In this section, the collision, trajectories, and deformation of two double-core compound droplets in a shear flow confined by the motion of two parallel walls with opposite velocities $\pm{U_o}$ and a shear rate of $\dot{\gamma} = 2U_o/H$, with $H$ being the vertical spacing between walls, are studied. Figure~\ref{fig:schematic-3} shows the schematic of a two dimensional simulation box with two initially circular compound droplets placed inside the computational domain. The shell droplets, both of radius $R_o$ and each having two core droplets of radius $R_i$, are placed in a continuous fluid between the walls.
The compound droplet pairs are separated by a horizontal distance of $\Delta X$ and a vertical offset of $\Delta Y$. The shell droplets are occupied by fluid $1$ with density $\rho_1$ and dynamic viscosity $\mu_1$. The core droplets, which are the same as the matrix fluid, contain fluid $2$ with density $\rho_2$ and dynamic viscosity $\mu_2$. The surface tension at each interface is denoted by $\sigma$. Given the scale of the droplets' size in our study, which is significantly smaller than the capillary length (i.e., $\sqrt(\sigma/(\Delta \rho g)$, where $g$ is gravitational acceleration), the influence of gravity has been neglected.

The dimensionless parameters that play critical roles in characterizing the coalescence of compound droplet pairs  are the Reynolds number ($Re =  \frac{\rho_2 \dot{\gamma }R_o^2}{\mu_2 }$, which quantifies the relative importance between the inertia force and the viscous force), the Capillary number ($Ca= \frac{\mu_2 \dot{\gamma }R_o}{\sigma}$, which shows the ratio of the viscous force to the surface tension force), density ratio ($\rho_{12}=\frac{\rho_1}{\rho_2}$), viscosity ratio ($\mu_{12}=\frac{\mu_1}{\mu_2}$), dimensionless vertical initial offset $(\frac{\Delta Y_o}{R_o})$, droplet radius ratio ($R_{r}= \frac{ R_o}{R_i}$), and the confinement ratio $(\frac{R_o}{H})$. The deformation of compound droplets (both core and shell) is expressed
by the Taylor deformation parameter, $D_T = \frac{A-B}{A+B}$, where $A$ and $B$ represent the major and minor axes of deformed droplets, respectively.
A summary of the detailed parameter set is provided in Table~\ref{tab:parameters} along with their values considered in this work. 


The reliability of our computational approach has been comprehensively validated in our previous works, specifically for the pairwise collisions of simple droplets and single-core compound droplets, as detailed in~\cite{al2022probing}  and~\cite{al2024hydrodynamic}. These studies demonstrated that our simulation algorithm provides results in great agreement with experimental  and theoretical expectations. In the current study, we extend this validated methodology to investigate the pairwise interactions of double-core compound droplets under confined shear flows, a problem that is novel and largely unexplored in the literature. Given the demonstrated reliability of our numerical model in analogous contexts and the lack of established benchmarks for this specific scenario, no additional validation is performed. Instead, our focus is on leveraging the proven strengths of our approach to provide pioneering insights into these complex interactions, offering a foundation for future experimental and theoretical explorations.

\begin{table}[hbt!]
\centering
	\caption{ System Parameters and Corresponding Values Utilized in the Simulations.} 
 \label{tab:parameters}
\setlength{\extrarowheight}{6pt}
\begin{tabular}{|l c  c|} 
 \hline
 Parameter & Definition & Value \\  
 \hline\hline
 Reynolds number & $Re = \frac{\rho_2\dot{\gamma }R_o^2}{\mu_2}$ & 1.0 \\ 
 Capillary number & $Ca = \frac{\mu_2 \dot{\gamma }R_o}{\sigma}$ & 0.05-0.10 \\
 confinement & $\frac{R_o}{H}$ & 0.40 \\
vertical Offset & $\frac{\Delta Y}{R_o}$ & 0.00-1.50 \\
 horizontal separation & $\frac{\Delta X}{R_o}$ & 2.50 \\ 
 radius ratio & $R _{r} = \frac{R _i}{R_o}$ & 0.375 \\
 density ratio & $\rho _{12} = \frac{\rho _1}{\rho _2}$ & 1.0-400.0 \\ 
 viscosity ratio & $\mu _{12} = \frac{\mu _1}{\mu _2}$ & 1.0-60.0 \\ 
\hline
\end{tabular} 
\end{table}


\subsection{Effect of Initial Vertical Offset and Ca number}\label{sec_5.2}

Building on insights from our previous research~\cite{al2024hydrodynamic}, which examined the effect of single-core droplet on the pair-wise collision of core-shell compound droplets in shear flows, we found that core droplets significantly influence the deformation and stability of the shell droplets. Surface tension, along with viscosity, plays a crucial role in determining how the droplet interfaces respond to different stress environments. The $Ca$ number serves as a key parameter to investigate the role of surface tension in these interactions. Additionally, the initial vertical offset between droplets has been identified in prior studies as a critical parameter influencing collision modes and deformation in the pairwise interactions of simple droplet pairs and single-core compound droplet pairs. In this section, we explore the interaction between the vertical initial offset ($\frac{\Delta Y}{R_o}=0-1.5$) and Ca ($Ca=0.05-0.1$), while keeping the viscosity and density ratios at unity, on the trajectories and collision modes of compound double core droplet pairs, which has not been addressed in previous investigations. 

\subsubsection{$Ca = 0.10$}

\begin{figure}[H]
          \centering
         
         \includegraphics[scale = 0.09]{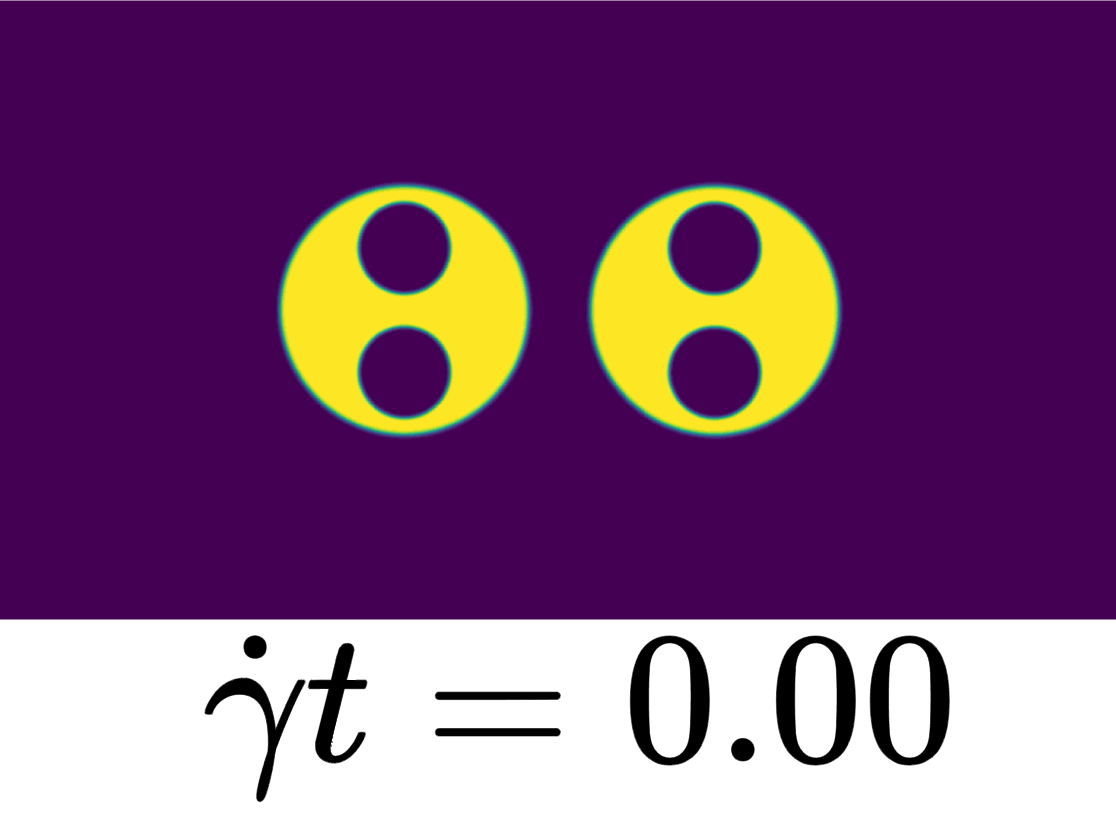}
		\includegraphics[scale = 0.09]{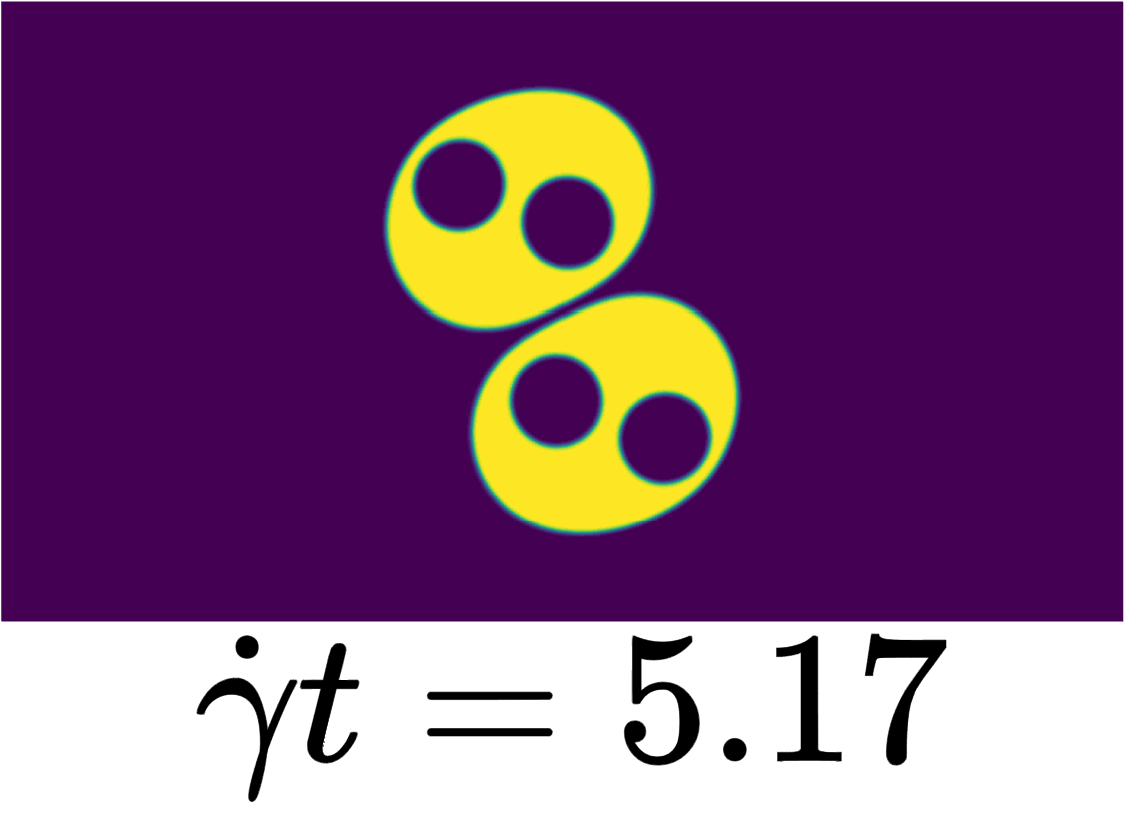}
		\includegraphics[scale = 0.09]{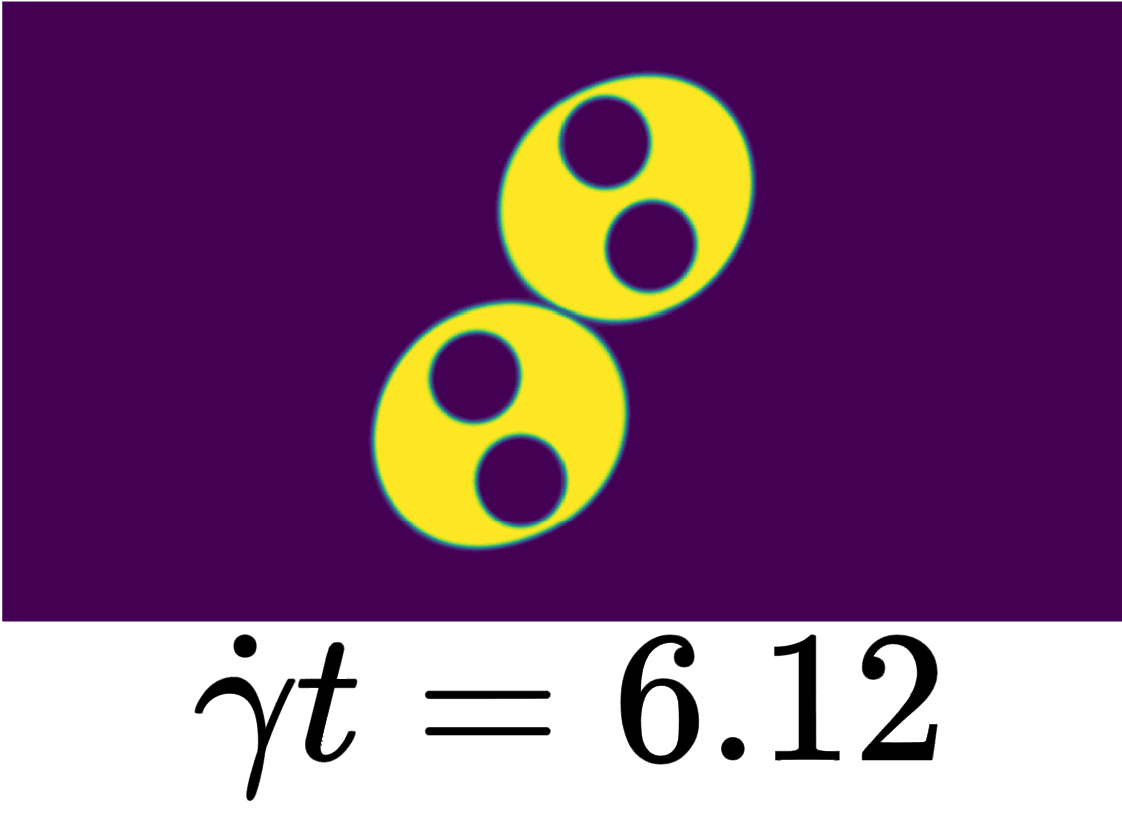}
		\includegraphics[scale = 0.09]{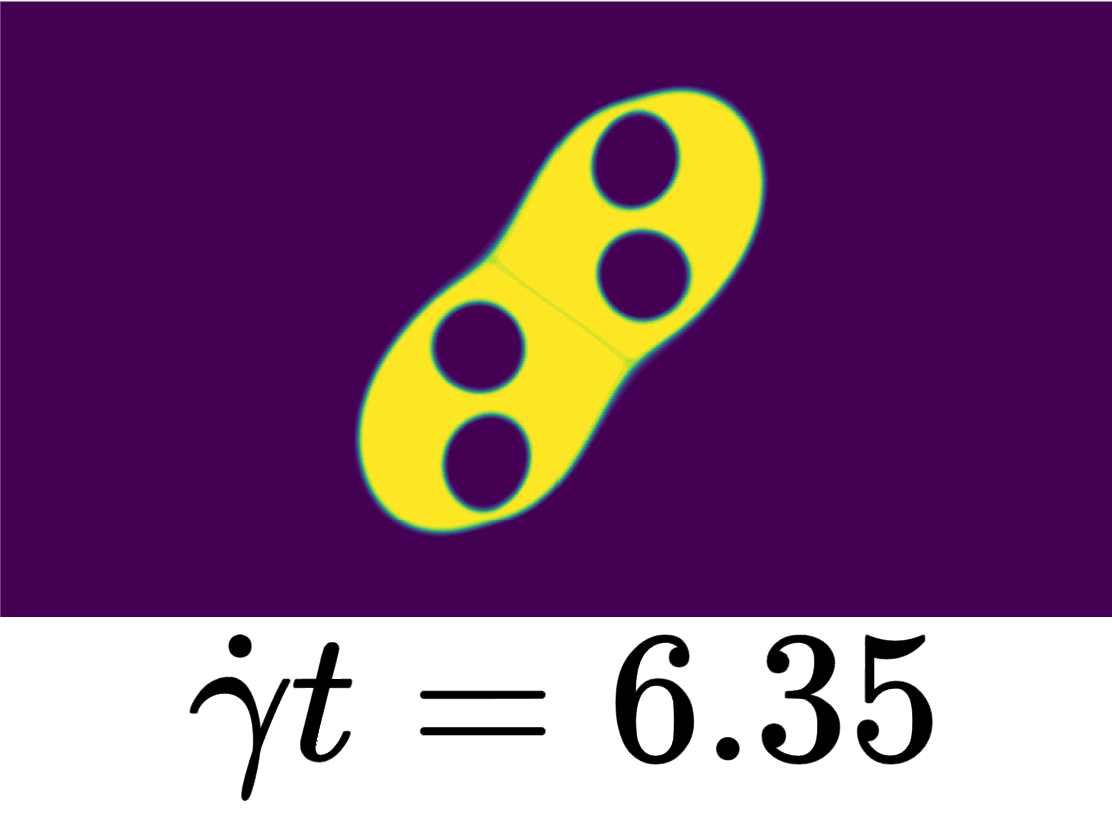}
		\includegraphics[scale = 0.09]{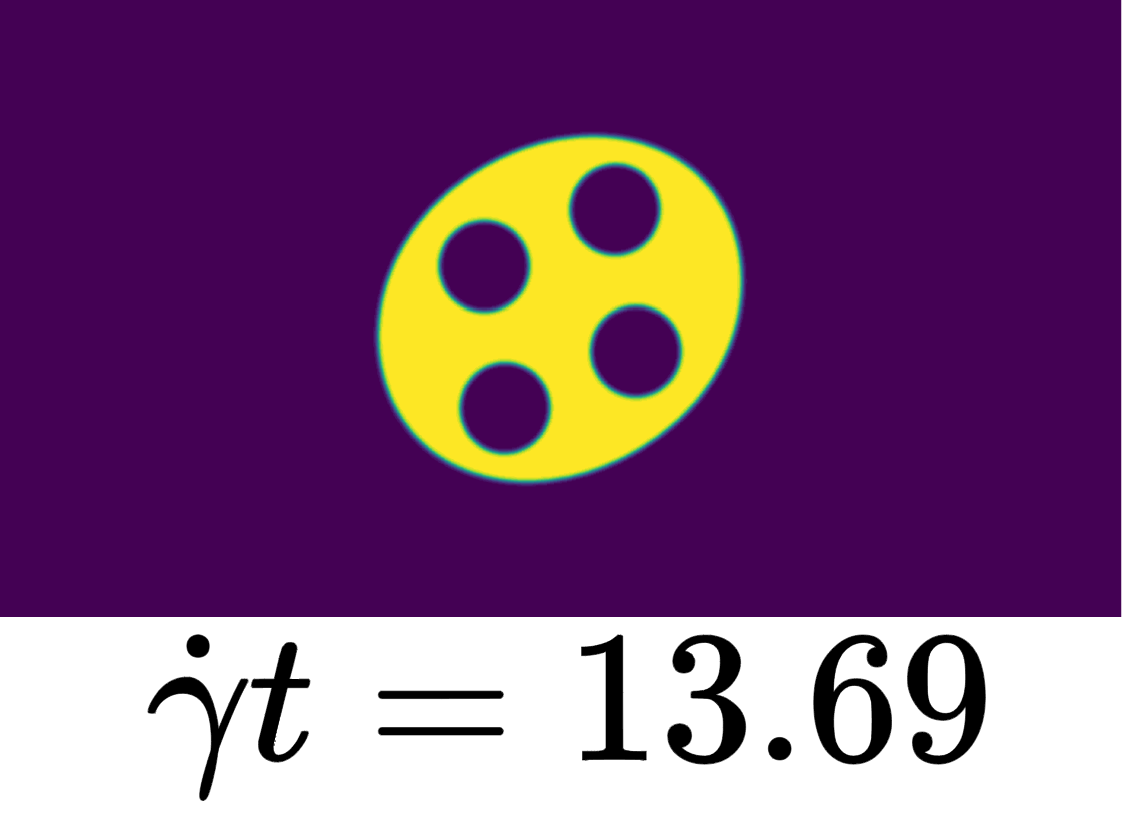} \\
        (a) $\frac{\Delta Y}{R_o}=0.00$~(Shell droplets coalesce)
		\\ \vspace{0.5cm}

        \includegraphics[scale = 0.09]{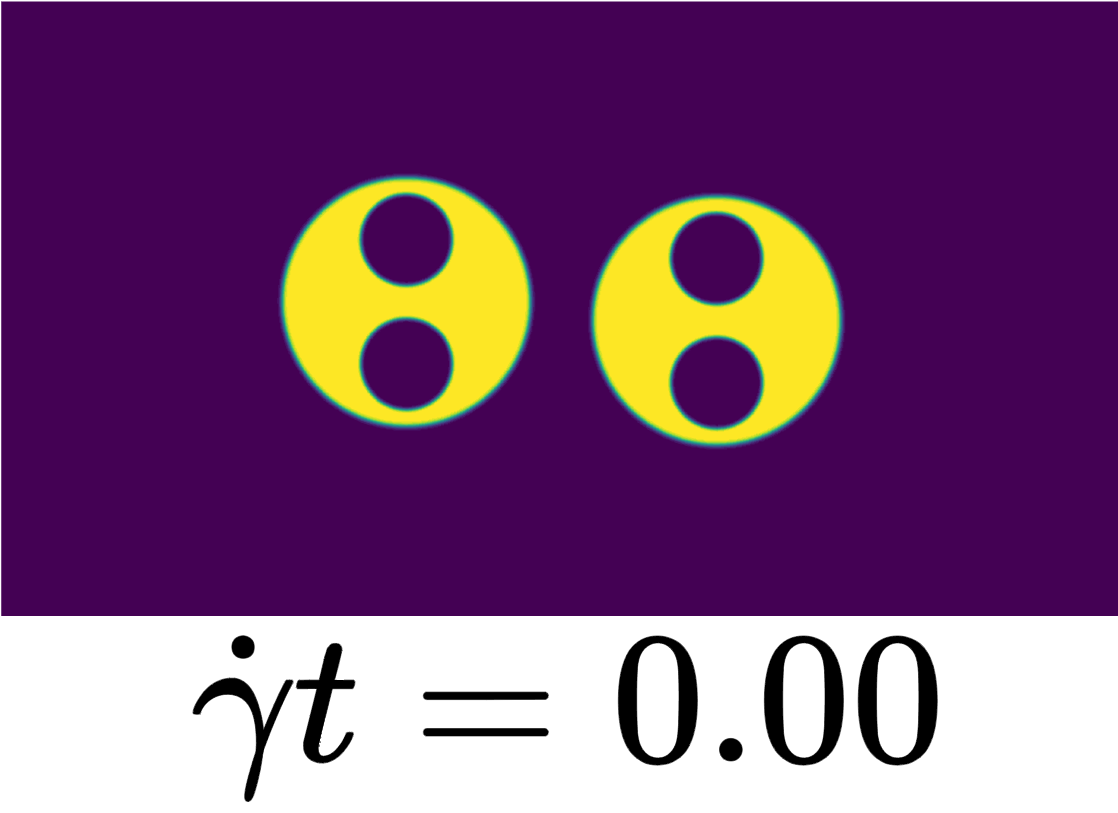}
            \includegraphics[scale = 0.09]{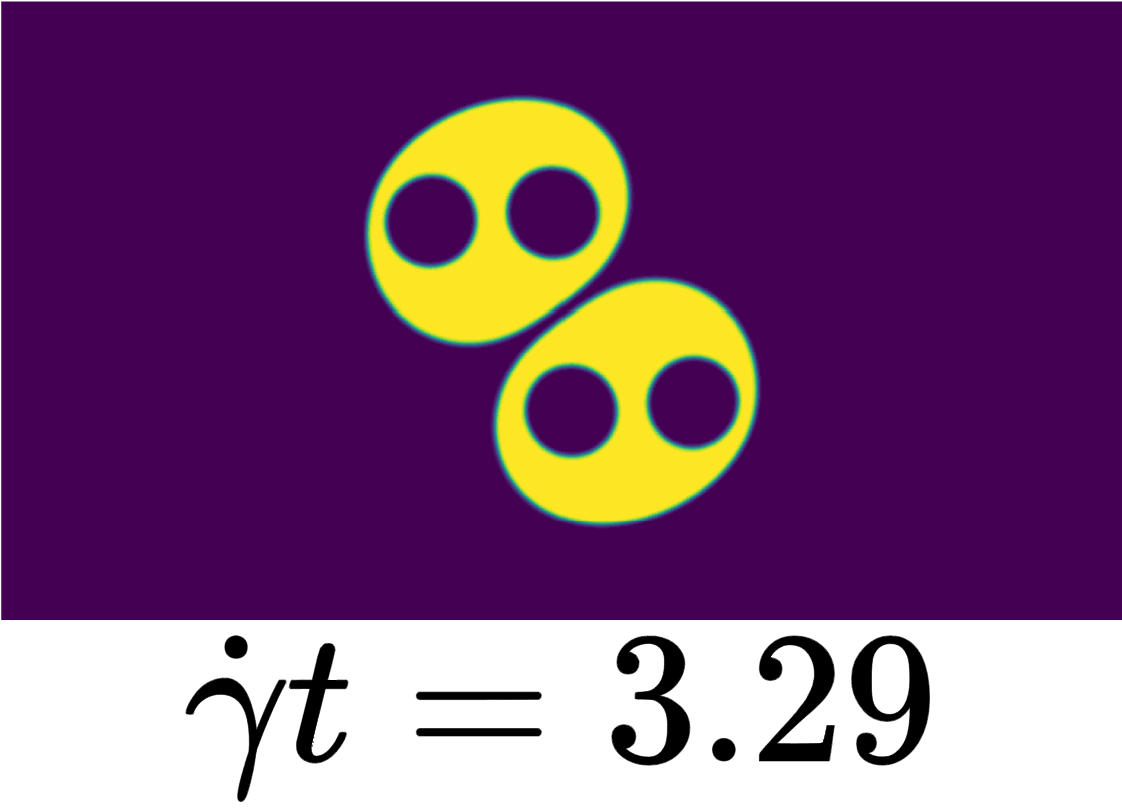}
		\includegraphics[scale = 0.09]{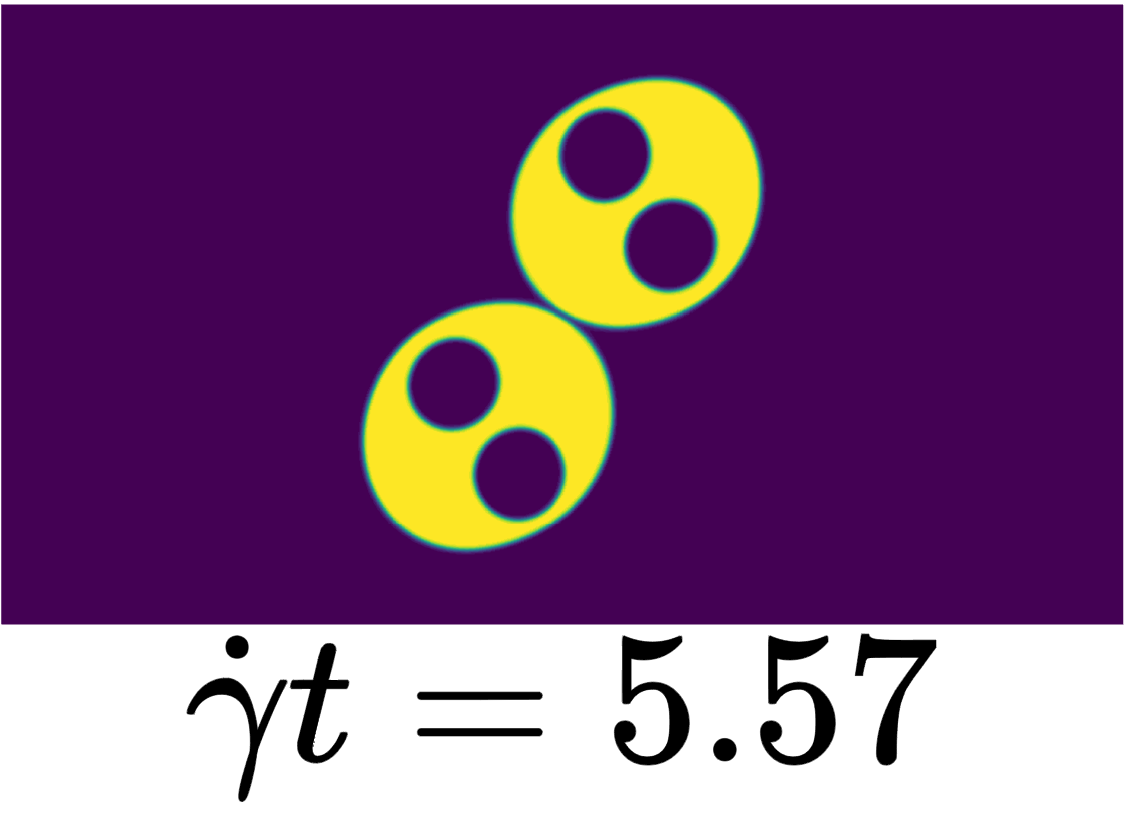}
		\includegraphics[scale = 0.09]{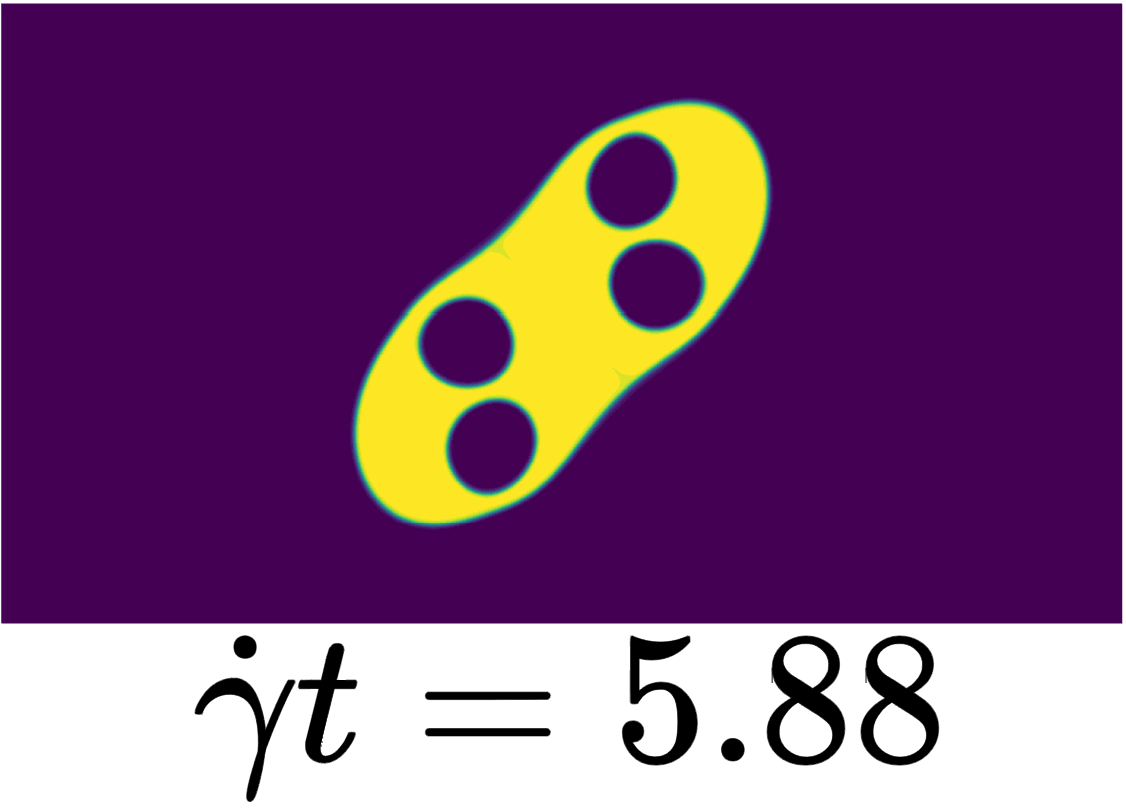}
		\includegraphics[scale = 0.09]{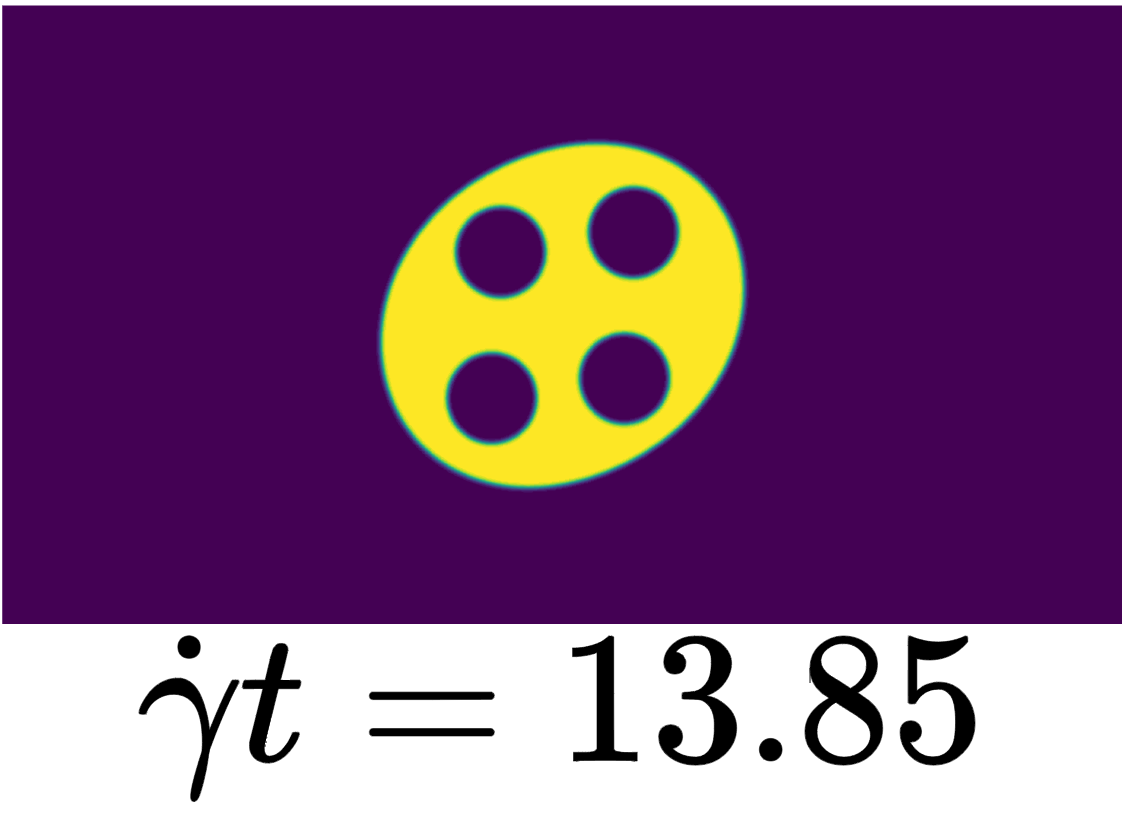}  \\
       (b)  $\frac{\Delta Y}{R_o}=0.15$~(Shell droplets coalesce)
  \vspace{0.5cm}

   \includegraphics[scale = 0.09]{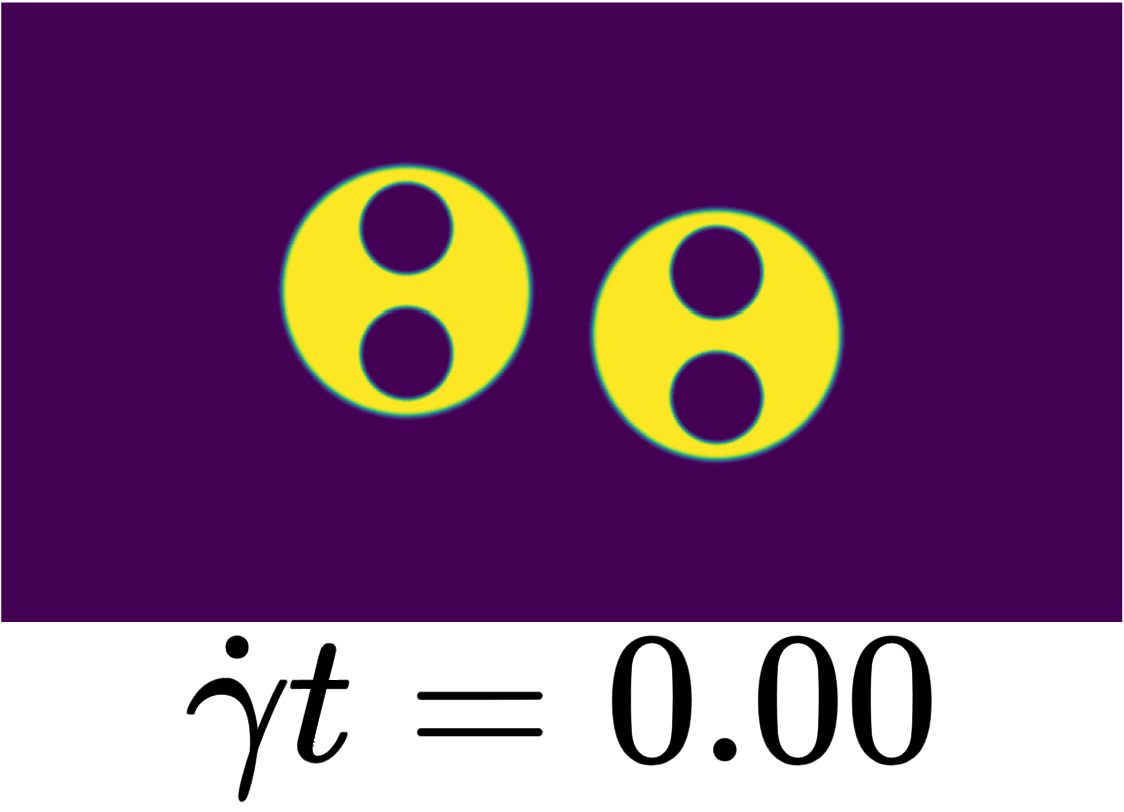}
		\includegraphics[scale = 0.09]{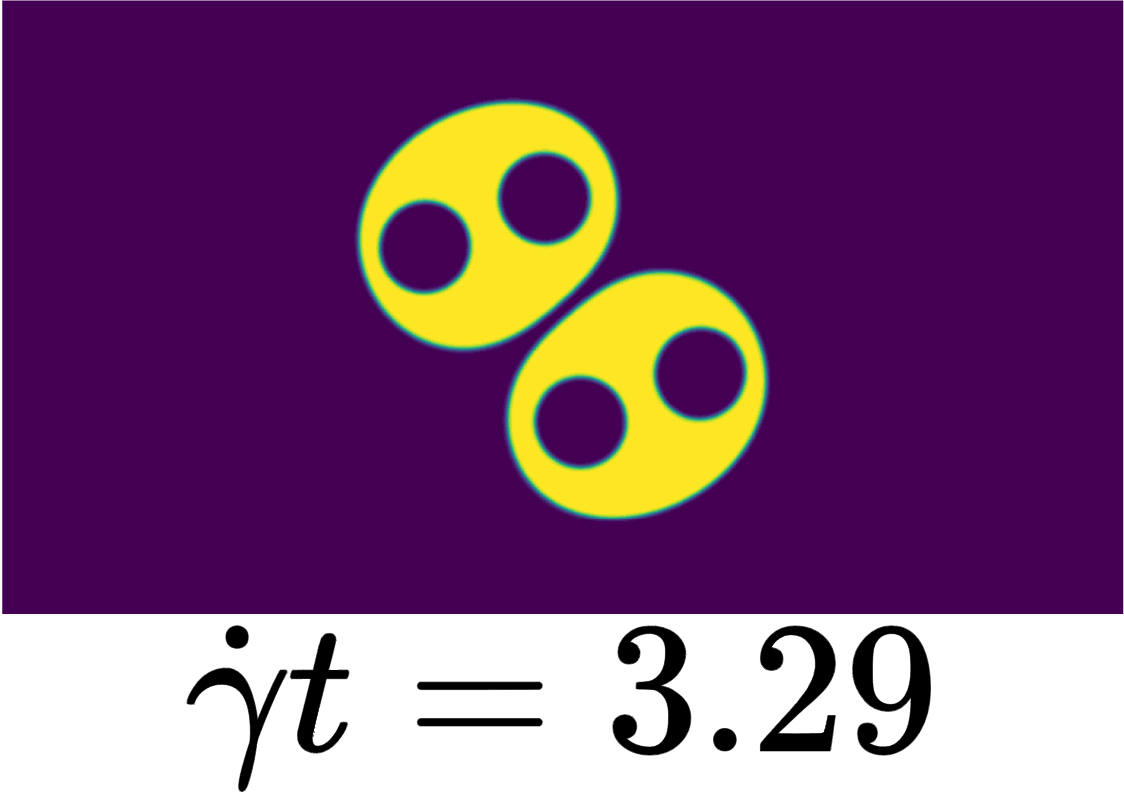}
		\includegraphics[scale = 0.09]{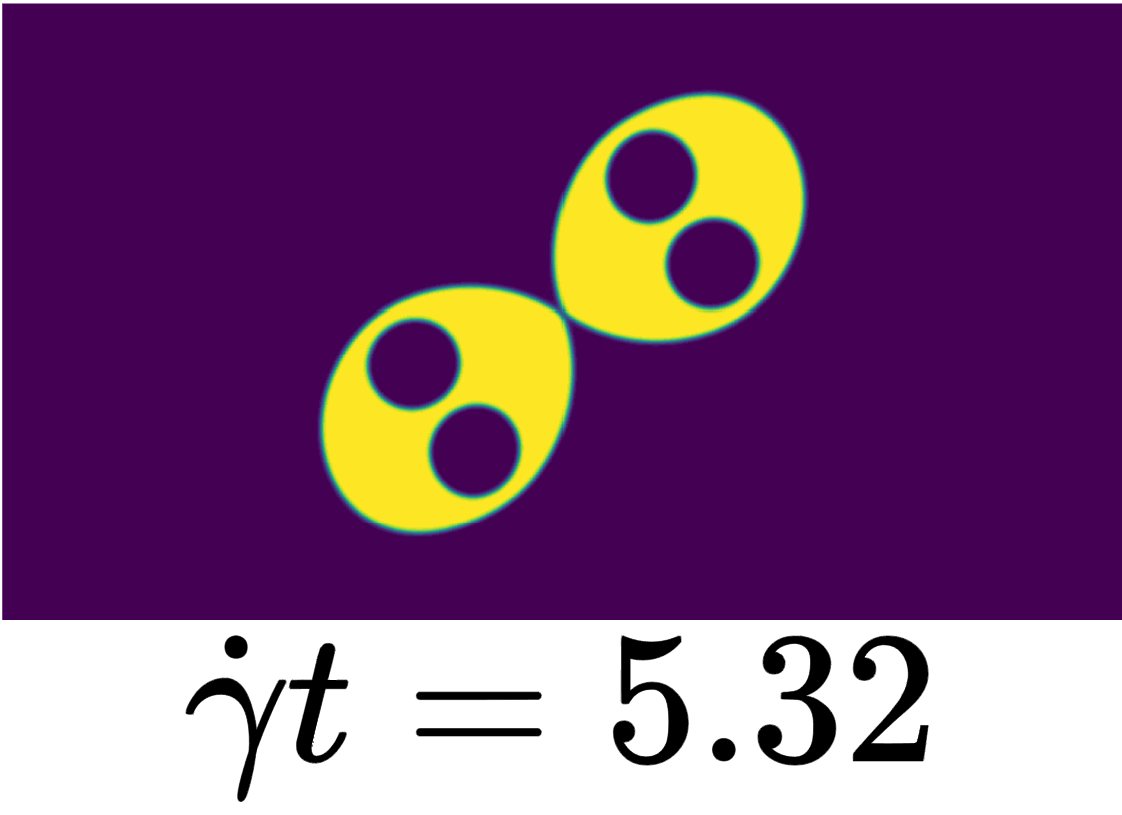}
		\includegraphics[scale = 0.09]{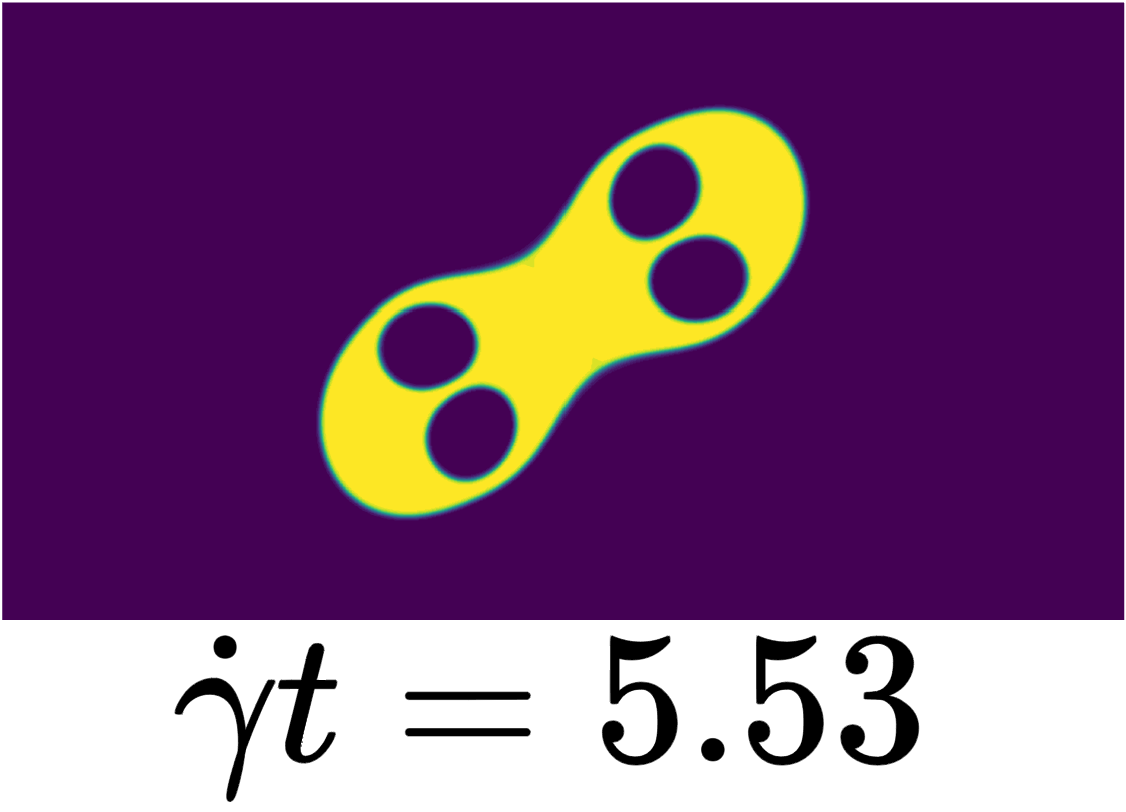}
		\includegraphics[scale = 0.09]{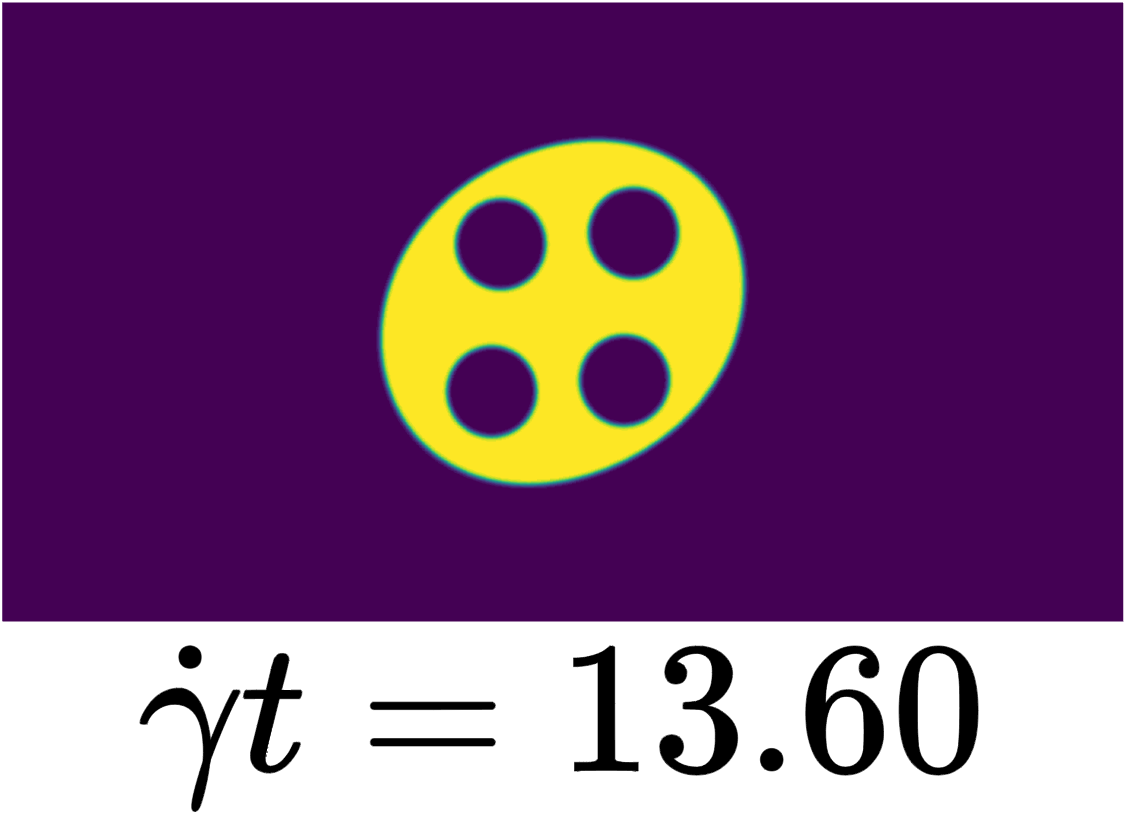} \\
        (c) $\frac{\Delta Y}{R_o}=0.35$~(Shell droplets coalesce)
		\\ \vspace{0.5cm}

  \includegraphics[scale = 0.09]{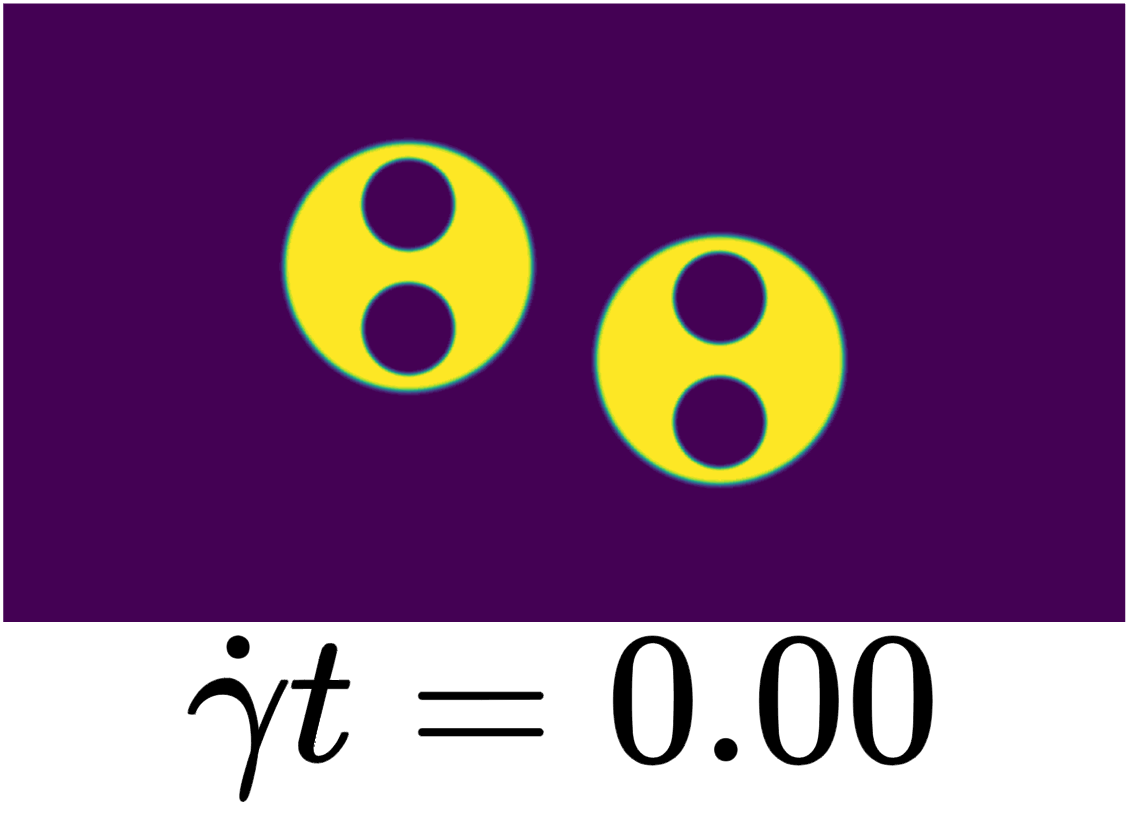}
		\includegraphics[scale = 0.09]{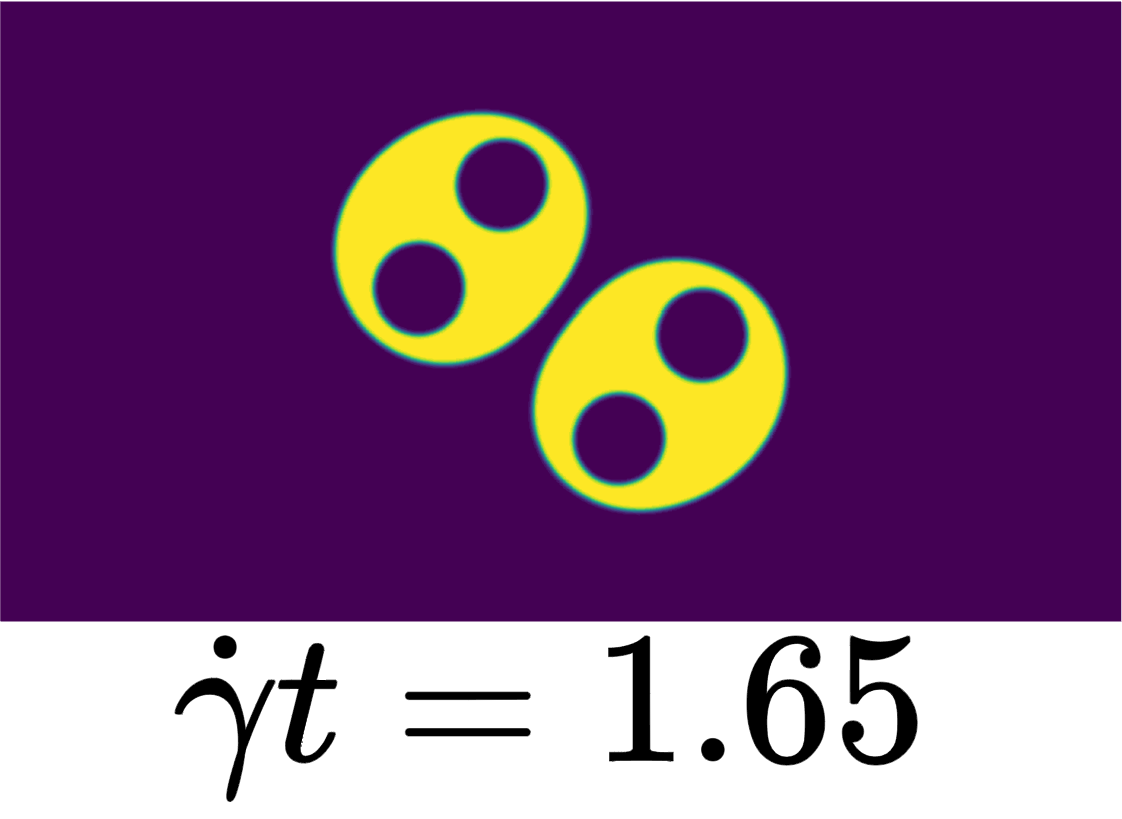}
		\includegraphics[scale = 0.09]{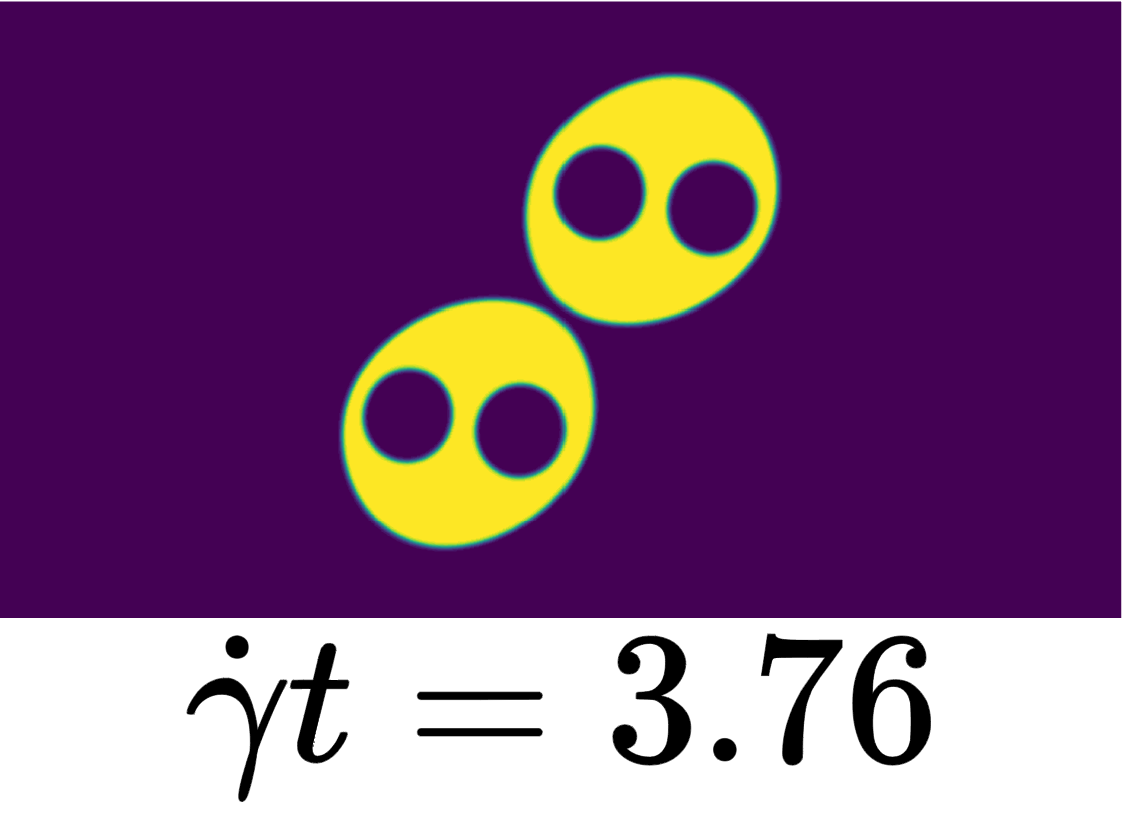}
		\includegraphics[scale = 0.09]{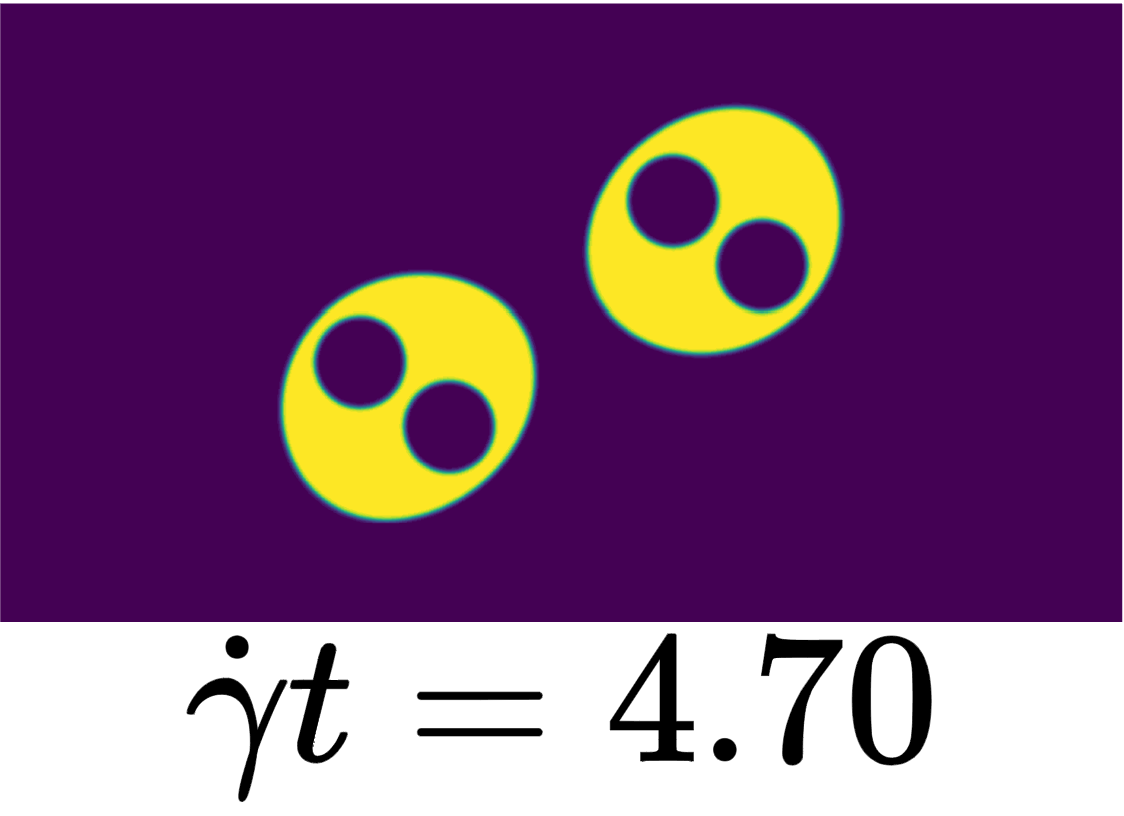}
		\includegraphics[scale = 0.09]{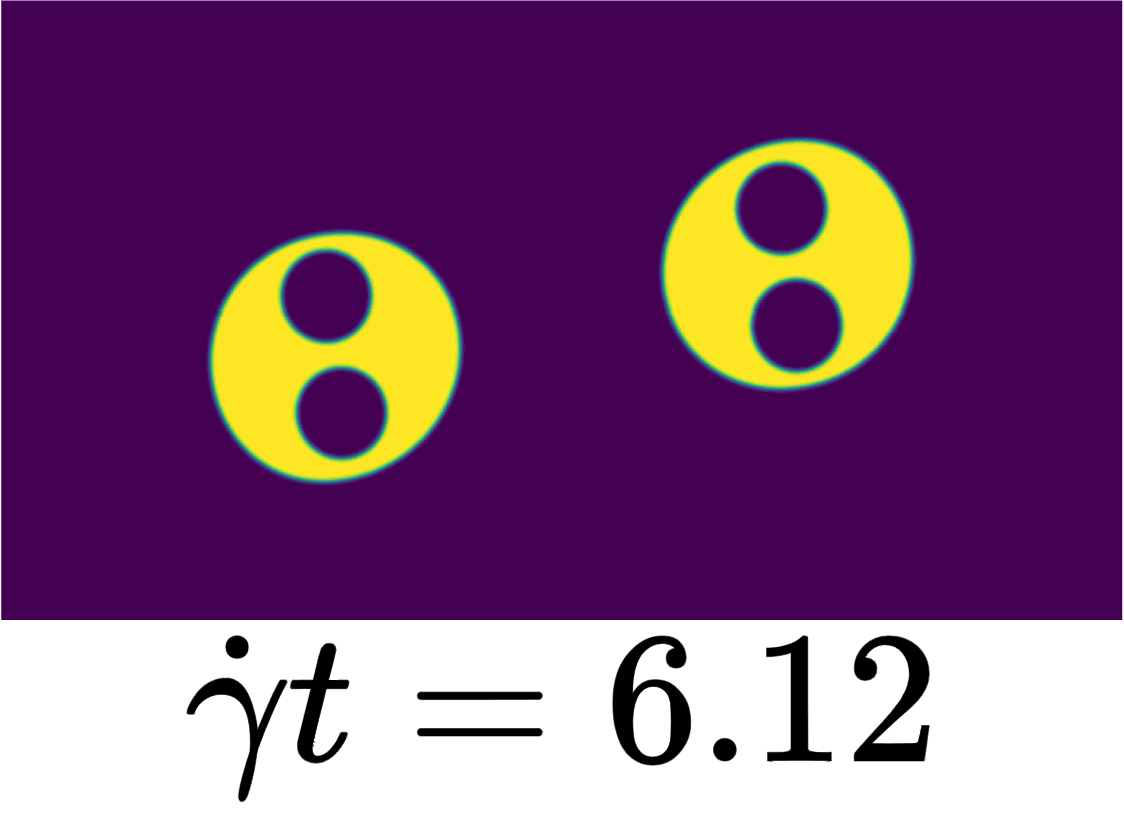} \\
        (d) $\frac{\Delta Y}{R_o}=0.75$~(Shell droplets pass over)
		\\ \vspace{0.5cm}

  \includegraphics[scale = 0.09]{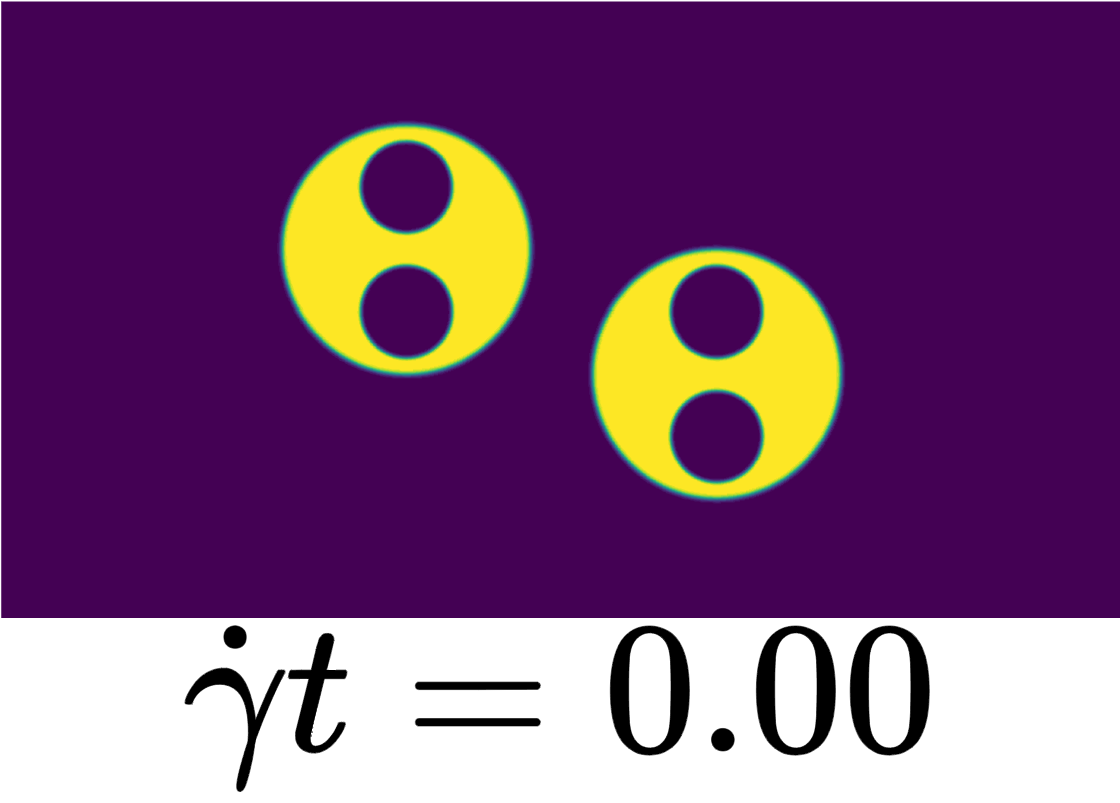}
		\includegraphics[scale = 0.09]{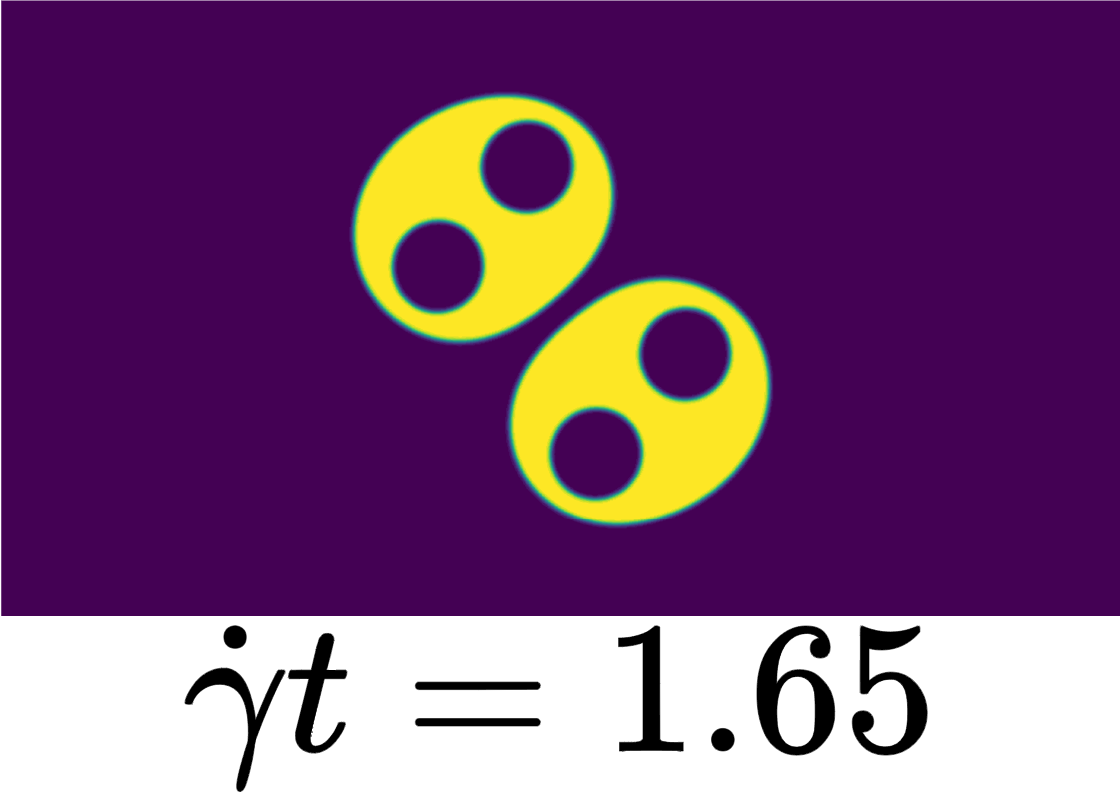}
		\includegraphics[scale = 0.09]{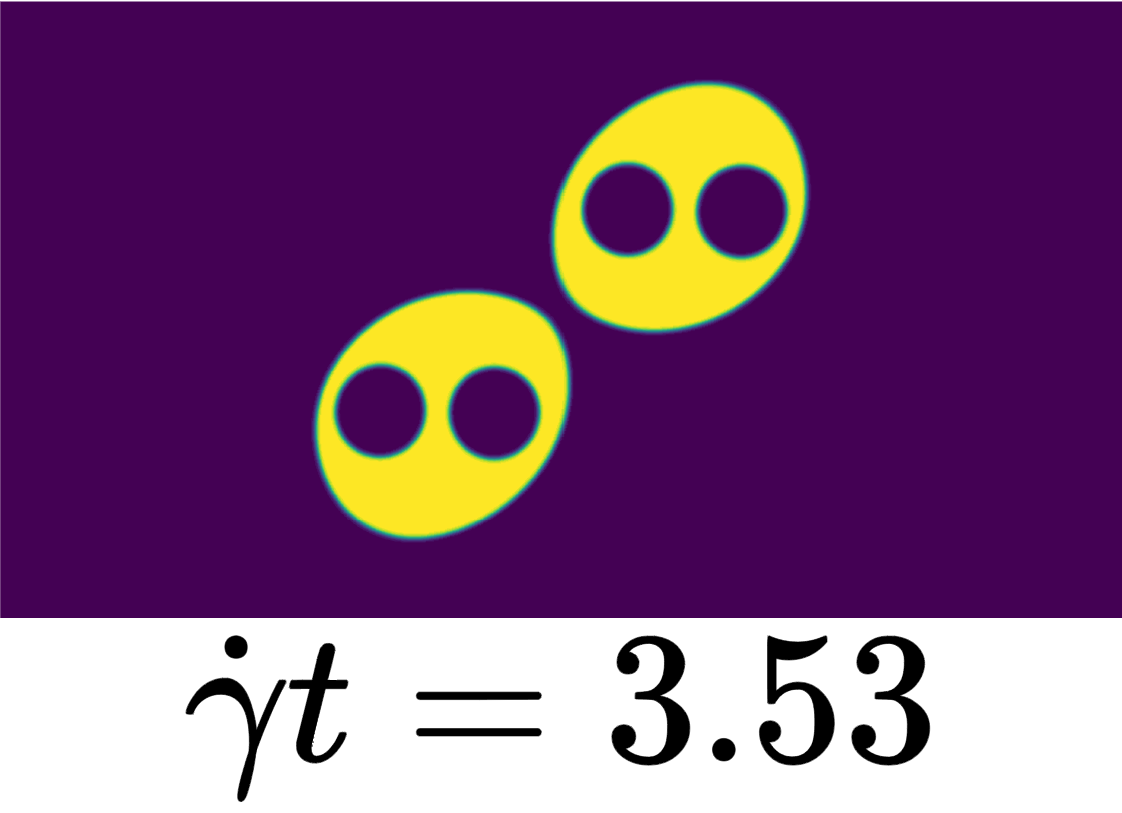}
		\includegraphics[scale = 0.09]{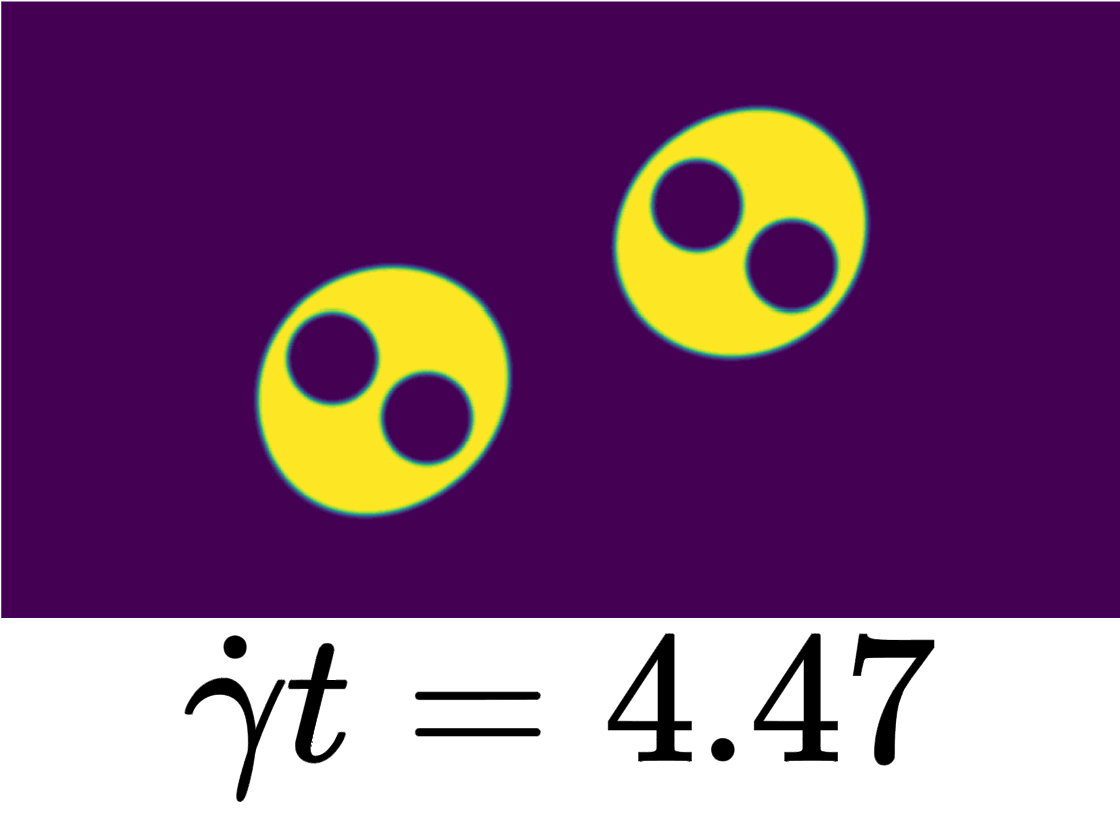}
		\includegraphics[scale = 0.09]{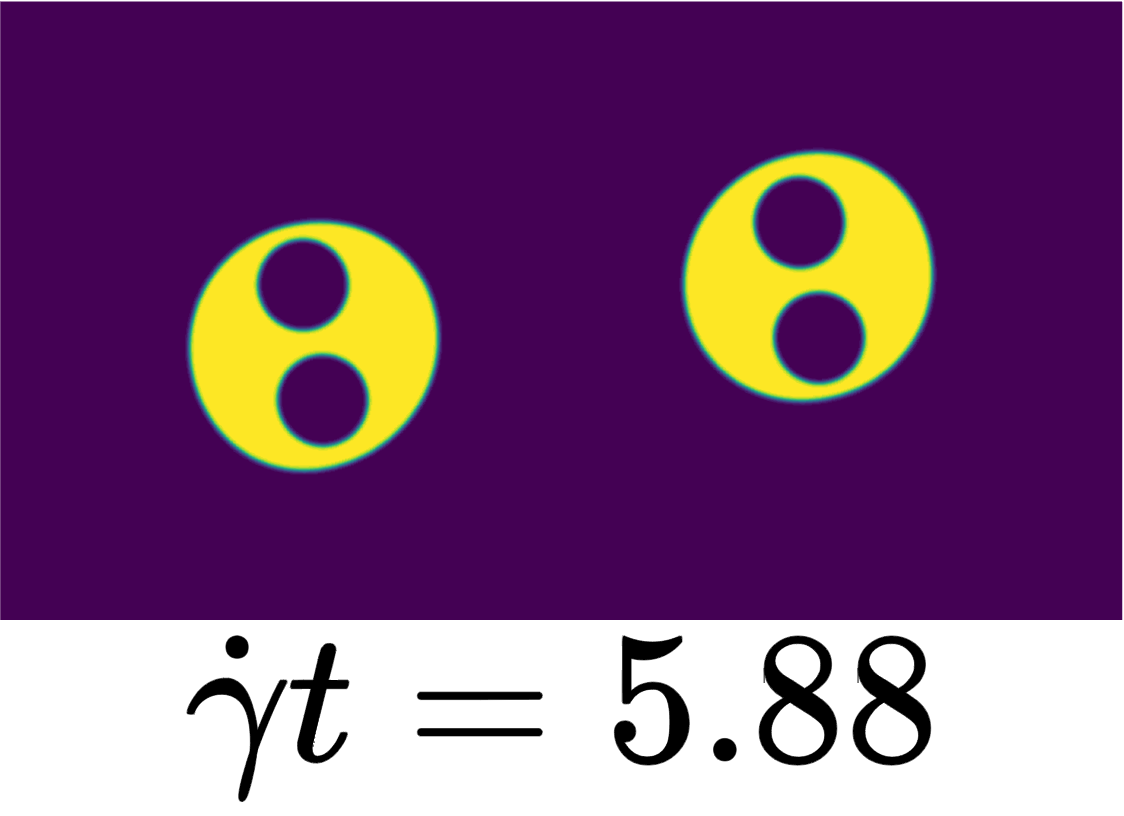} \\
        (e) $\frac{\Delta Y}{R_o}=1.00$~(Shell droplets pass over)
		\\ \vspace{0.5cm}

  \includegraphics[scale = 0.09]{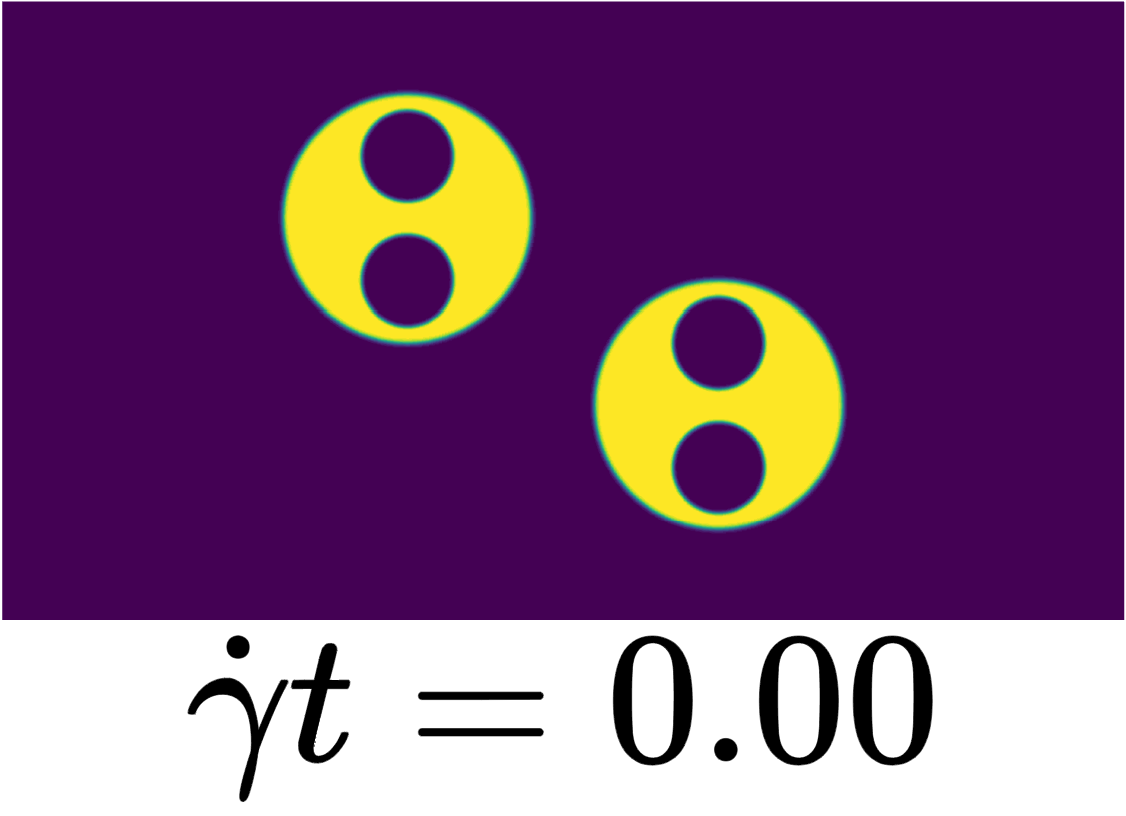}
		\includegraphics[scale = 0.09]{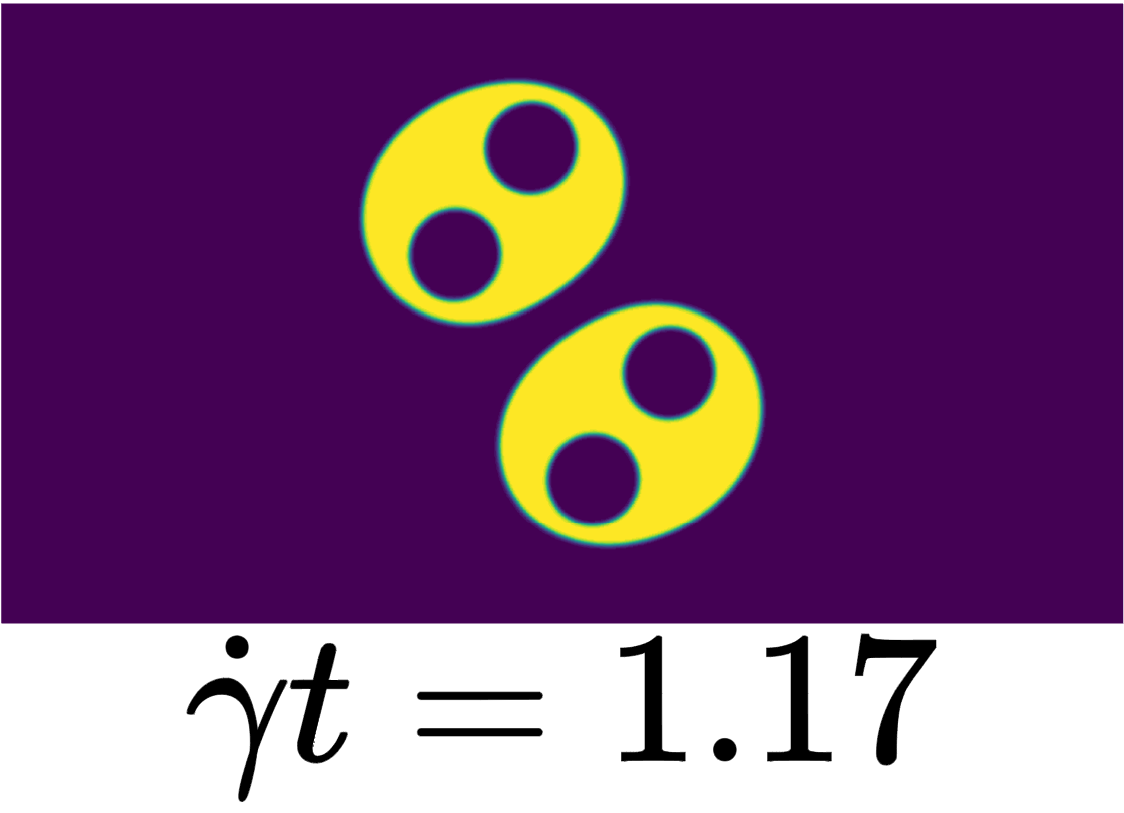}
		\includegraphics[scale = 0.09]{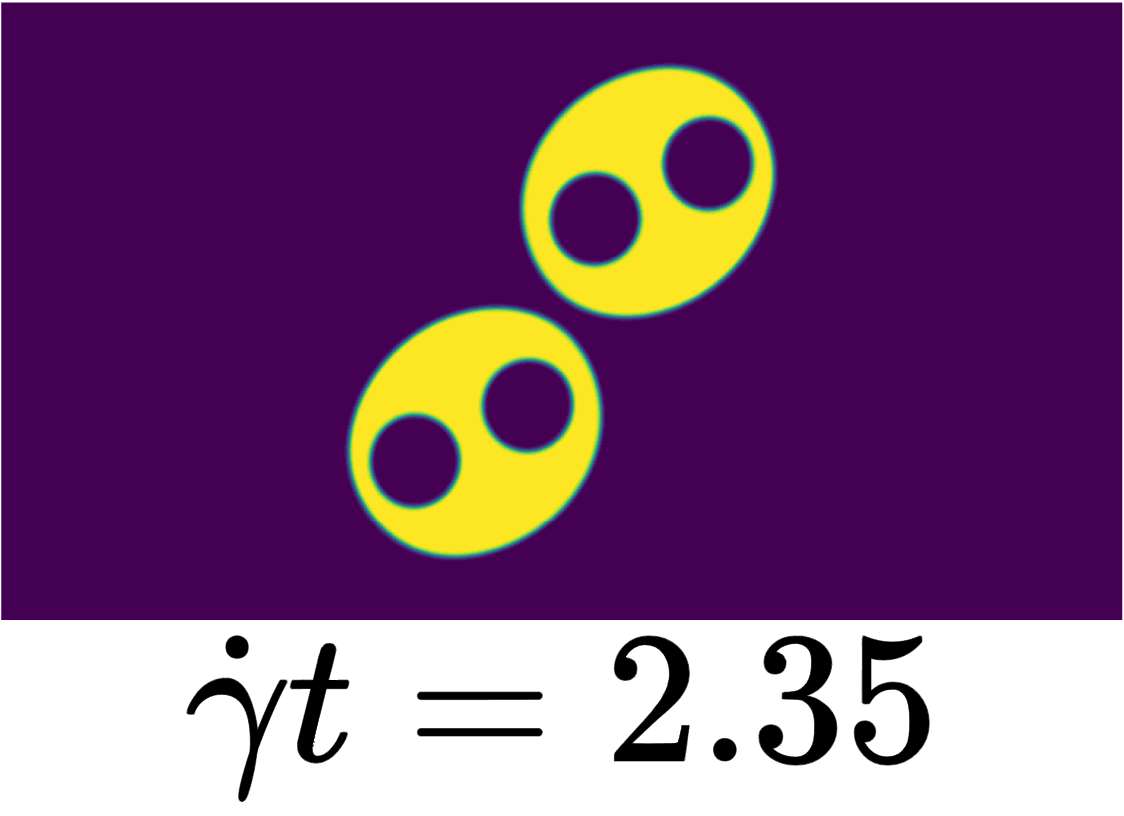}
		\includegraphics[scale = 0.09]{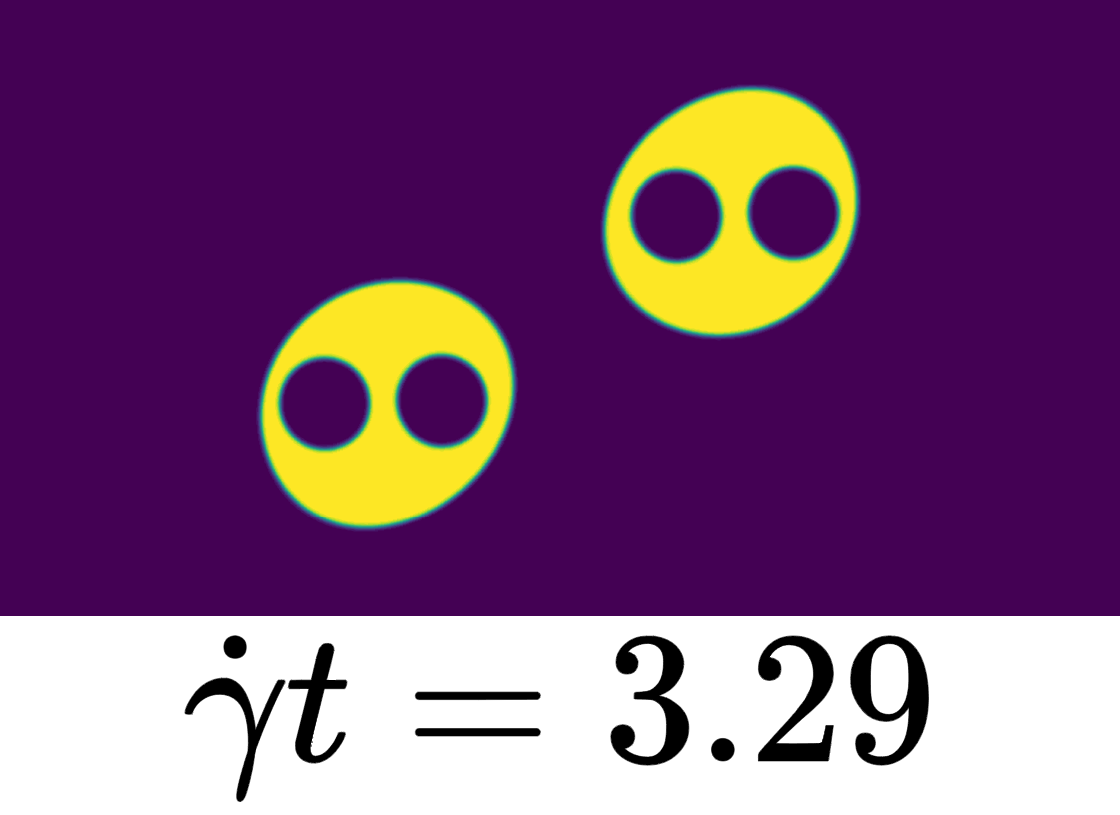}
		\includegraphics[scale = 0.09]{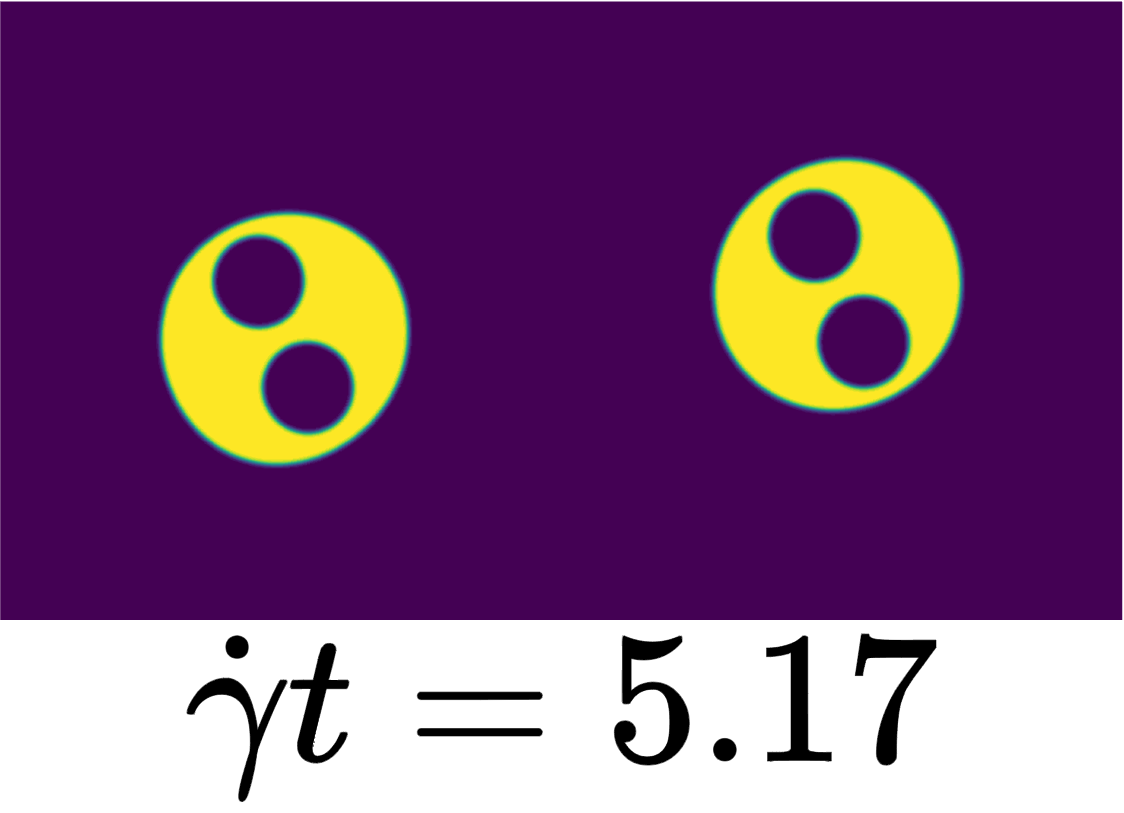} \\
        (f) $\frac{\Delta Y}{R_o}=1.50$~(Shell droplets pass over)
		\\ \vspace{0.5cm}
  
 	\caption{Time-lapse images of double-core compound droplet collision for six different initial vertical offset ($\frac{\Delta Y}{R_o}=0.00, 0.15, 0.35, 0.75, 1.00$, and $1.50$) with $Ca=0.10$, $Re=0.10$, $\rho_{12}=1.0$, $\mu_{12}=1$, $\Delta X_o/R_o = 2.50,$ and $R_o/H = 0.40.$} 
	\label{fig:Double-core-Ca-0.10-varying-offset-snap}
\end{figure}

We begin our analysis with the high value of $Ca=0.1$ and six different vertical initial offsets ($\frac{\Delta Y}{R_o}=0.00, 0.15, 0.35, 0.75, 1.00$, and $1.50$), while keeping all other parameters fixed. Figure~\ref{fig:Double-core-Ca-0.10-varying-offset-snap} depicts time-lapse images capturing critical instances during the evolution of each case. These images offer a comprehensive view of how varying initial offsets influences the behavior of multicore compound droplets. For the cases with the three lower initial vertical offsets ($\frac{\Delta Y}{R_o}=0.00, 0.15$, and $0.35$), it has been observed that the shell droplet pairs coalesce. 

Despite the coalescence of the outer shells, the core droplets inside remain distinct and continue their motion. The motion of core droplets is characterized by a planetary-like rotation within the newly formed, larger coalesced shell droplet. This behavior highlights the ability of the core droplets to maintain their integrity and independent motion even after the outer shells have merged. The internal dynamics of these core droplets are influenced by the confined environment of the coalesced shell, which sustains their rotational movement. In contrast, for the cases with the three higher initial vertical offsets  ($\frac{\Delta Y}{R_o}=0.75, 1.00$, and $1.5$), the shell droplets do not coalesce but instead pass over each other. During this pass-over motion, the core droplets inside the shell droplets exhibit complex movements influenced by the motion of the shell droplets and the shear forces exerted by the surrounding continuous fluid. The interaction between the shell droplets and the shear flow induces both translational and rotational movements in the core droplets.

\begin{figure}[H]
          \centering

         \includegraphics[scale = 0.085]{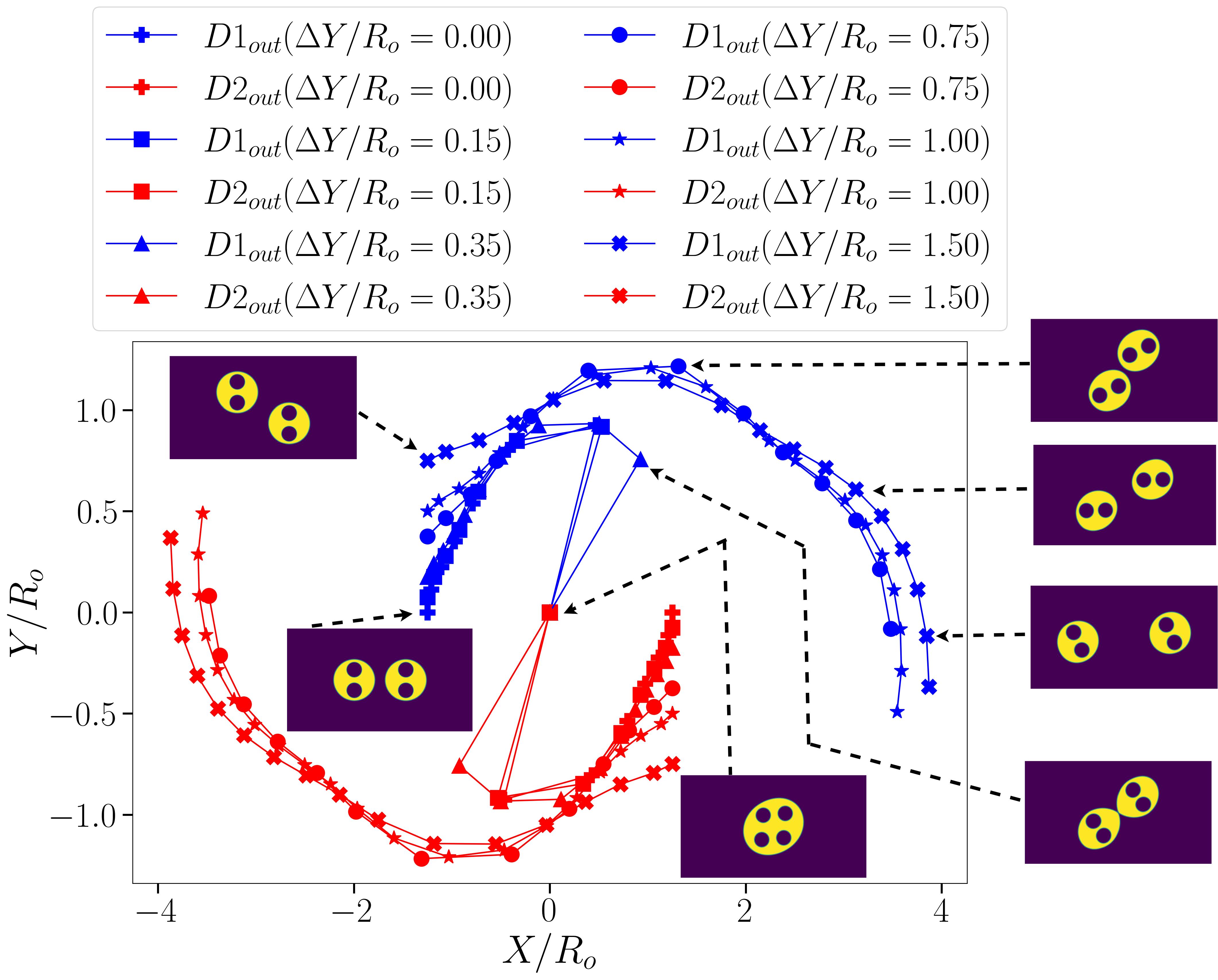} \\
          \vspace{0.45cm}

	\caption{Trajectories of the shell droplet pairs ($D1_{out}$ and $D2_{out}$) with the six different initial vertical offset ($\frac{\Delta Y}{R_o}=0.00, 0.15, 0.35, 0.75, 1.00$, and $1.50$) for $Ca=0.10$.}
	\label{fig:Double-core-Ca-0.10-varrying-offset-outer-trajectory}
\end{figure}

To gain better understanding into the motion of the shell droplets,  trajectory plots have been generated accompanied by key snapshots at various critical instances, as illustrated in Figure~\ref{fig:Double-core-Ca-0.10-varrying-offset-outer-trajectory}. These plots and snapshots provide a broader view of the shell droplets' motion under different initial vertical offsets, highlighting the distinct outcomes of pass-over and coalescence events. For the cases where the shell droplets pass over each other, the trajectory plots reveal that the droplets tend to move towards the vertical center of the domain after they completely pass over. In scenarios where the shell droplets coalesce, the trajectory plots show that the center of the newly formed shell droplet settles at the domain's center. This consistent central positioning is a direct consequence of the linear and symmetric nature of the shear flow, which facilitates the coalesced droplet's movement towards equilibrium in the middle of the domain. The snapshots accompanying these trajectory plots further illustrate that the newly formed coalesced droplet adopts an ellipsoidal shape. 

\begin{figure}[H]
          \centering
          
        \includegraphics[scale = 0.08]{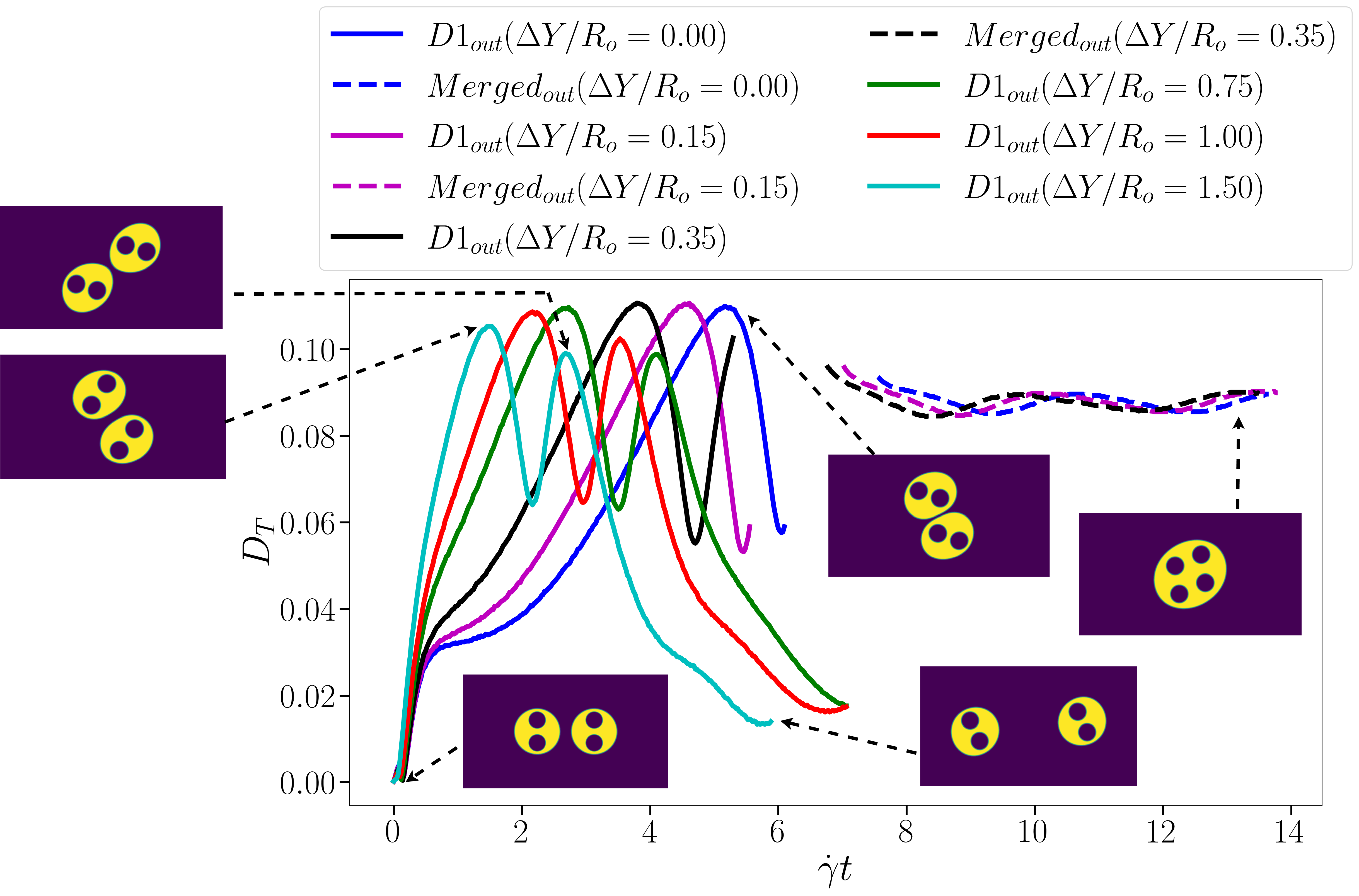} \\
    \vspace{0.45cm}
    
	\caption{Deformation quantification of the shell droplet before and after coalescence ($D1_{out}$ and $Merged_{out}$) with the six different initial vertical offset ($\frac{\Delta Y}{R_o}=0.00, 0.15, 0.35, 0.75, 1.00$, and $1.50$) for the $Ca$ of $0.10$.}
	\label{fig:Double-core-Ca-0.10-varrying-offset-outer-deform}
\end{figure}

\begin{figure}[H]
    \centering
    \begin{tabular}{cc}
        \includegraphics[scale = 0.2]{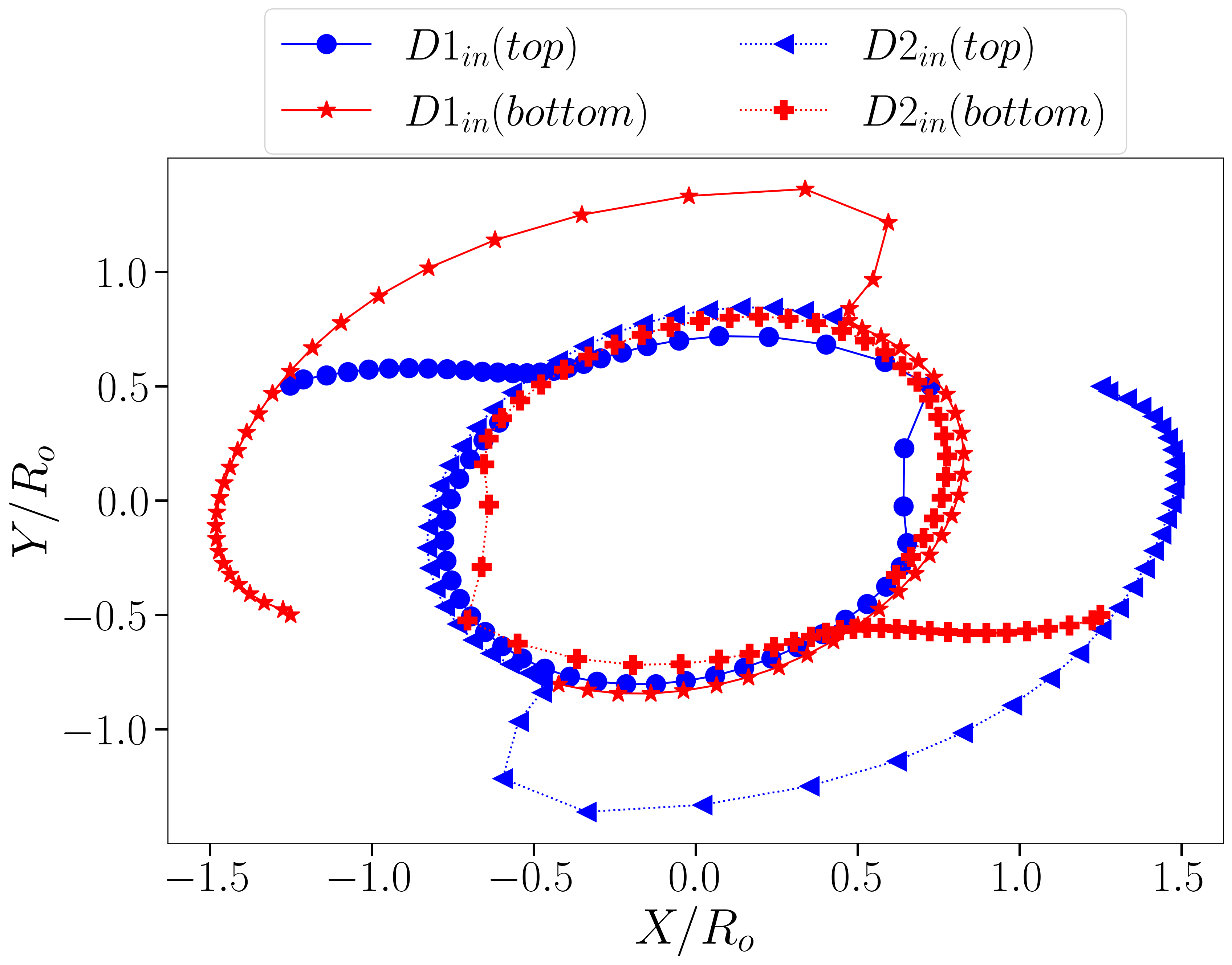} & \includegraphics[scale = 0.2]{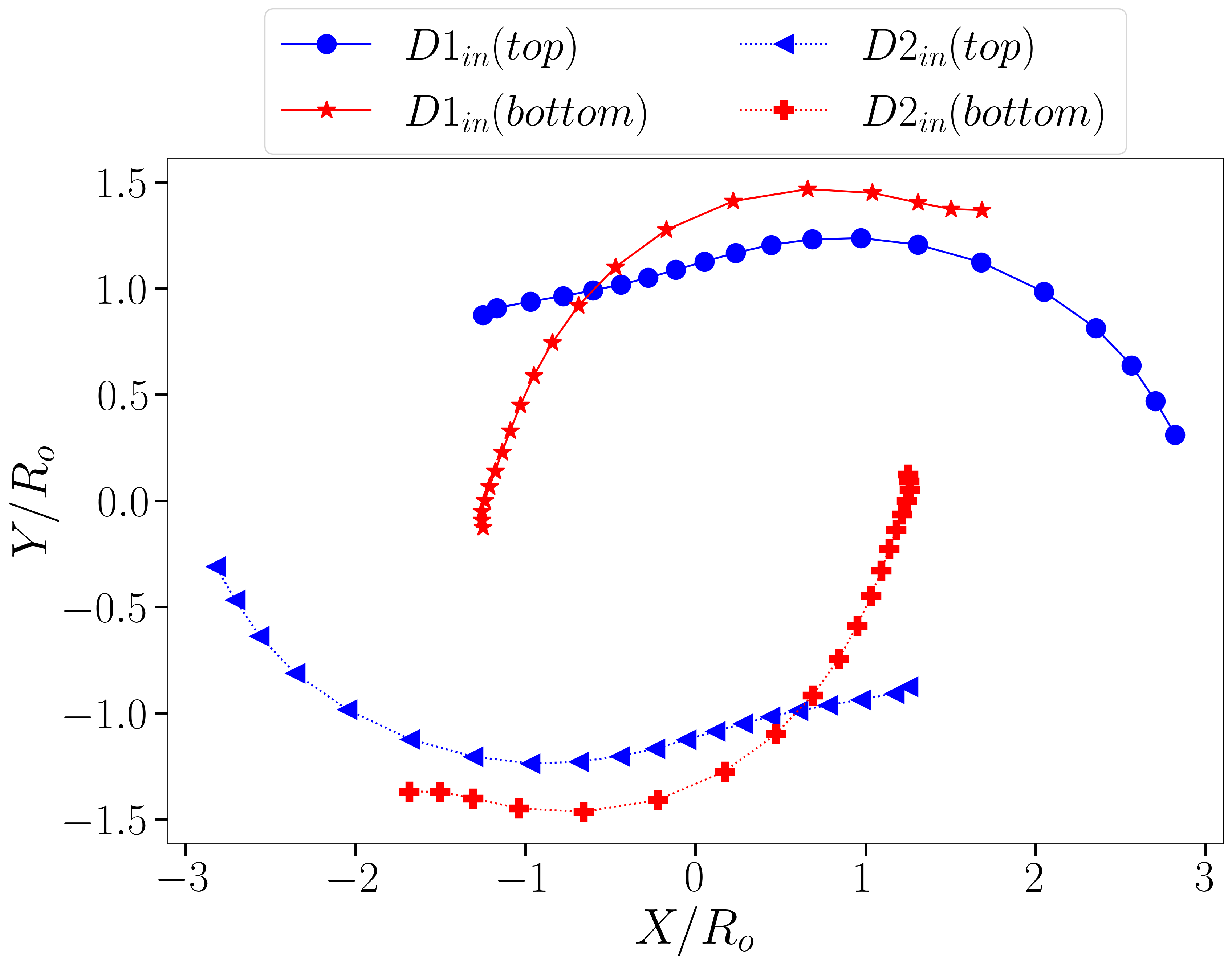} \\
        (a) $\Delta Y_o/R_o = 0.00$  &
        (d) $\Delta Y_o/R_o = 0.75$  \\
        \includegraphics[scale = 0.2]{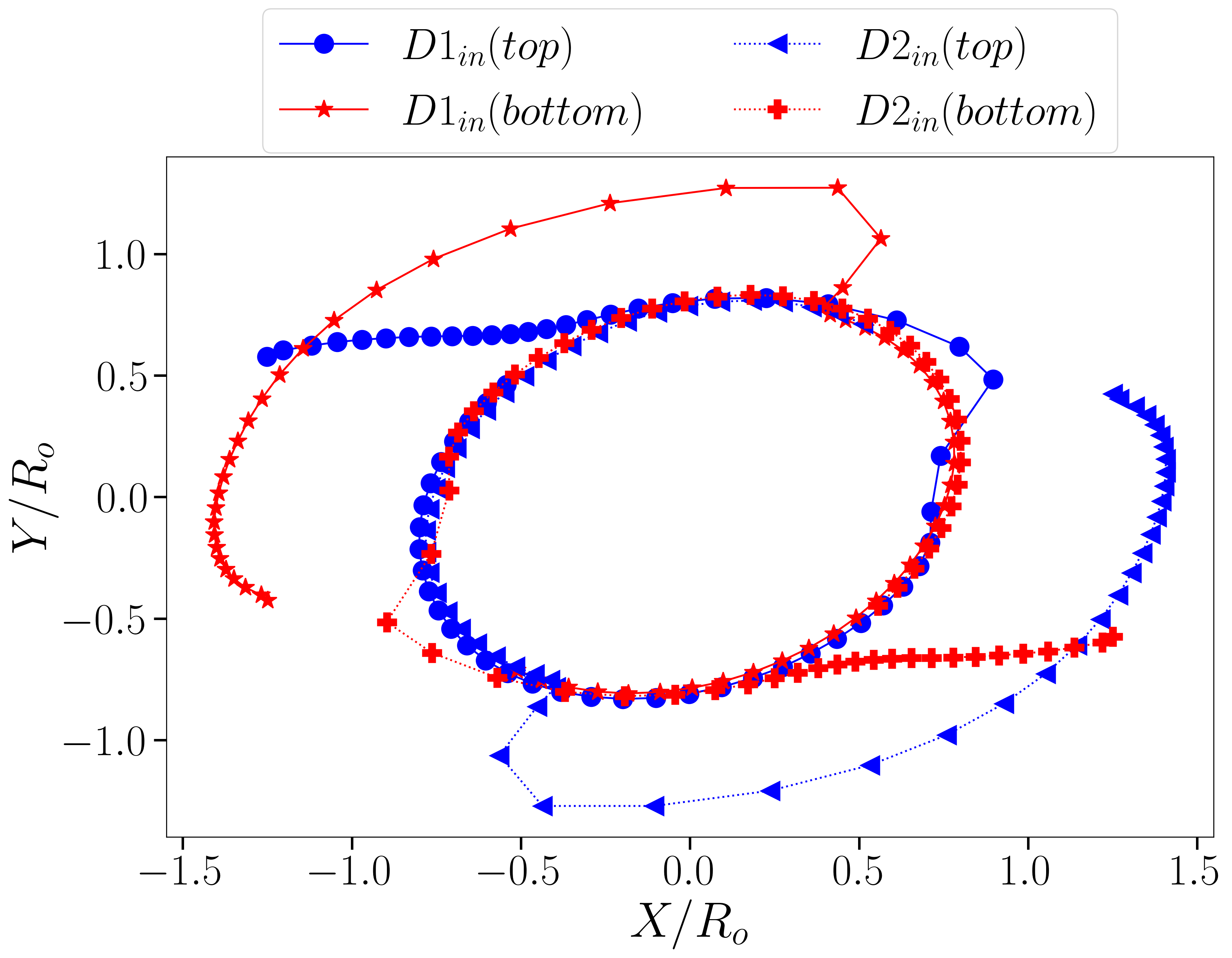} & \includegraphics[scale = 0.2]{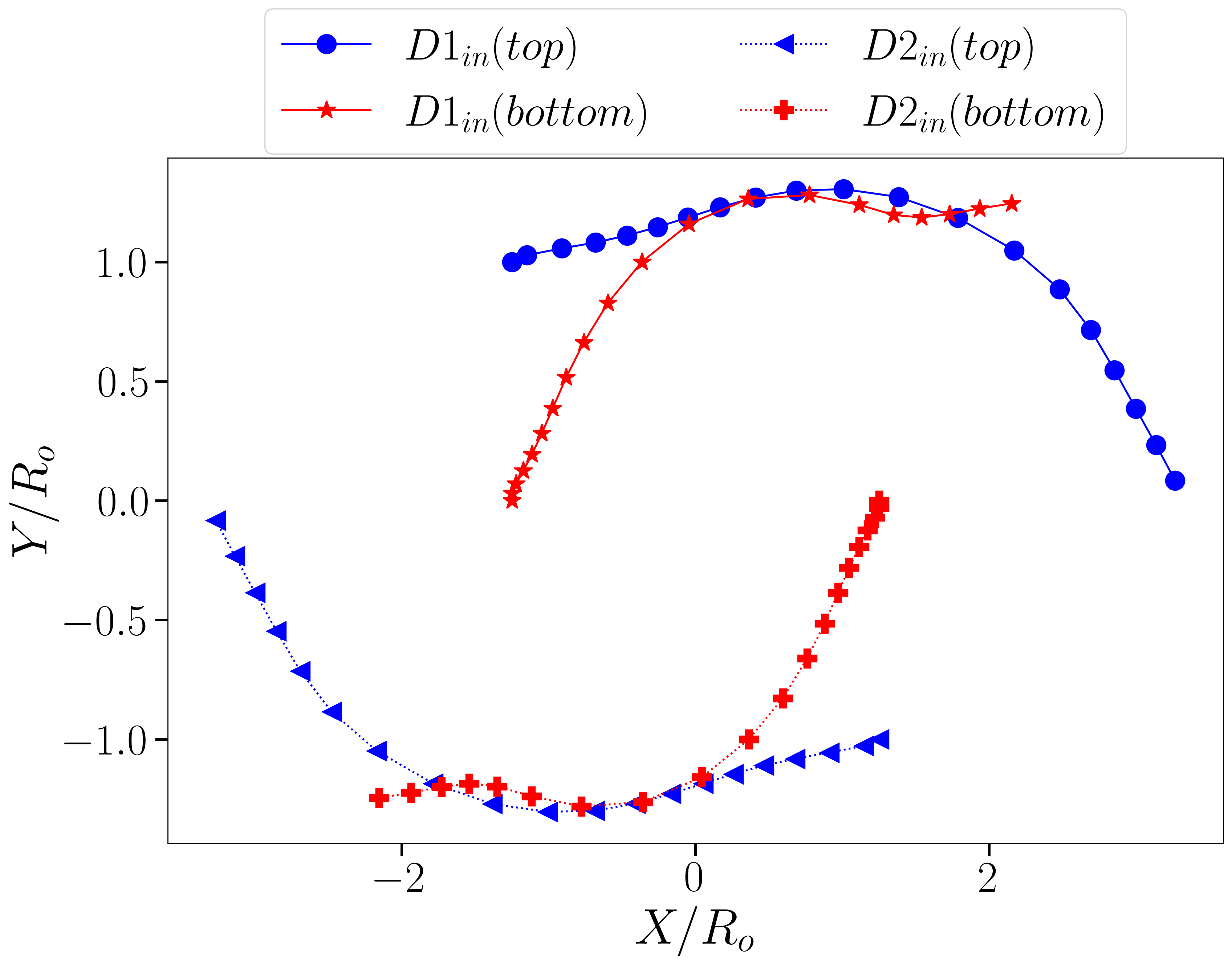} \\
        (b) $\Delta Y_o/R_o = 0.15$  &
        (e) $\Delta Y_o/R_o = 1.00$  \\
        \includegraphics[scale = 0.2]{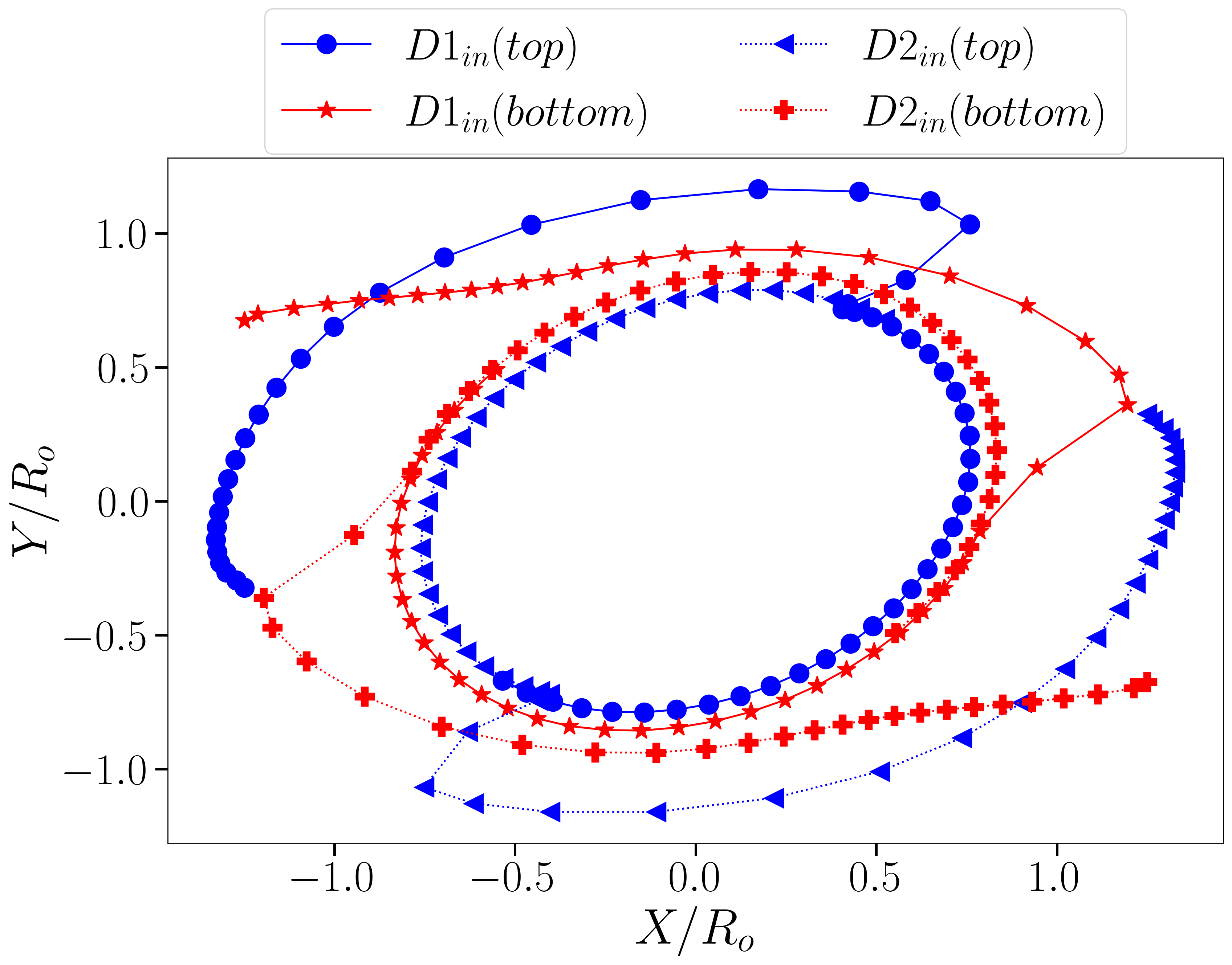} & \includegraphics[scale = 0.2]{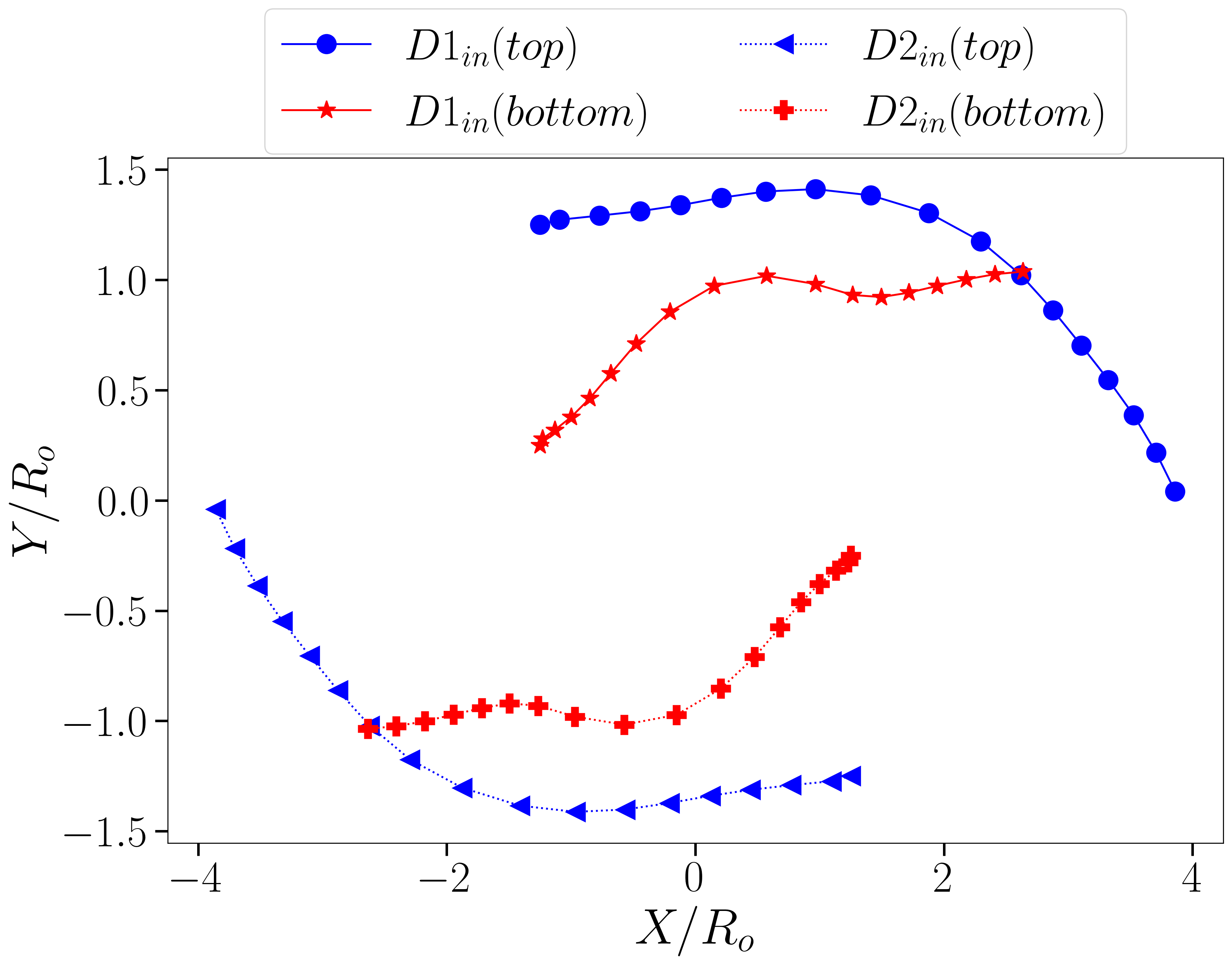} \\
        (c) $\Delta Y_o/R_o = 0.35$  &
        (f) $\Delta Y_o/R_o = 1.50$  \\
    \end{tabular}

    \caption{Trajectories of the core droplets ($D1_{in}(top)$, $D1_{in}(bottom)$, $D2_{in}(top)$ and $D2_{in}(bottom)$) for the six different initial vertical offset ($\frac{\Delta Y}{R_o}=0.00, 0.15, 0.35, 0.75, 1.00$, and $1.50$) between shell droplets with $Ca=0.10$.}
    \label{fig:Double-core-Ca-0.10-varrying-offset-inner-trajectory}
\end{figure}

This shape formation can be attributed to the continuous influence of the shear flow, which stretches the droplet along the flow direction, resulting in an ellipsoidal configuration. The stress dynamics of the shell droplets in this case study is influenced by both external shear flow and internal droplet interactions. The stress environment experienced by the shell droplets is effectively reflected in their deformation, which can be quantitatively analyzed throughout their evolution. Similar to the trajectory plots for the six different initial vertical offsets with $Ca=0.10$, the deformation of the shell droplets has been quantified and displayed in Figure~\ref{fig:Double-core-Ca-0.10-varrying-offset-outer-deform}. Due to the symmetry of the shear flow and the geometry of the system, only the deformation of the left shell droplet ($D1_{out}$) is considered for analysis. 
In cases where the shell droplets coalesce, the deformation of the resulting coalesced shell droplet ($Merged_{out}$) is also included in the plot. The deformation curves reveal critical insights into the stress dynamics and the behavior of the droplets under varying conditions. For the cases resulting in coalescence, the deformation of the shell droplets begins to rise under the influence of the shear flow. The deformation reaches its peak during the collision and initial attempt to pass over, as depicted in Figure~\ref{fig:Double-core-Ca-0.10-varrying-offset-outer-deform}. As the coalescence process initiates, the shell droplets assume highly irregular shapes, making it challenging to define their deformation quantitatively. However, once the coalesced shell droplet stabilizes into a regular form, the deformation can be recalculated. 
It is observed that the deformation stabilizes and aligns with the formation of an ellipsoidal shape, with only slight fluctuations. These minor fluctuations in deformation are attributed to the rotation of the internal core droplets. The internal dynamics, particularly the rotational motion of the core droplets, create periodic variations in the stress distribution on the shell droplet interface, resulting in slight bumps. These bumps are reflected as small fluctuations in the deformation plot. The rise and peak in deformation during collision and coalescence highlight the significant stress experienced by the droplets. The subsequent stabilization into an ellipsoidal shape indicates that the coalesced droplet reaches an equilibrium state under the symmetric shear flow. It is also notable that for the pass-over cases, the shell droplet shows two peak in it's deformation curve. The second peak indicates another rise on deformation at the point when the shell droplets starts to separate from each other. The reason behind this increase is due to a low pressure region generated between the shell droplet which has been explained earlier. To further understand the behavior of core droplets in terms of their trajectory and deformation, we have generated similar plots as for the shell droplets. Figure~\ref{fig:Double-core-Ca-0.10-varrying-offset-inner-trajectory} illustrates the trajectories of the core droplets for the six different initial vertical offset cases. For cases resulting in coalescence (i.e., Figure~\ref{fig:Double-core-Ca-0.10-varrying-offset-inner-trajectory}(a-c)), the trajectories of the core droplets reveal a more complex motion compared to the shell droplets. Despite the intricate motion, it can be clearly observed that the core droplets tend to follow nearly circular paths even after the shell droplets coalesce. This circular motion is indicative of the phenomena where the core droplets continue to interact and rotate around each other in a planetary-like manner. In contrast, for the pass-over cases (i.e., Figure~\ref{fig:Double-core-Ca-0.10-varrying-offset-inner-trajectory}(d-f)), the trajectories of the core droplets exhibit both translational and rotational movements. As the shell droplets pass over each other, the core droplets undergo significant repositioning due to the combined effects of the shell droplets' motion and the shear flow. This interaction results in the core droplets flipping their initial positions by the end of the pass-over process. The translational movement is primarily influenced by the overall drift of the shell droplets, while the rotational movement is driven by the internal interactions and the shear forces acting within the continuous fluid. 

\begin{figure}[H]
    \centering
    \begin{tabular}{cc}
        \includegraphics[scale = 0.17]{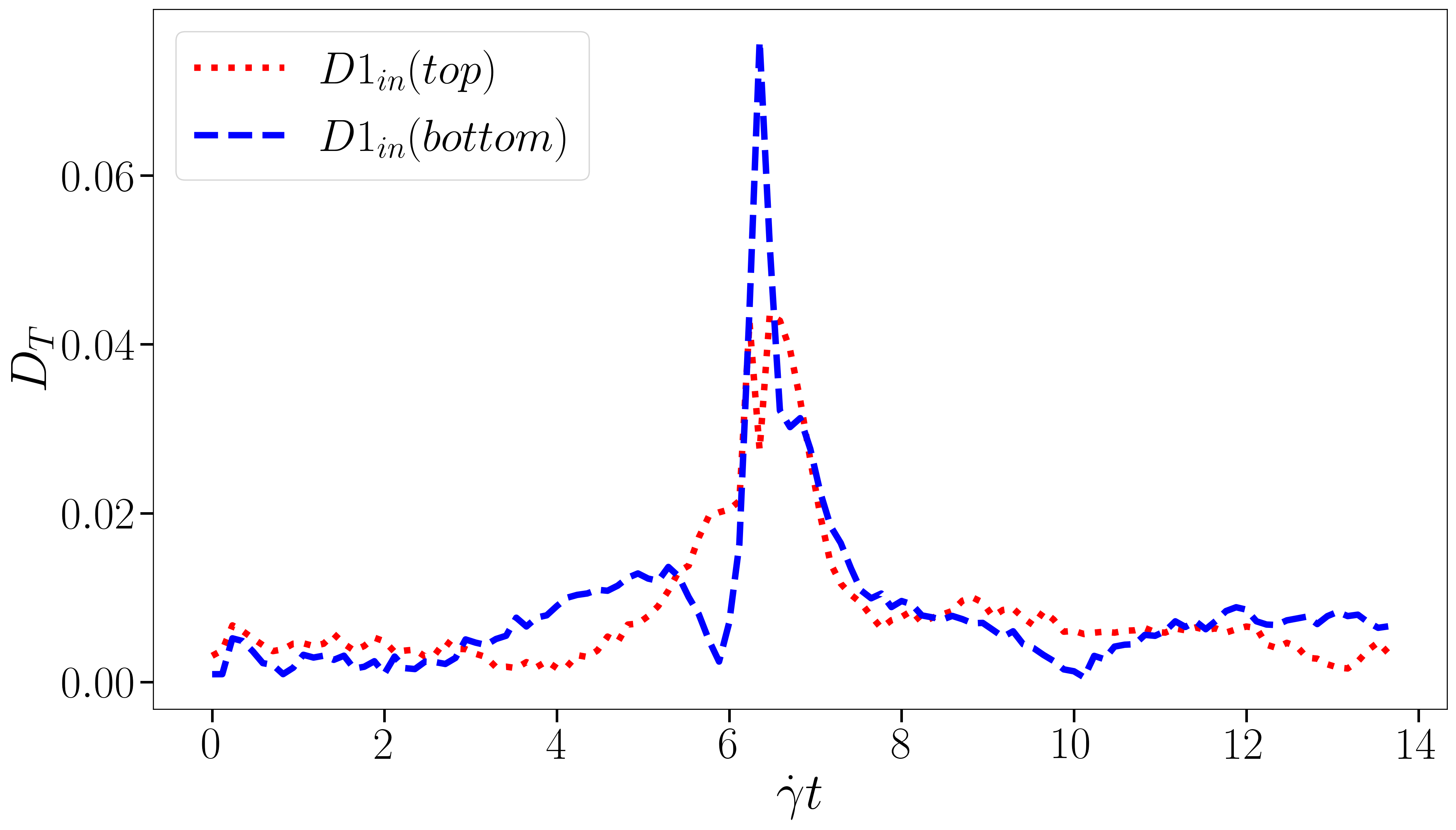} & \includegraphics[scale = 0.17]{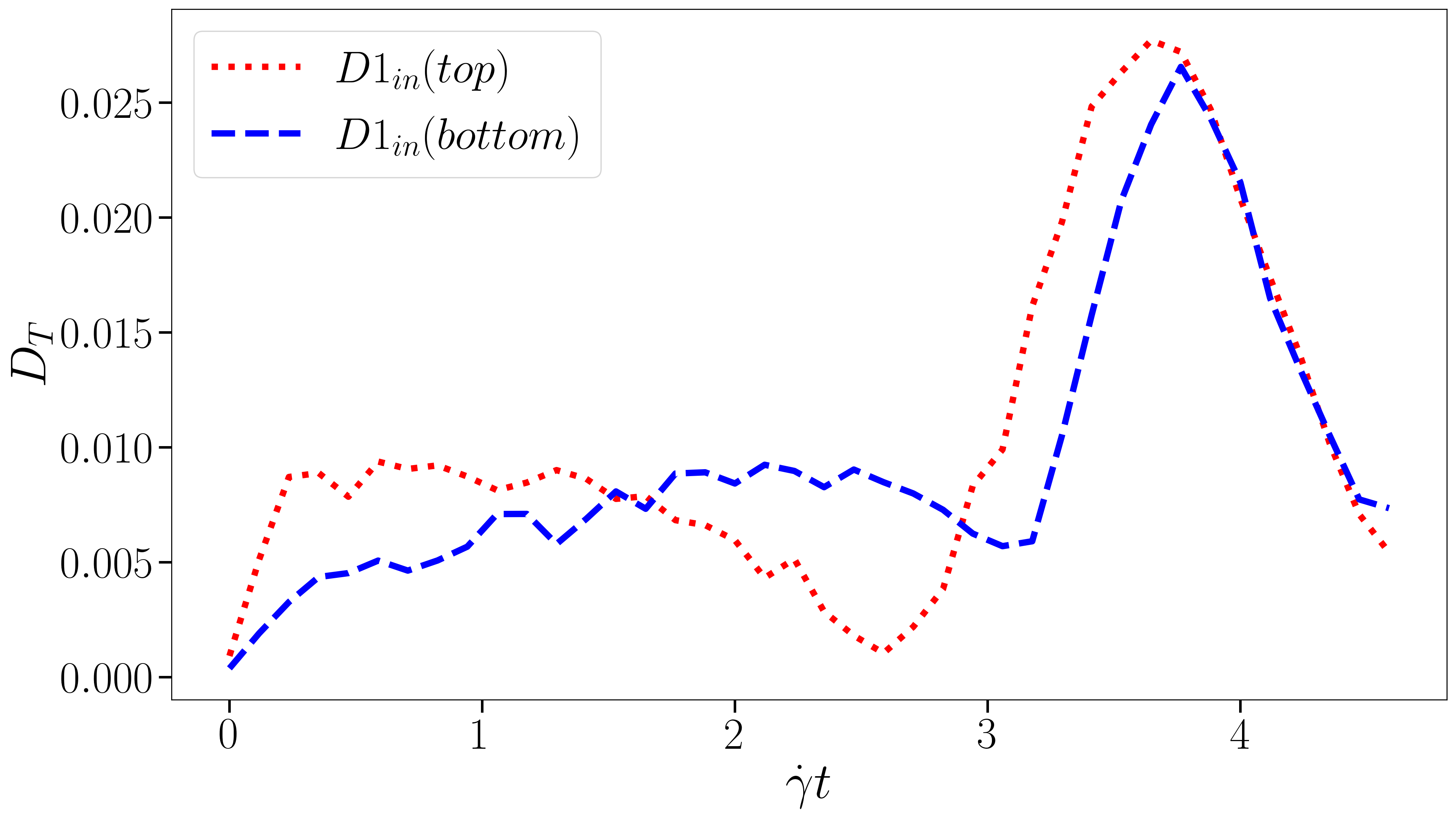} \\
        (a) $\Delta Y_o/R_o = 0.00$  &
        (d) $\Delta Y_o/R_o = 0.75$  \\
        \includegraphics[scale = 0.17]{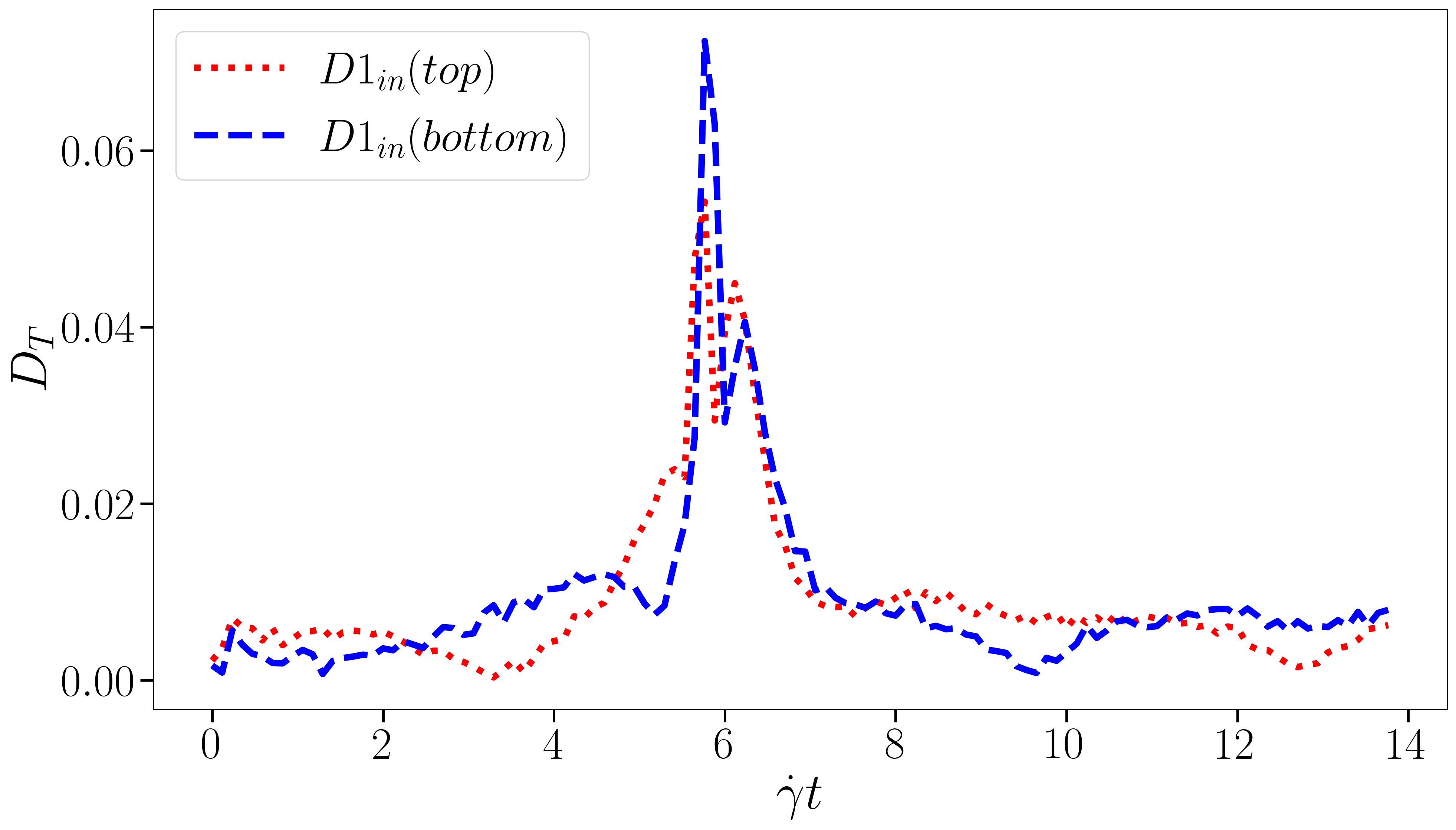} & \includegraphics[scale = 0.17]{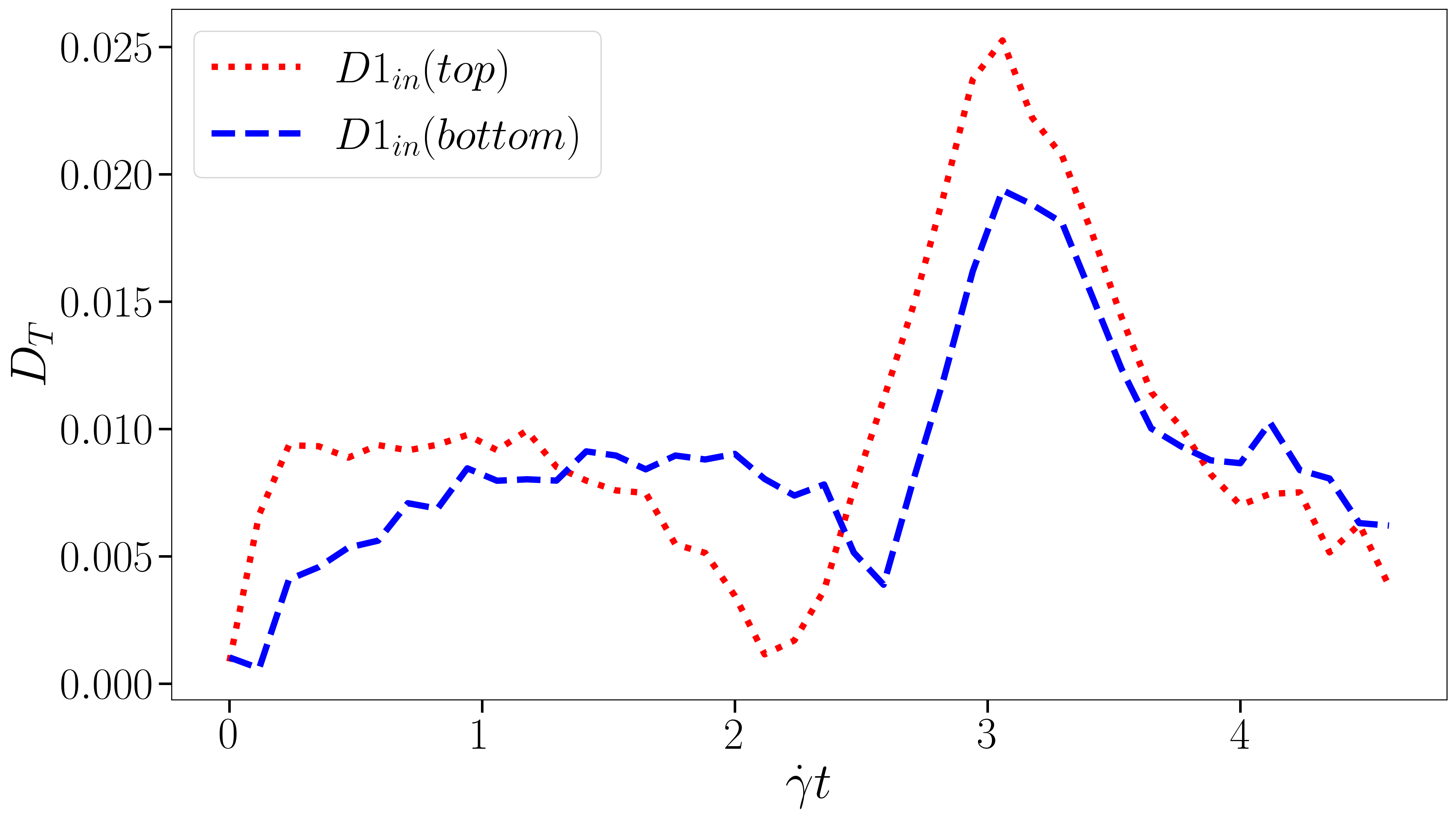} \\
        (b) $\Delta Y_o/R_o = 0.15$  &
        (e) $\Delta Y_o/R_o = 1.00$  \\
        \includegraphics[scale = 0.17]{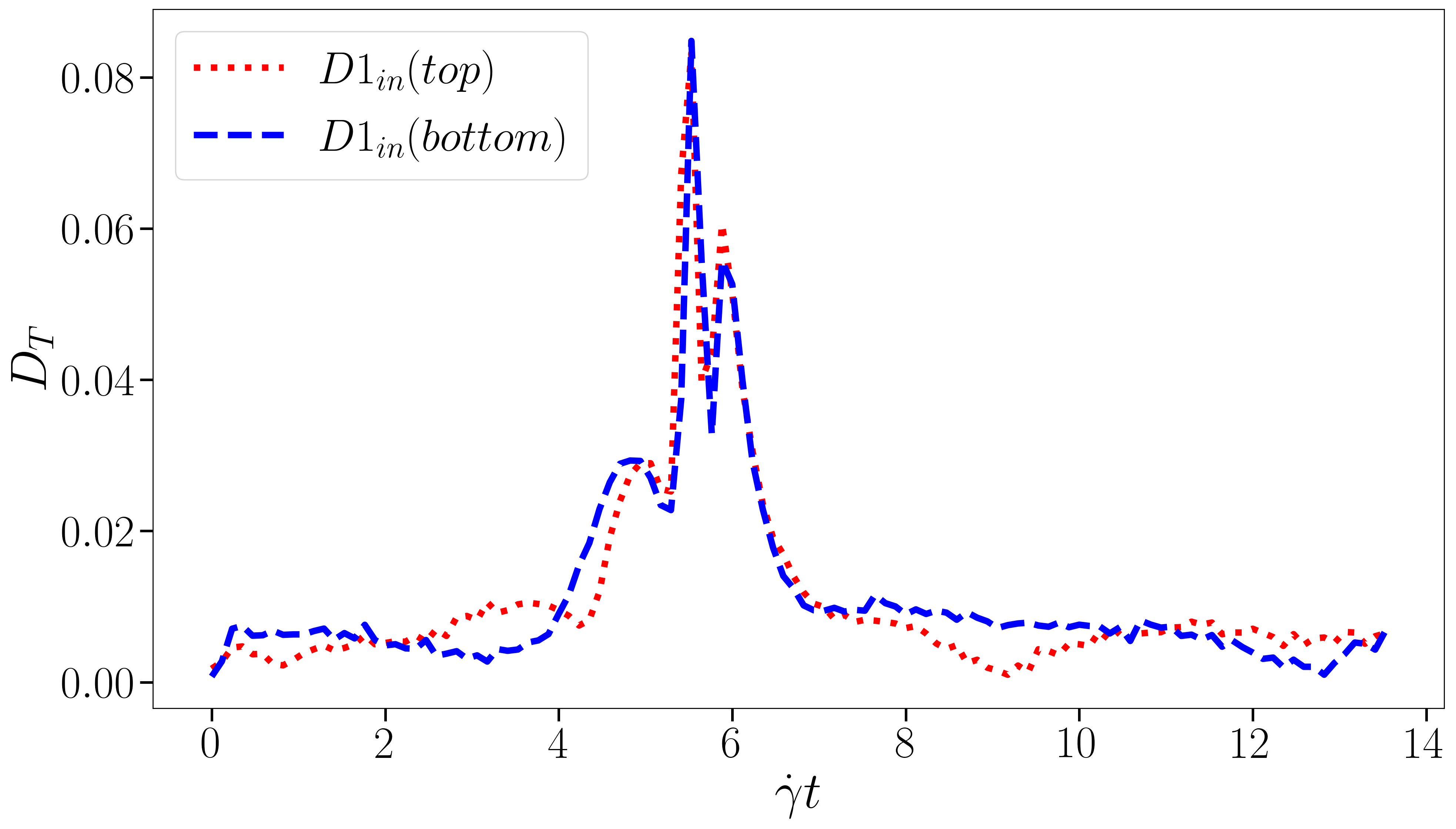} & \includegraphics[scale = 0.17]{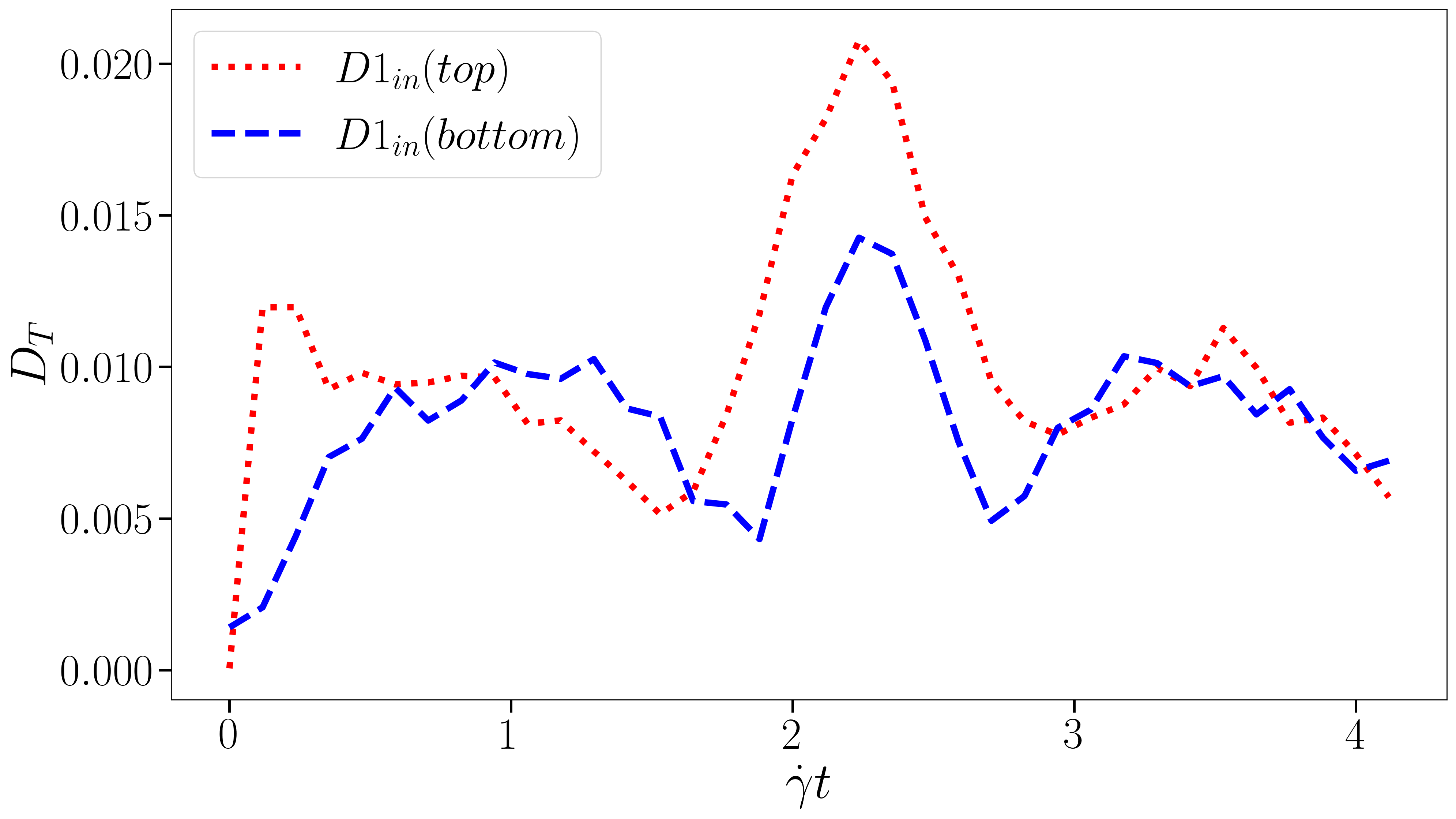} \\
        (c) $\Delta Y_o/R_o = 0.35$  &
        (f) $\Delta Y_o/R_o = 1.50$  \\
    \end{tabular}

    \caption{Deformation quantification of the core droplets ($D1_{in}(top)$ and $D1_{in}(bottom)$) for the six different initial vertical offset ($\frac{\Delta Y}{R_o}=0.00, 0.15, 0.35, 0.75, 1.00$, and $1.50$) in between shell droplets with $Ca=0.10$.}
    \label{fig:Double-core-Ca-0.10-varrying-offset-inner-deform}
\end{figure}

The trajectory plots highlight the different behaviors of the core droplets depending on whether the shell droplets coalesce or pass over each other. In the coalescence scenarios, the nearly circular paths suggest a stable, rotating system within the merged shell droplet. This stability is maintained by the balance of forces within the coalesced structure. Conversely, the pass-over scenarios demonstrate the dynamic nature of the system, where the core droplets are subject to more complex and variable stress environments, leading to their eventual repositioning and rotation. It can be observed from the deformation plots for core droplets in these simulation cases, as depicted in Figure~\ref{fig:Double-core-Ca-0.10-varrying-offset-inner-deform}, that the core droplets experience peaks in deformation during their evolution, closely aligning with the instances when the shell droplets undergo higher deformation. This correlation indicates that the stress environment within the shell droplets directly influences the deformation behavior of the core droplets. Apart from these peaks, the deformation of the core droplets remains relatively low and exhibits minor fluctuations. This stability in deformation can be attributed to the smaller size of the core droplets. The core droplets, being encapsulated within the shell droplets, are somewhat shielded from the direct shear forces acting on the outer interface. As a result, they maintain a more stable shape with less deformation, except during moments of significant stress transfer from the shell droplet.


\subsubsection{$Ca = 0.07$}

\begin{figure}[H]
          \centering

         \includegraphics[scale = 0.09]{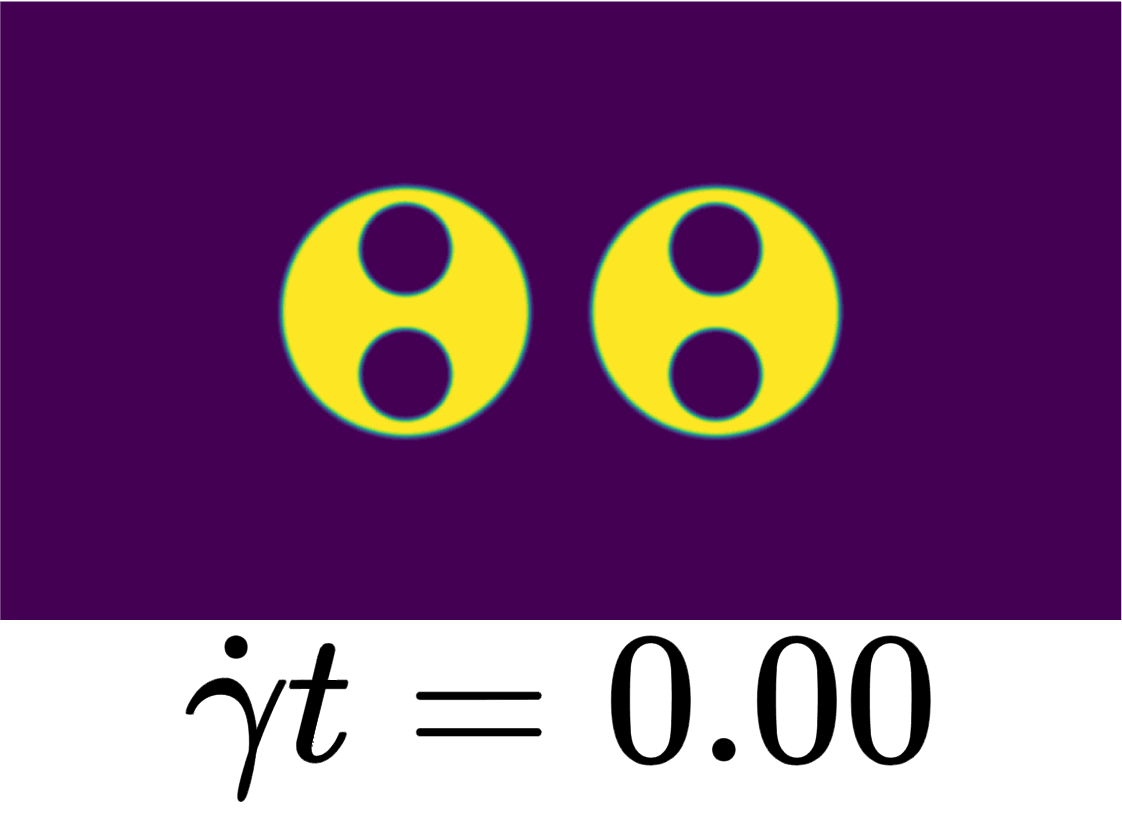}
		\includegraphics[scale = 0.09]{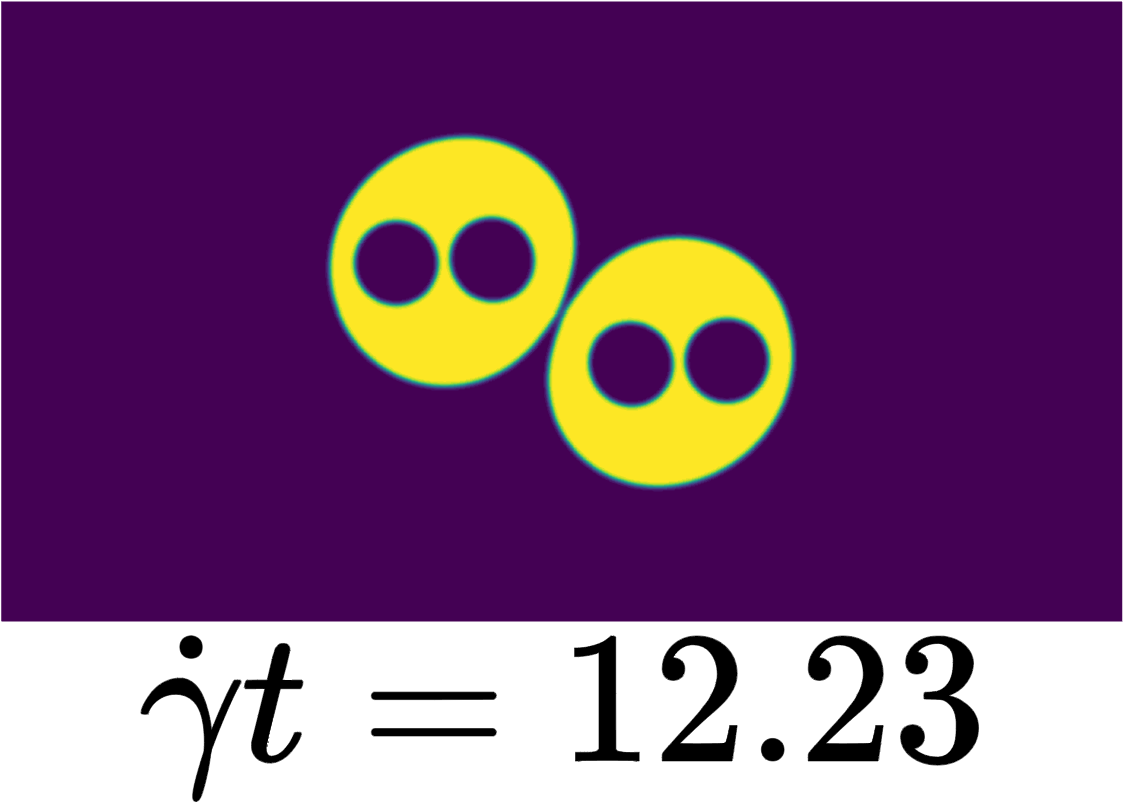}
		\includegraphics[scale = 0.09]{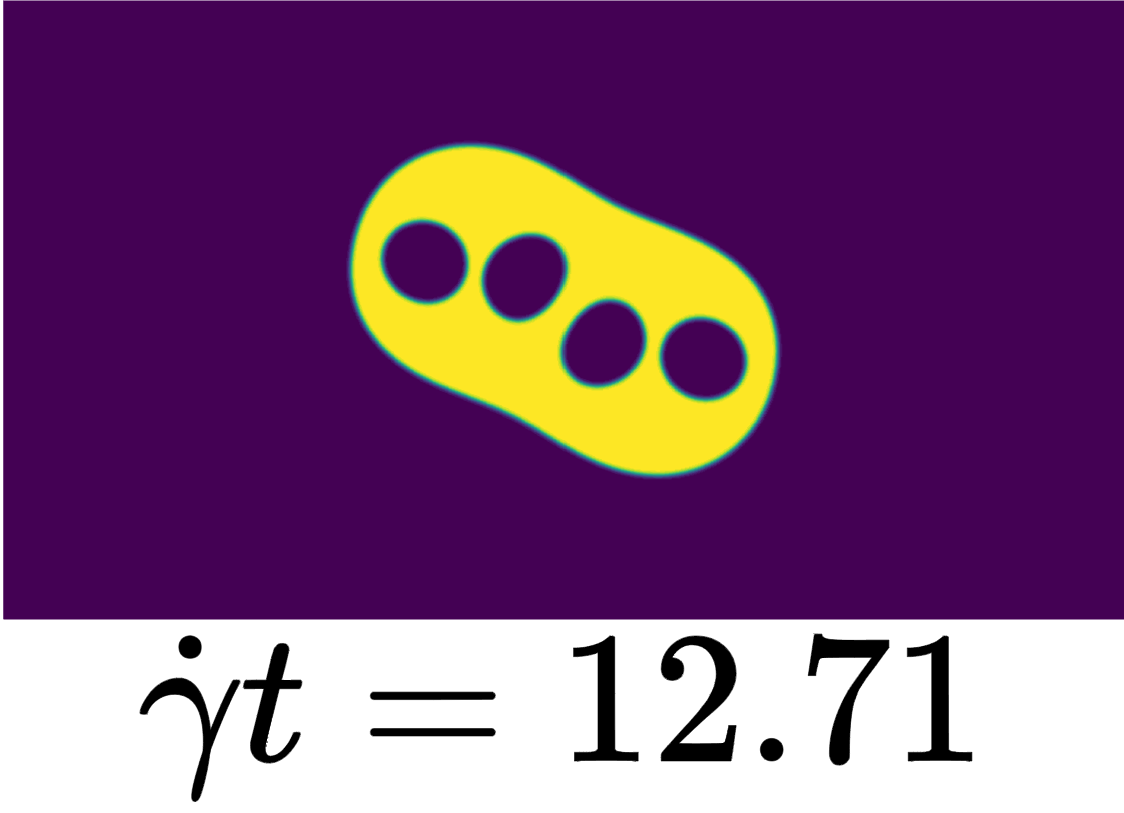}
		\includegraphics[scale = 0.09]{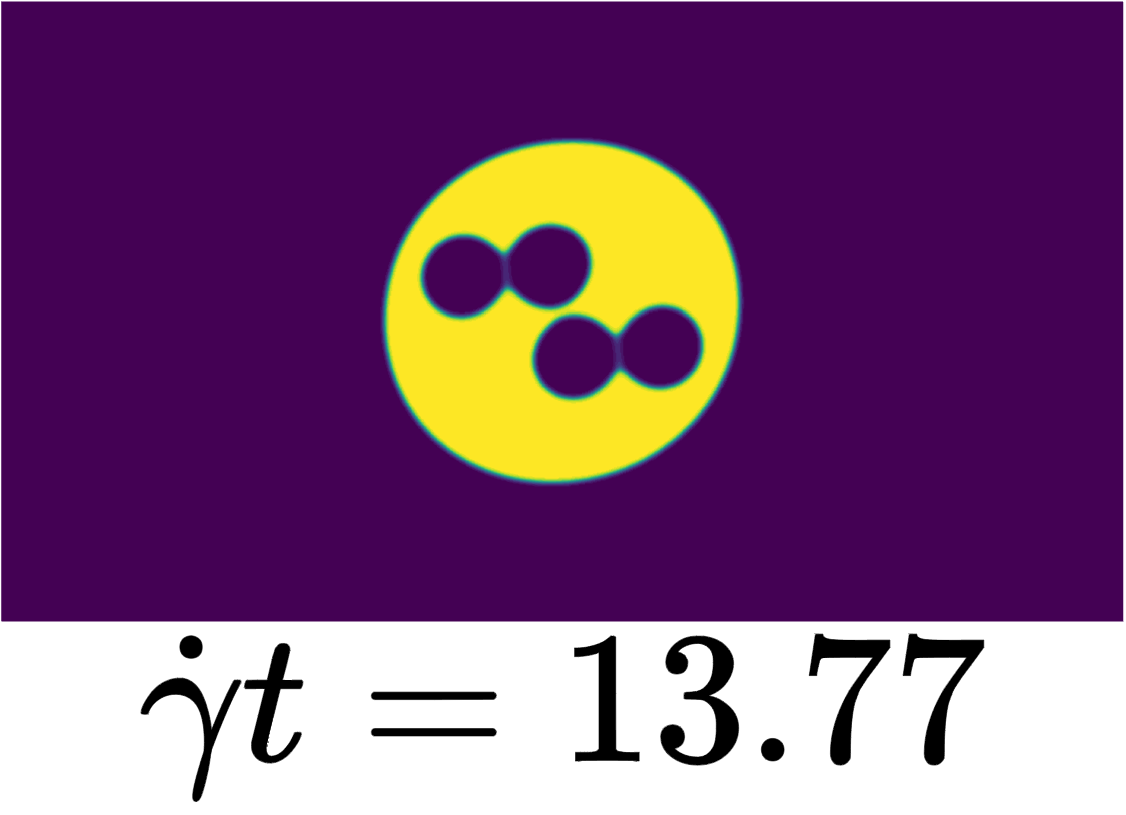}
		\includegraphics[scale = 0.09]{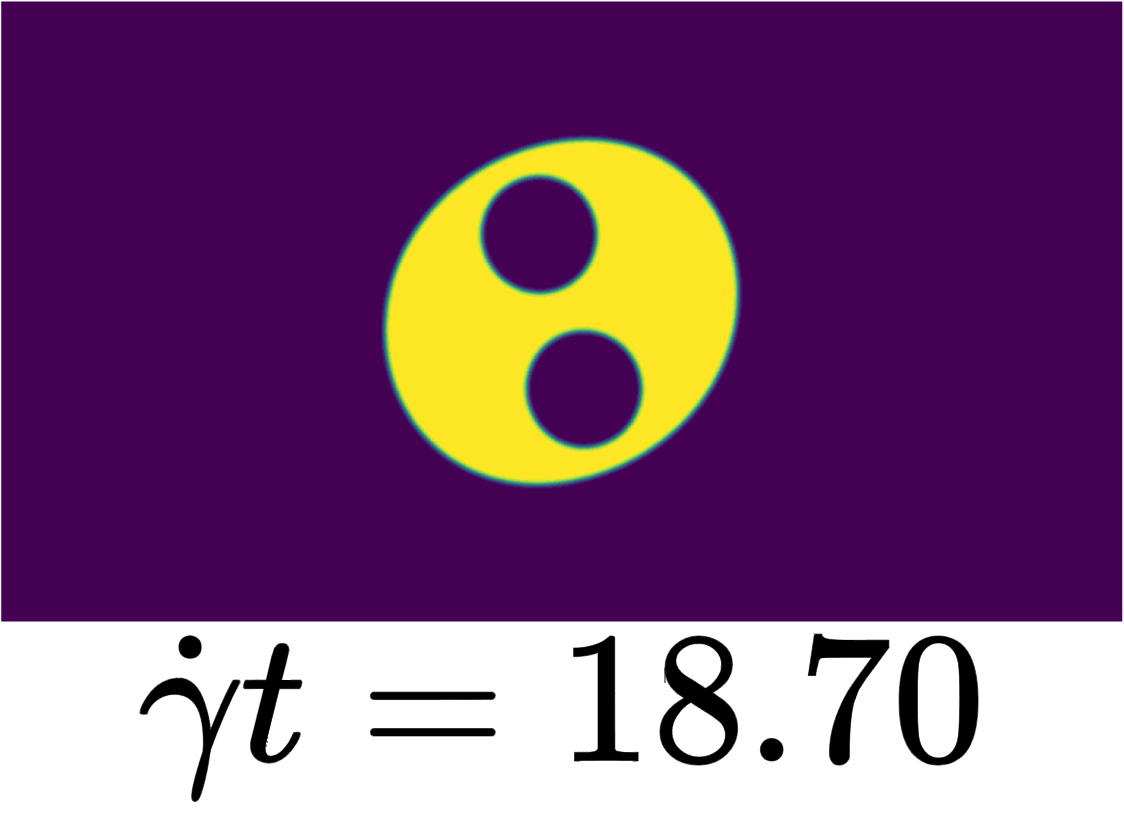} \\
        (a) $\frac{\Delta Y}{R_o}=0.00$~(Shell droplets coalesce, core droplets coalesce)
		\\ \vspace{0.5cm}

        \includegraphics[scale = 0.09]{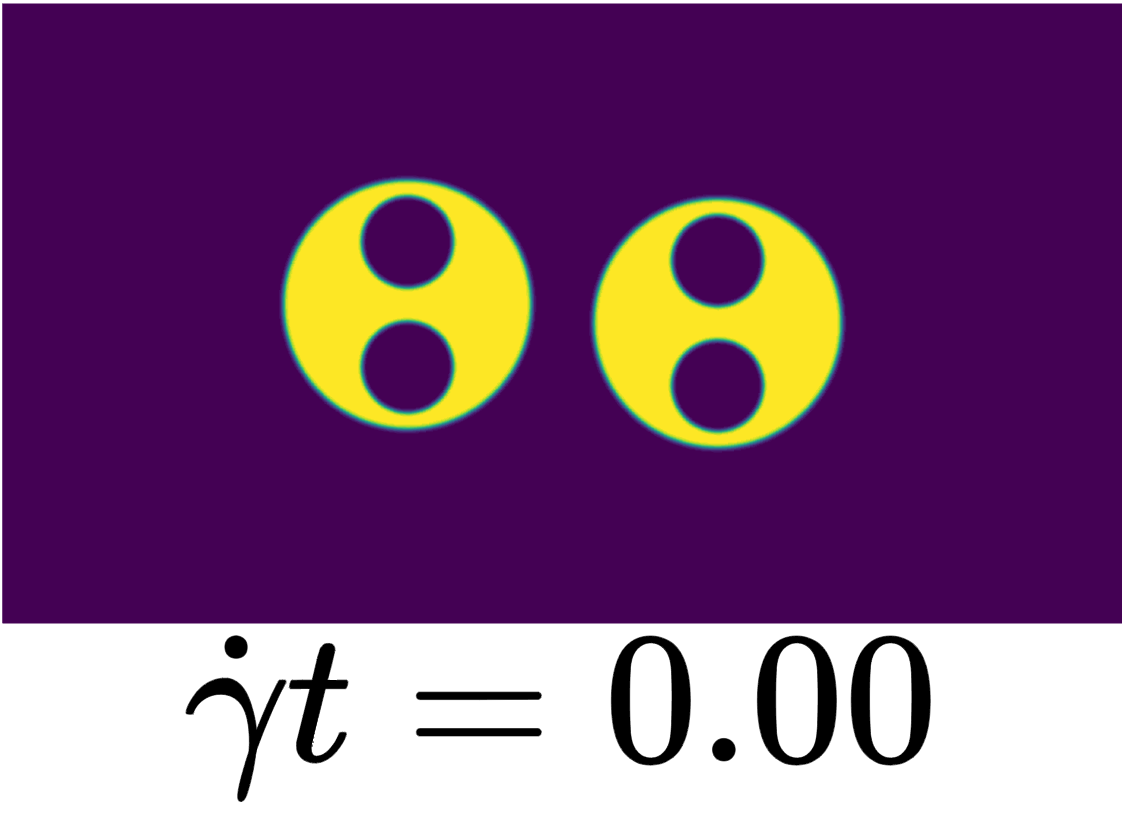}
            \includegraphics[scale = 0.09]{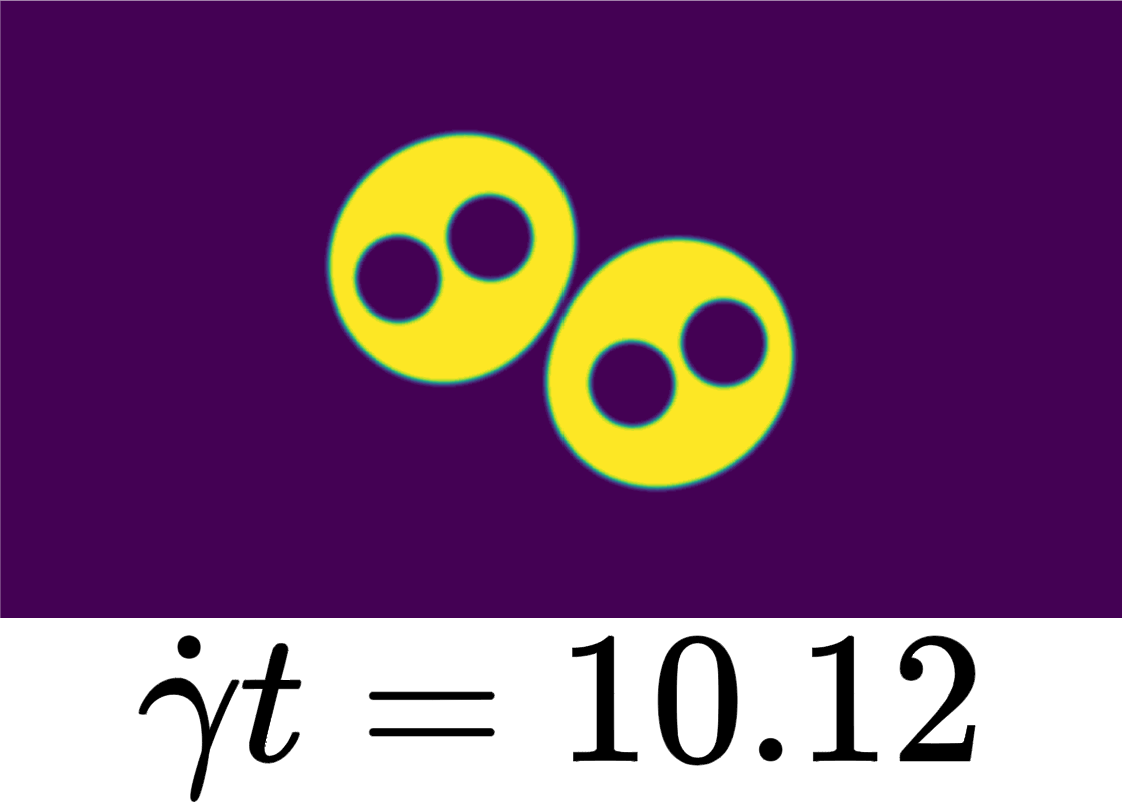}
		\includegraphics[scale = 0.09]{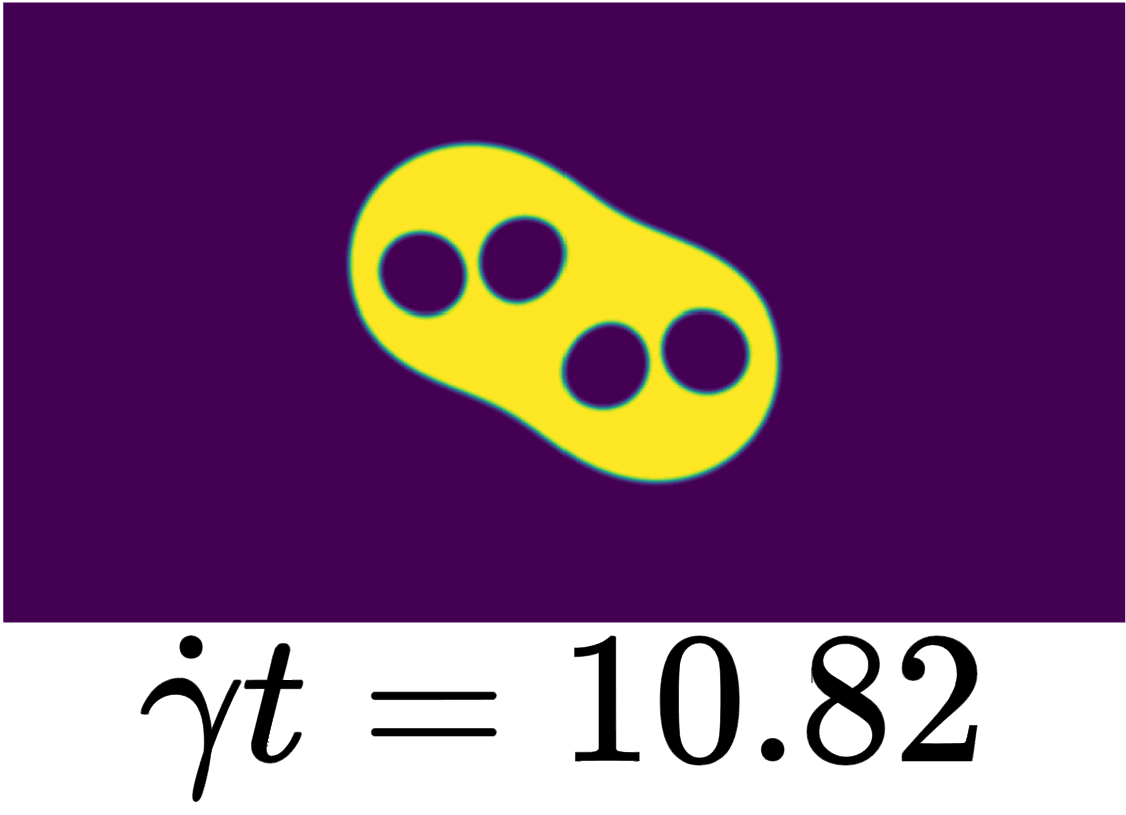}
		\includegraphics[scale = 0.09]{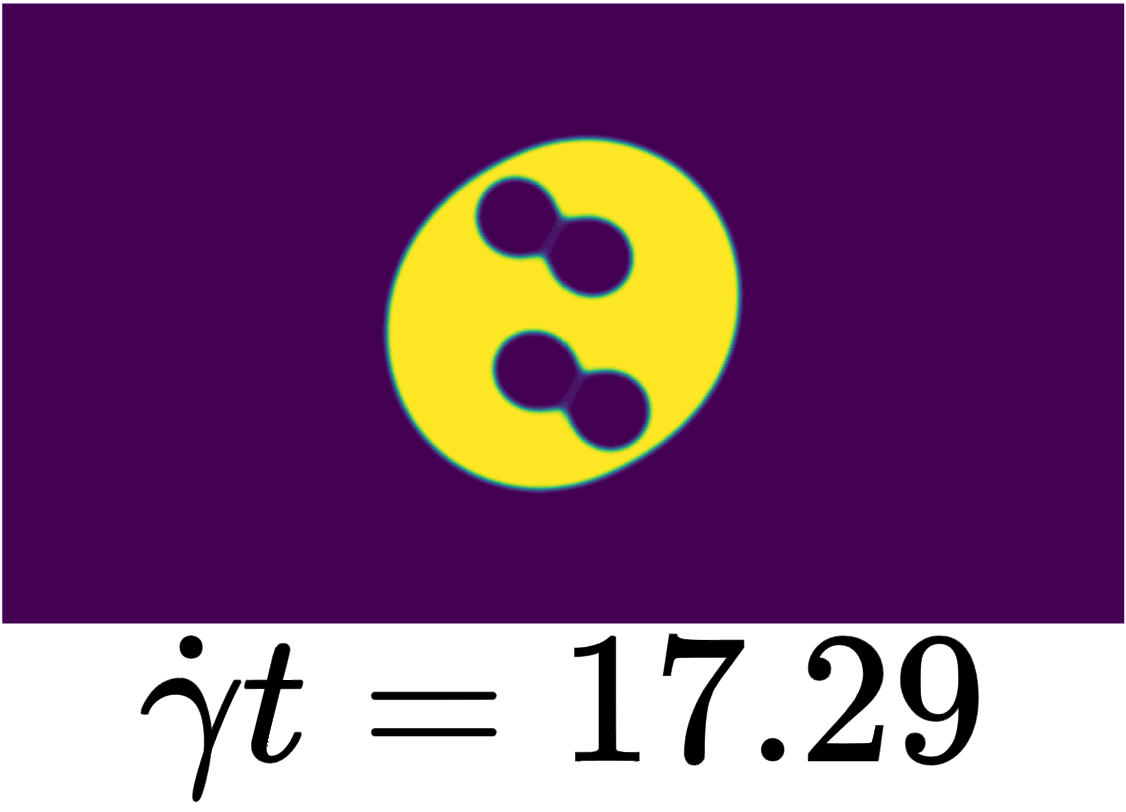}
		\includegraphics[scale = 0.09]{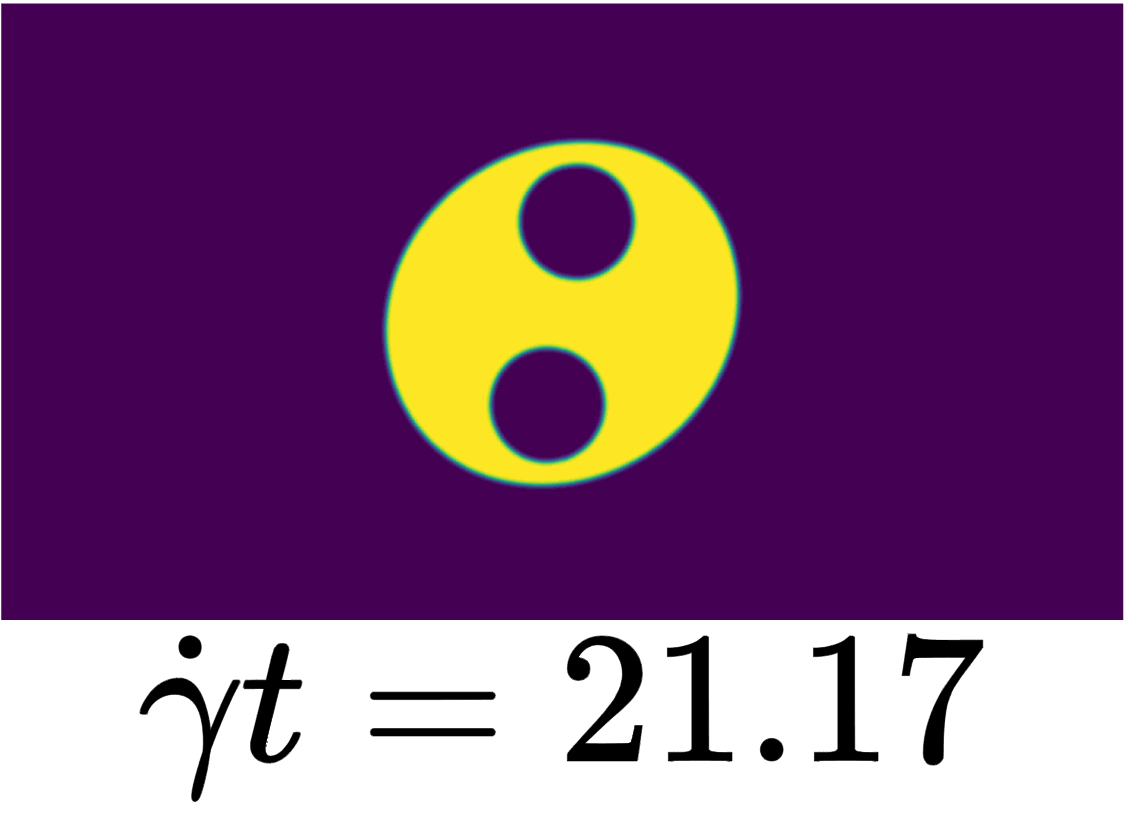}  \\
       (b)  $\frac{\Delta Y}{R_o}=0.15$~(Shell droplets coalesce, core droplets coalesce)
  \vspace{0.5cm}

   \includegraphics[scale = 0.09]{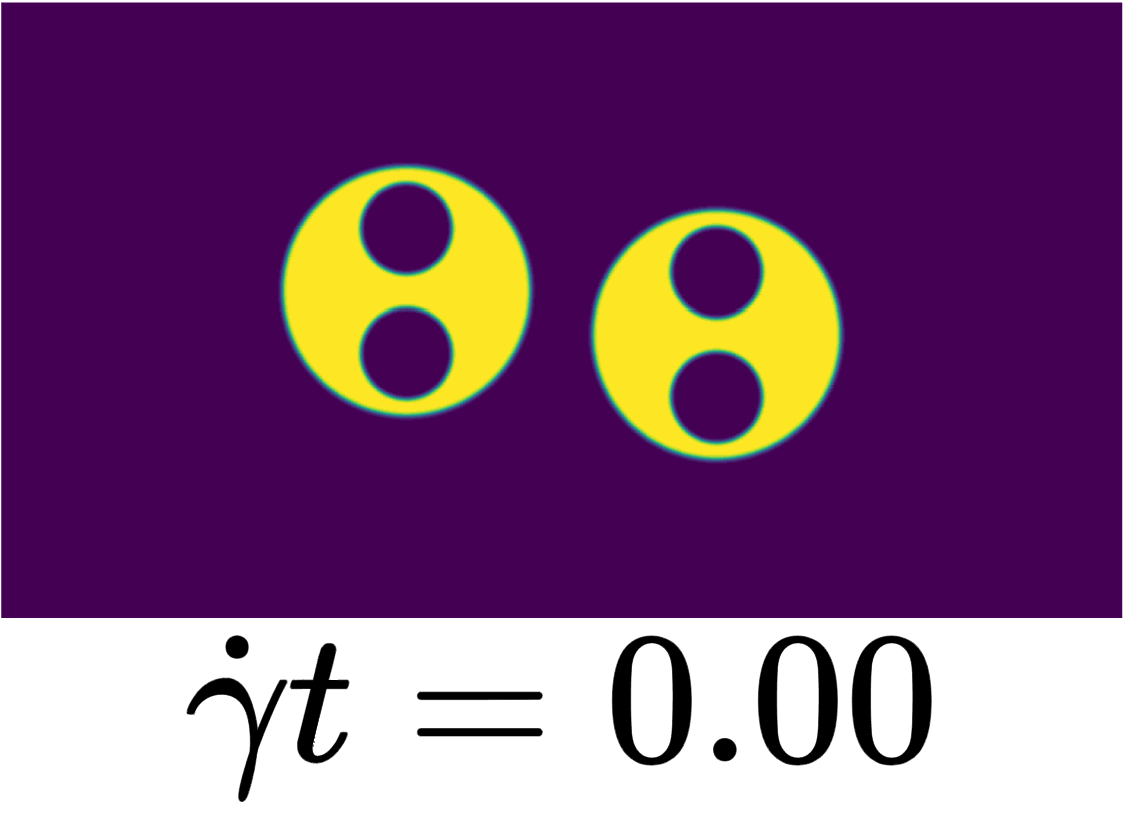}
		\includegraphics[scale = 0.09]{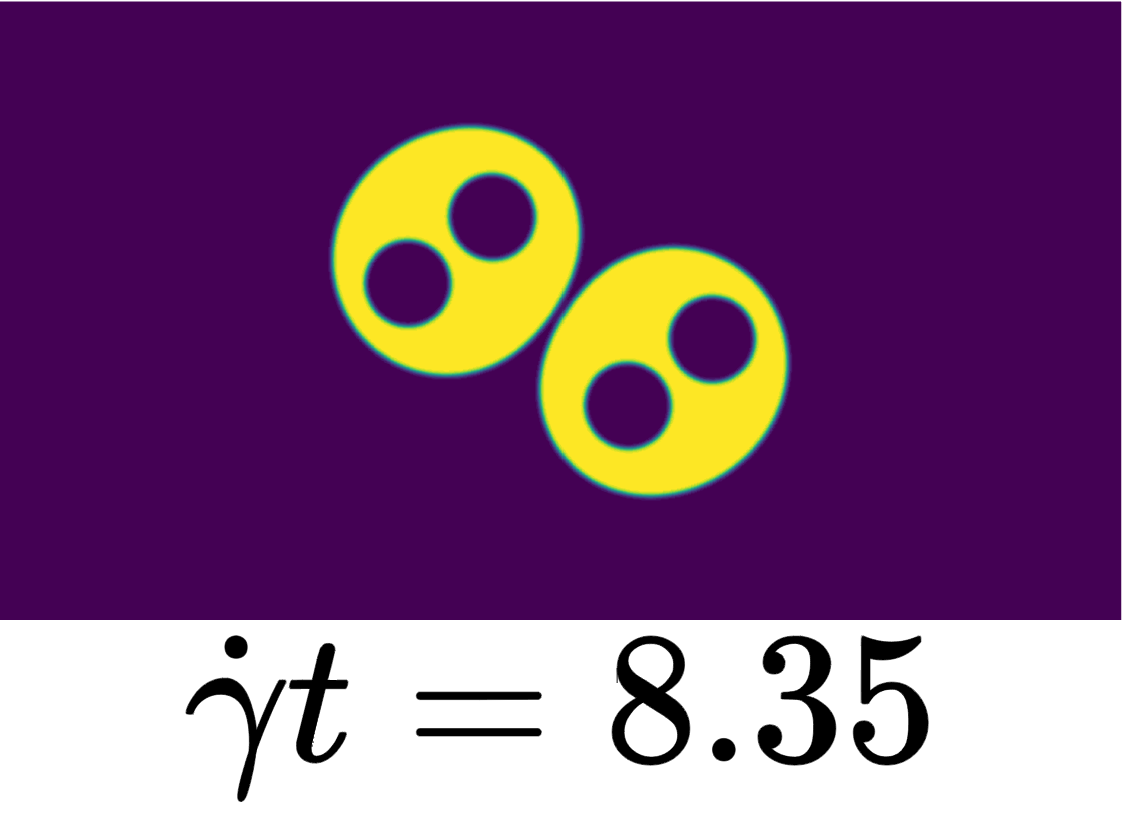}
		\includegraphics[scale = 0.09]{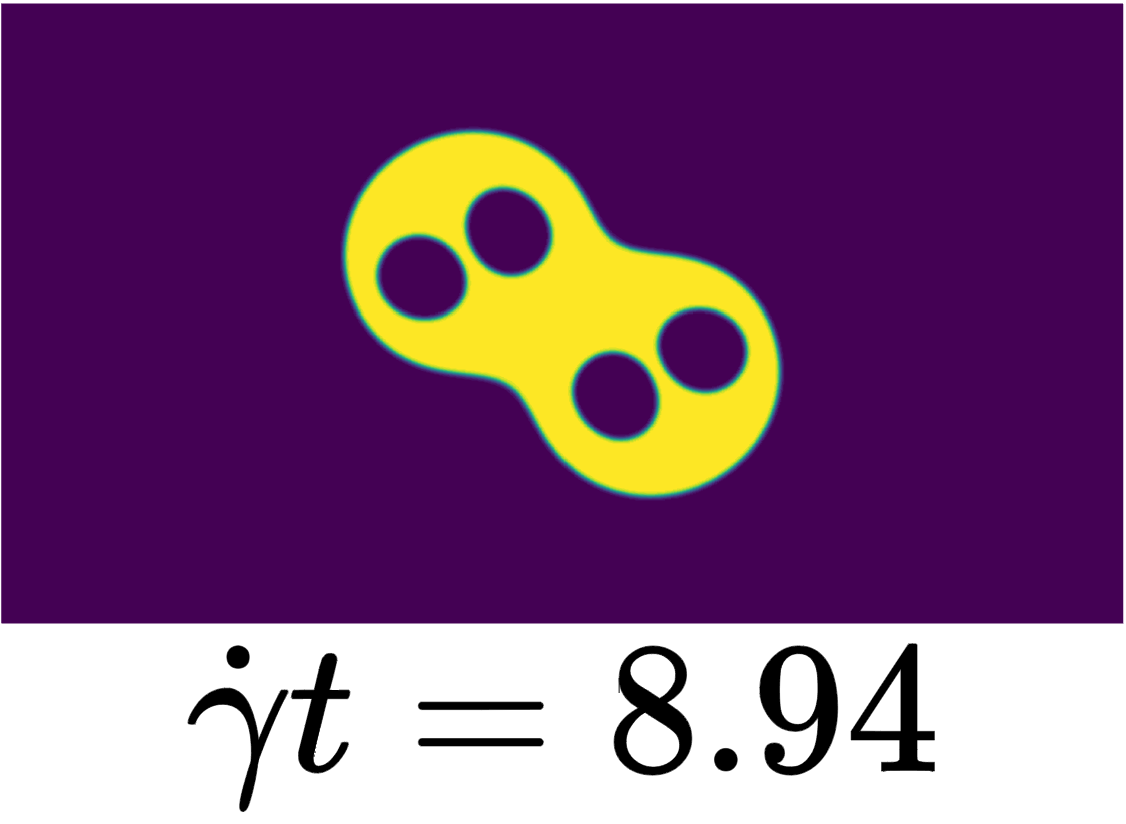}
		\includegraphics[scale = 0.09]{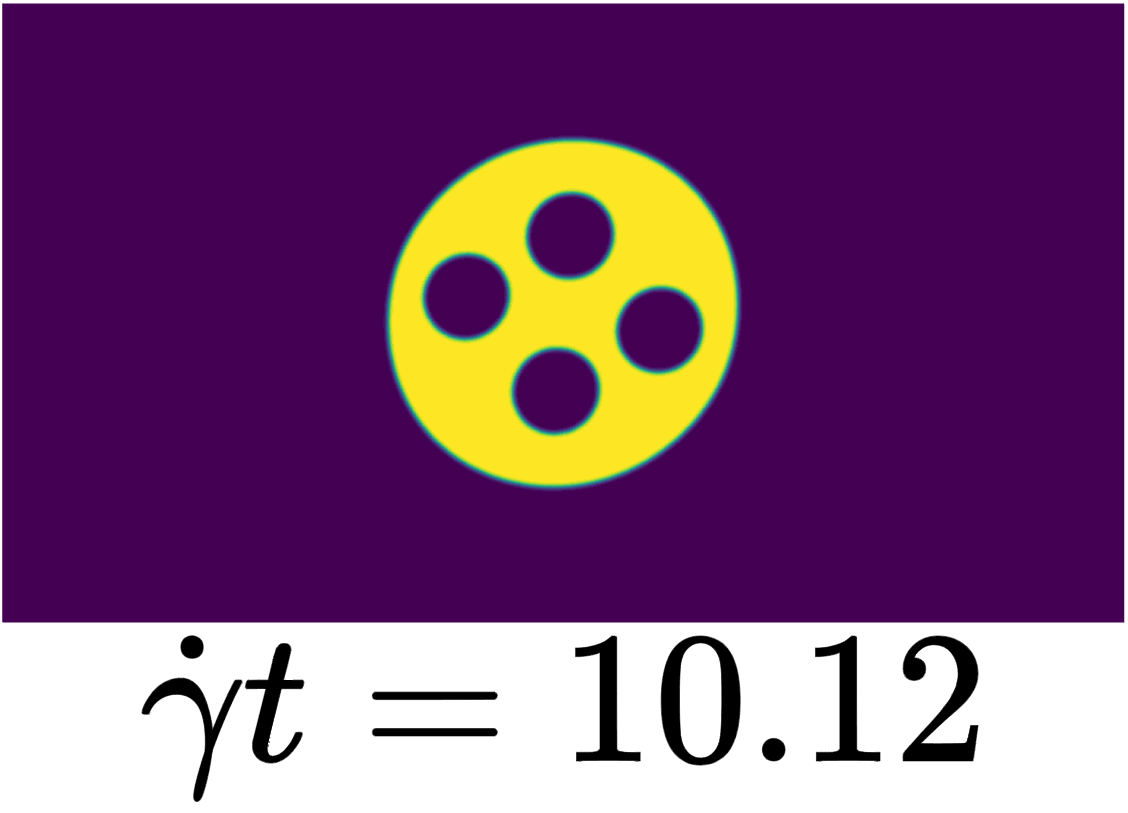}
		\includegraphics[scale = 0.09]{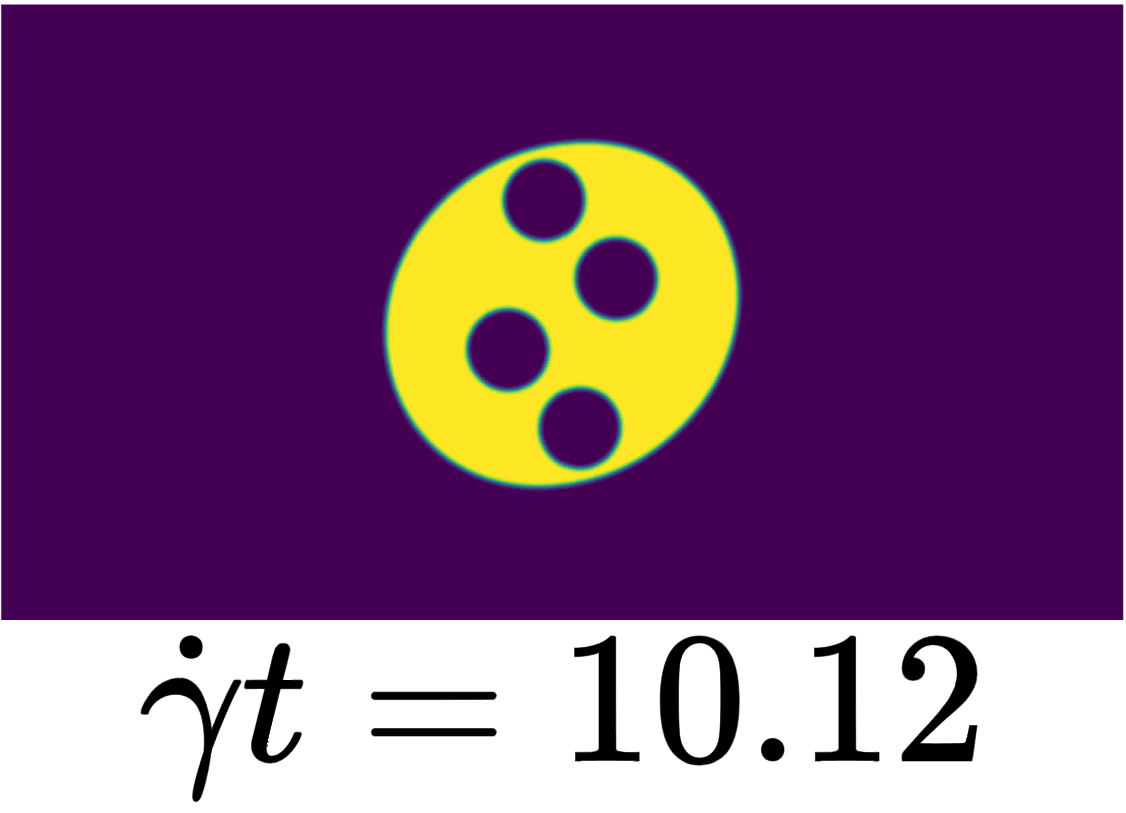} \\
        (c) $\frac{\Delta Y}{R_o}=0.35$~(Only shell droplets coalesce)
		\\ \vspace{0.5cm}

  \includegraphics[scale = 0.09]{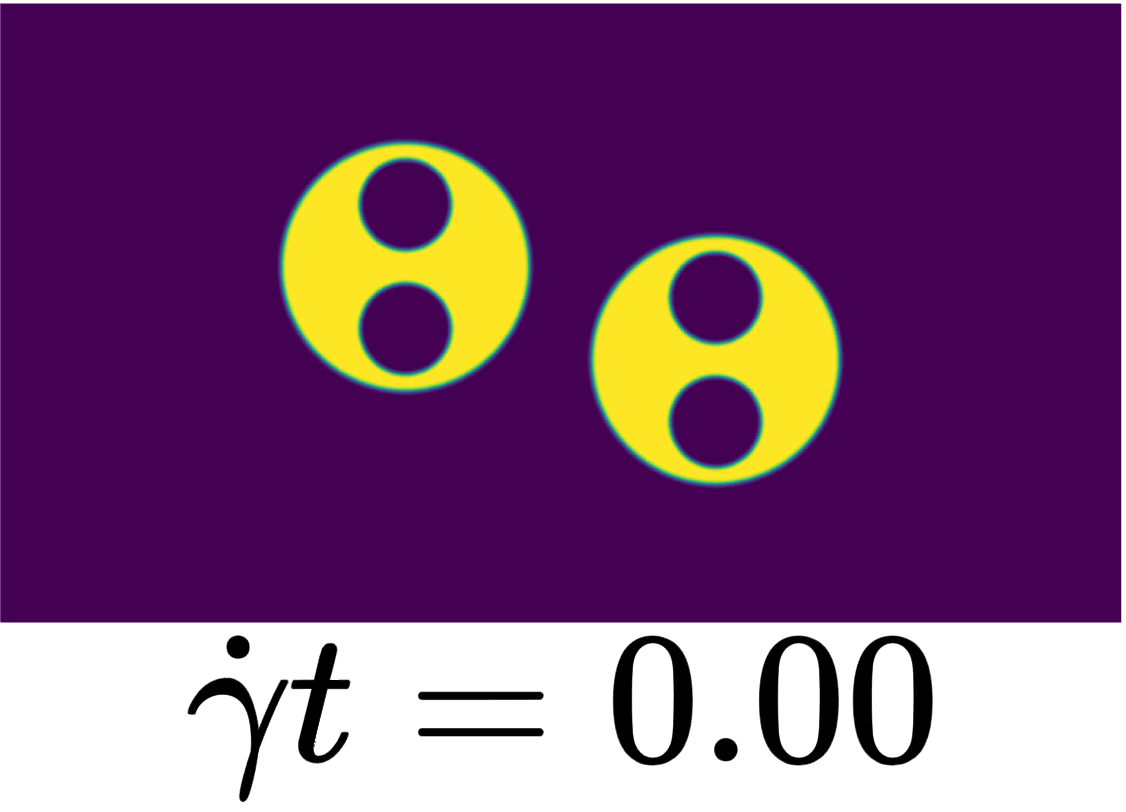}
		\includegraphics[scale = 0.09]{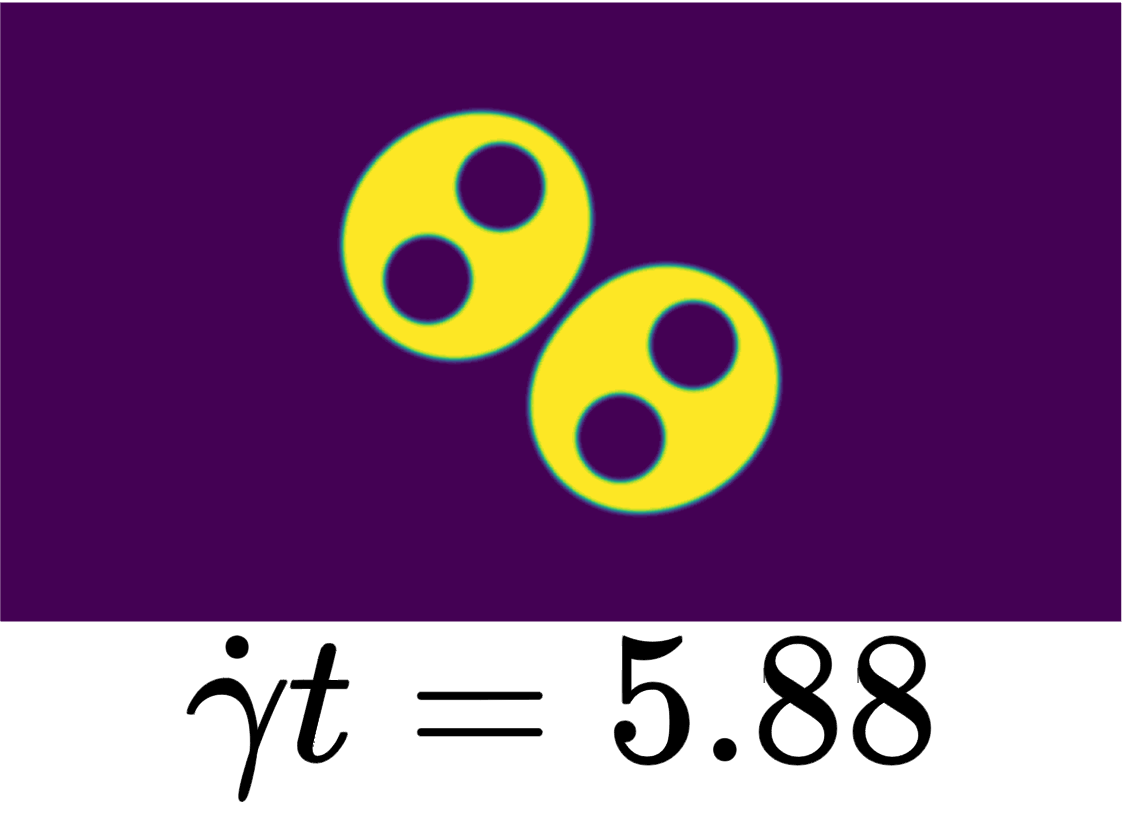}
		\includegraphics[scale = 0.09]{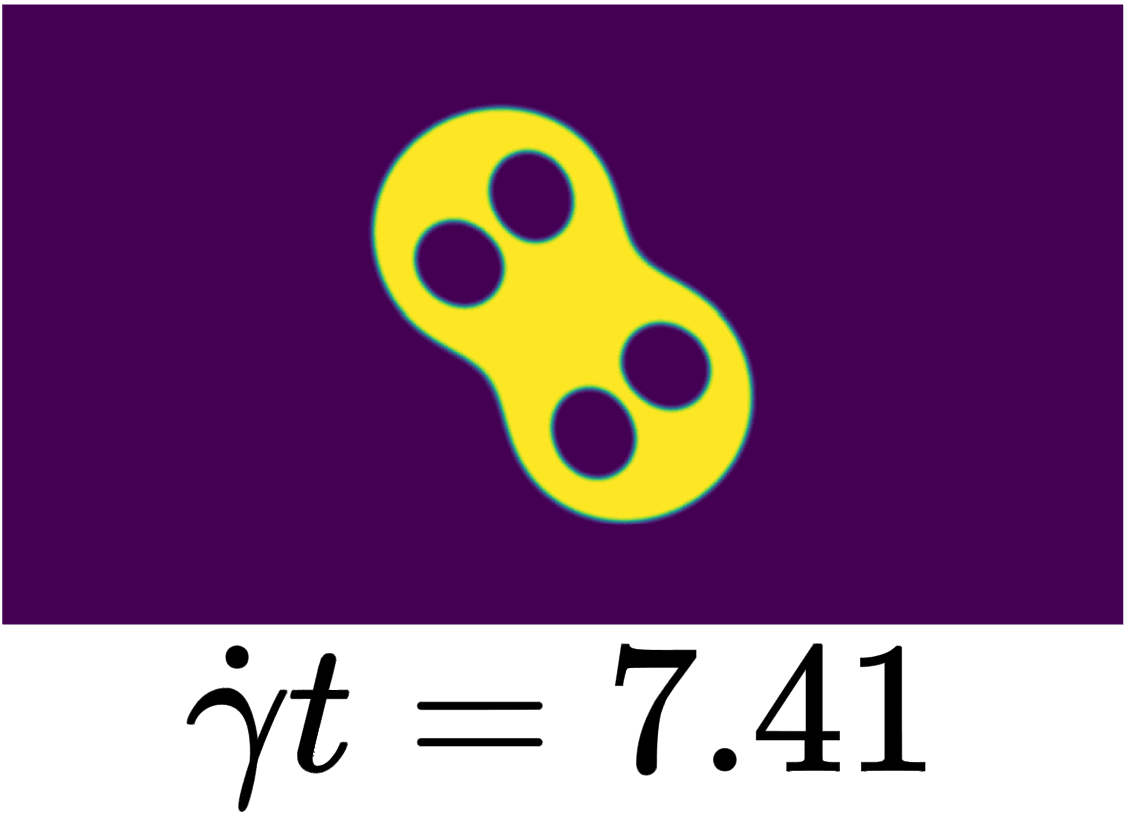}
		\includegraphics[scale = 0.09]{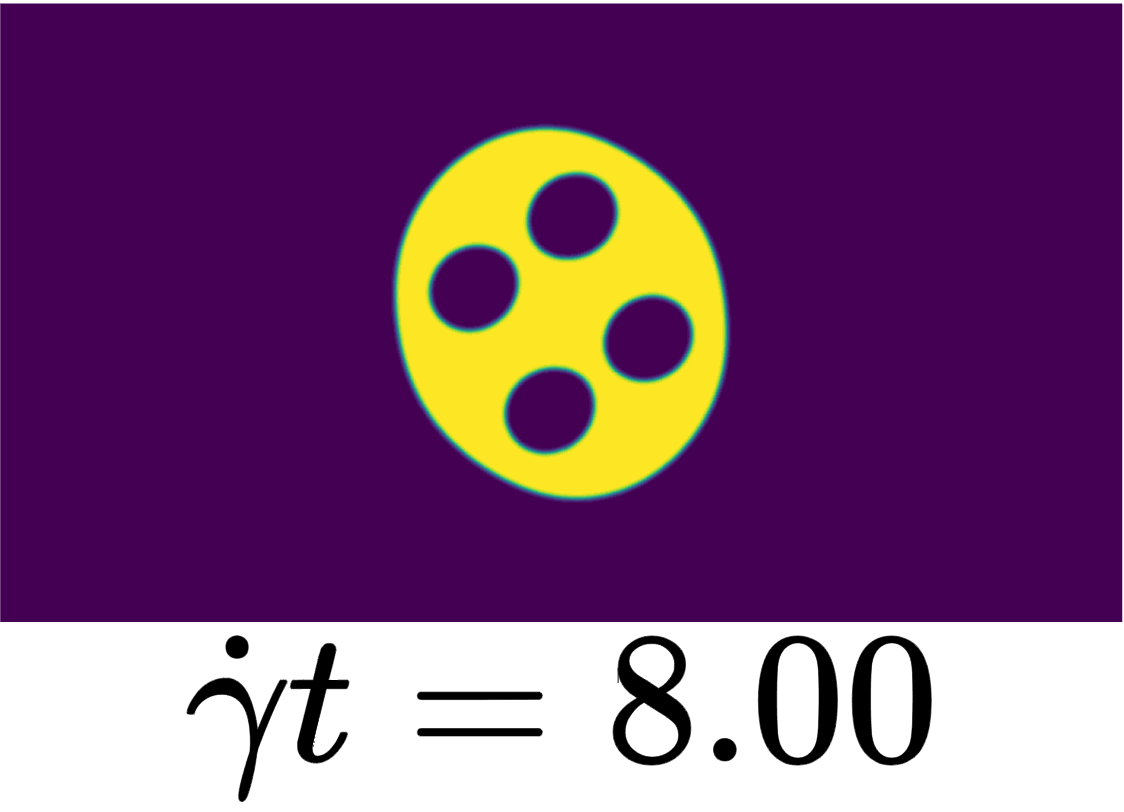}
		\includegraphics[scale = 0.09]{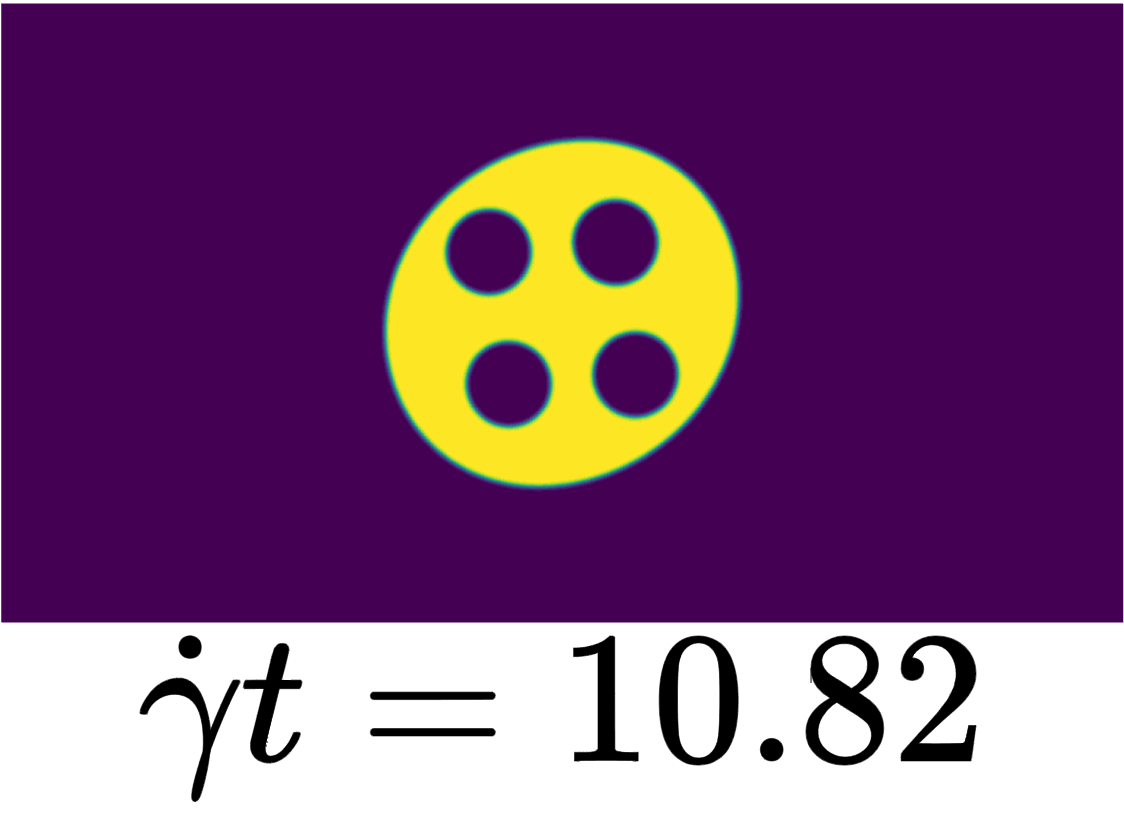} \\
        (d) $\frac{\Delta Y}{R_o}=0.75$~(Only shell droplets coalesce)
		\\ \vspace{0.5cm}

  \includegraphics[scale = 0.09]{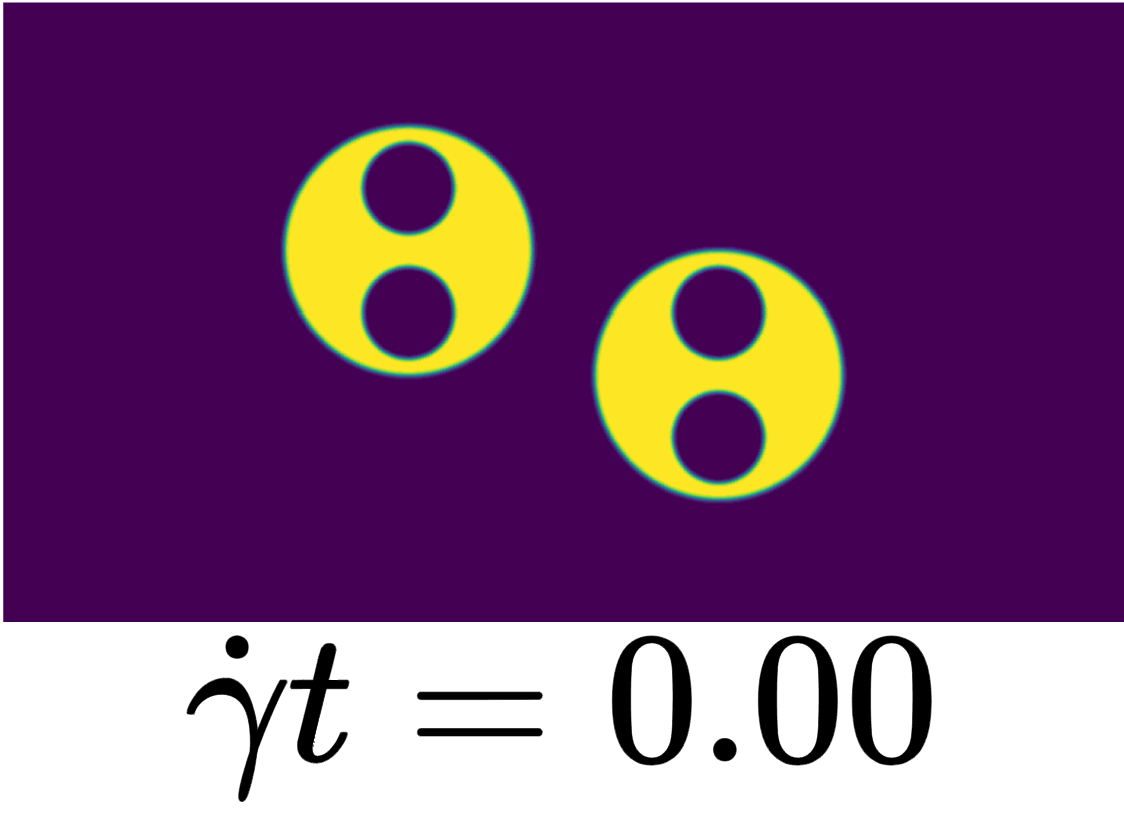}
		\includegraphics[scale = 0.09]{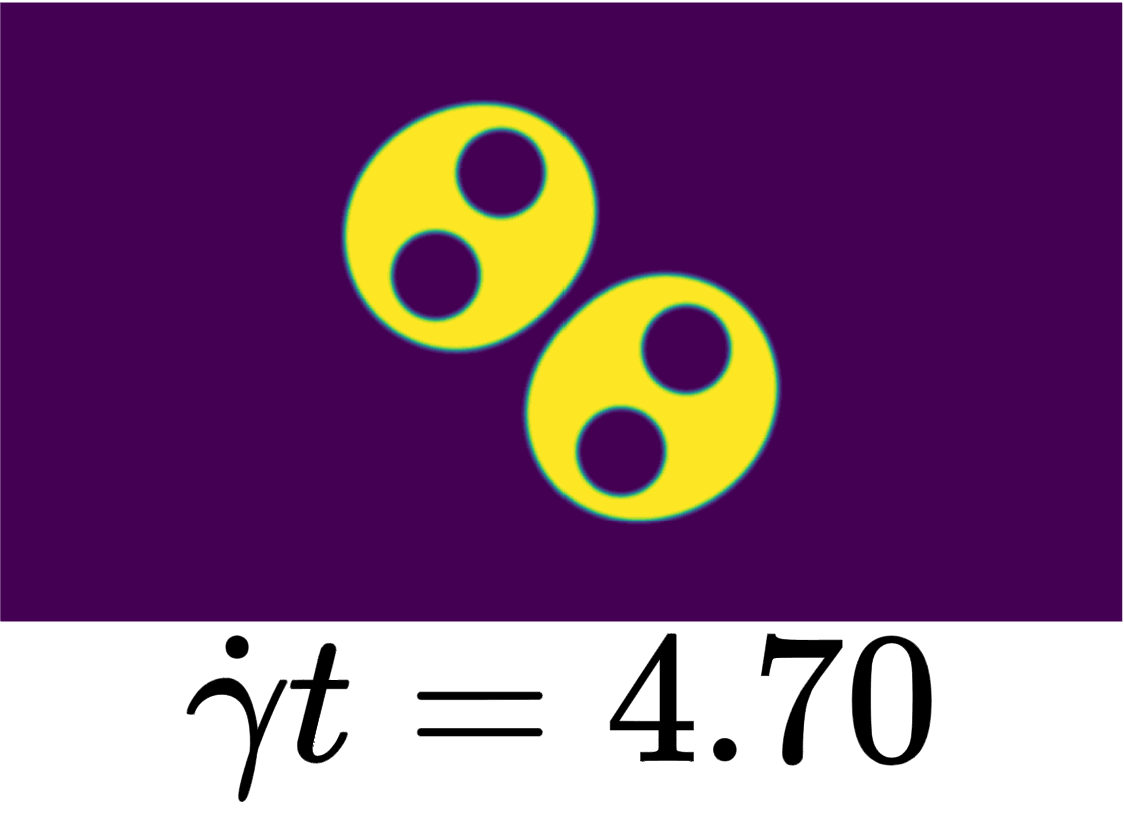}
		\includegraphics[scale = 0.09]{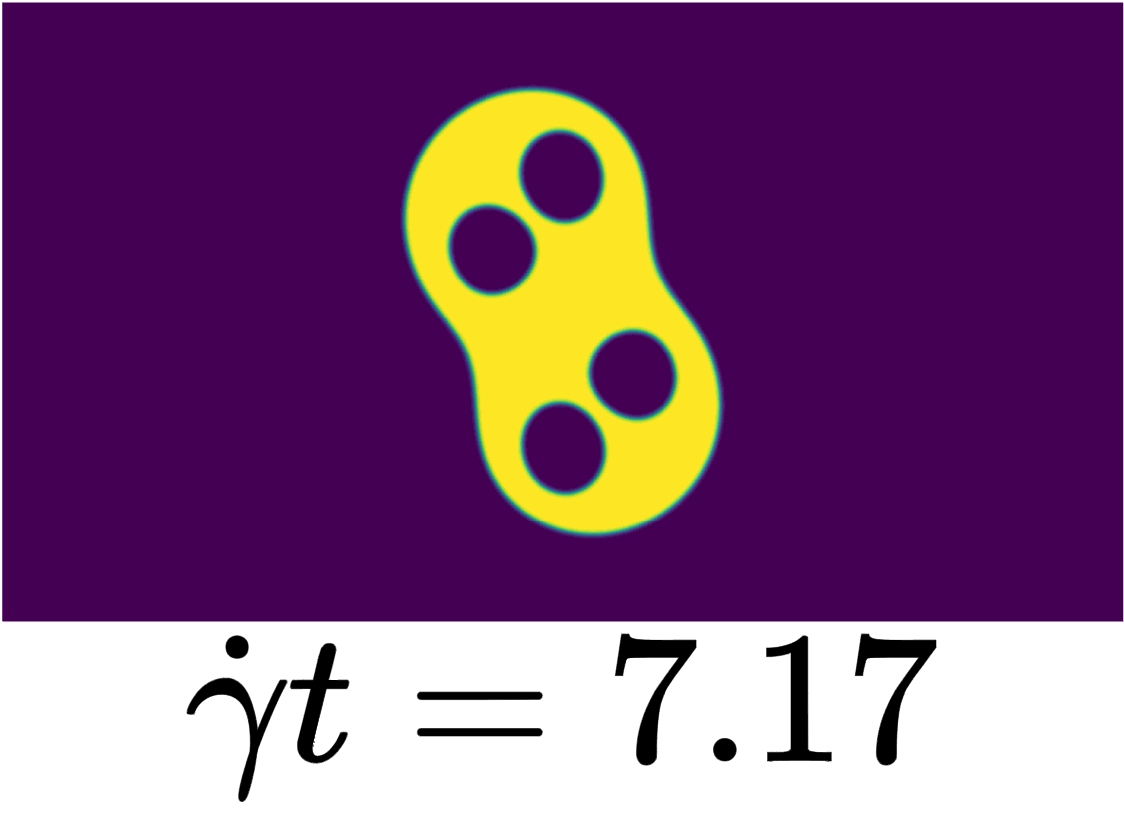}
		\includegraphics[scale = 0.09]{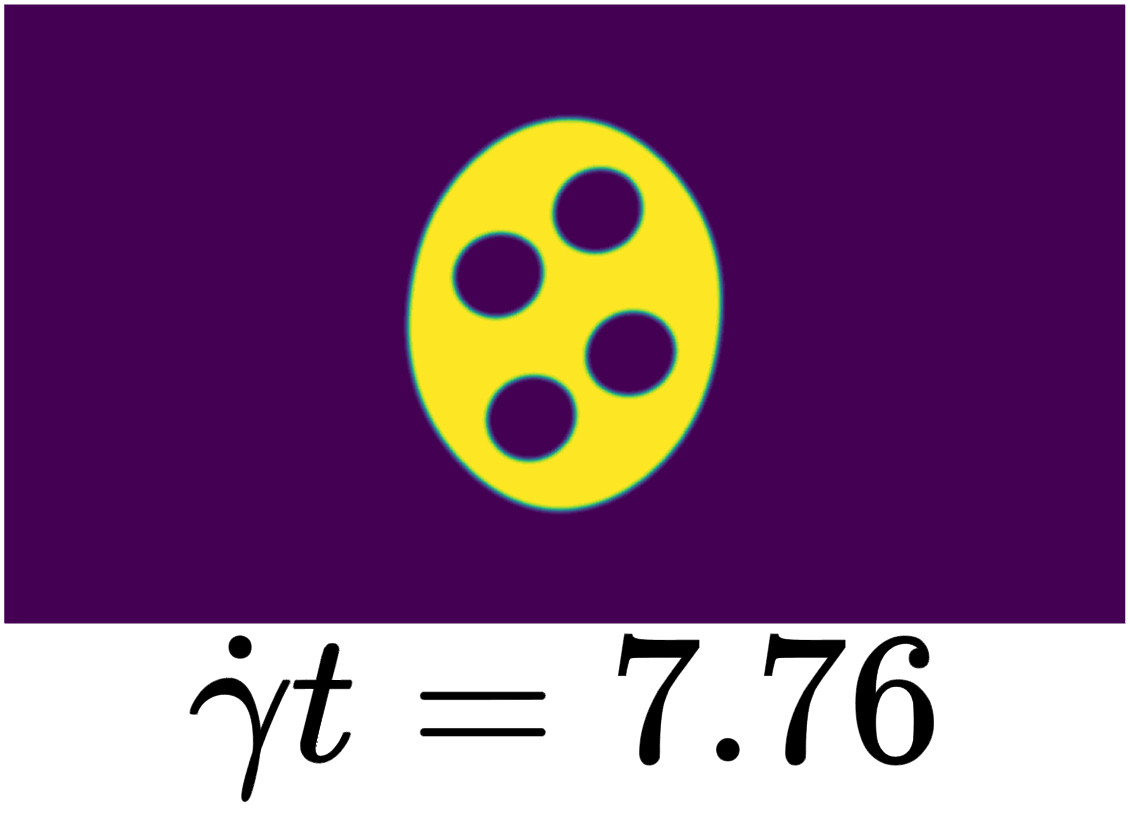}
		\includegraphics[scale = 0.09]{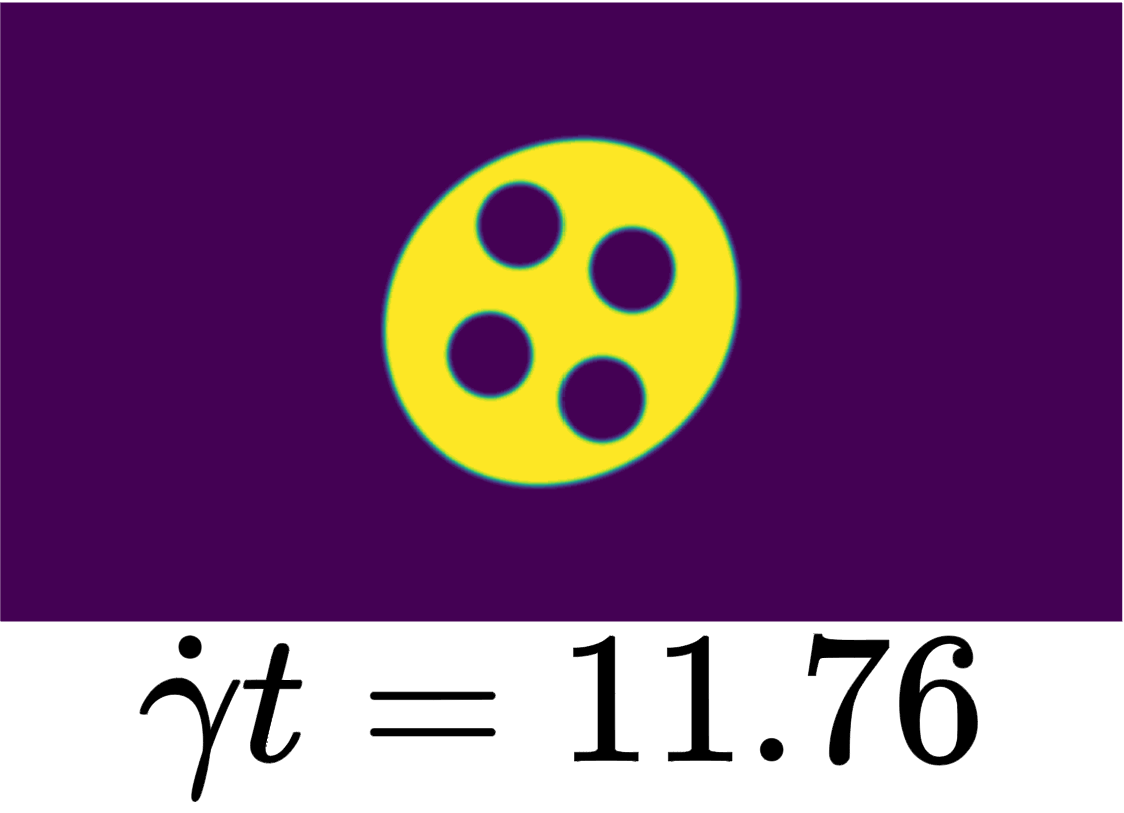} \\
        (e) $\frac{\Delta Y}{R_o}=1.00$~(Only shell droplets coalesce)
		\\ \vspace{0.5cm}

  \includegraphics[scale = 0.09]{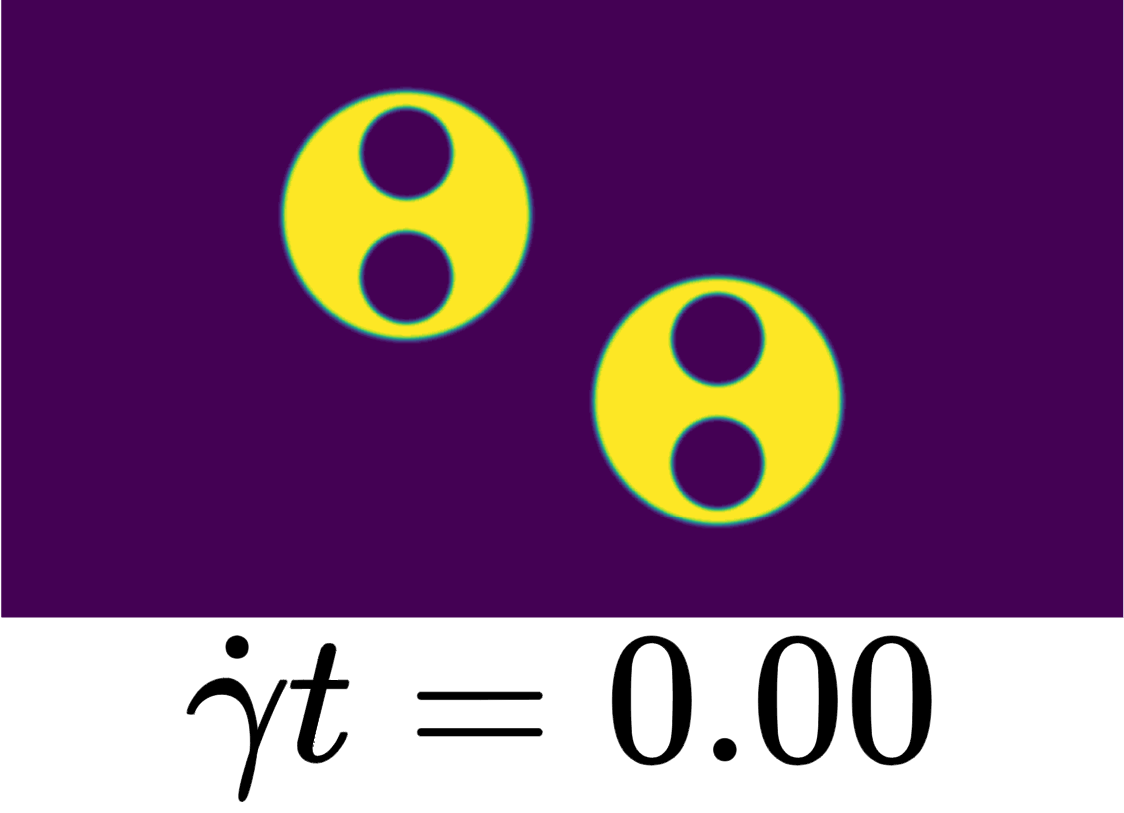}
		\includegraphics[scale = 0.09]{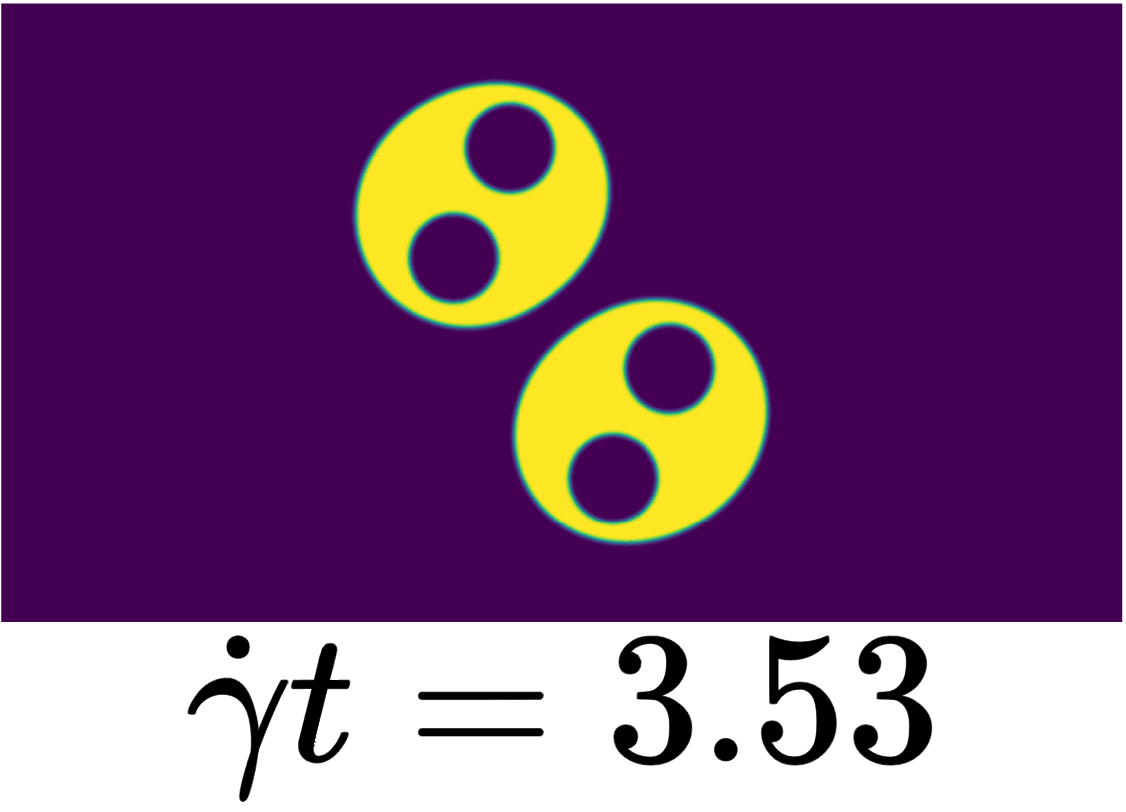}
		\includegraphics[scale = 0.09]{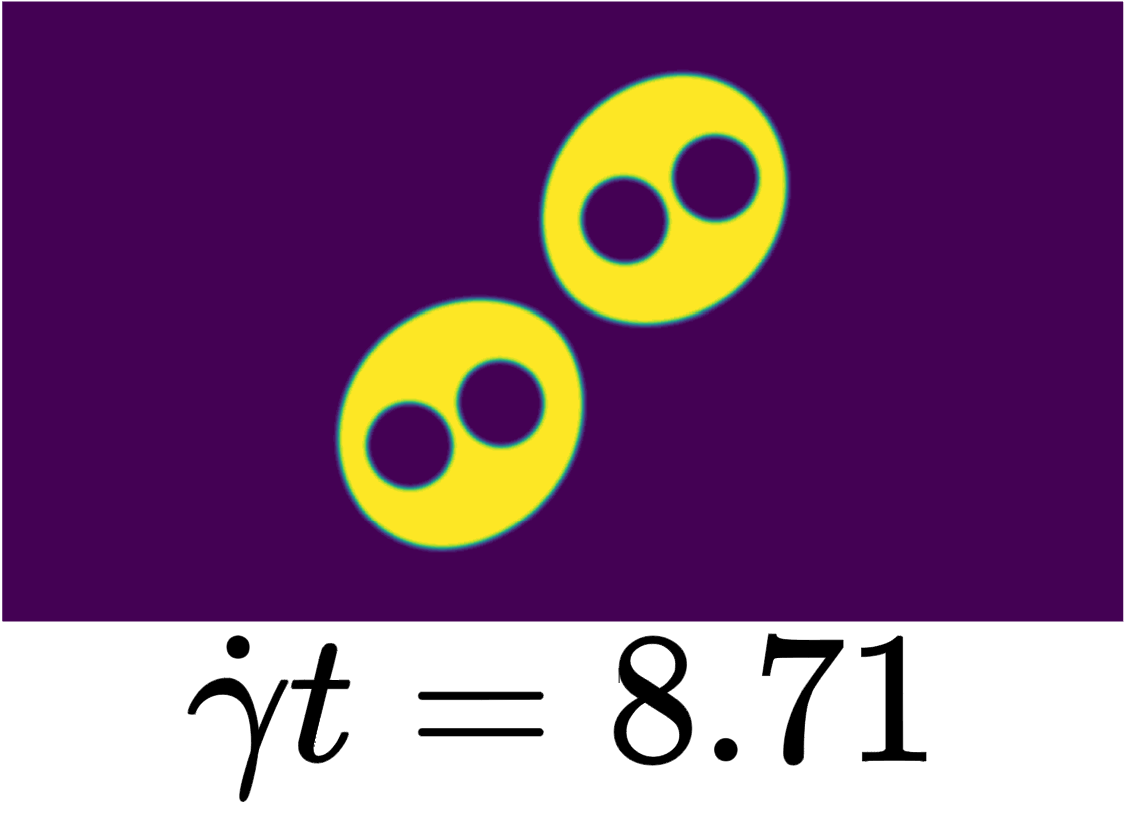}
		\includegraphics[scale = 0.09]{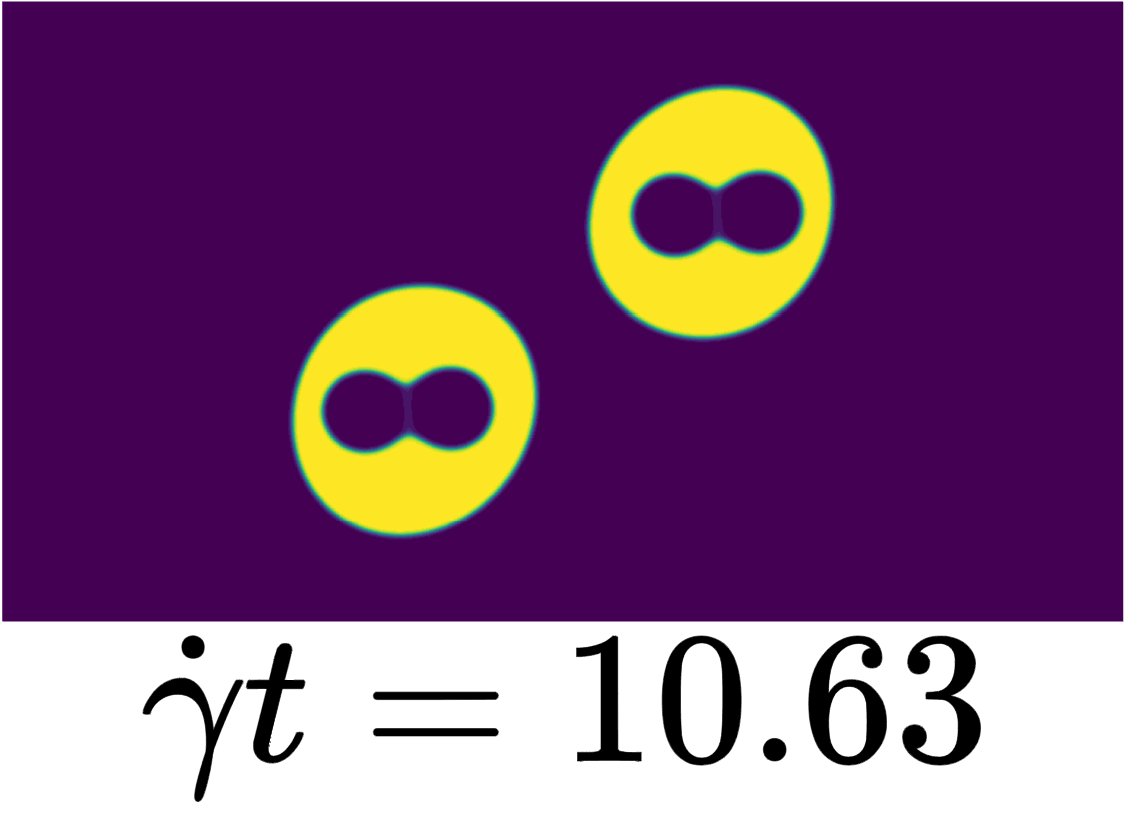}
		\includegraphics[scale = 0.09]{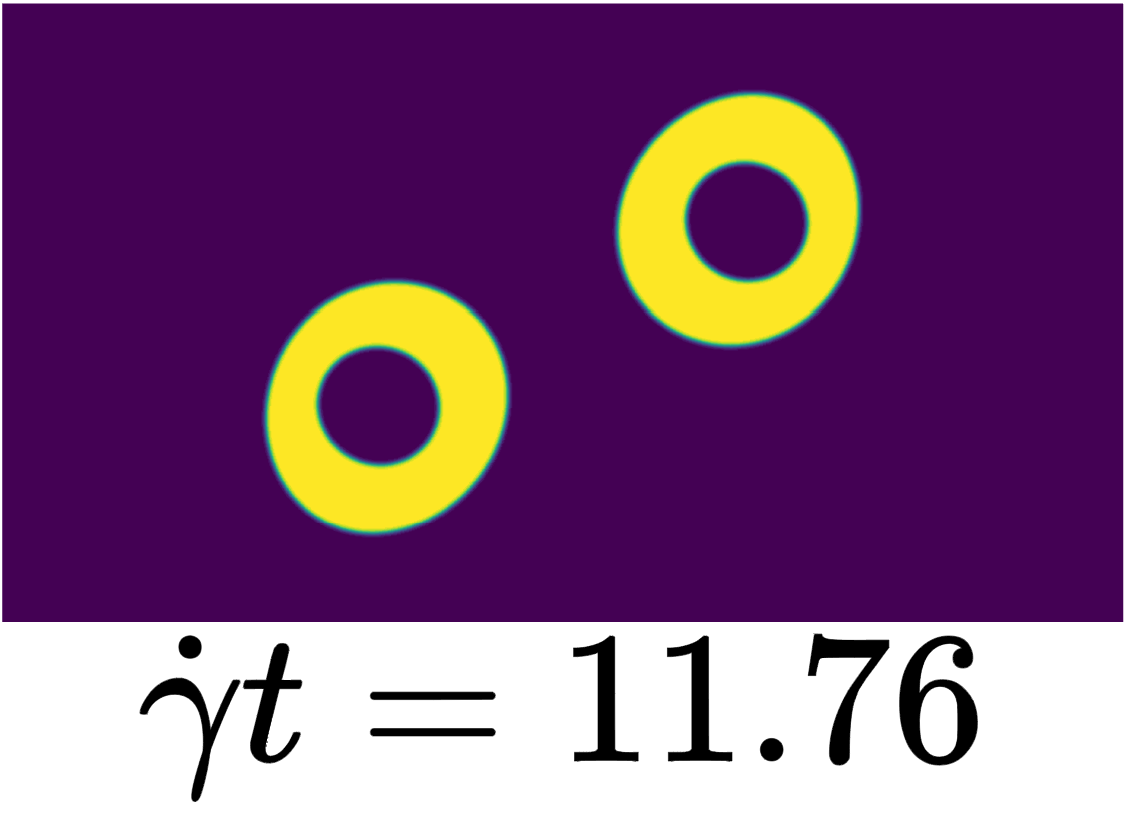} \\
        (f) $\frac{\Delta Y}{R_o}=1.50$~(Shell droplets pass over, core droplets coalesce)
		\\ \vspace{0.5cm}
  
 	\caption{Time-lapse images of double-core compound droplet collision for six different initial vertical offset ($\frac{\Delta Y}{R_o}=0.00, 0.15, 0.35, 0.75, 1.00$, and $1.50$) with $Ca=0.07$, $Re=0.10$, $\rho_{12}=1.0$, $\mu_{12}=1$, $\Delta X_o/R_o = 2.50,$ and $R_o/H = 0.40.$} 
	\label{fig:Double-core-Ca-0.07-varying-offset-snap}
\end{figure}

In Section~\ref{sec_5.2}(1), the outcomes of double-core compound droplet collisions at a high value of Ca (i.e., $0.10$) have been examined, revealing two distinct phenomena: coalescence of the shell droplets or pass-over motion of the shell droplets, while the core droplets remain separated in both scenarios. Here, simulation results have been presented for the same analysis but with a slightly lower $Ca$ value of $0.07$. Here, the primary objective is to determine if changes in surface tension, as indicated by the reduced $Ca$, influence the collision outcomes and dynamics of the droplets. In Figure~\ref{fig:Double-core-Ca-0.07-varying-offset-snap}, the collision outcomes are captured through critical snapshots at various stages of evolution for different initial vertical offset cases. These outcomes can be categorized into three distinct scenarios: (1) coalescence of both the shell and core droplets, (2) coalescence of only the shell droplets, and (3) pass-over of the shell droplets with coalescence of the core droplets. Notably, at this reduced Ca value of 0.07, core droplet coalescence is observed, which was absent for the case of high Ca (i.e., $Ca=0.10$). This significant difference underscores the effect of the $Ca$ value on the coalescence phenomena. The reduction in $Ca$, corresponding to an increase in surface tension (keeping viscosity fixed), appears to facilitate the merging of the core droplets even as the shell droplets exhibit varied outcomes. 

\begin{figure}[H]
          \centering

         \includegraphics[scale = 0.085]{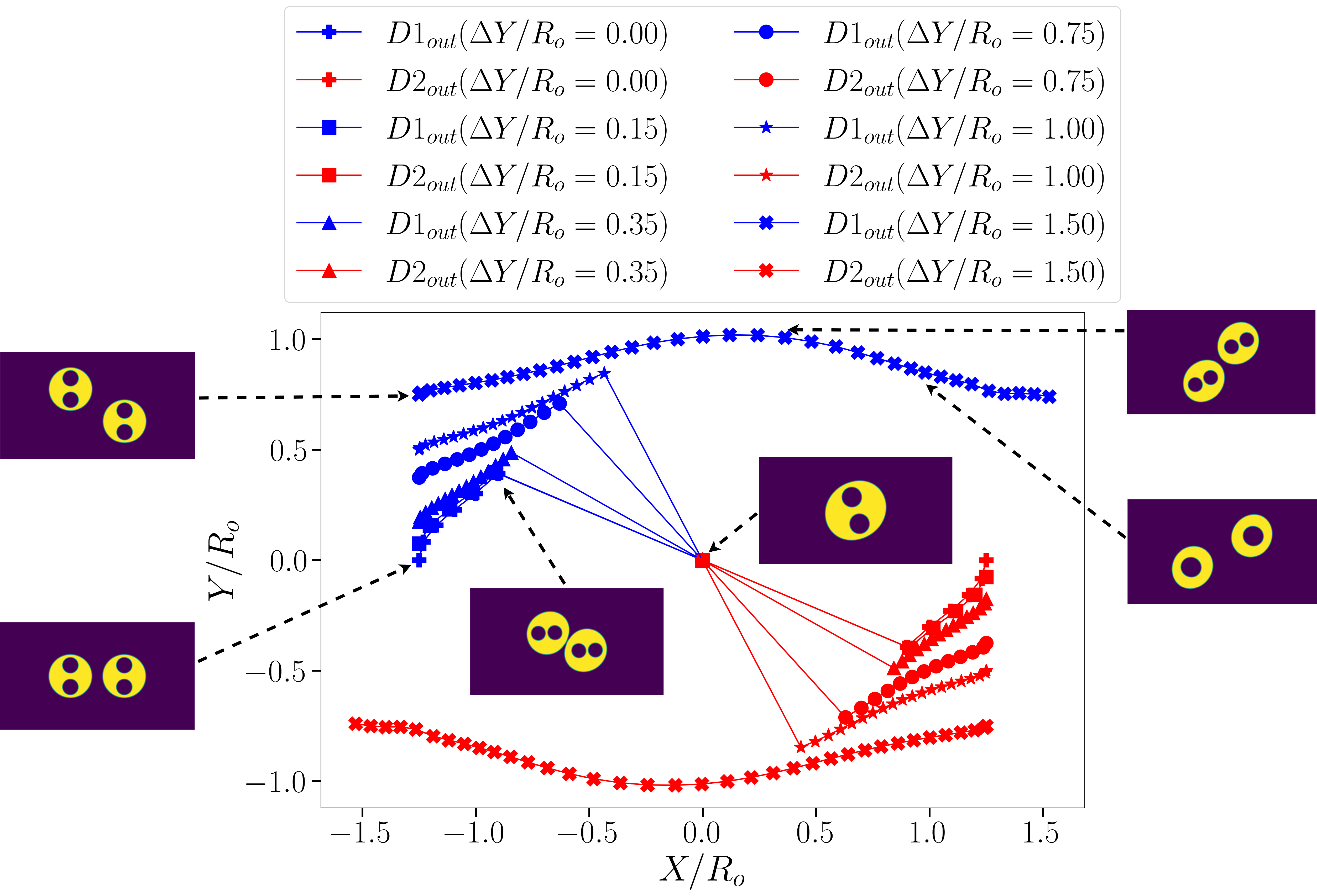} \\
          \vspace{0.45cm}

	\caption{Trajectories of the shell droplet pairs ($D1_{out}$ and $D2_{out}$) for the six different initial vertical offset ($\frac{\Delta Y}{R_o}=0.00, 0.15, 0.35, 0.75, 1.00$, and $1.50$) with $Ca=0.07$.}
	\label{fig:Double-core-Ca-0.07-varrying-offset-outer-trajectory}
\end{figure}

The trajectory of the shell droplets has been plotted and illustrated in Figure~\ref{fig:Double-core-Ca-0.07-varrying-offset-outer-trajectory}, accompanied by key snapshots to better understand the situations at critical instances. The behavior of the shell droplets in terms of their trajectories for both coalescence and pass-over cases at $Ca=0.07$ exhibits a consistent pattern compared to those observed at $Ca=0.10$, suggesting that the primary factors influencing the trajectories of the shell droplets are the initial vertical offset and the imposed shear flow. 

\begin{figure}[H]
          \centering
          
        \includegraphics[scale = 0.075]{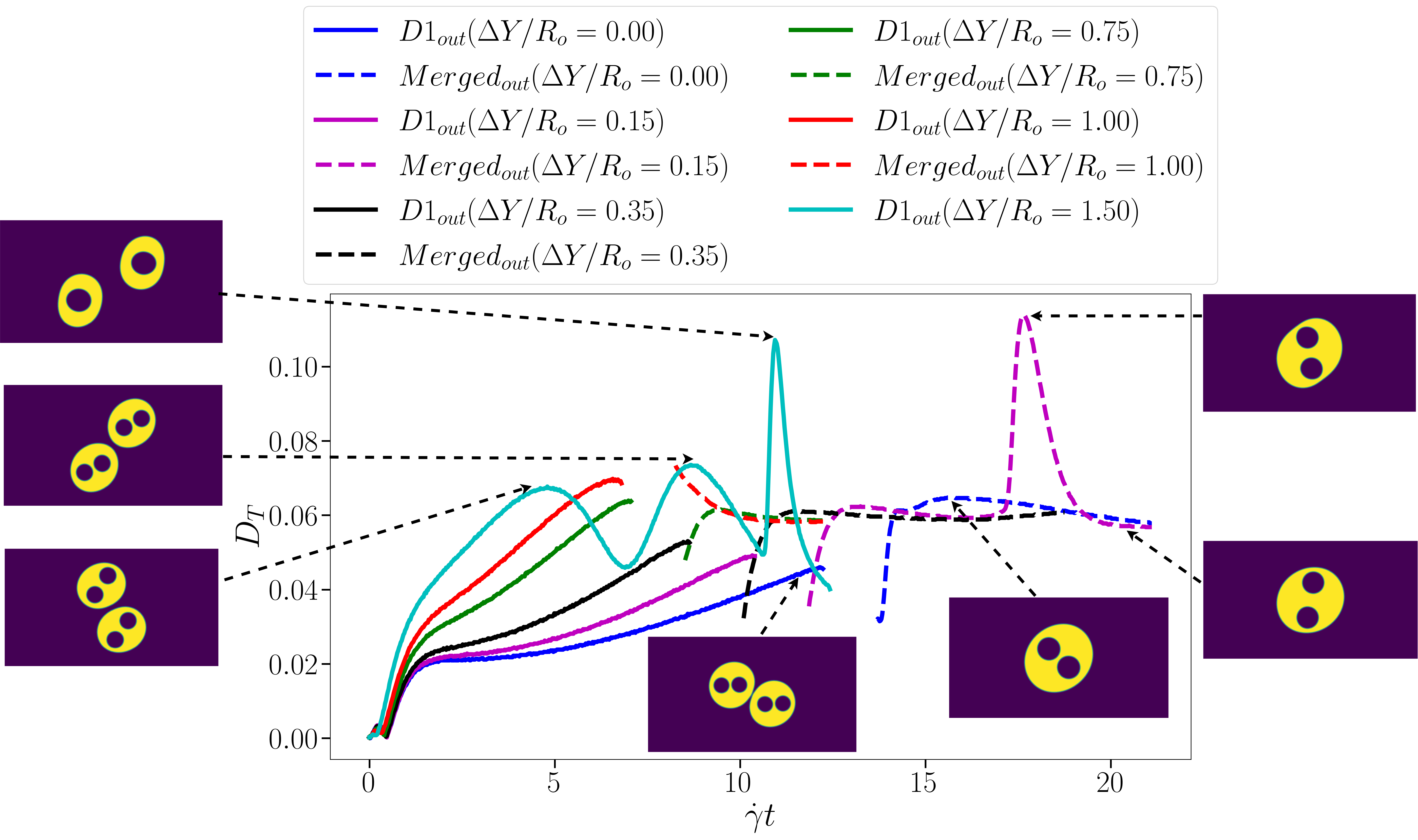} \\
    \vspace{0.45cm}
    
	\caption{Deformation quantification of the shell droplet before and after coalescence ($D1_{out}$ and $Merged_{out}$) for the six different initial vertical offset ($\frac{\Delta Y}{R_o}=0.00, 0.15, 0.35, 0.75, 1.00$, and $1.50$) with $Ca=0.07$.}
	\label{fig:Double-core-Ca-0.07-varrying-offset-outer-deform}
\end{figure}

Figure~\ref{fig:Double-core-Ca-0.07-varrying-offset-outer-deform} depicts the deformation of the shell droplet both before and after the coalescence of shell or core droplets. The deformation plots reveal trends similar to those observed for $Ca=0.10$, but with a noticeable sudden increase in the deformation curves when the core droplets coalesce inside the shell droplets, for both the shell droplet coalescence and pass-over cases. The coalescence of the core droplets induces a slight instability in the shape of the shell droplet, which manifests as a spike in the deformation curve. 

Interestingly, among the two cases of shell droplet coalescence accompanied by core droplet coalescence (i.e., $\frac{\Delta Y}{R_o}=0.00, 0.15$), only the case with $\frac{\Delta Y}{R_o}=0.15$ shows this increase in the deformation curve. We observed that in the case without the spike, where $\frac{\Delta Y}{R_o}=0.0$, the coalescence of the core droplets occurs shortly after the shell droplet coalesces. This timing absorbed the effect of core droplet coalescence on the shell droplet's deformation curve, as the newly formed shell droplet displays deformation rise to some extent by the time the core droplets coalesce.

\begin{figure}[H]
    \centering
    \begin{tabular}{cc}
        \includegraphics[scale = 0.18]{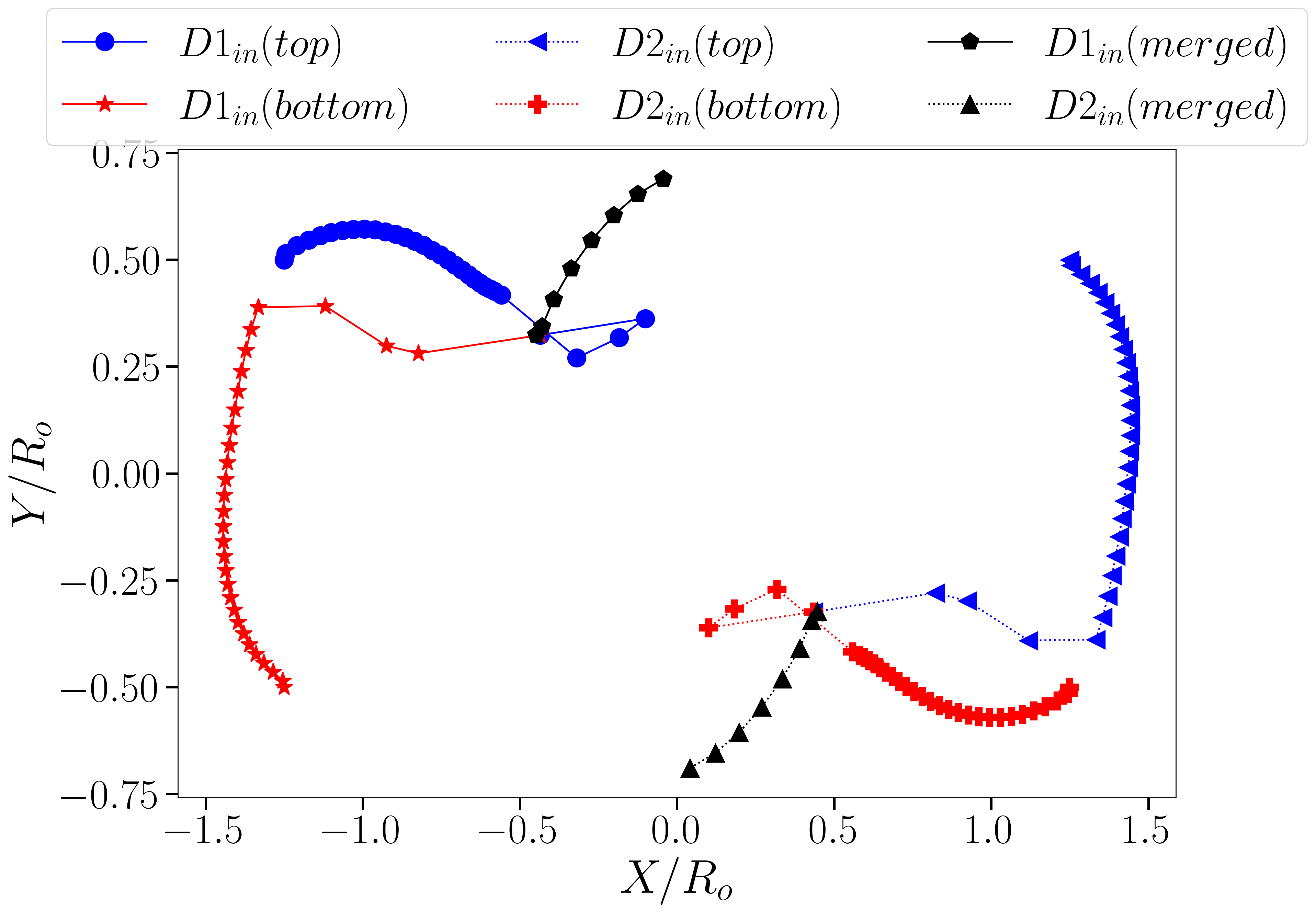} & \includegraphics[scale = 0.18]{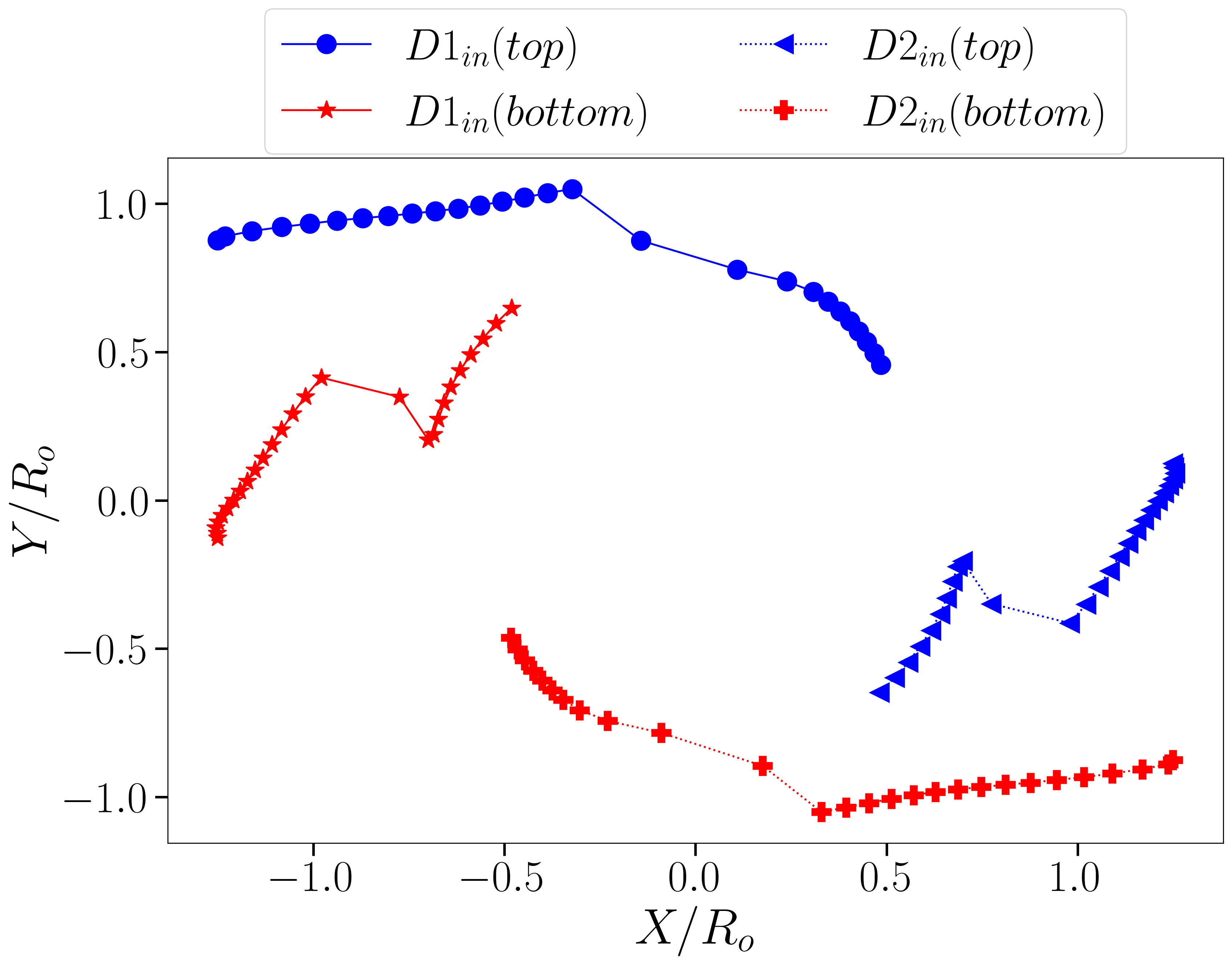} \\
        (a) $\Delta Y_o/R_o = 0.00$  &
        (d) $\Delta Y_o/R_o = 0.75$  \\
        \includegraphics[scale = 0.18]{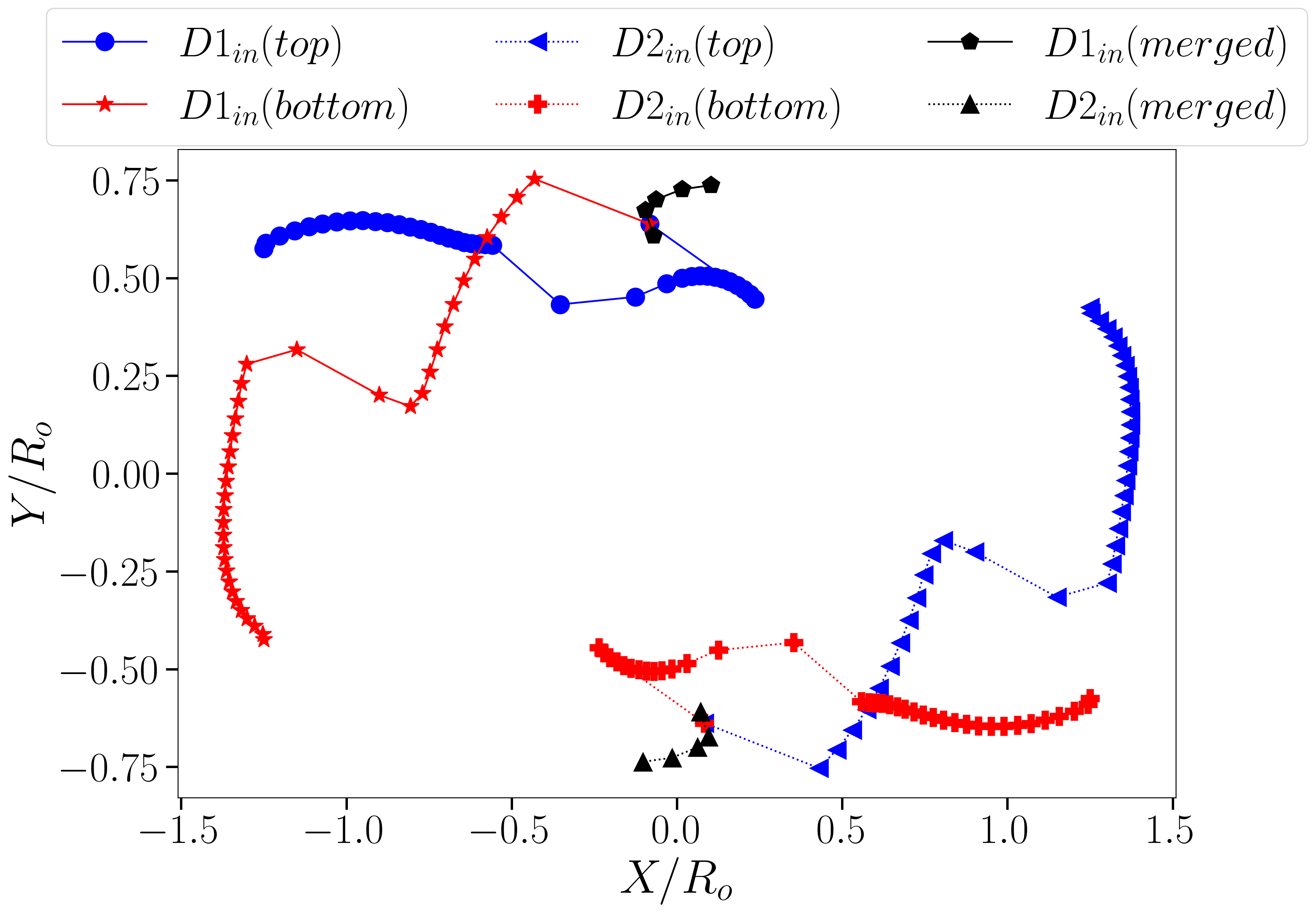} & \includegraphics[scale = 0.18]{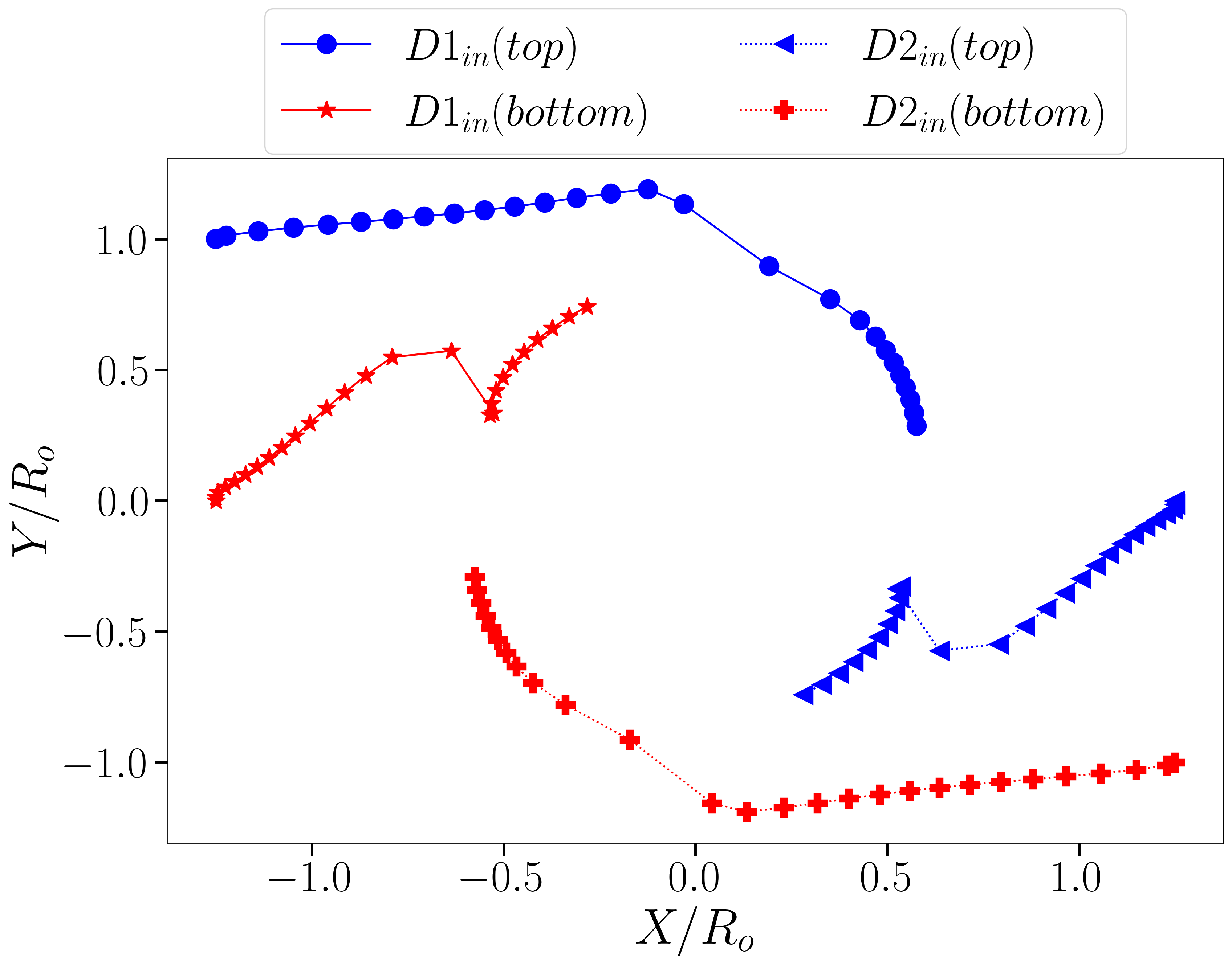} \\
        (b) $\Delta Y_o/R_o = 0.15$  &
        (e) $\Delta Y_o/R_o = 1.00$  \\
        \includegraphics[scale = 0.18]{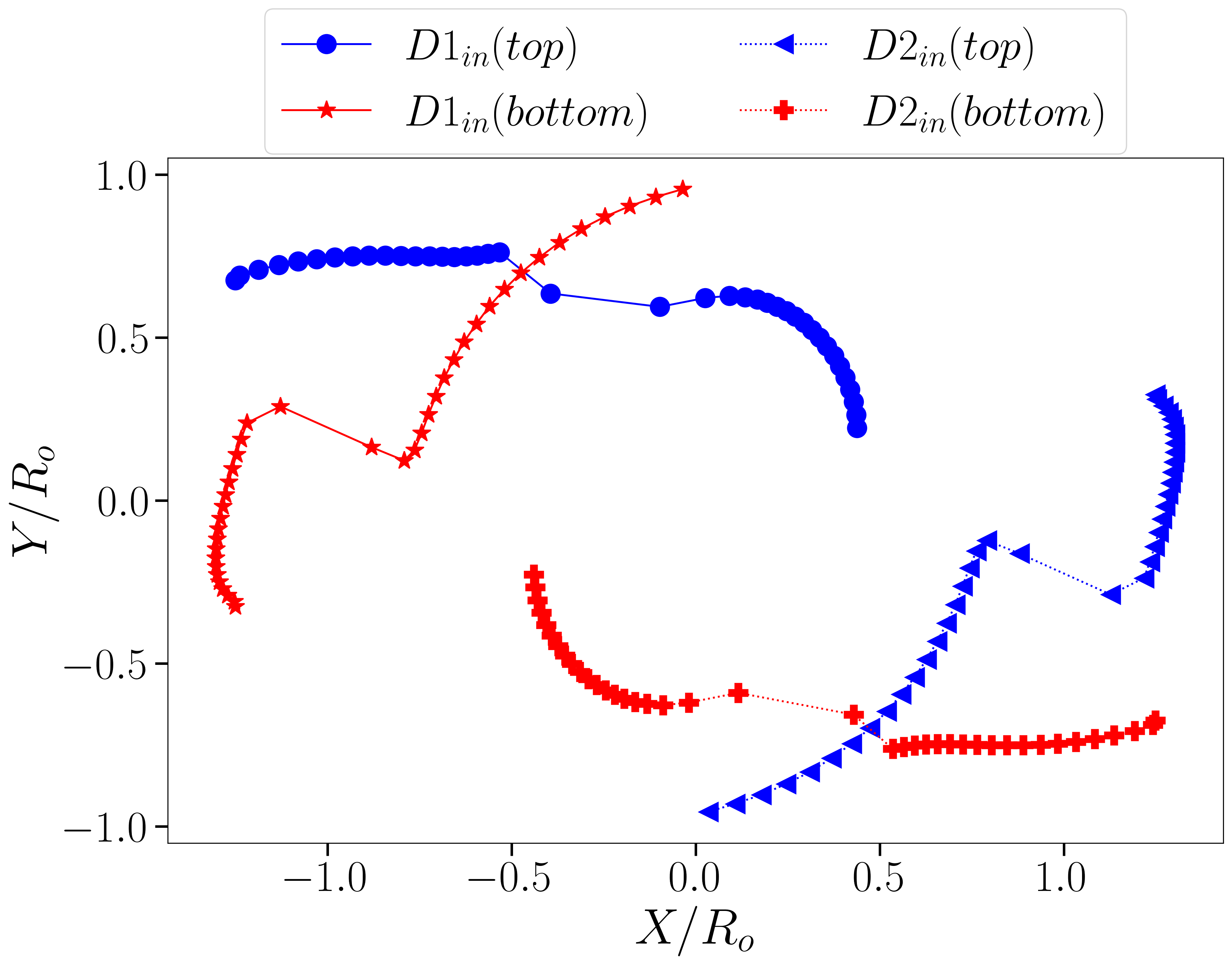} & \includegraphics[scale = 0.18]{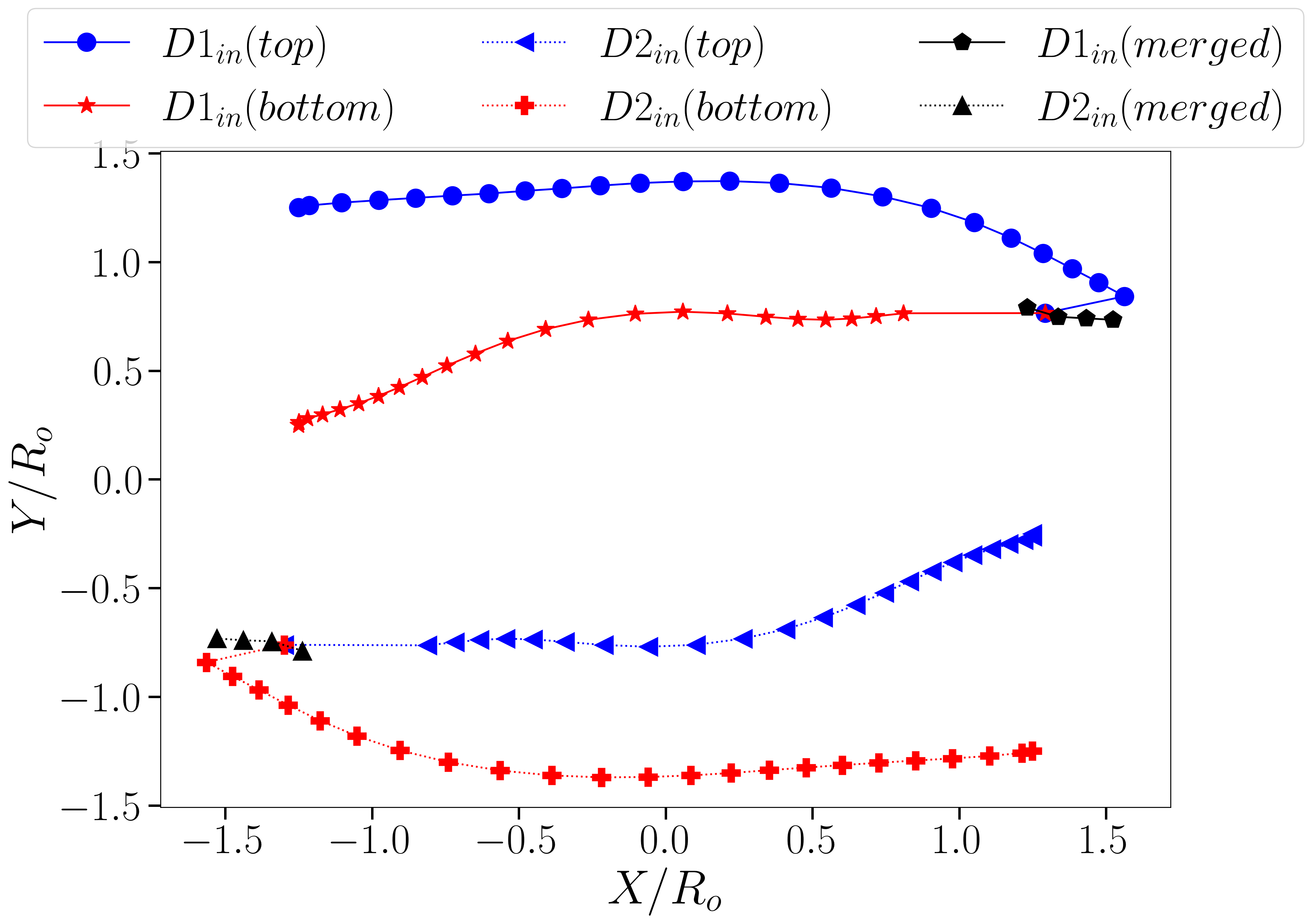} \\
        (c) $\Delta Y_o/R_o = 0.35$  &
        (f) $\Delta Y_o/R_o = 1.50$  \\
    \end{tabular}

    \caption{Trajectories of the core droplets ($D1_{in}(top)$, $D1_{in}(bottom)$, $D2_{in}(top)$ and $D2_{in}(bottom)$) for the six different initial vertical offset ($\frac{\Delta Y}{R_o}=0.00, 0.15, 0.35, 0.75, 1.00$, and $1.50$) in between shell droplets with $Ca=0.07$.}
    \label{fig:Double-core-Ca-0.07-varrying-offset-inner-trajectory}
\end{figure}

To observe the behavior of core droplets in terms of their movement, trajectory plots have been generated and are illustrated in Figure~\ref{fig:Double-core-Ca-0.07-varrying-offset-inner-trajectory}. Despite slight variations, the general trend for the core droplets is to rotate in a planetary-like motion in cases where the shell droplets coalesce (Figure~\ref{fig:Double-core-Ca-0.07-varrying-offset-inner-trajectory}(c-e)). Even when the core droplets also coalesce (Figure~\ref{fig:Double-core-Ca-0.07-varrying-offset-inner-trajectory}(a,b)), they continue to follow a circular path inside the coalesced shell droplet. This indicates that the core droplets maintain their rotational movement despite the merging of the outer shells. In the scenario where the shell droplets pass over each other but the core droplets inside each shell coalesce (Figure~\ref{fig:Double-core-Ca-0.07-varrying-offset-inner-trajectory}(f)), the coalesced core droplets remain positioned centrally within the shell droplet and move along with it. This central positioning highlights the balance of forces and the stability of the coalesced core droplet within the moving shell droplet.

\begin{figure}[H]
    \centering
    \begin{tabular}{cc}
        \includegraphics[scale = 0.17]{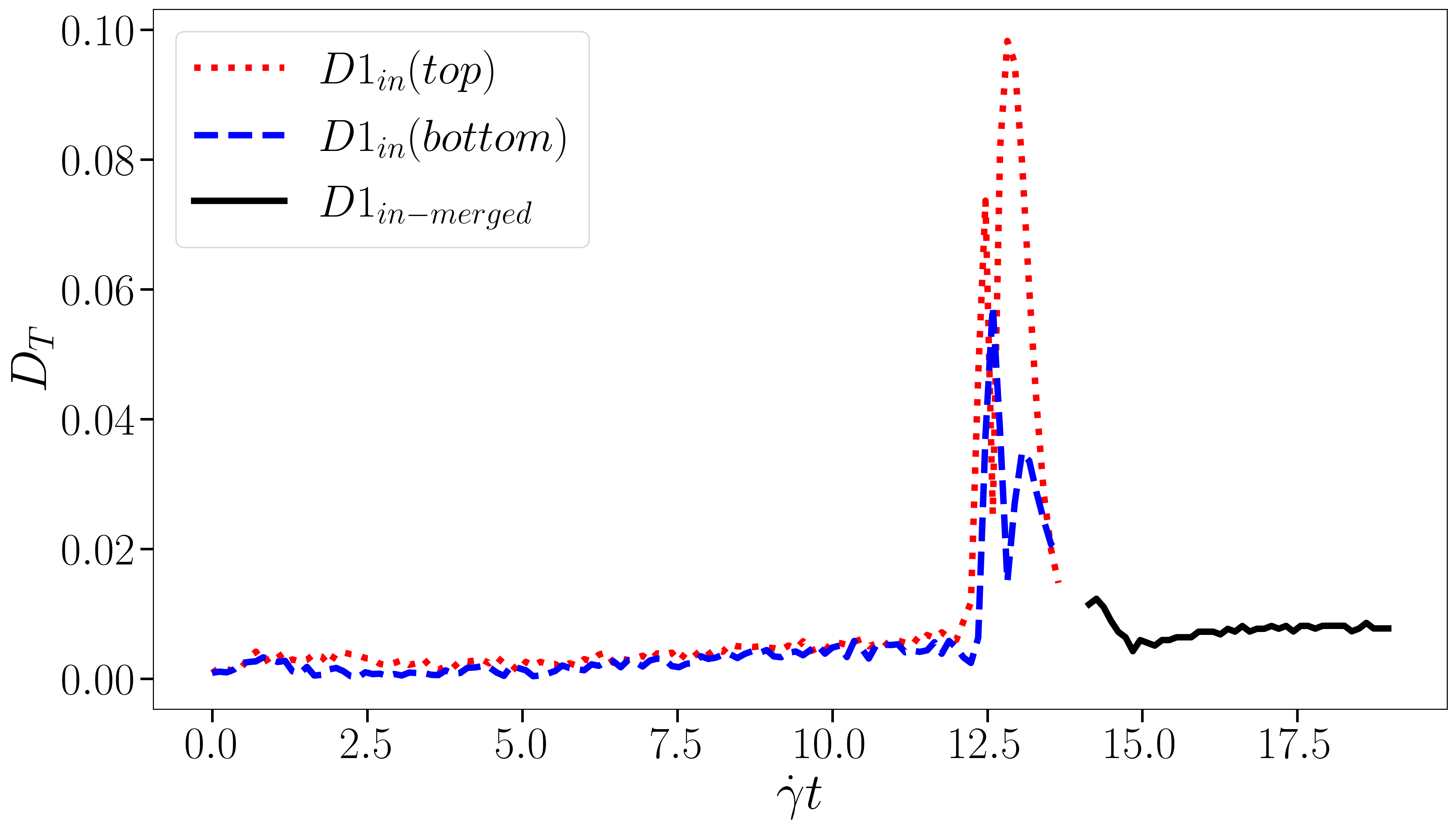} & \includegraphics[scale = 0.17]{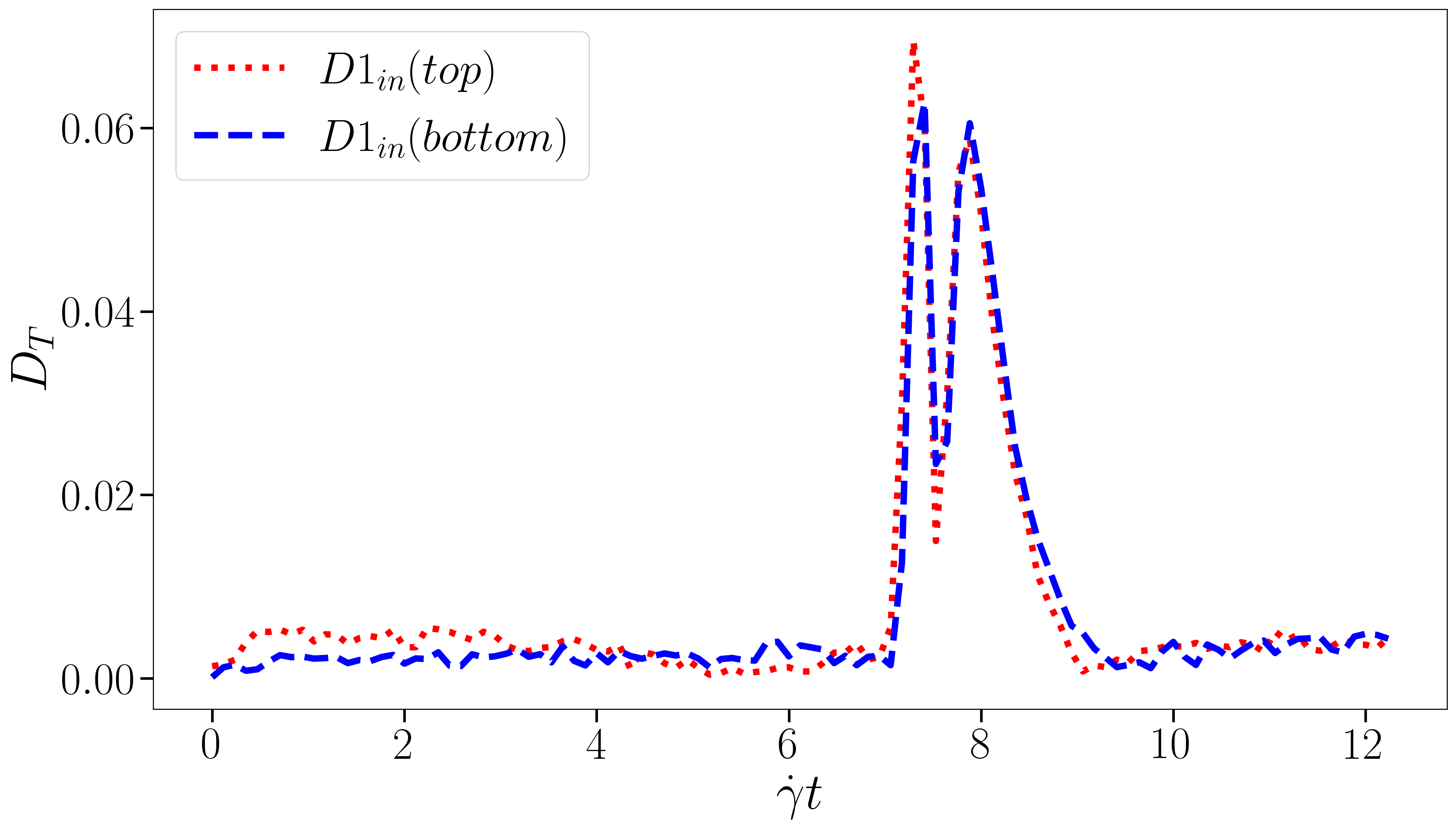} \\
        (a) $\Delta Y_o/R_o = 0.00$  &
        (d) $\Delta Y_o/R_o = 0.75$  \\
        \includegraphics[scale = 0.17]{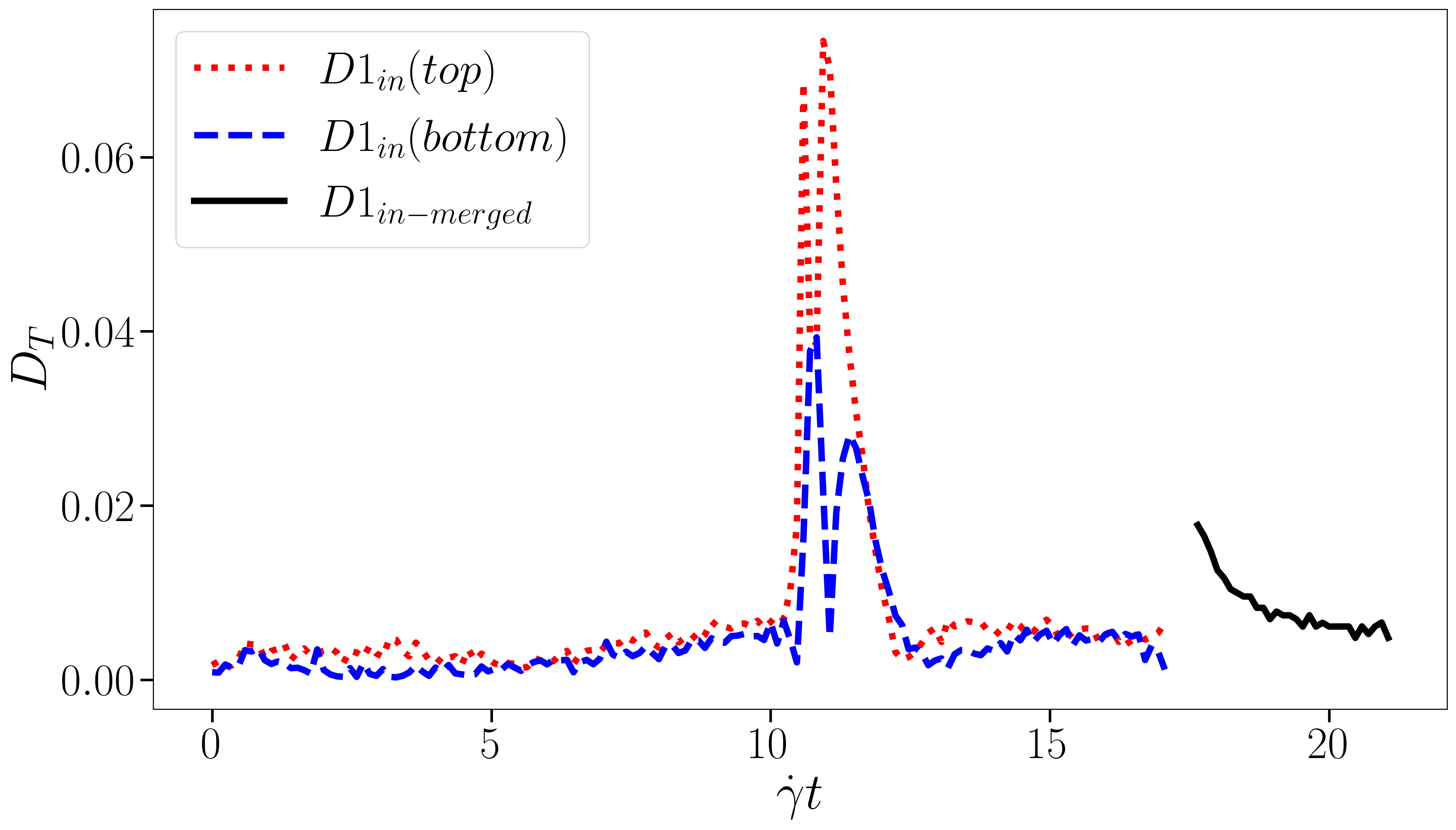} & \includegraphics[scale = 0.17]{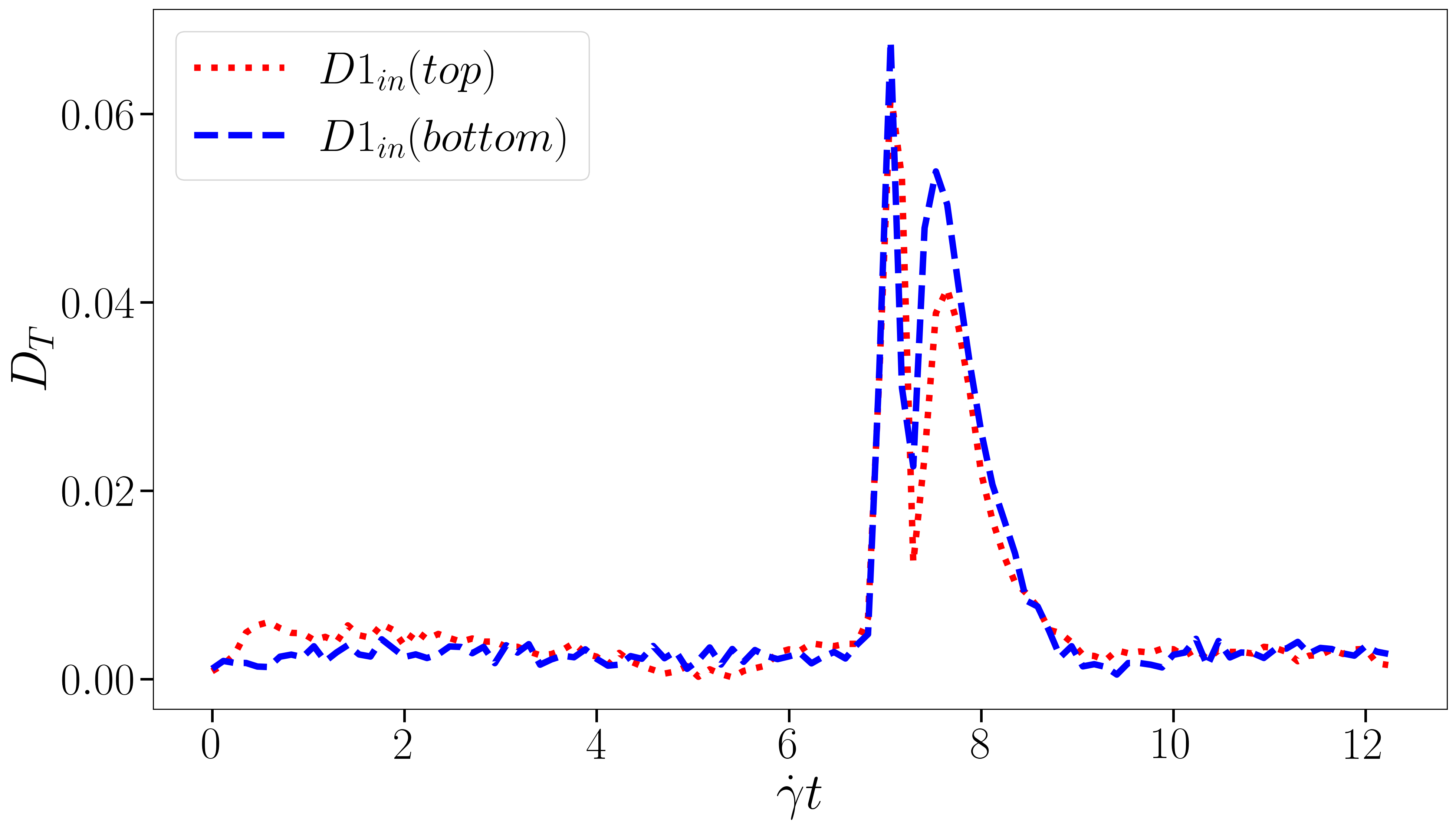} \\
        (b) $\Delta Y_o/R_o = 0.15$  &
        (e) $\Delta Y_o/R_o = 1.00$  \\
        \includegraphics[scale = 0.17]{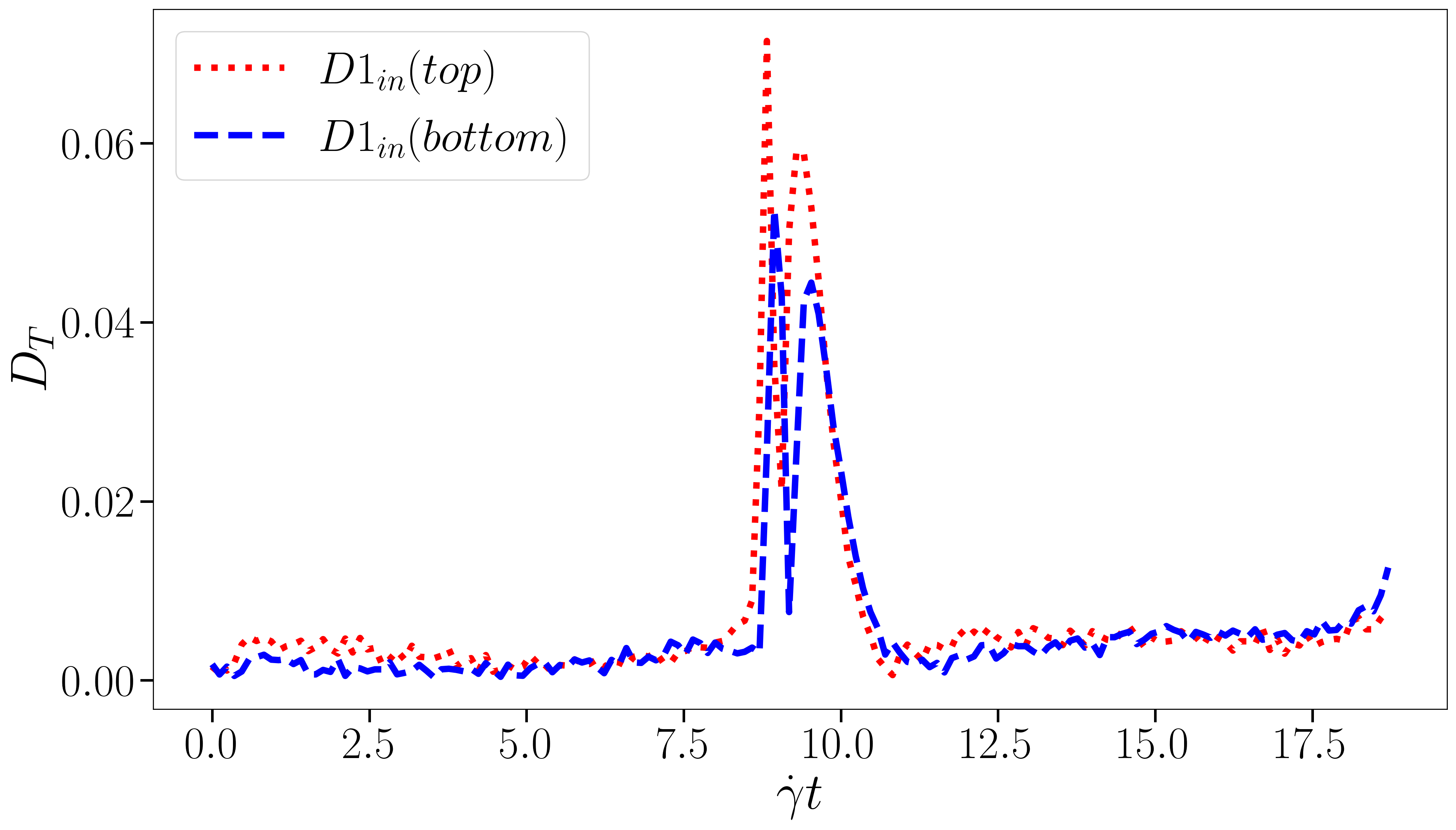} & \includegraphics[scale = 0.17]{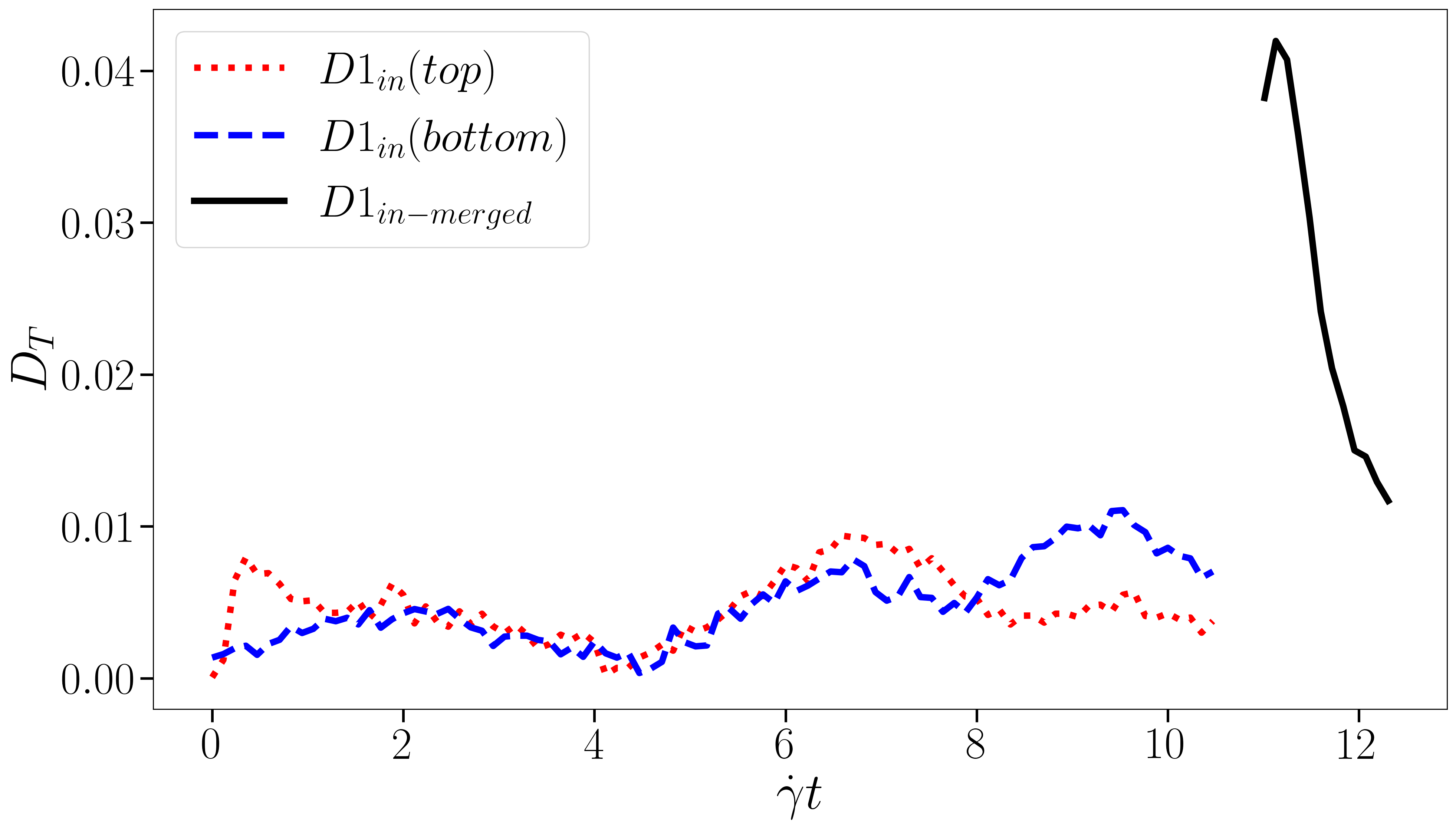} \\
        (c) $\Delta Y_o/R_o = 0.35$  &
        (f) $\Delta Y_o/R_o = 1.50$  \\
    \end{tabular}

    \caption{Deformation quantification of the core droplets ($D1_{in}(top)$, $D1_{in}(bottom)$, $D2_{in}(top)$ and $D2_{in}(bottom)$) for the six different initial vertical offset ($\frac{\Delta Y}{R_o}=0.00, 0.15, 0.35, 0.75, 1.00$, and $1.50$) in between shell droplets with $Ca=0.07$.}
    \label{fig:Double-core-Ca-0.07-varrying-offset-inner-deform}
\end{figure}

Examining the deformation plots for the core droplets, as displayed in Figure~\ref{fig:Double-core-Ca-0.07-varrying-offset-inner-deform}, a couple of key features has been found. One common characteristic observed across most cases is the presence of two spikes in the deformation curves. The first spike occurs during the collision of the shell droplets, coinciding with the rise in deformation of the shell droplets themselves. 

This initial spike highlights the immediate impact of the collision on the core droplets. The second spike in the deformation curve of the core droplets appears during the coalescence phase of the shell droplets. This spike indicates the time when the shell droplets experience fluctuating deformation as they merge. This phase introduces additional stress on the core droplets, reflected in their increased deformation. For cases where the core droplets coalesce inside the shell droplets (Figure~\ref{fig:Double-core-Ca-0.07-varrying-offset-inner-deform}(a,b)), an additional increase in deformation is observed. This spike occurs as the core droplets merge, leading to a temporary instability in their shape. However, this instability is short-lived, as the deformation of the merged core droplets soon stabilizes and they attain a steady shape.
\subsubsection{$Ca = 0.05$}

\begin{figure}[H]
          \centering

         \includegraphics[scale = 0.09]{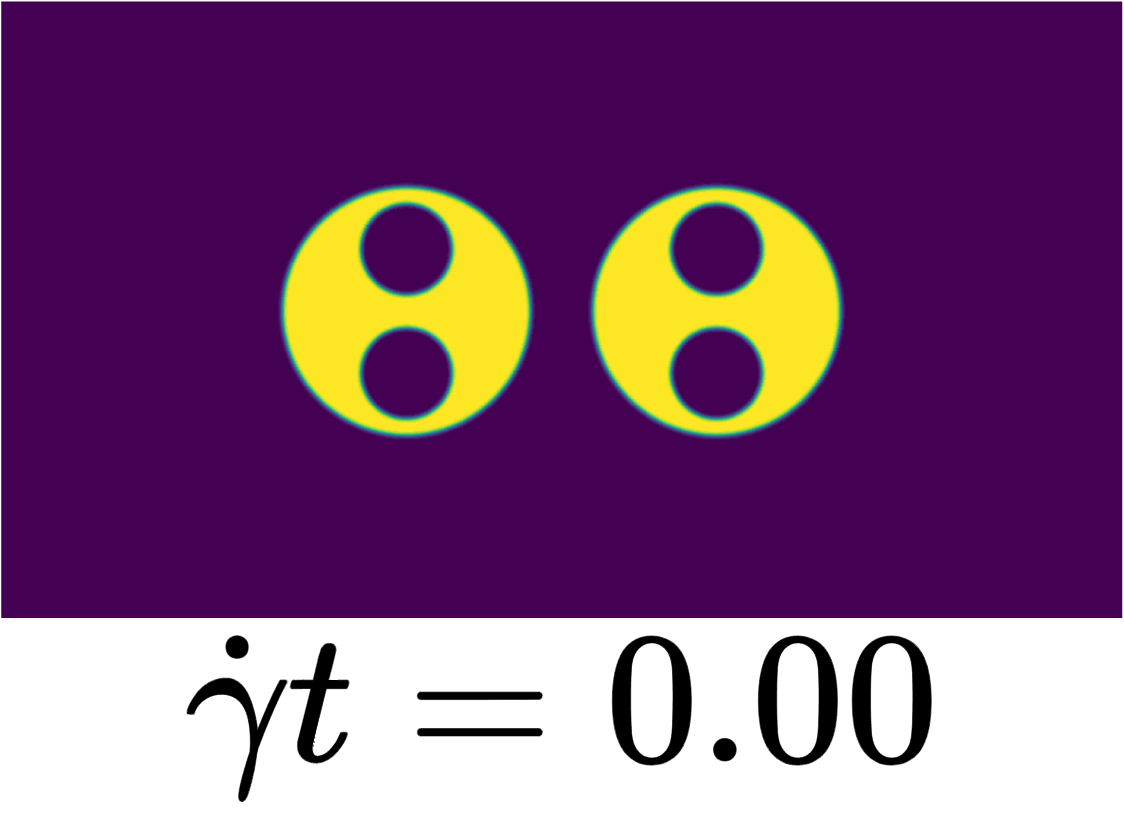}
		\includegraphics[scale = 0.09]{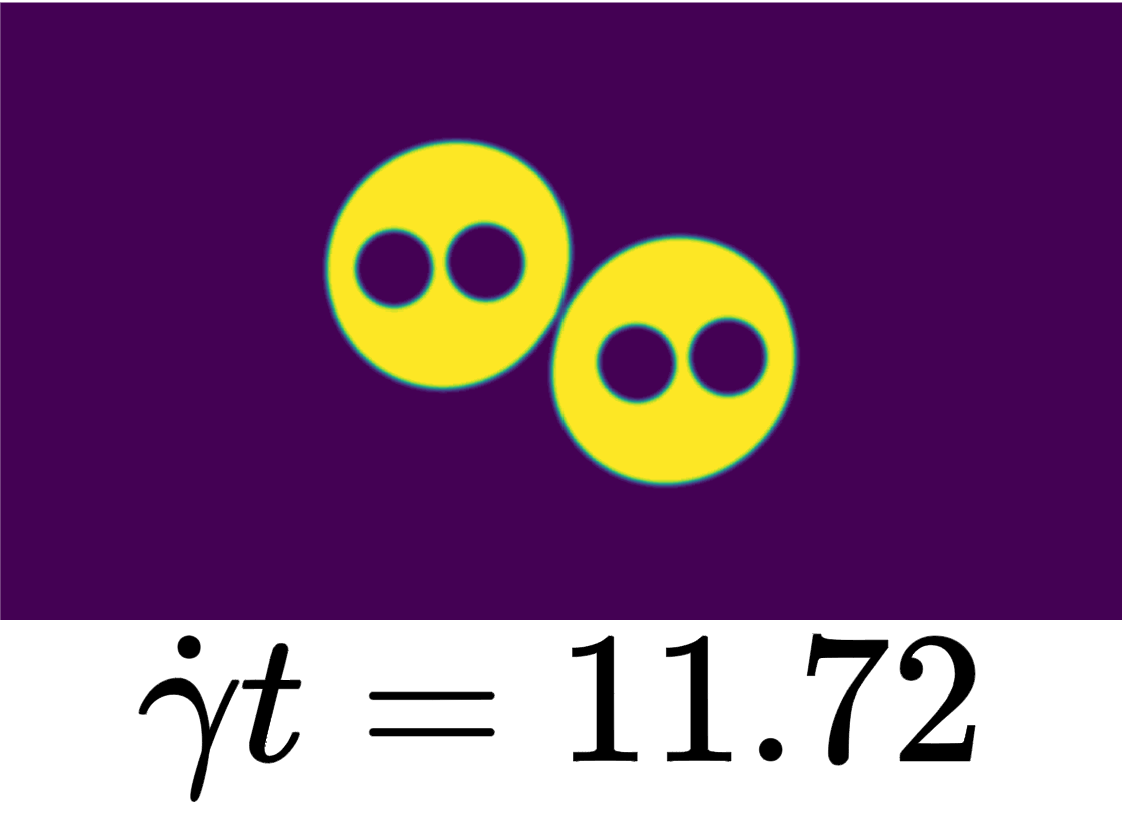}
		\includegraphics[scale = 0.09]{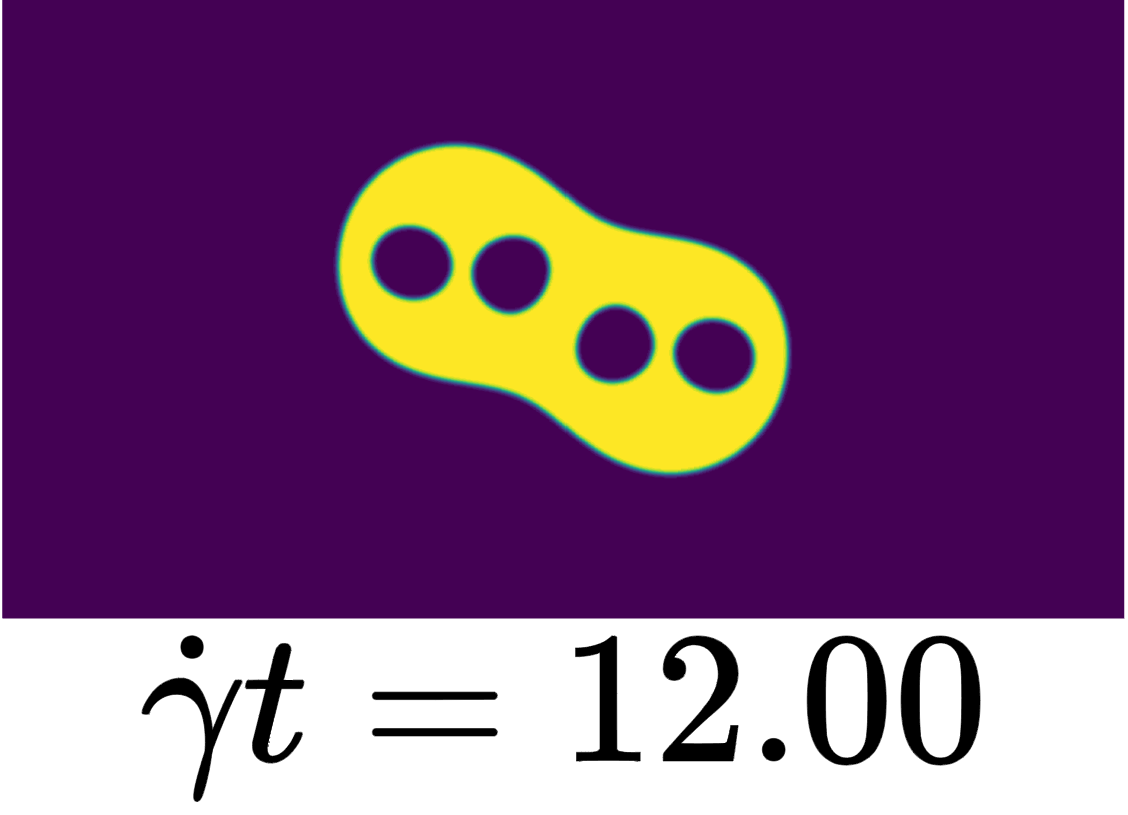}
		\includegraphics[scale = 0.09]{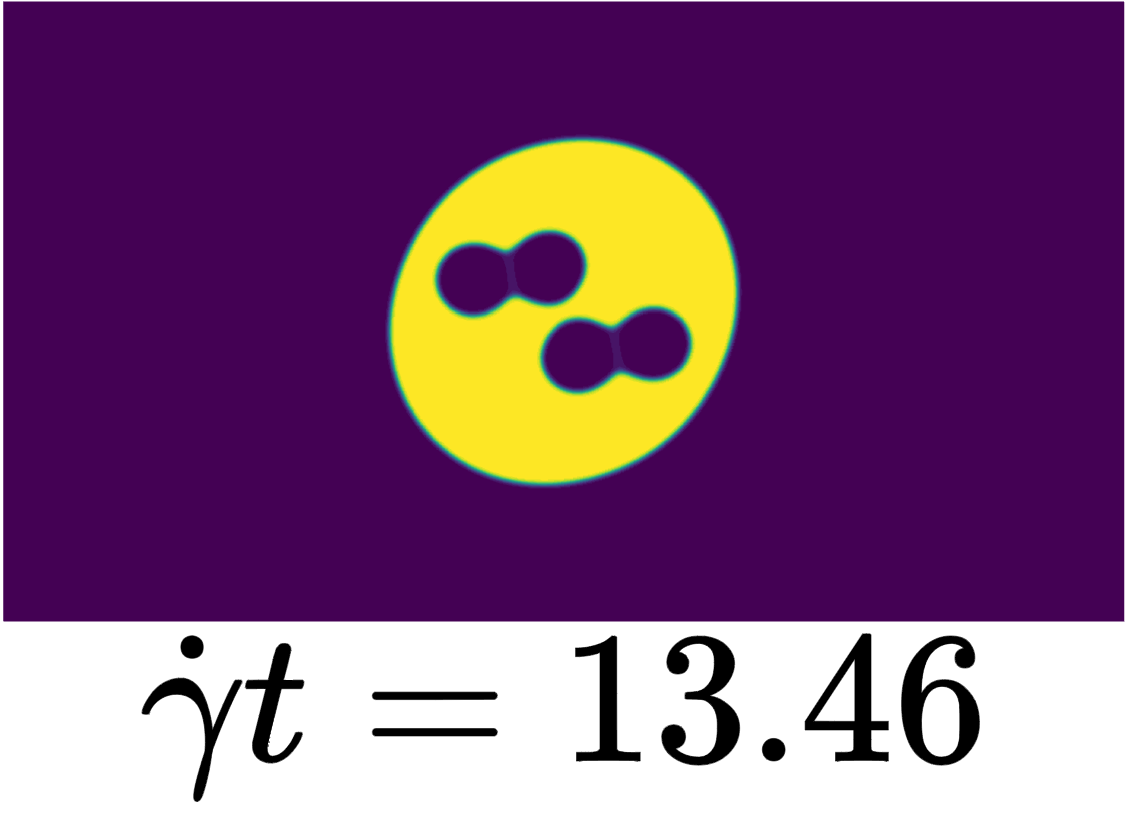}
		\includegraphics[scale = 0.09]{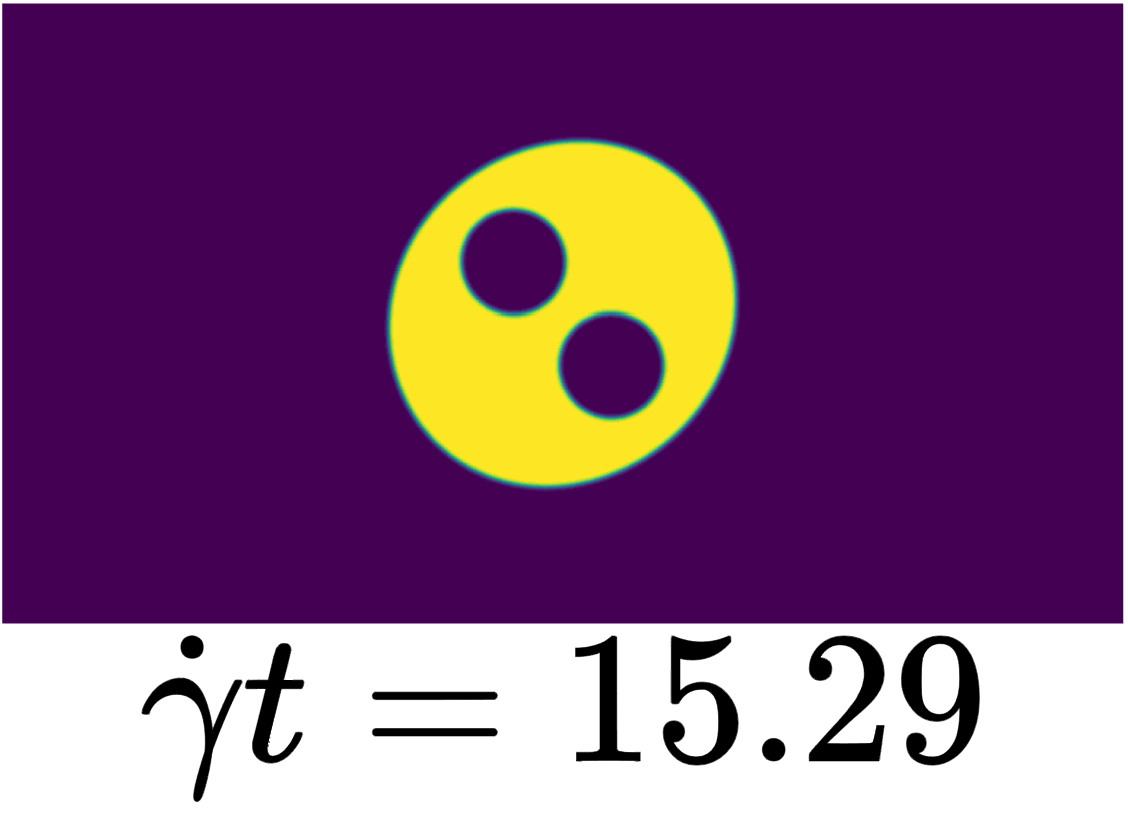} \\
        (a) $\frac{\Delta Y}{R_o}=0.00$ ~(Shell droplet coalesce, core droplet coalesce)
		\\ \vspace{0.5cm}

        \includegraphics[scale = 0.09]{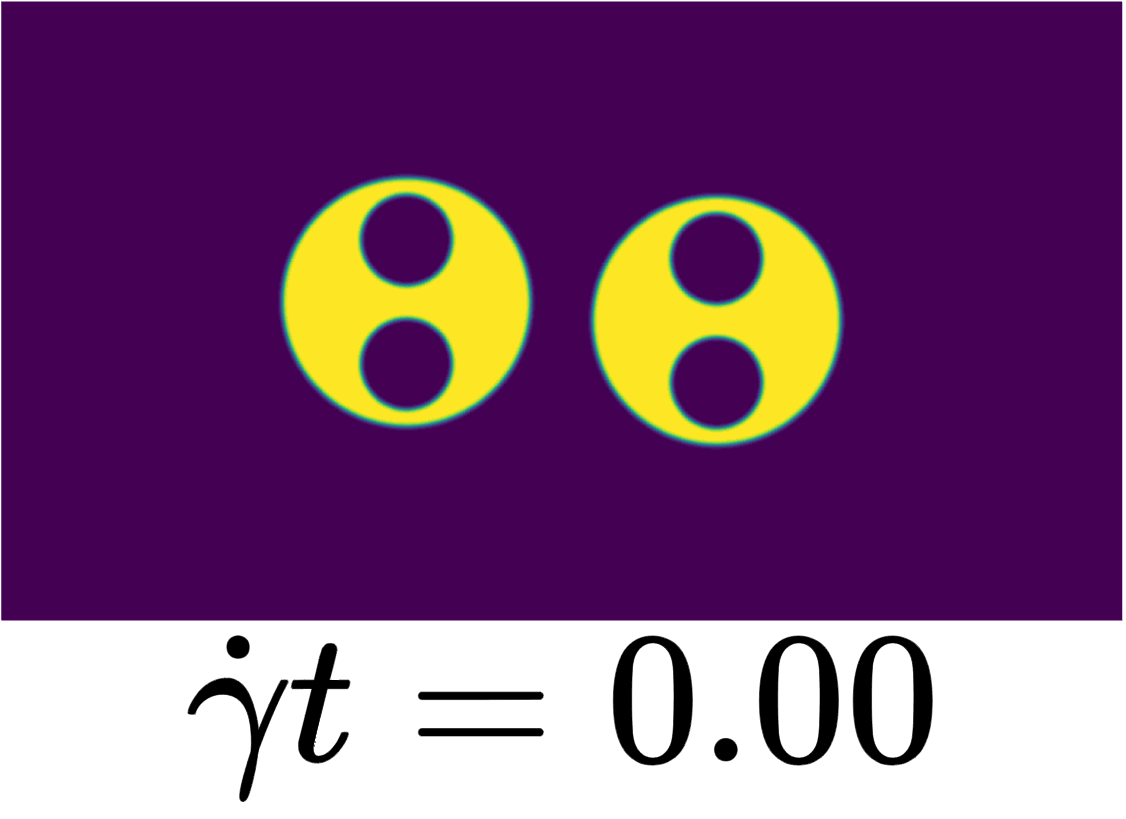}
            \includegraphics[scale = 0.09]{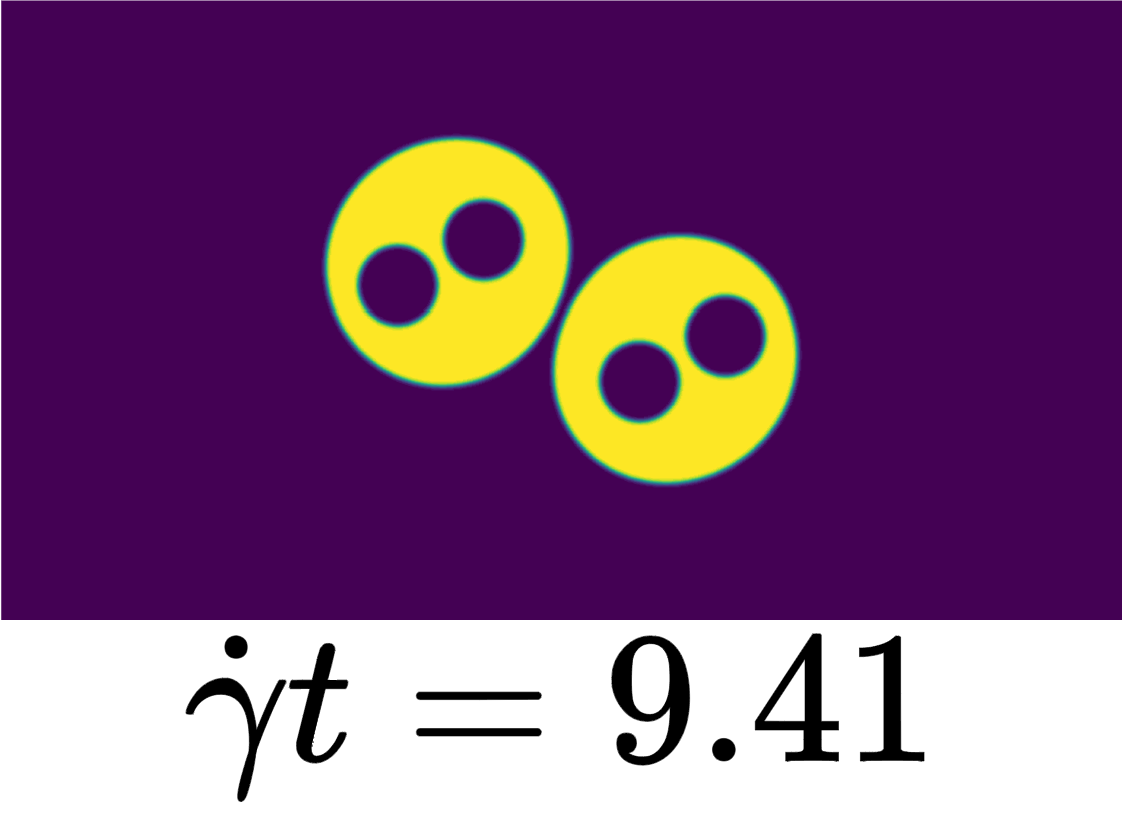}
		\includegraphics[scale = 0.09]{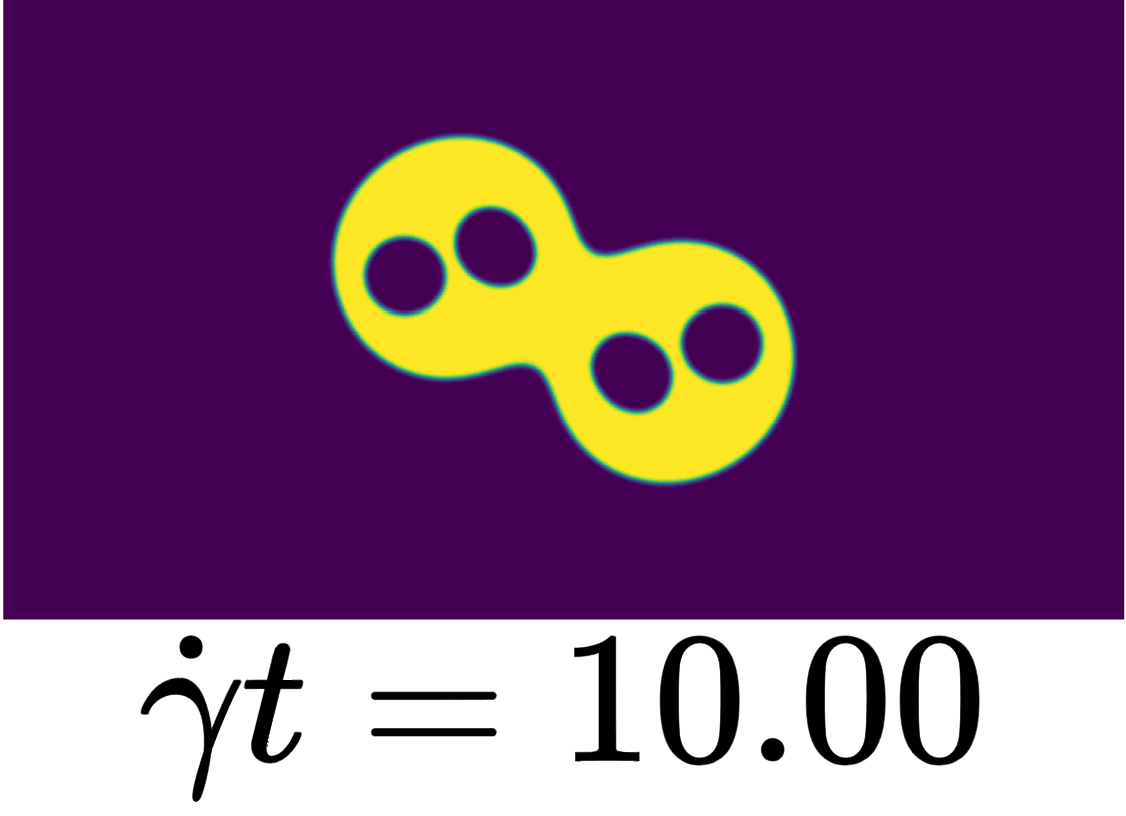}
		\includegraphics[scale = 0.09]{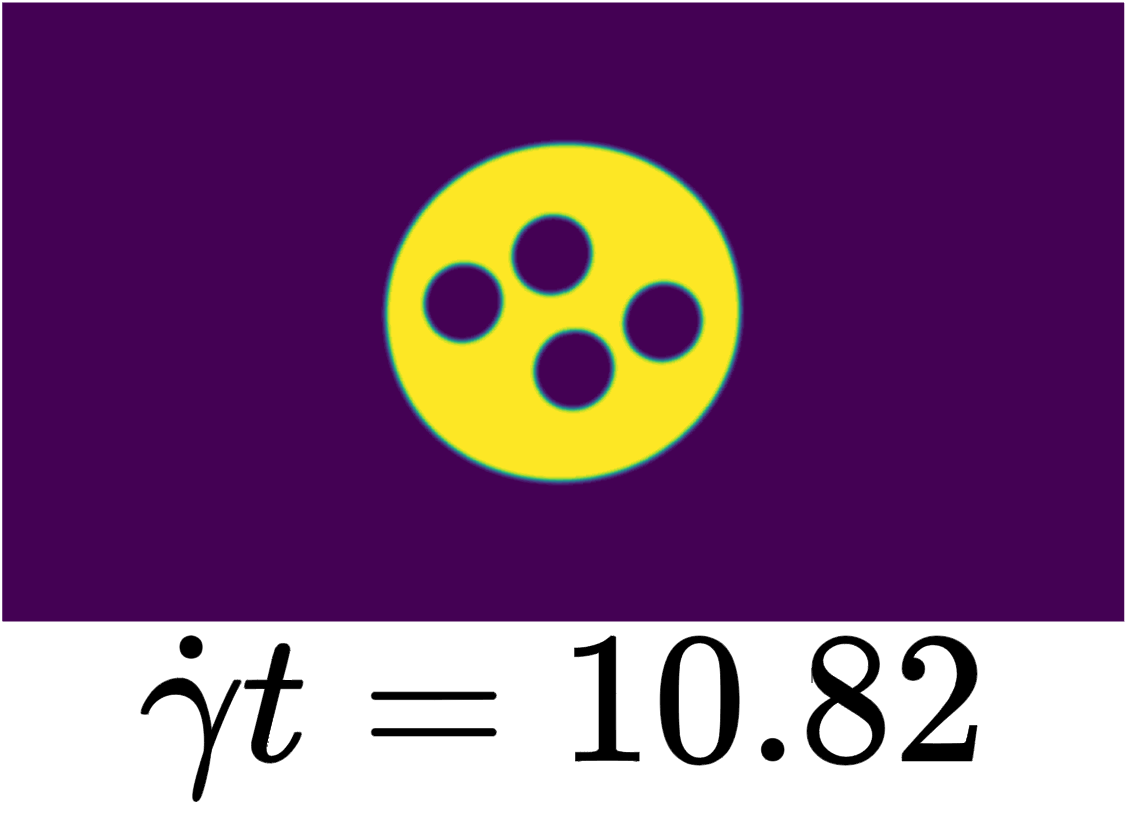}
		\includegraphics[scale = 0.09]{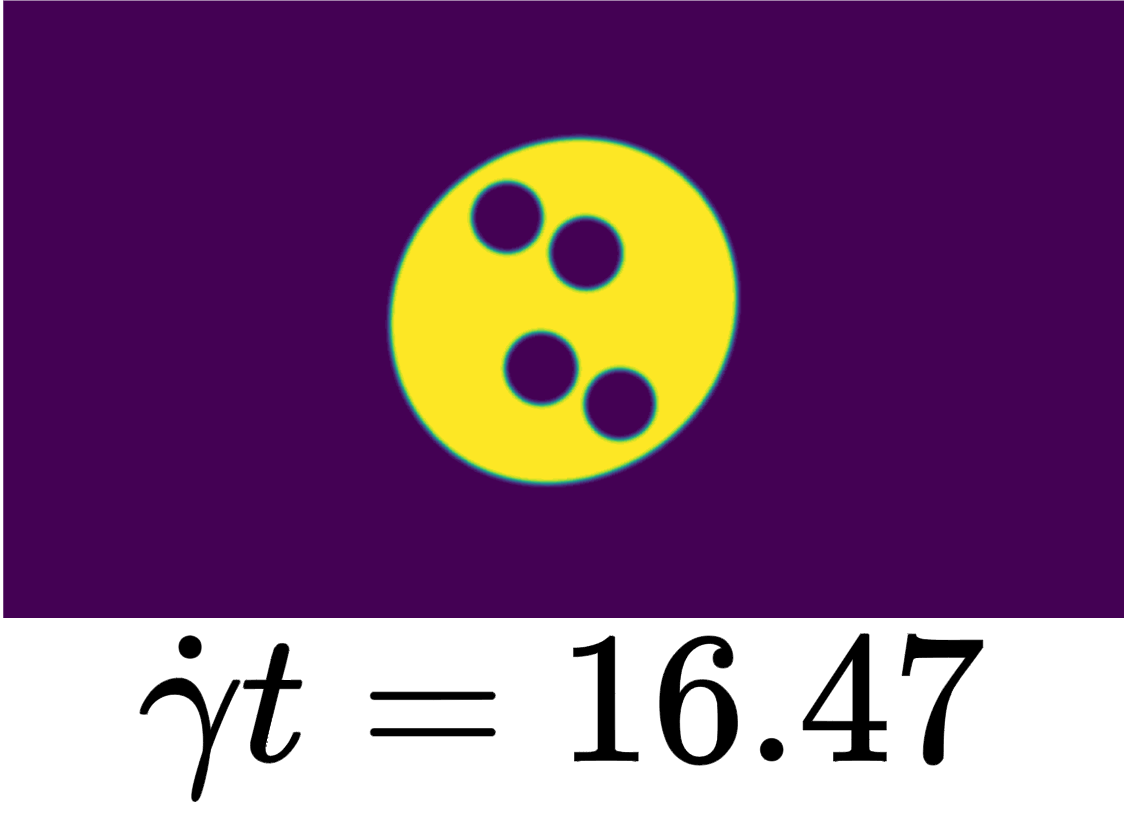}  \\
       (b)  $\frac{\Delta Y}{R_o}=0.15$~(Only shell droplet coalesce)
  \vspace{0.5cm}

   \includegraphics[scale = 0.09]{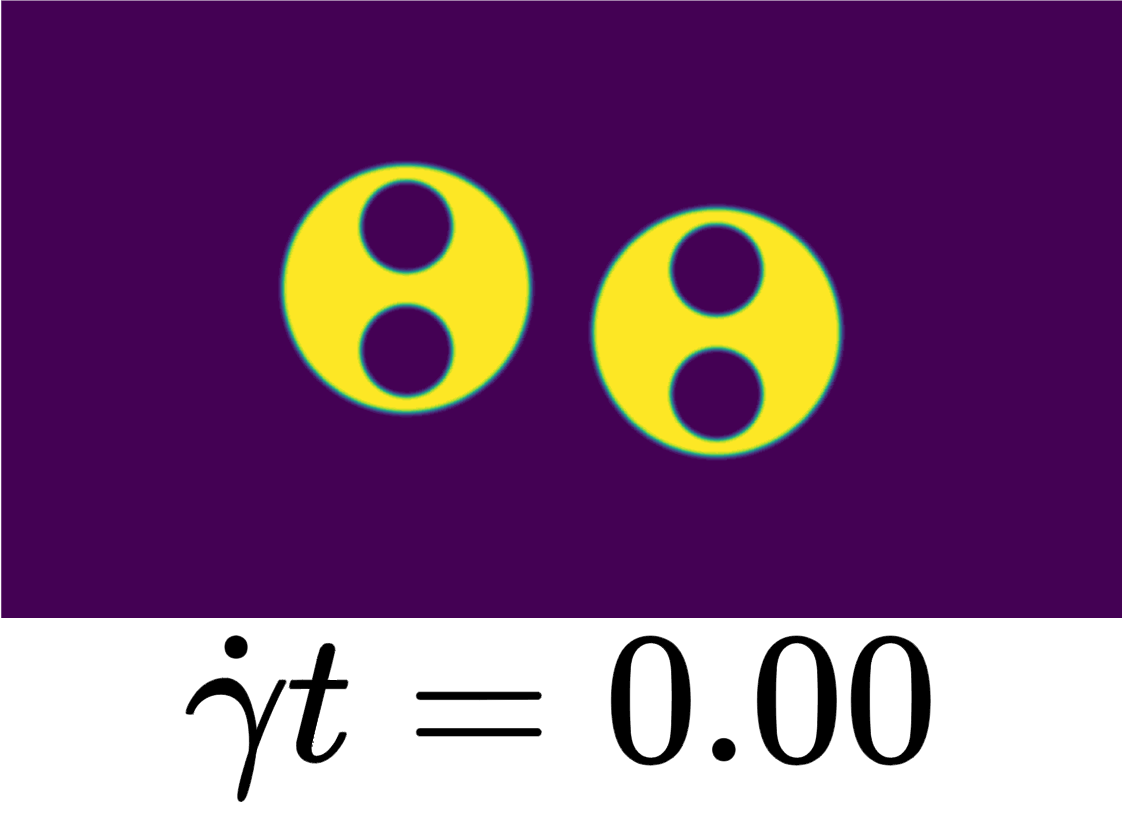}
		\includegraphics[scale = 0.09]{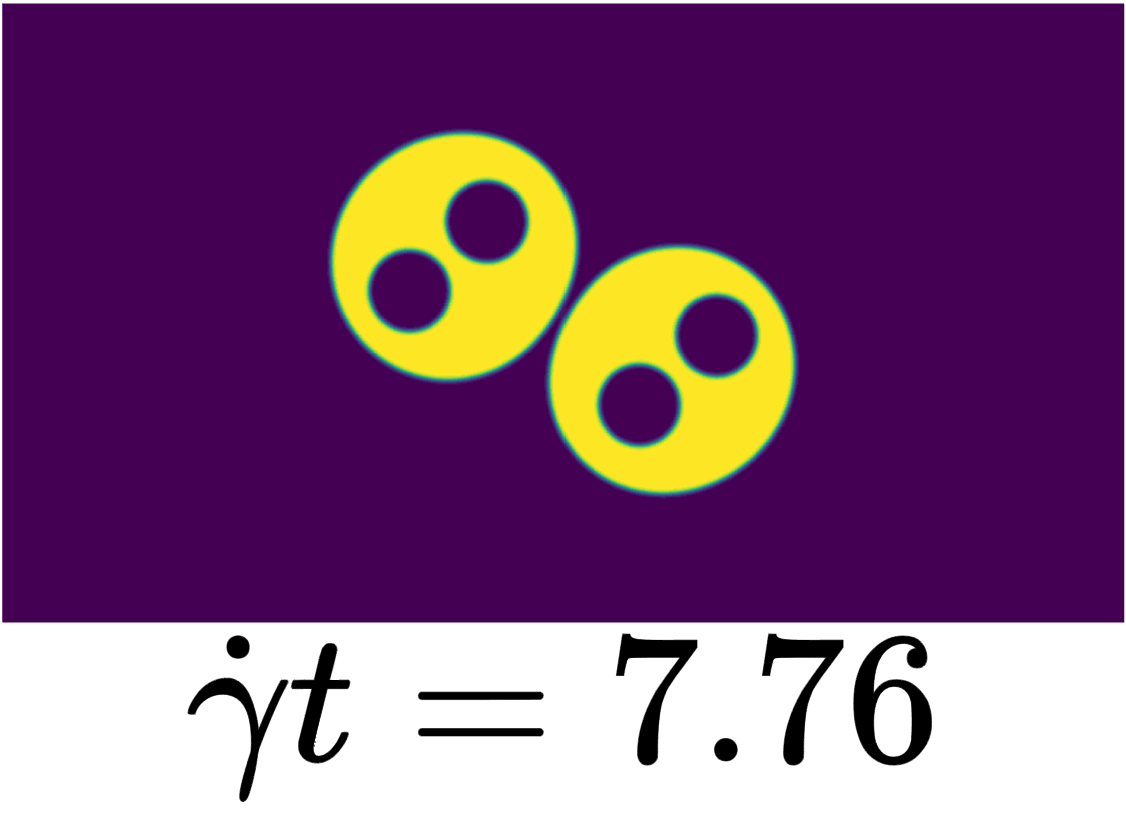}
		\includegraphics[scale = 0.09]{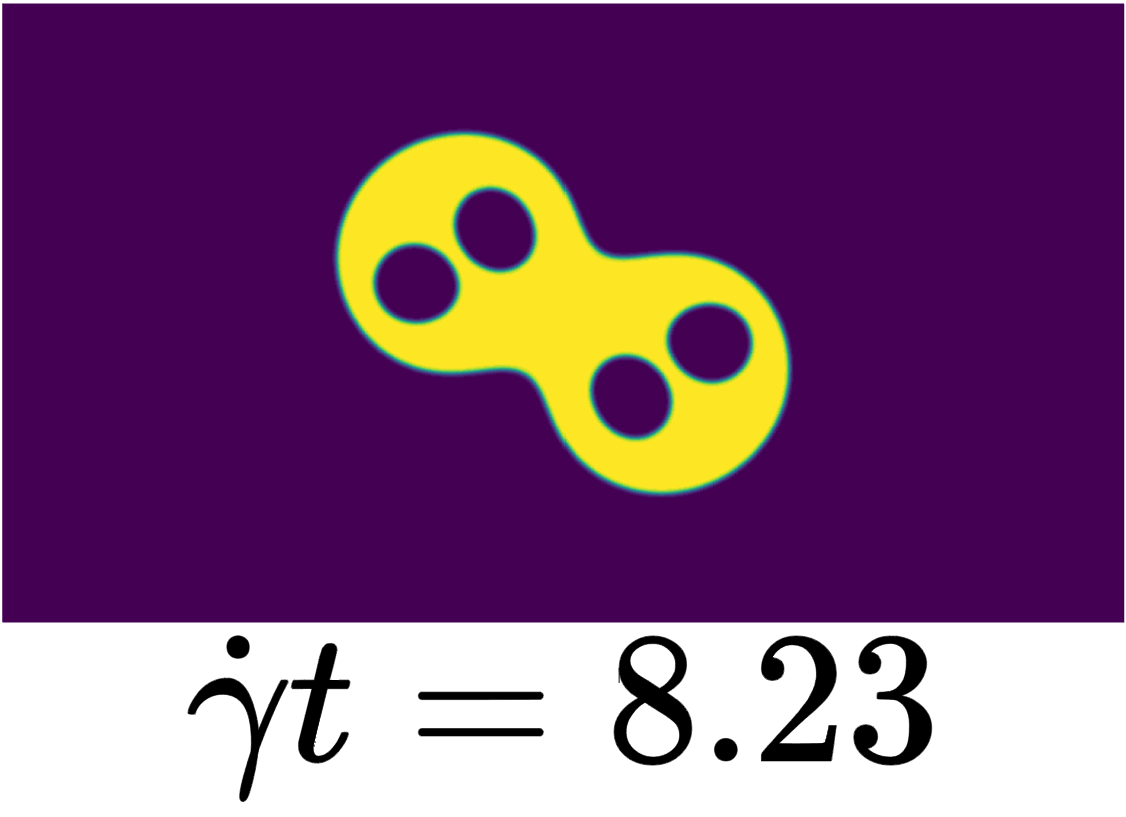}
		\includegraphics[scale = 0.09]{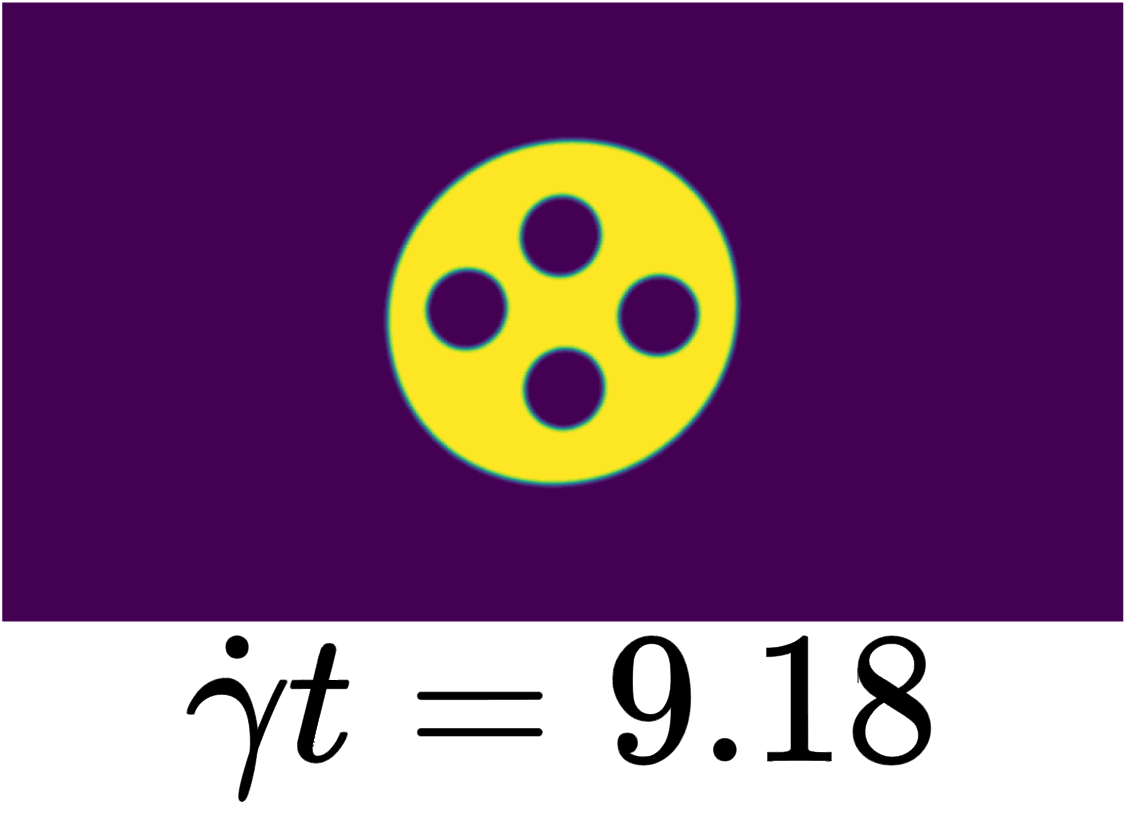}
		\includegraphics[scale = 0.09]{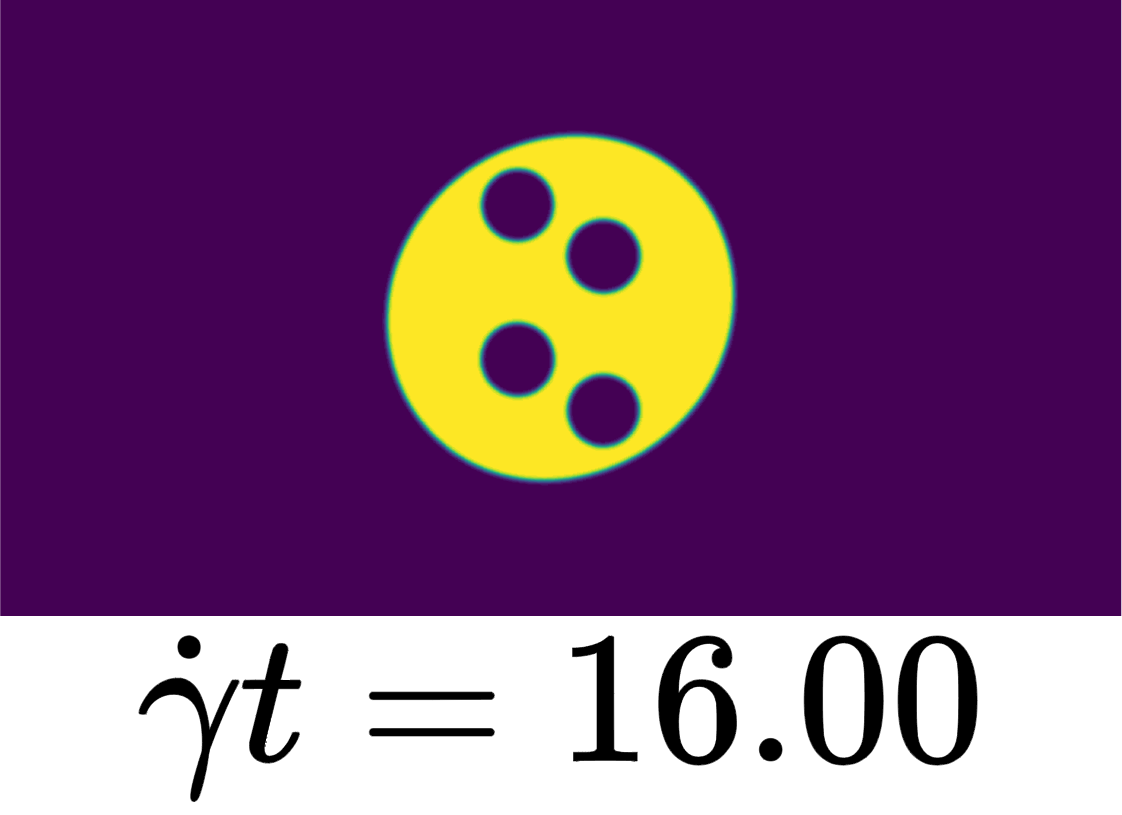} \\
        (c) $\frac{\Delta Y}{R_o}=0.35$~(Only shell droplet coalesce)
		\\ \vspace{0.5cm}

  \includegraphics[scale = 0.09]{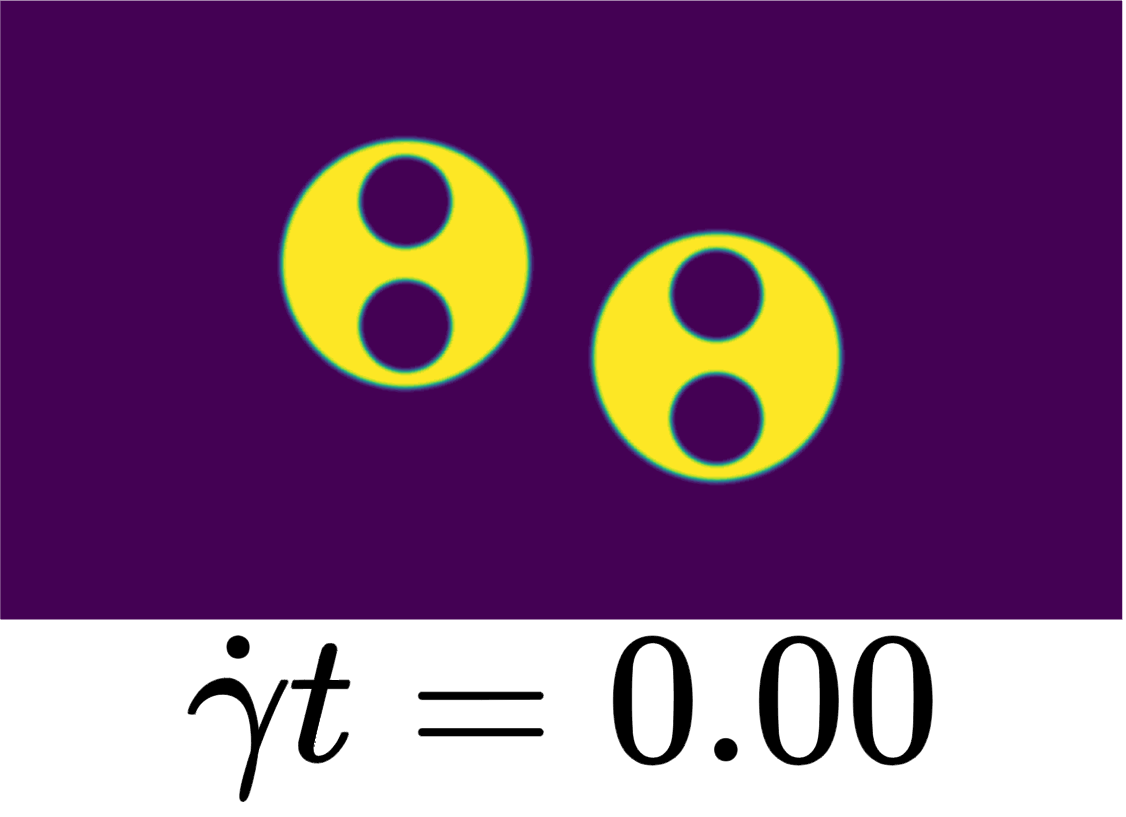}
		\includegraphics[scale = 0.09]{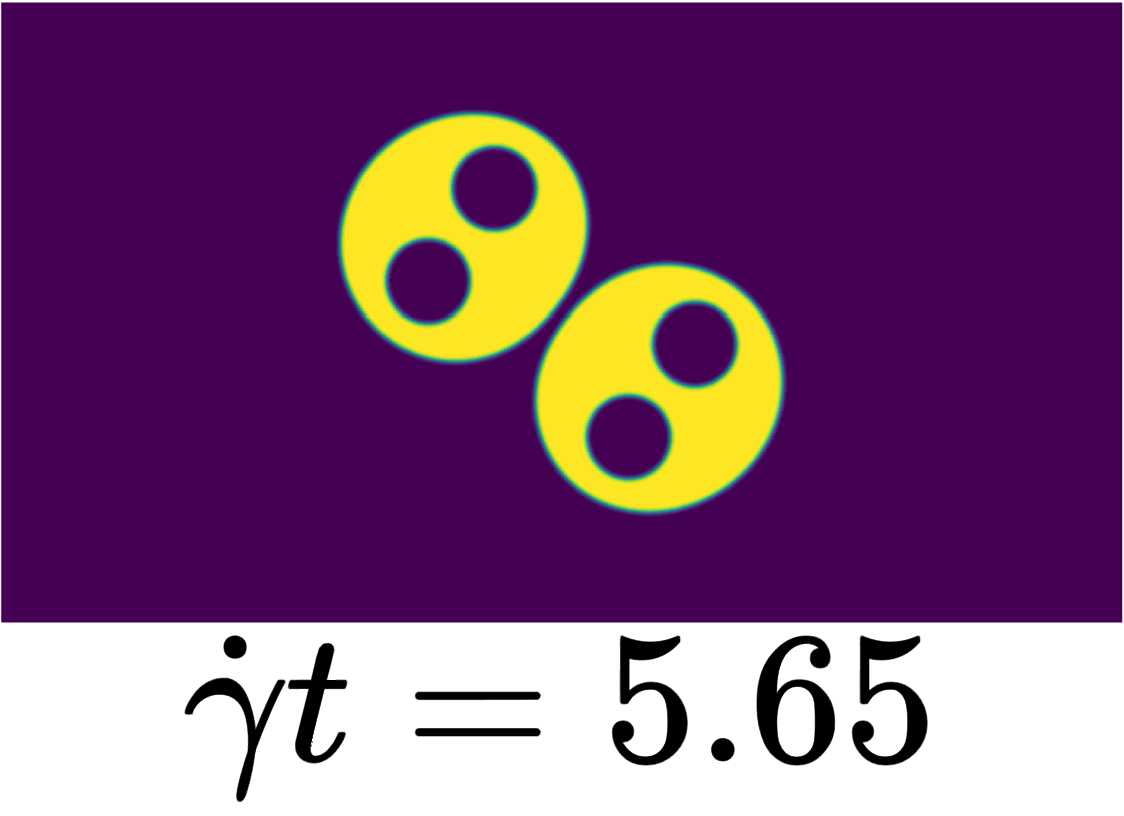}
		\includegraphics[scale = 0.09]{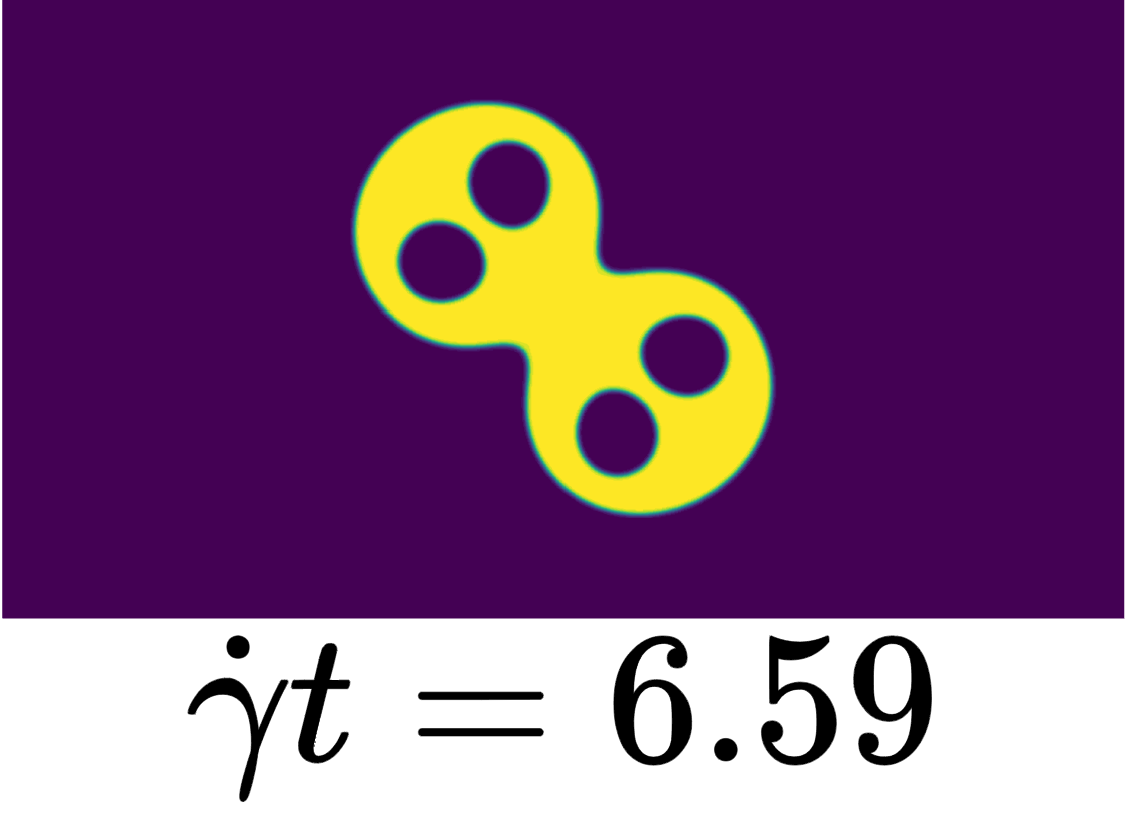}
		\includegraphics[scale = 0.09]{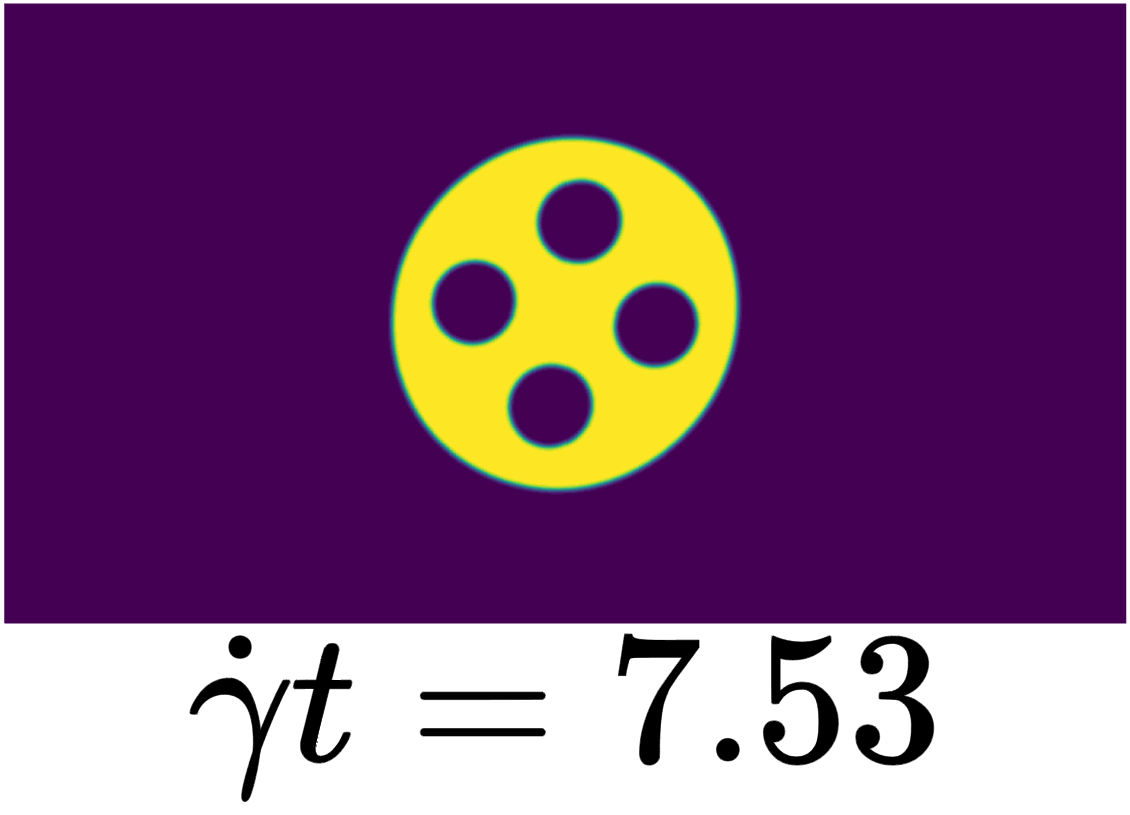}
		\includegraphics[scale = 0.09]{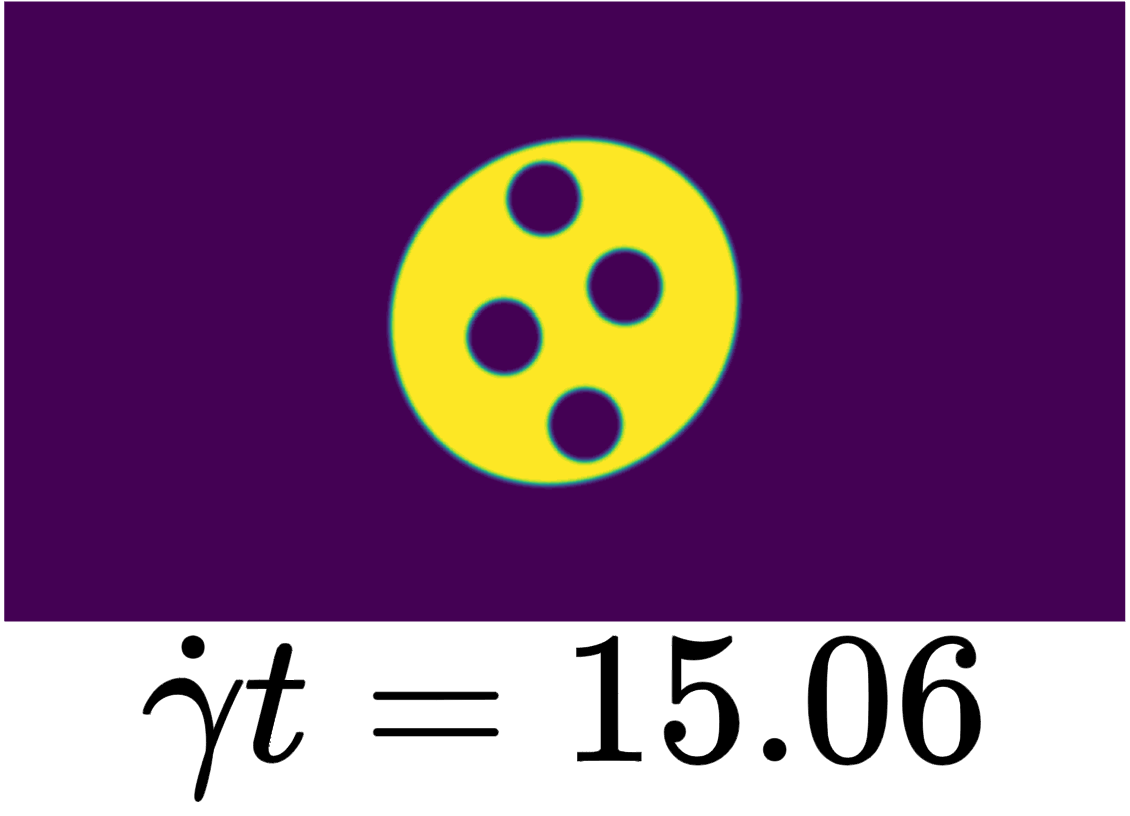} \\
        (d) $\frac{\Delta Y}{R_o}=0.75$~(Only shell droplet coalesce)
		\\ \vspace{0.5cm}

  \includegraphics[scale = 0.09]{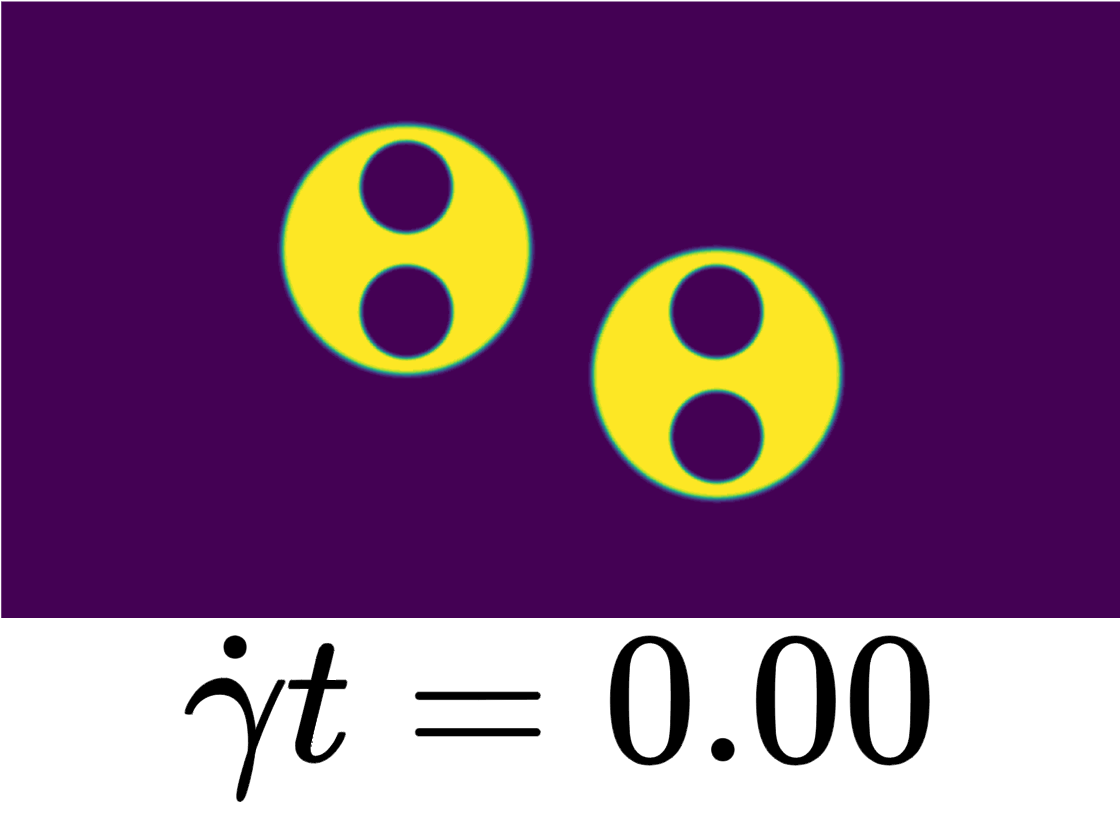}
		\includegraphics[scale = 0.09]{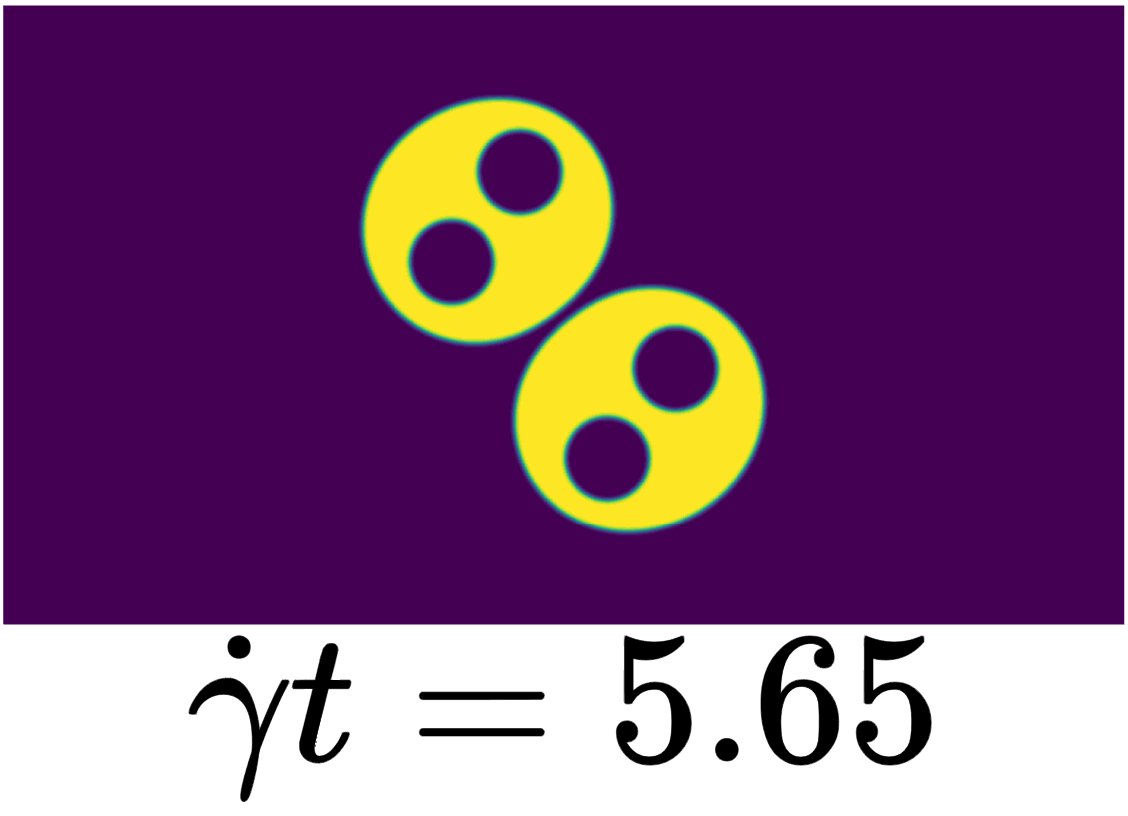}
		\includegraphics[scale = 0.09]{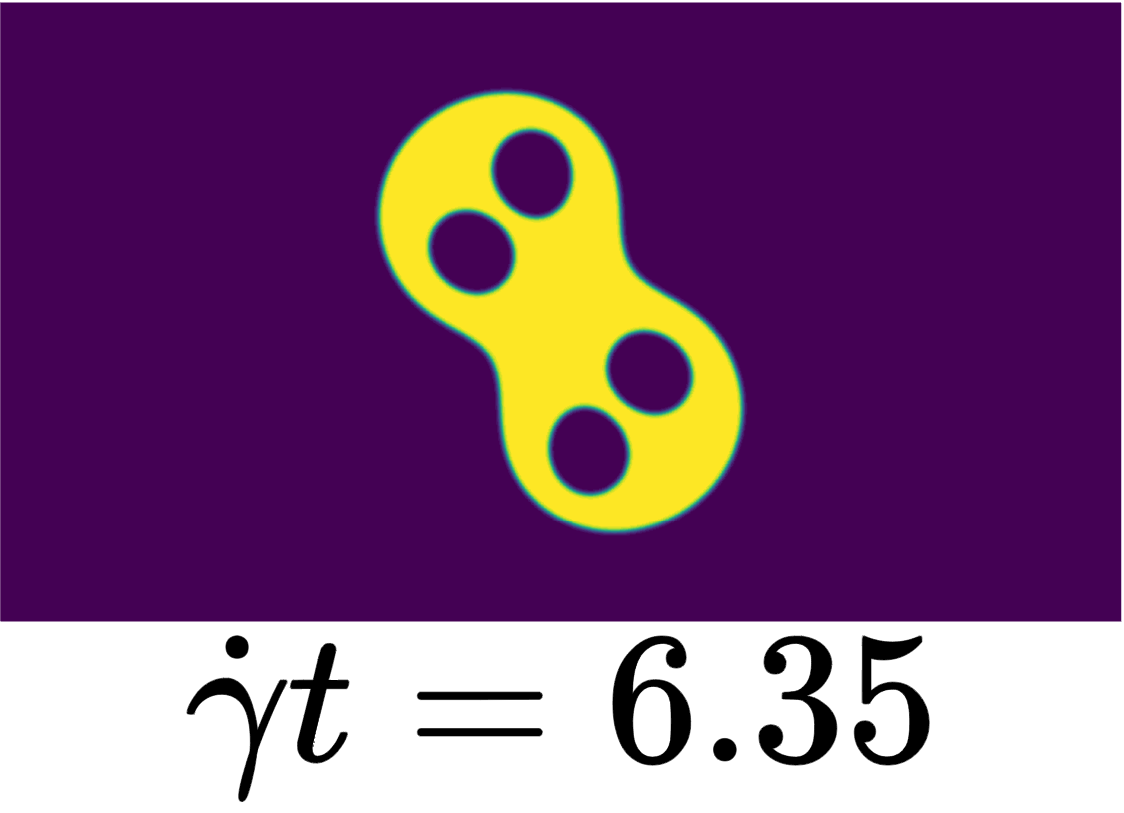}
		\includegraphics[scale = 0.09]{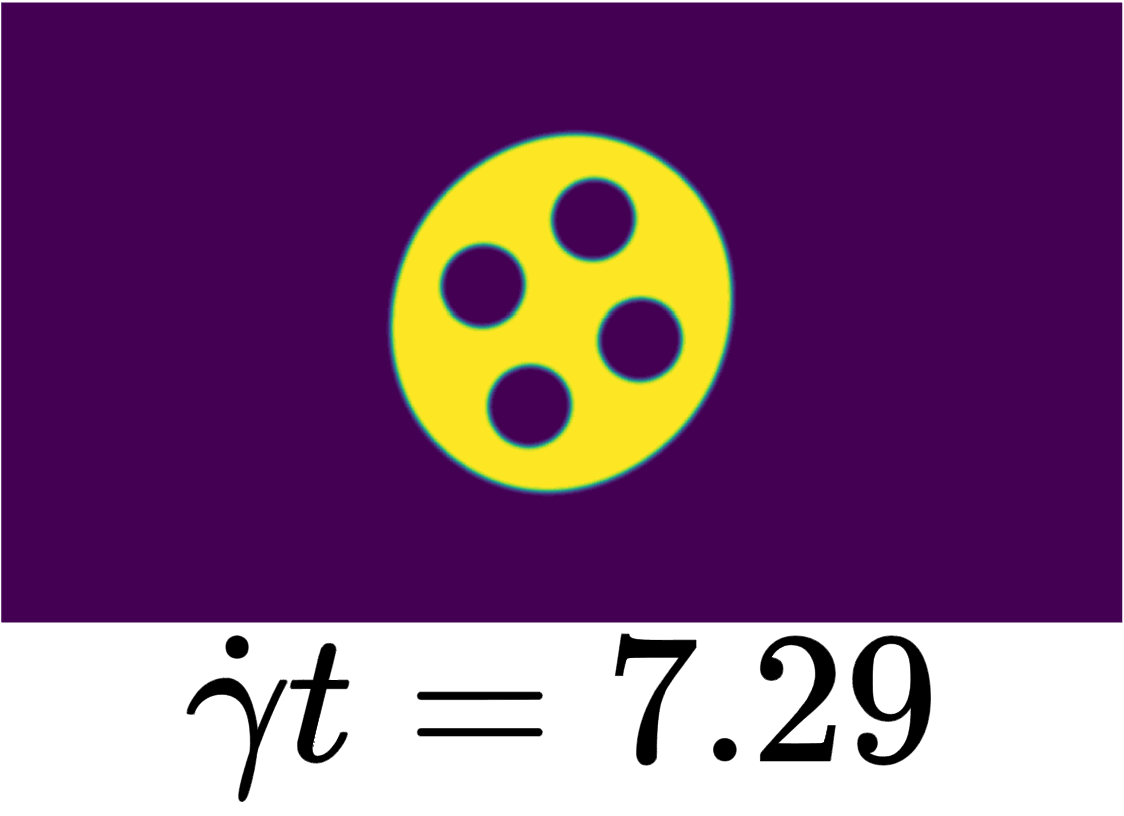}
		\includegraphics[scale = 0.09]{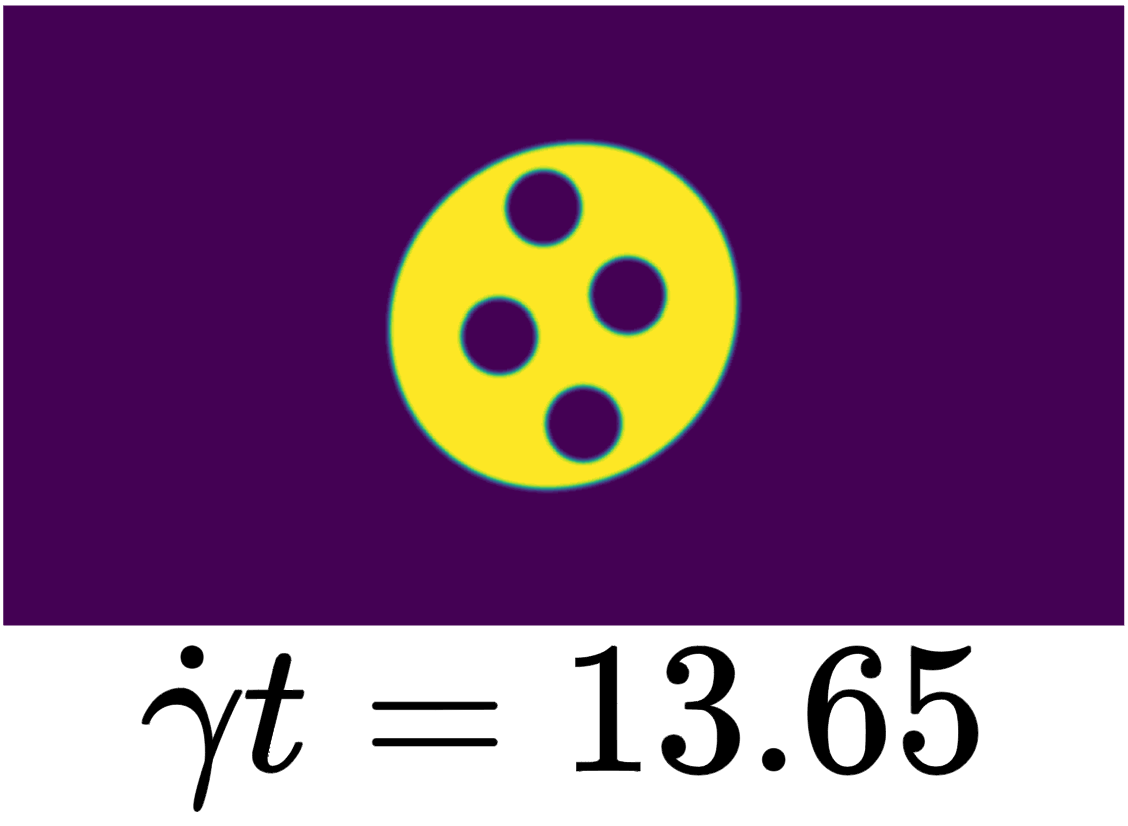} \\
        (e) $\frac{\Delta Y}{R_o}=1.00$~(Only shell droplet coalesce)
		\\ \vspace{0.5cm}

  \includegraphics[scale = 0.09]{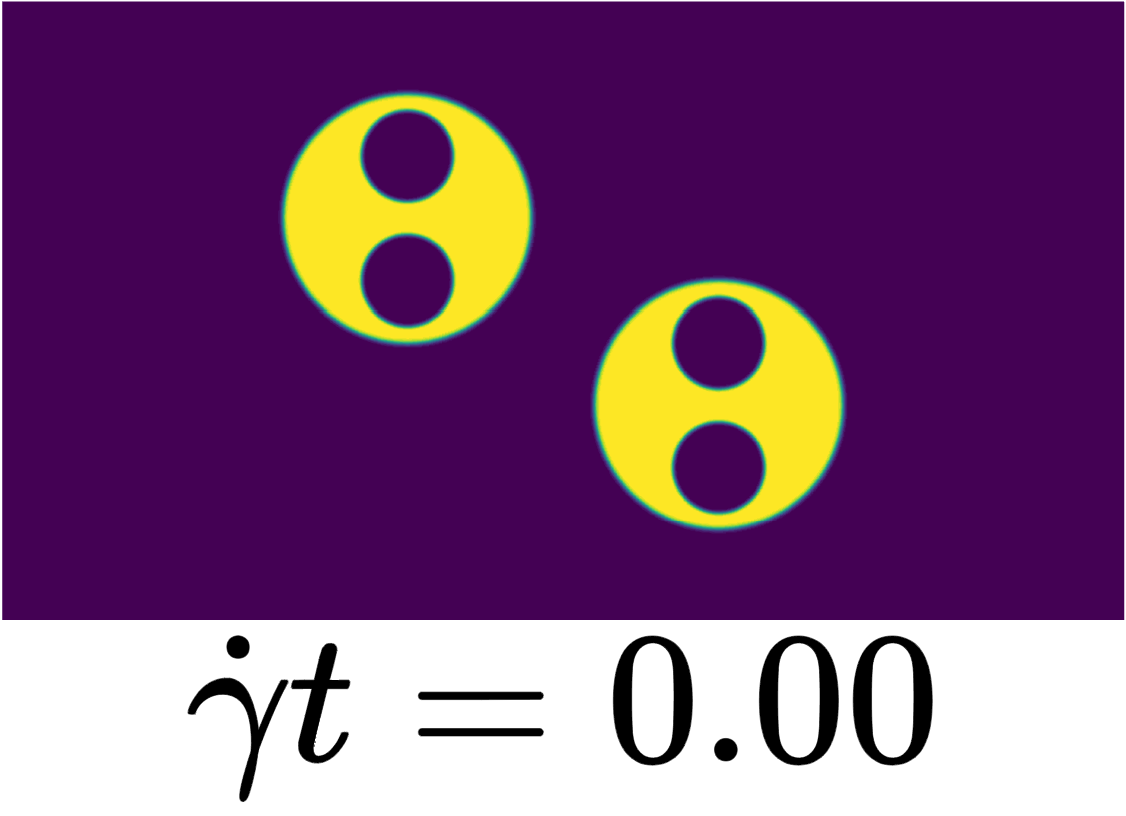}
		\includegraphics[scale = 0.09]{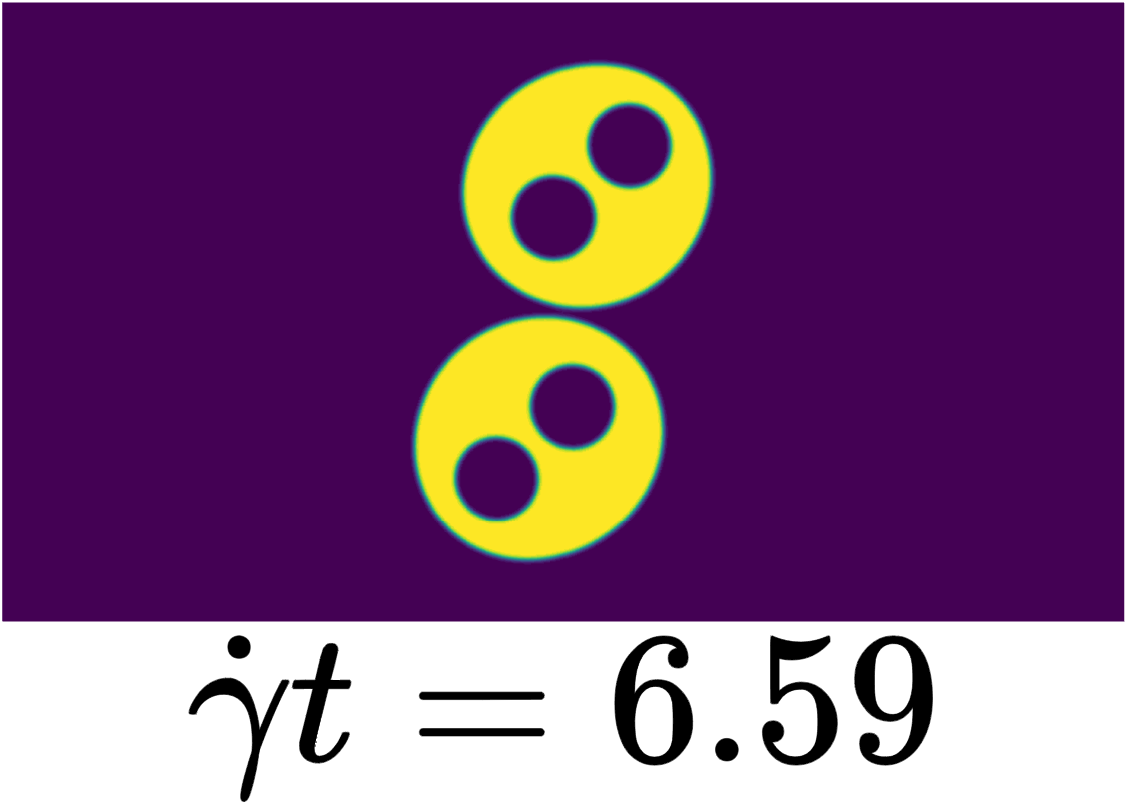}
		\includegraphics[scale = 0.09]{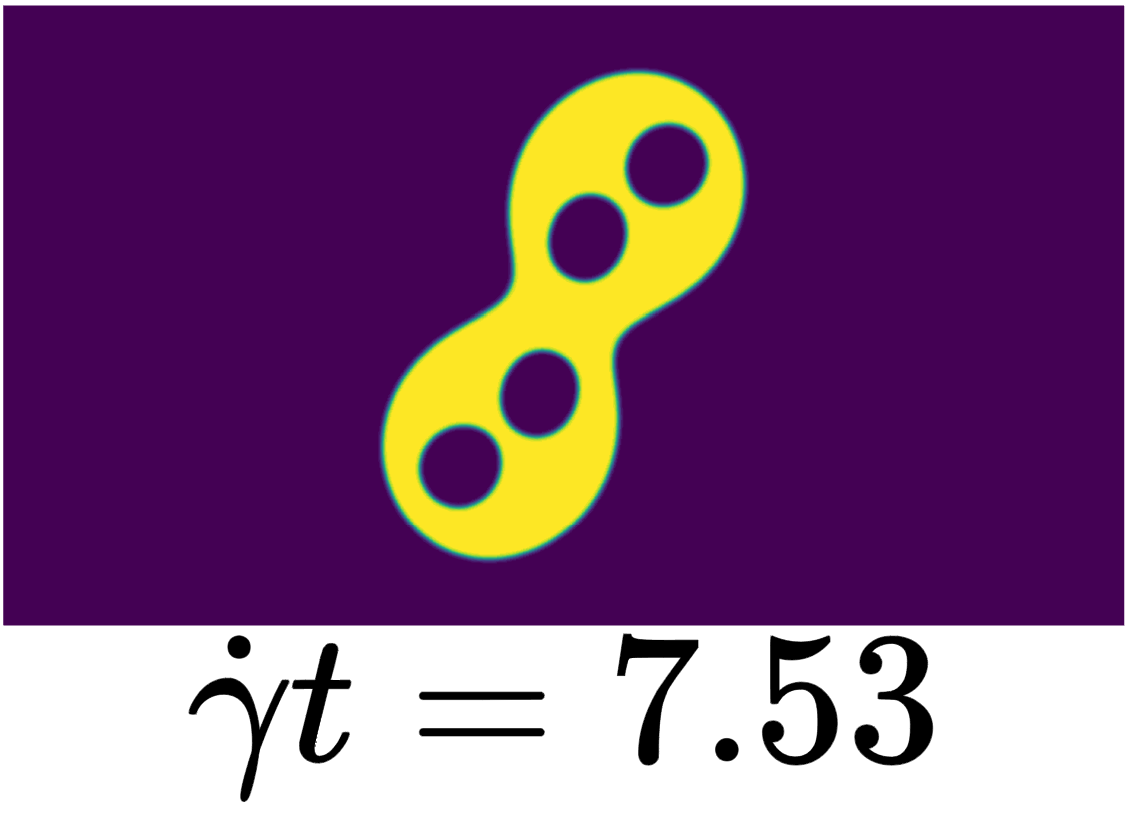}
		\includegraphics[scale = 0.09]{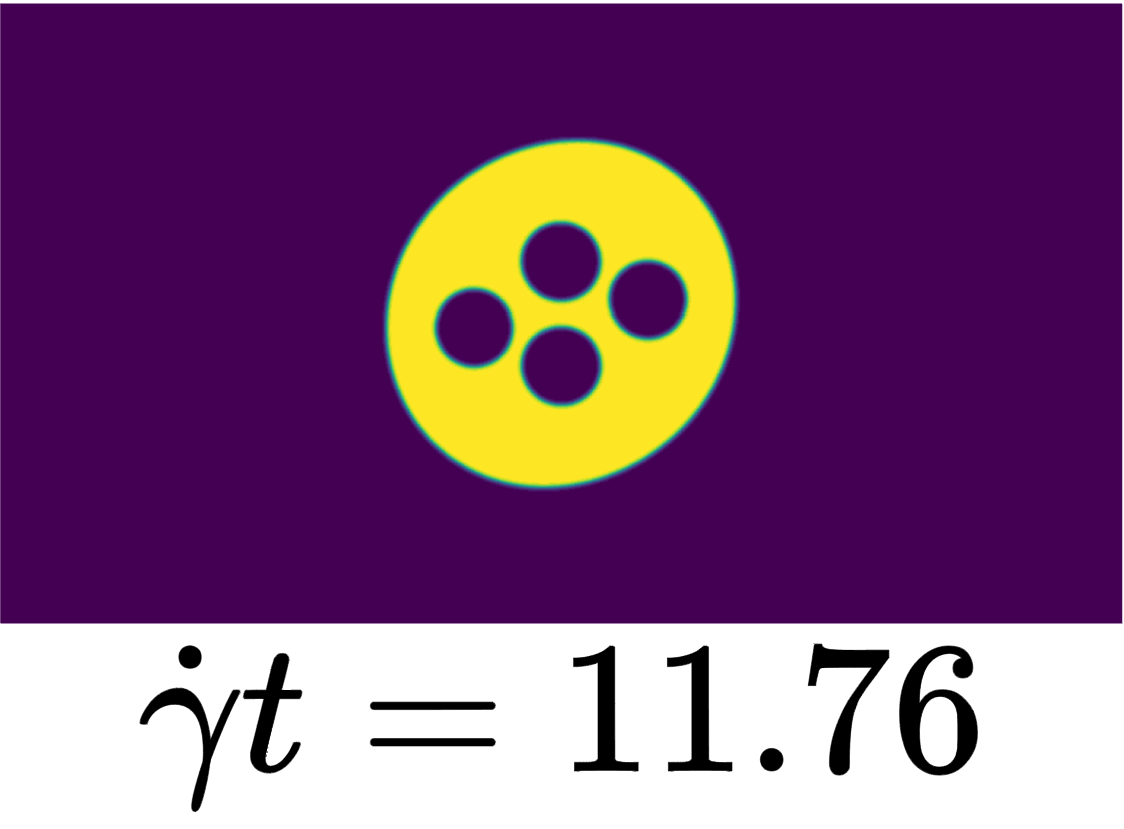}
		\includegraphics[scale = 0.09]{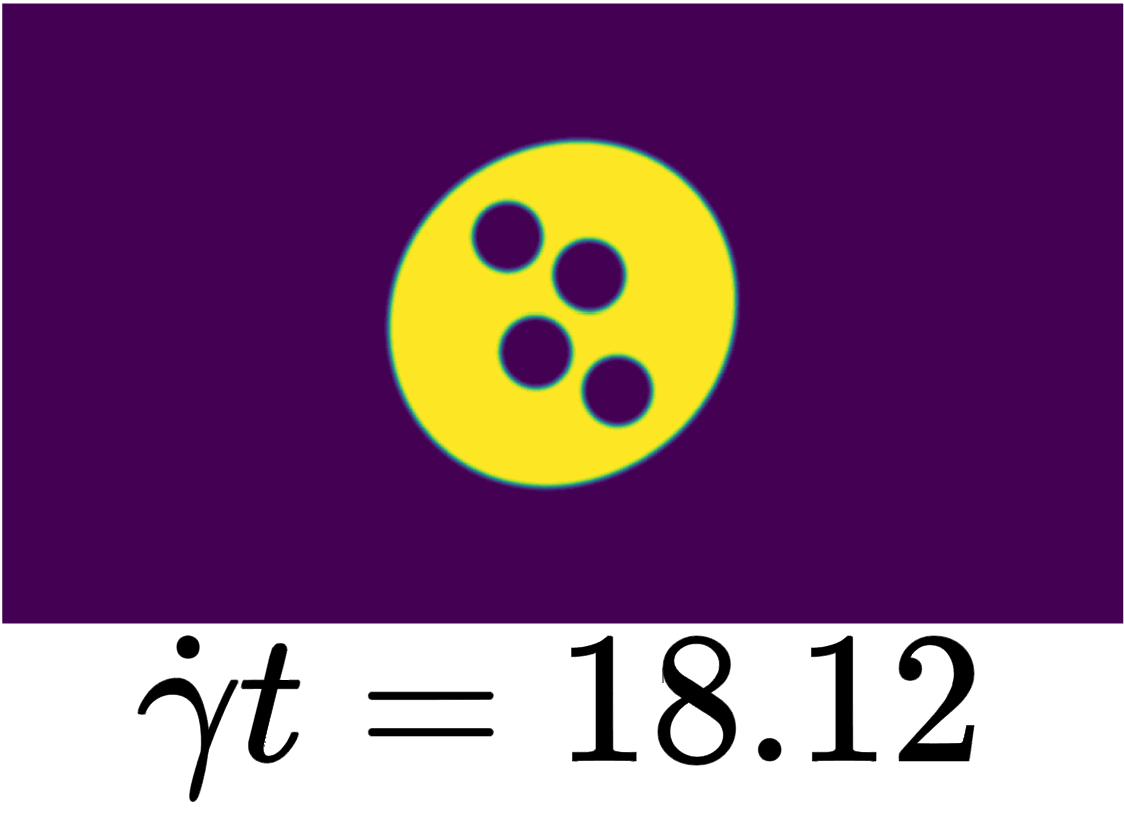} \\
        (f) $\frac{\Delta Y}{R_o}=1.50$~(Only shell droplet coalesce)
		\\ \vspace{0.5cm}
  
 	\caption{Time-lapse images of double-core compound droplet collision for six different initial vertical offset ($\frac{\Delta Y}{R_o}=0.00, 0.15, 0.35, 0.75, 1.00$, and $1.50$) with $Ca=0.05$, $Re=0.10$, $\rho_{12}=1$, $\mu_{12}=1$, $\Delta X_o/R_o = 2.50,$ and $R_o/H = 0.40.$} 
	\label{fig:Double-core-Ca-0.05-varying-offset-snap}
\end{figure}

\begin{figure}[H]
          \centering

         \includegraphics[scale = 0.085]{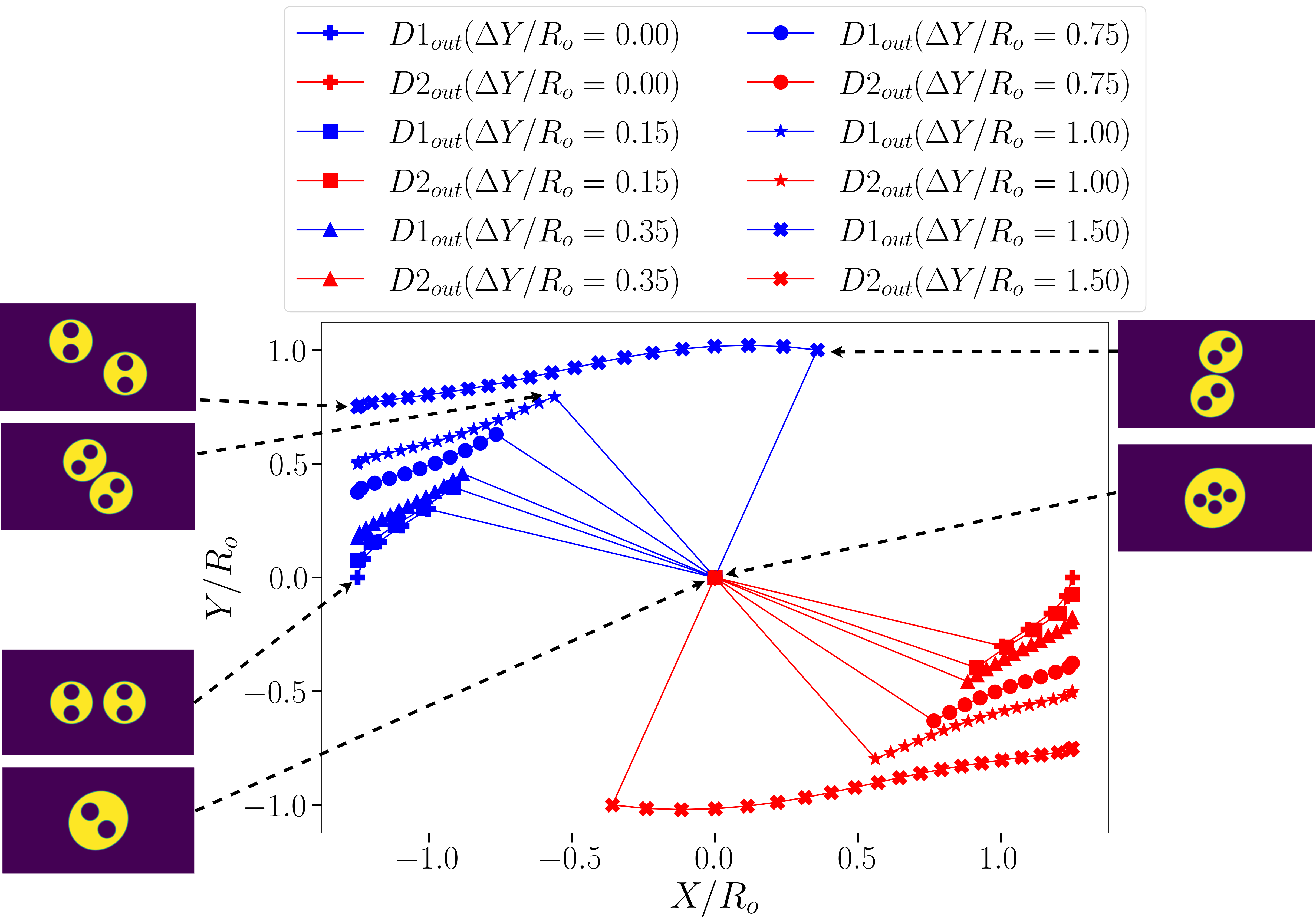} \\
          \vspace{0.45cm}

	\caption{Trajectories of the shell droplet pairs ($D1_{out}$ and $D2_{out}$) for six different initial vertical offset ($\frac{\Delta Y}{R_o}=0.00, 0.15, 0.35, 0.75, 1.00$, and $1.50$) with $Ca=0.05$.}
	\label{fig:Double-core-Ca-0.05-varrying-offset-outer-trajectory}
\end{figure}

In this section, the results from simulations with various initial vertical offsets ($\frac{\Delta Y}{R_o}=0.00, 0.15, 0.35, 0.75, 1.00$, and $1.50$), but with a reduced $Ca$ of $0.05$ will be discussed. The aim is to determine if further decrease in the $Ca$ value alters the collision outcomes or affects the dynamics of the droplets. Similar to the previous analyses at higher $Ca$ values, the evolution of the compound droplet collisions has been illustrated using key snapshots for all six initial vertical offset cases in Figure~\ref{fig:Double-core-Ca-0.05-varying-offset-snap}. For all initial vertical offset cases, it has been observed that the shell droplets coalesce. Notably, in one of the cases with $\frac{\Delta Y}{R_o}=0.00$, the core droplets also coalesce. Although the timing of the shell droplet coalescence varies depending on the initial offset, their behavior post-coalescence remains consistent. After coalescence, the core droplets continue their planetary-like motion inside the merged shell droplet. This consistent coalescence across varying initial offsets suggests that a lower Ca value enhances the likelihood of shell droplet coalescence.
The trajectories of the shell droplets in the cases with $Ca=0.05$ remain very consistent and similar to those in the previously analyzed cases with $Ca$ values of $0.10$ and $0.07$. This consistency is clearly observed in Figure~\ref{fig:Double-core-Ca-0.05-varrying-offset-outer-trajectory}, where the shell droplets’ trajectories have been plotted. 
\begin{figure}[H]
          \centering
          
        \includegraphics[scale = 0.075]{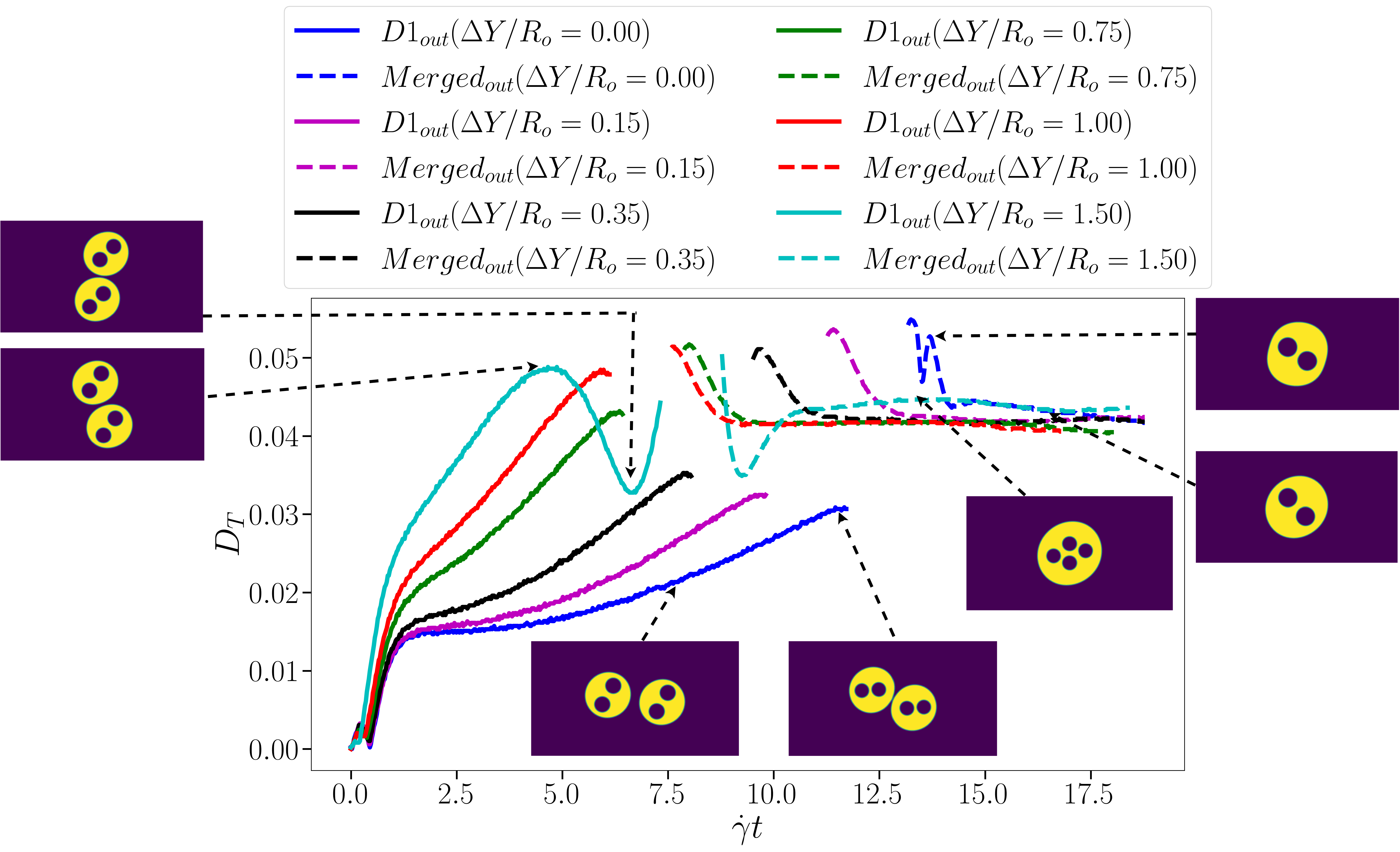} \\
    \vspace{0.45cm}
    
	\caption{Deformation quantification of the shell droplet before and after coalescence ($D1_{out}$ and $Merged_{out}$) for six different initial vertical offset ($\frac{\Delta Y}{R_o}=0.00, 0.15, 0.35, 0.75, 1.00$, and $1.50$) with $Ca=0.05$.}
	\label{fig:Double-core-Ca-0.05-varrying-offset-outer-deform}
\end{figure}

For cases with higher initial vertical offsets, the droplets travel further in the shear flow direction before coalescing compared to cases with lower initial vertical offsets. However, in all cases, the shell droplets eventually coalesce.

The deformation plots depicted in Figure~\ref{fig:Double-core-Ca-0.05-varrying-offset-outer-deform} shows consistent trends with the previously analyzed coalescence cases. As the shell droplets approach collision, there is a noticeable rise in deformation. Following coalescence, the deformation stabilizes as the newly formed shell droplet adopts an ellipsoidal shape. However, it is notable that the presence of multiple core droplets influences the final shape of the coalesced shell droplet. Instead of forming a perfect ellipsoid, the coalesced droplet appears to be somewhat slanted ovaloid. The core droplets, continuing their rotational motion inside the coalesced shell droplet, create internal stresses that distort the shell droplet’s shape.

\begin{figure}[H]
    \centering
    \begin{tabular}{cc}
        \includegraphics[scale = 0.18]{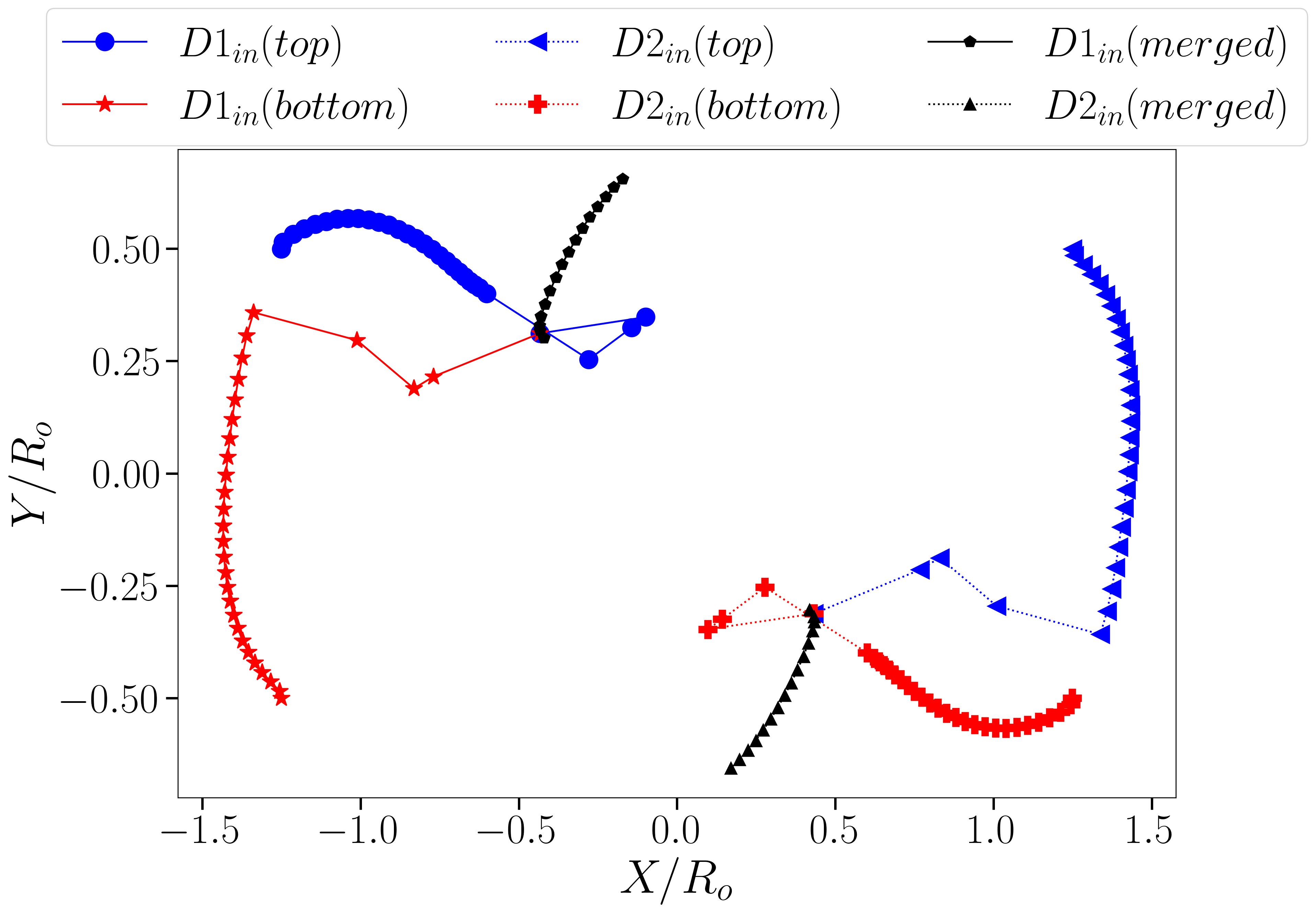} & \includegraphics[scale = 0.18]{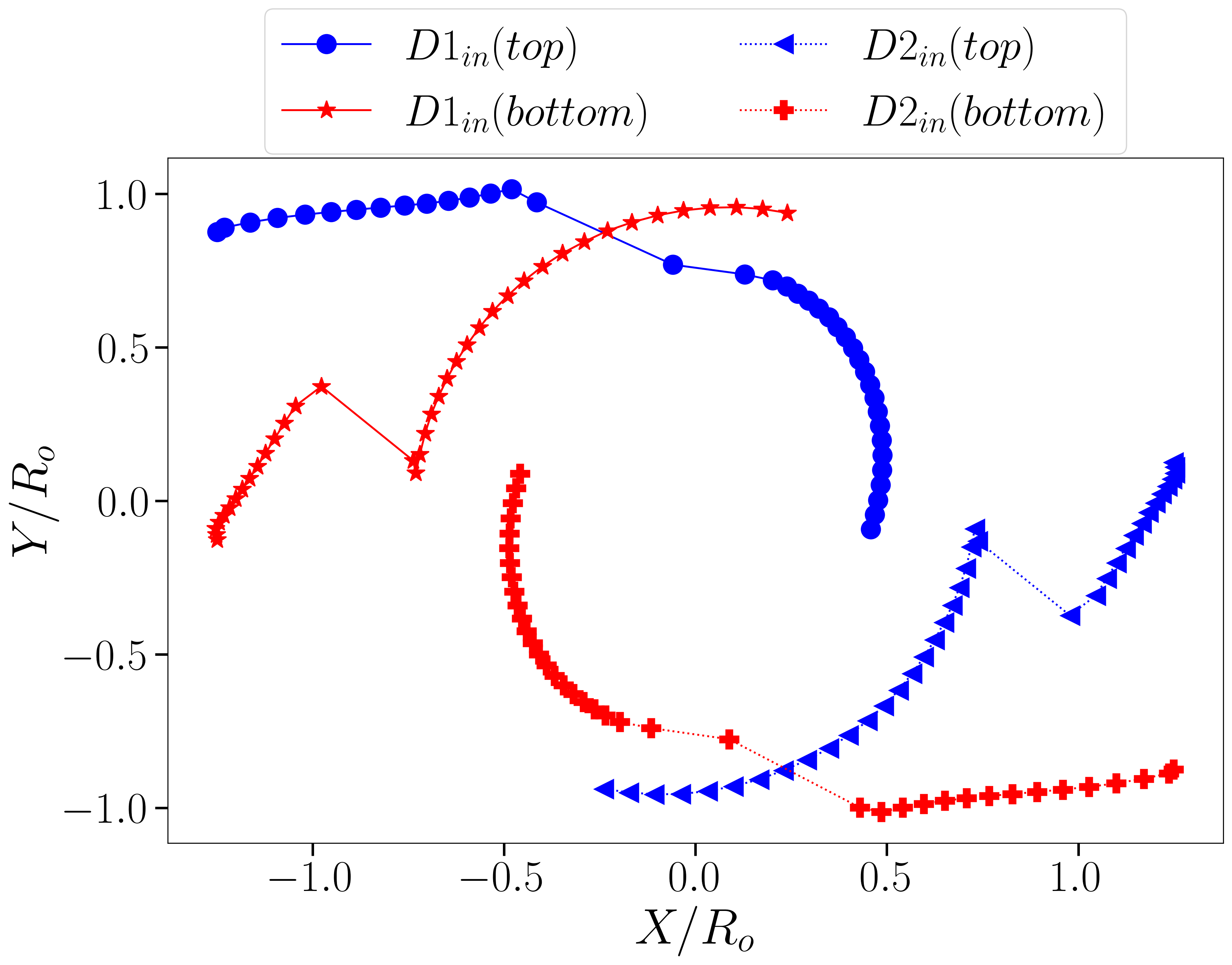} \\
        (a) $\Delta Y_o/R_o = 0.00$  &
        (d) $\Delta Y_o/R_o = 0.75$  \\
        \includegraphics[scale = 0.18]{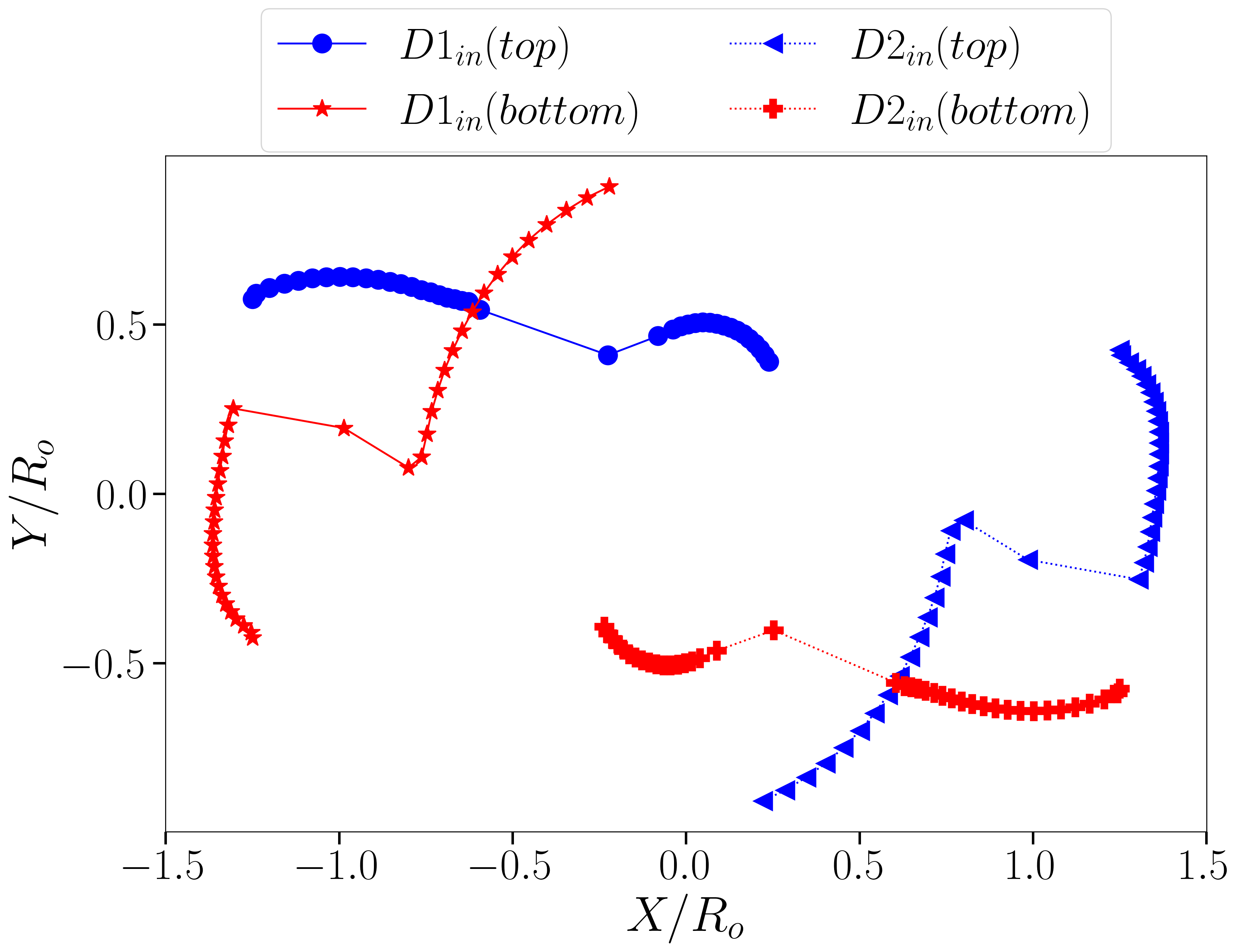} & \includegraphics[scale = 0.18]{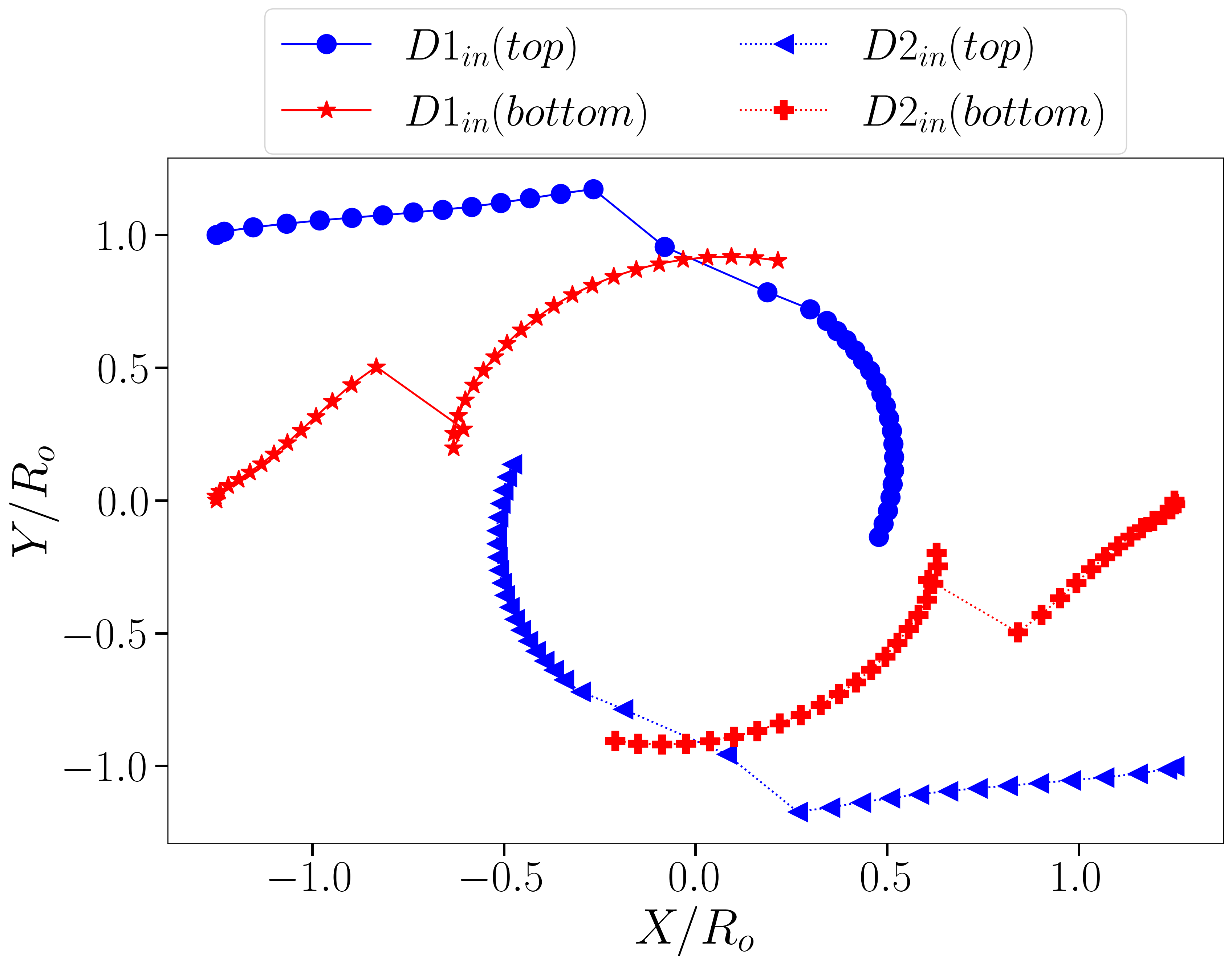} \\
        (b) $\Delta Y_o/R_o = 0.15$  &
        (e) $\Delta Y_o/R_o = 1.00$  \\
        \includegraphics[scale = 0.18]{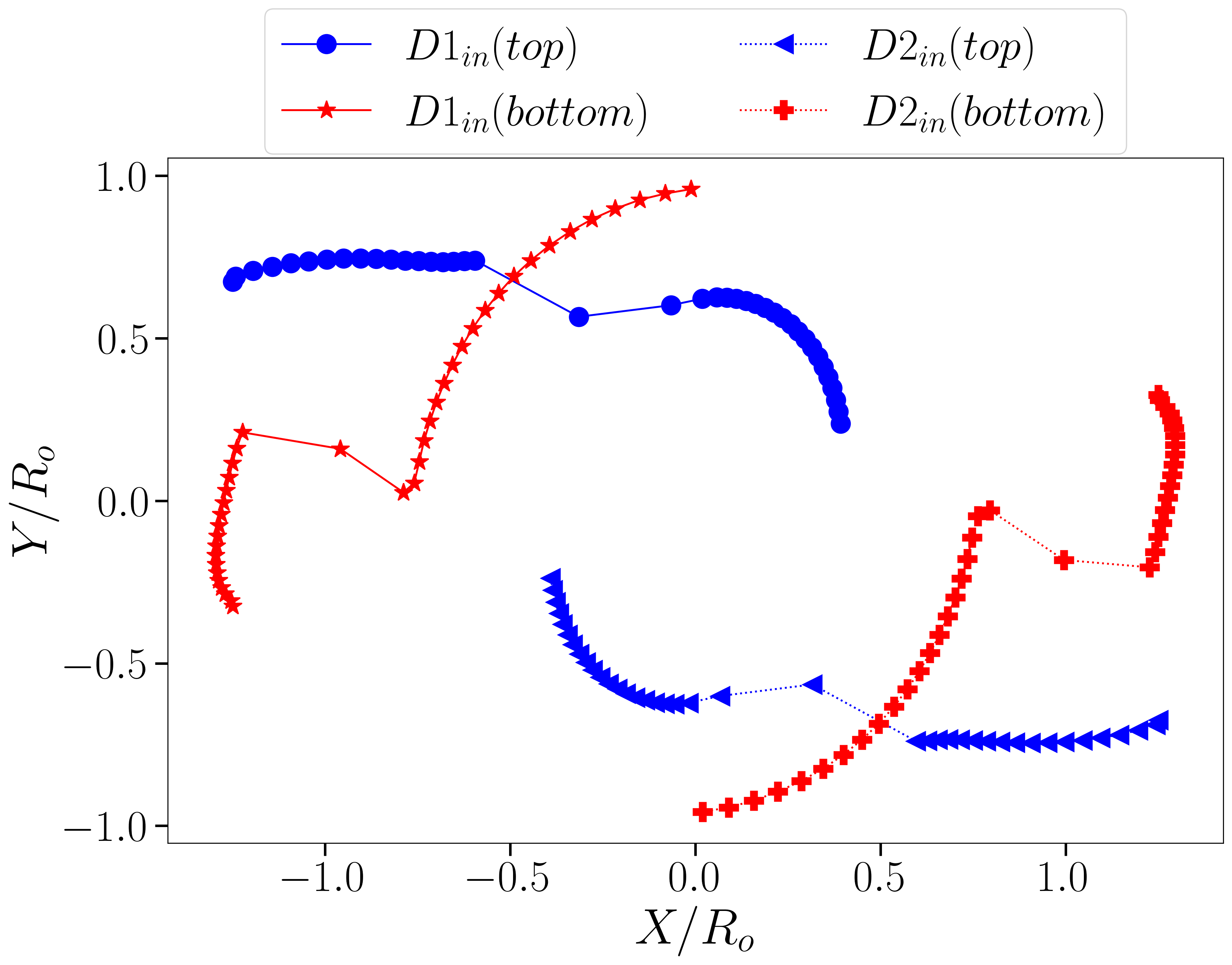} & \includegraphics[scale = 0.18]{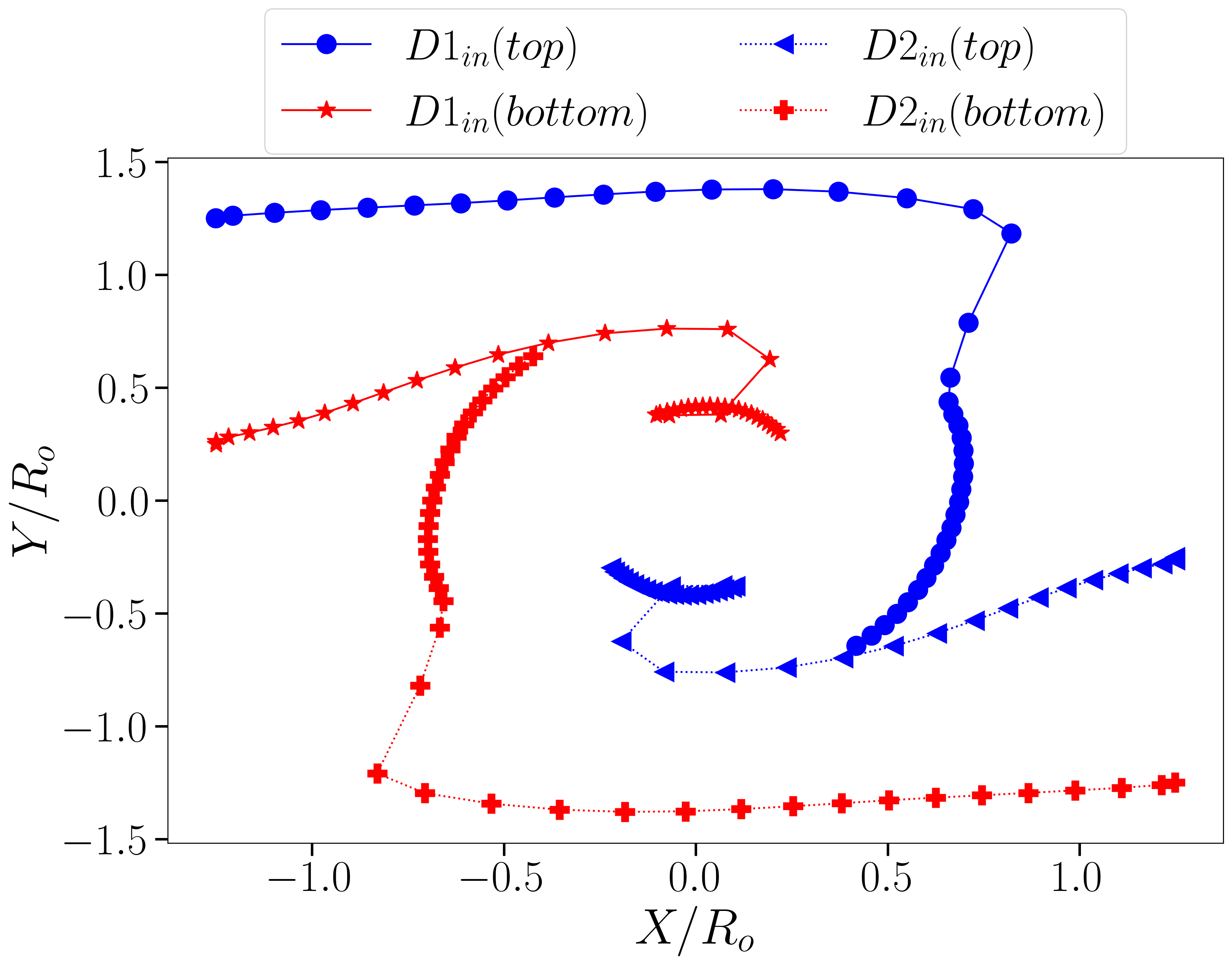} \\
        (c) $\Delta Y_o/R_o = 10.35$  &
        (f) $\Delta Y_o/R_o = 1.50$  \\
    \end{tabular}

    \caption{Trajectories of the core droplets ($D1_{in}(top)$, $D1_{in}(bottom)$, $D2_{in}(top)$ and $D2_{in}(bottom)$) for the six different initial vertical offset ($\frac{\Delta Y}{R_o}=0.00, 0.15, 0.35, 0.75, 1.00$, and $1.50$) in between shell droplets with $Ca=0.05$.}
    \label{fig:Double-core-Ca-0.05-varrying-offset-inner-trajectory}
\end{figure}

Based on the trajectory plots presented in Figure~\ref{fig:Double-core-Ca-0.05-varrying-offset-inner-trajectory}, the behavior of core droplets exhibits consistency throughout the various $Ca$ number examined in this study. Before shell droplet coalescence, core droplets maintain a rotational motion within their respective shells, influenced by the surrounding shear flow and encapsulation within the shell droplet. Post-coalescence of the shell droplets, the multiple core droplets converge into a circular path within the newly formed, merged shell droplet. This behavior suggests that despite variations in capillary number ($Ca=0.05$ in this case), the core droplets consistently undergo rotational and translational motions dictated by the overall dynamics of the coalesced shell droplet for all the shell droplet coalescence cases.

\begin{figure}[H]
    \centering
    \begin{tabular}{cc}
        \includegraphics[scale = 0.17]{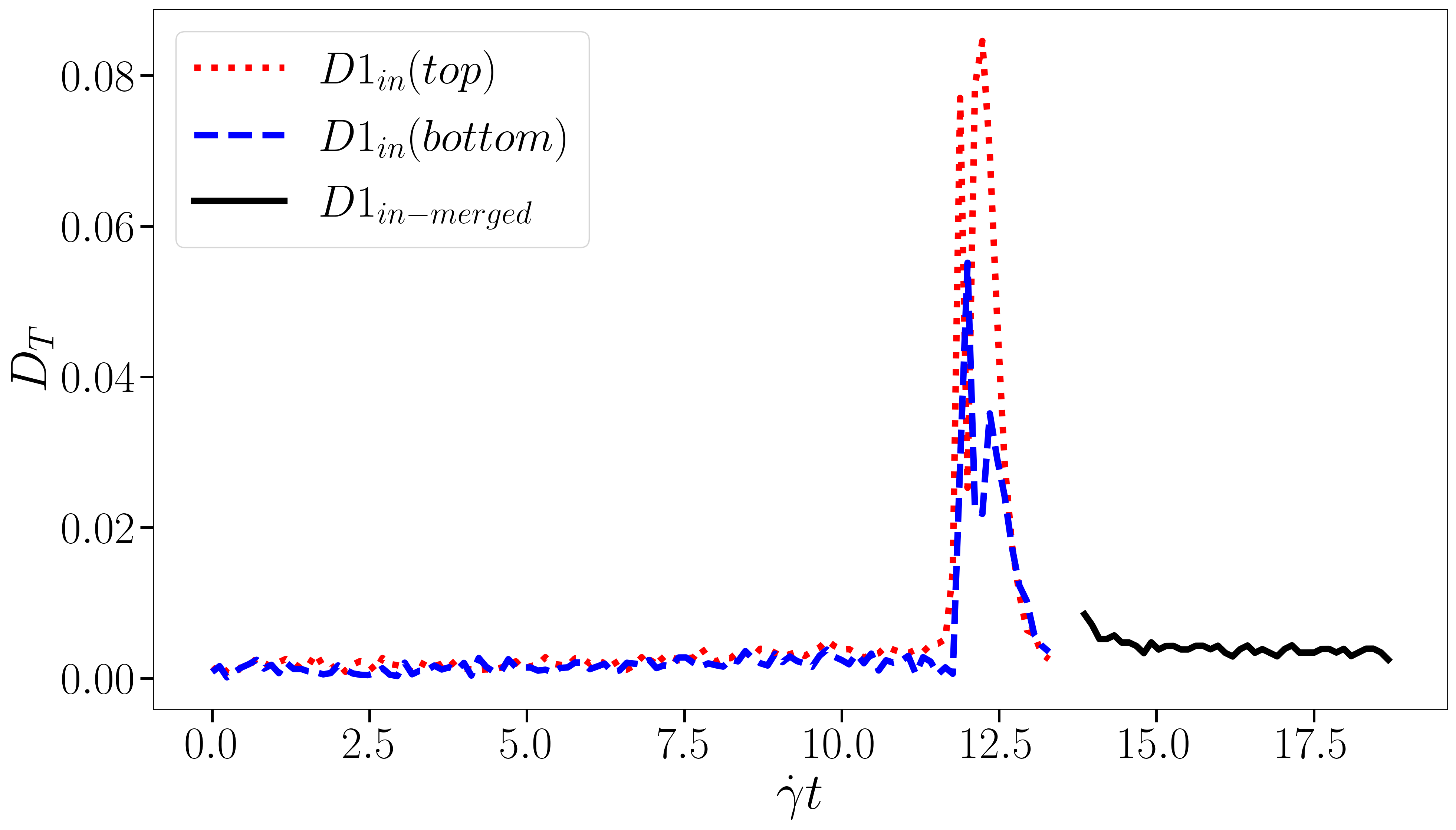} & \includegraphics[scale = 0.17]{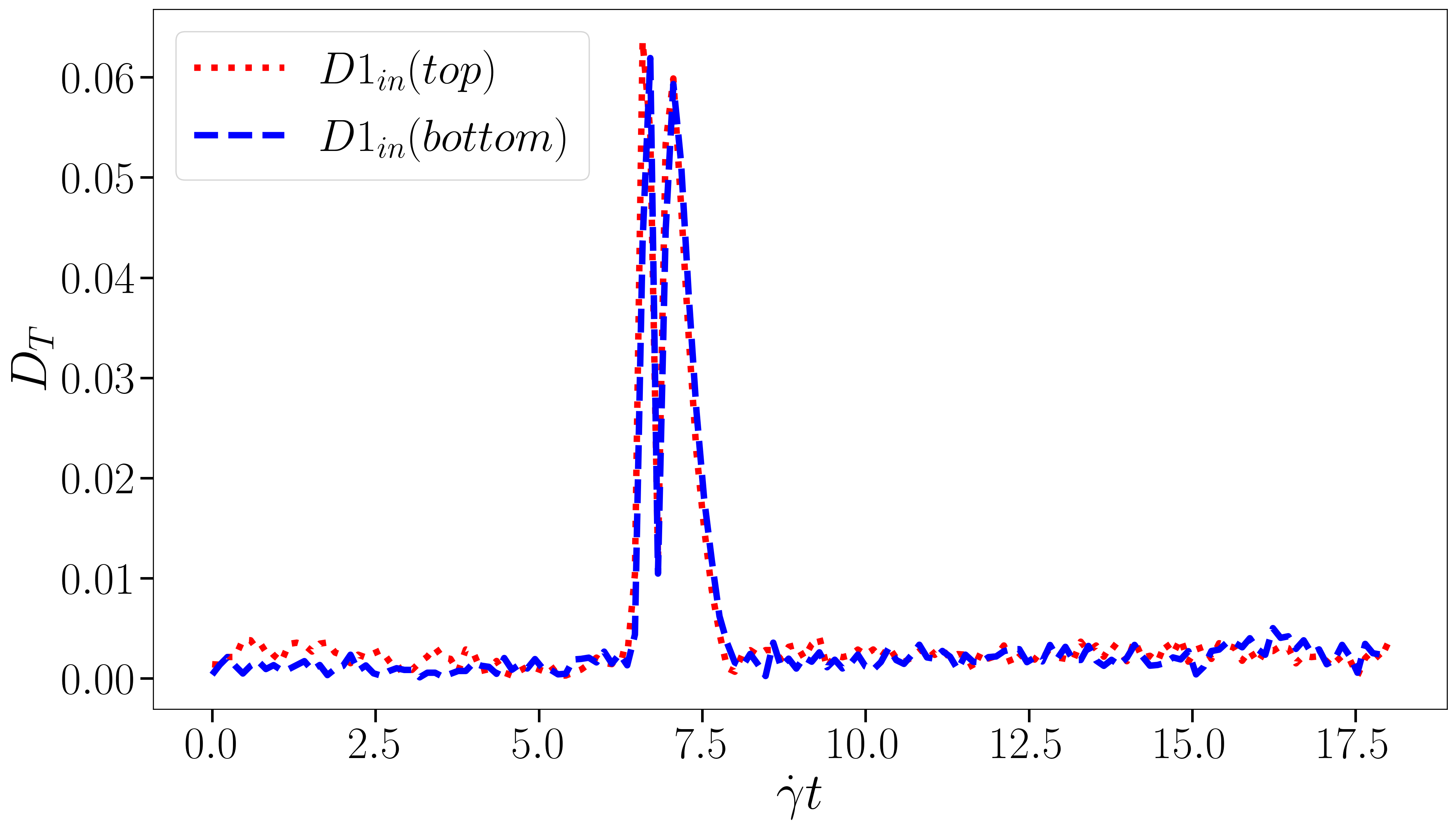} \\
        (a) $\Delta Y_o/R_o = 0.00$  &
        (d) $\Delta Y_o/R_o = 0.75$  \\
        \includegraphics[scale = 0.17]{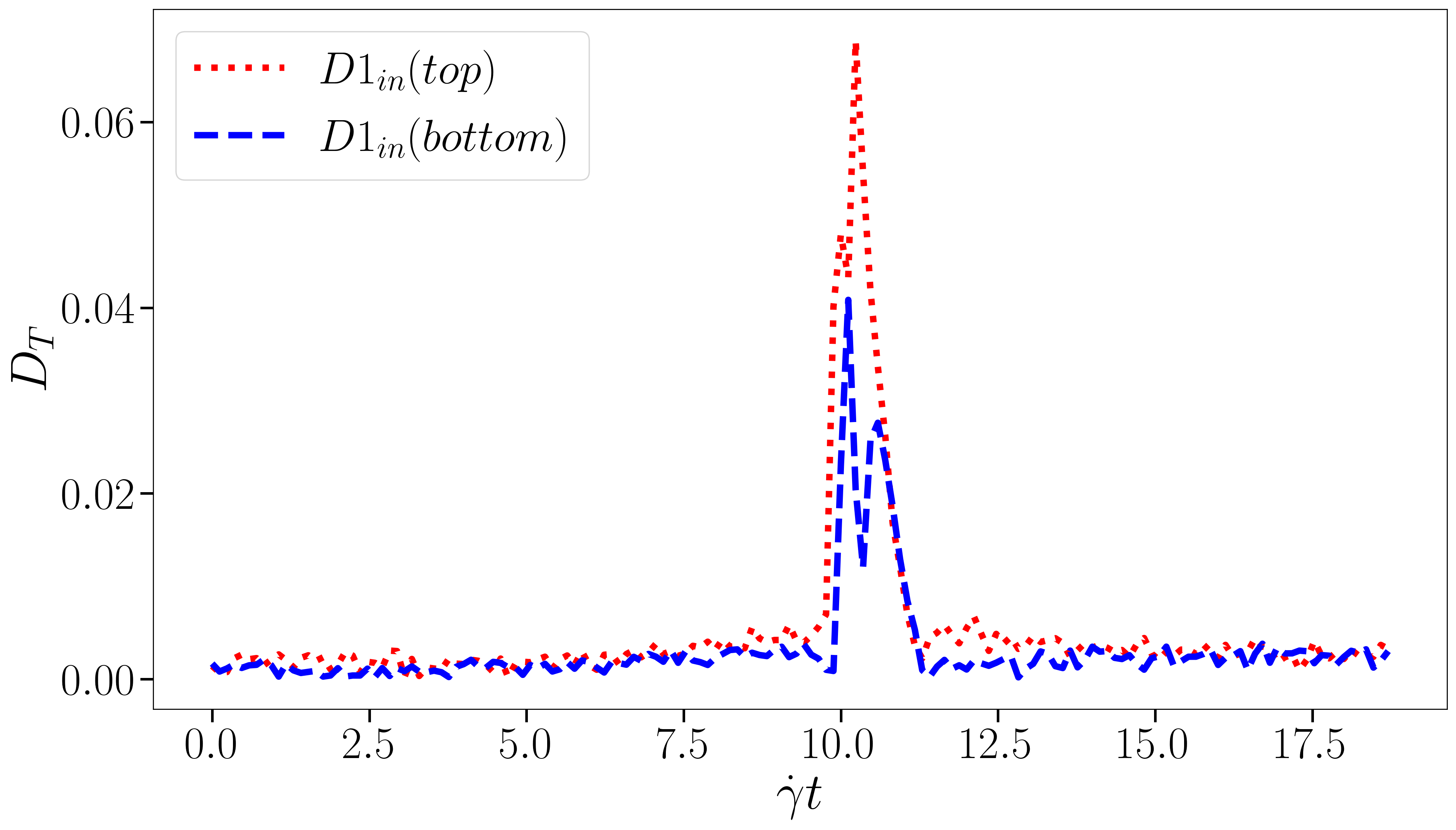} & \includegraphics[scale = 0.17]{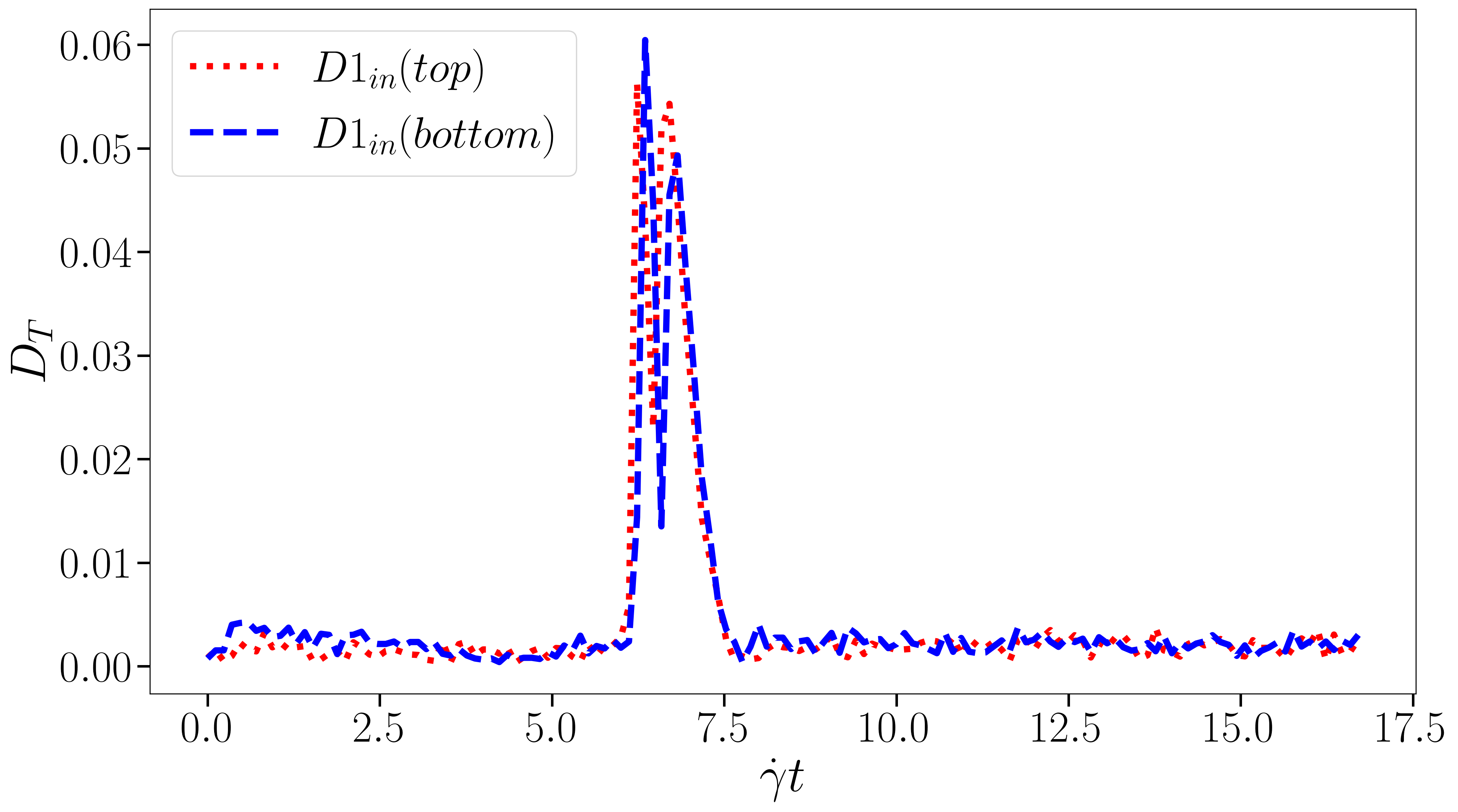} \\
        (b) $\Delta Y_o/R_o = 0.15$  &
        (e) $\Delta Y_o/R_o = 1.00$  \\
        \includegraphics[scale = 0.17]{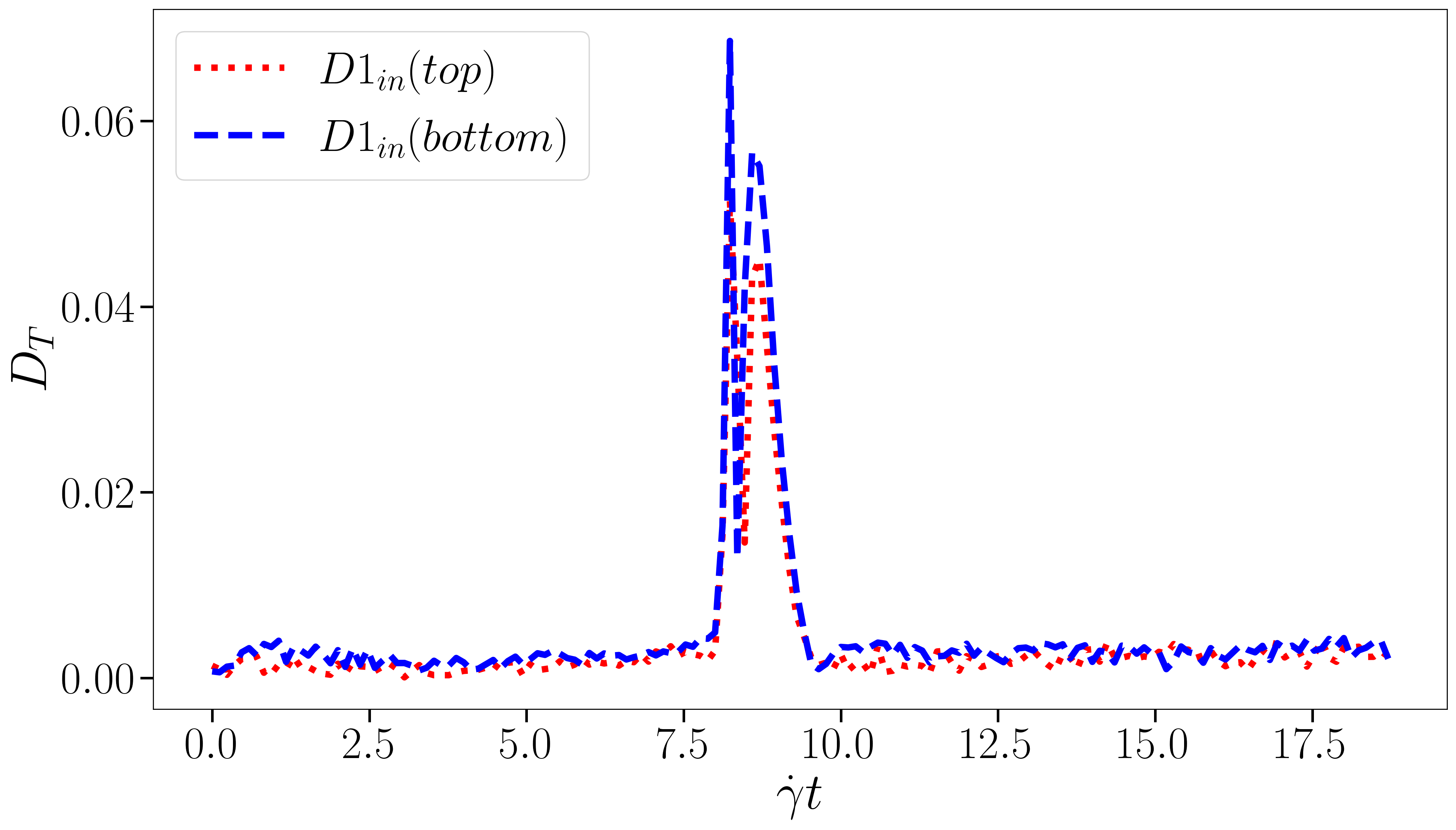} & \includegraphics[scale = 0.17]{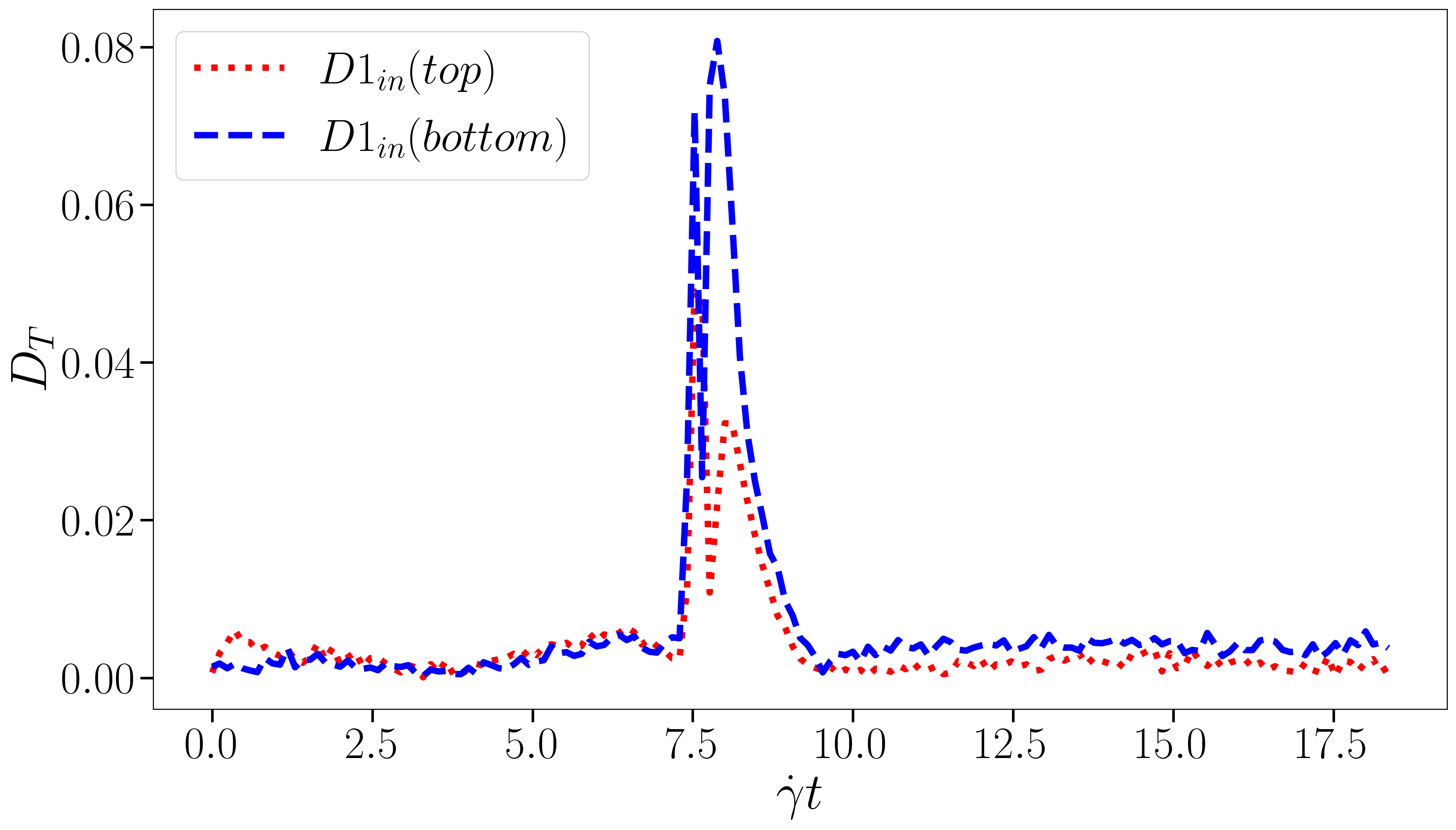} \\
        (c) $\Delta Y_o/R_o = 0.35$  &
        (f) $\Delta Y_o/R_o = 1.50$  \\
    \end{tabular}

    \caption{Deformation quantification of the core droplets ($D1_{in}(top)$, $D1_{in}(bottom)$, $D2_{in}(top)$ and $D2_{in}(bottom)$) for the six different initial vertical offset ($\frac{\Delta Y}{R_o}=0.00, 0.15, 0.35, 0.75, 1.00$, and $1.50$) in between shell droplets with $Ca=0.05$.}
    \label{fig:Double-core-Ca-0.05-varrying-offset-inner-deform}
\end{figure}

Based on the deformation plots shown in Figure~\ref{fig:Double-core-Ca-0.05-varrying-offset-inner-deform}, the core droplets exhibit consistent behavior with two distinct spikes in their deformation curves, similar to observations discussed previously in Section~\ref{sec_5.2}(2). These spikes coincide with critical instances during the collision and coalescence phases of the shell droplets. Interestingly, while in most cases both top and bottom core droplets show peaks in deformation at nearly the same times, there are cases where one core droplet deforms more significantly than the other. 

This variability in deformation can be attributed to the specific timing of shell droplet coalescence and the resulting positions of the core droplets within the shell at those moments. For instance, when the shell droplet coalesces, the internal arrangement of core droplets may vary, leading to different exposure to deformation-inducing forces such as shear and pressure gradients within the shell droplet. Such variations underscore the dynamic interaction between multiple core droplets and the evolving shape of the shell droplet during collision dynamics, highlighting the complexity and sensitivity of these systems to initial conditions and internal dynamics.

The study highlights the significant interplay between $Ca$ number and the initial vertical offset in determining the collision outcomes of double-core compound droplets within a confined shear flow. At higher $Ca$ value ($Ca=0.10$), the surface tension is relatively low compared to the shear forces. This allows the droplets to deform more easily and move past each other, leading to pass-over events at larger initial vertical offsets. For lower offsets, the reduced initial distance facilitates coalescence of the shell droplets, although the core droplets remain separated due to their rotational motion and insufficient force to overcome the internal separation. As $Ca$ decreases to $0.07$, surface tension becomes more influential. At this intermediate level, the competition between surface tension and shear forces results in varied outcomes. At lower offsets, both shell and core droplets coalesce due to increased interactions and the inability of the shear forces to keep them separated. As the offset increases, the shell droplets still tend to coalesce, but the core droplets, having more space and influenced by internal fluid dynamics, remain separated in most cases. With the lowest $Ca$ value ($Ca=0.05$), surface tension dominates, leading to a higher tendency for coalescence of the shell droplets. The initial vertical offset further influences these dynamics by altering the proximity and interaction time between the droplets. Understanding these behaviors is crucial for applications in microfluidics and emulsion technologies, where precise control over droplet interactions can enhance process efficiency and outcomes.

\subsection{Effect of Initial Vertical Offset, Density Ratio, and Viscosity Ratio}\label{sec_5.3}

In earlier research found in the literature, the influence of density ratios and viscosity ratios greater than unity has not been extensively studied, particularly in scenarios involving the collision of compound droplets containing two core droplets each. Recognizing this gap in the literature, we aim to explore and understand the behavior of such systems through our simulations. In this section, we present our findings on the effects of varying higher density ratios (specifically $100$, $200$, and $400$) and viscosity ratios (specifically $24$, $42$, and $60$) under conditions of different vertical offset (i.e, $\frac{\Delta Y}{R_o}=0.00-1.50$) between the shell droplets. These specific ratios were selected to examine the interplay between these two critical parameters and their impact on droplet dynamics. To ensure consistency and relevance, we maintained the $Re$ at $1.0$, the $Ca$ at $0.10$, the radius ratio ($R _{r} = \frac{R _i}{R_o}$) at $0.375$, and the confinement ($\frac{R_o}{H}$) at $0.40$ throughout our simulations. The choice of these parameters was determined through careful trials to ensure accurate physical simulations, avoiding any unexpected behaviors, and ensuring that the simulation time remains feasible within our computational resource constraints.

\subsubsection{High vertical initial offset: $\frac{\Delta Y}{R_o}=1.5$}
\begin{figure}[H]
    \centering
    \begin{tabular}{cc}
        \includegraphics[scale = 0.20]{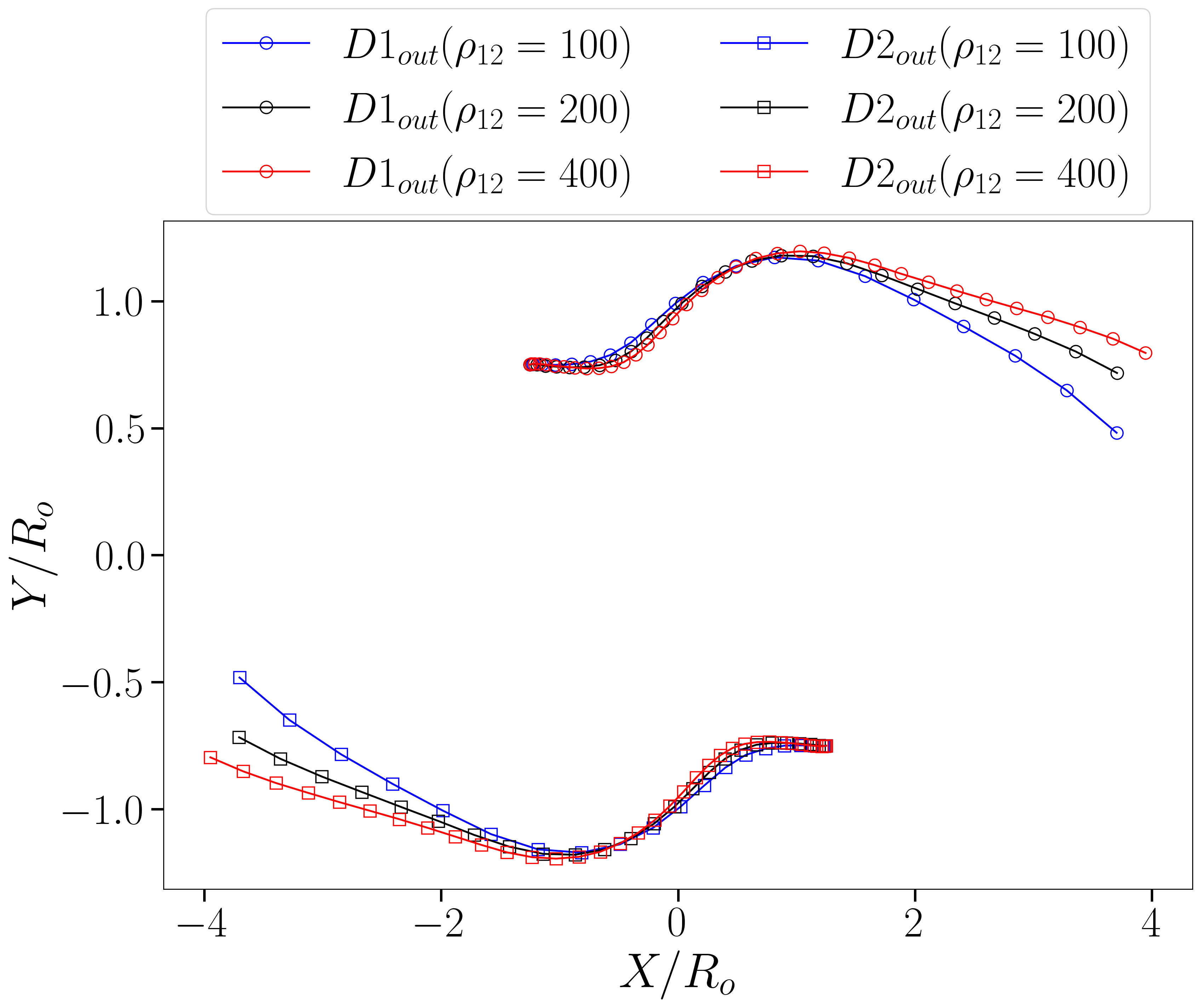} & \includegraphics[scale = 0.20]{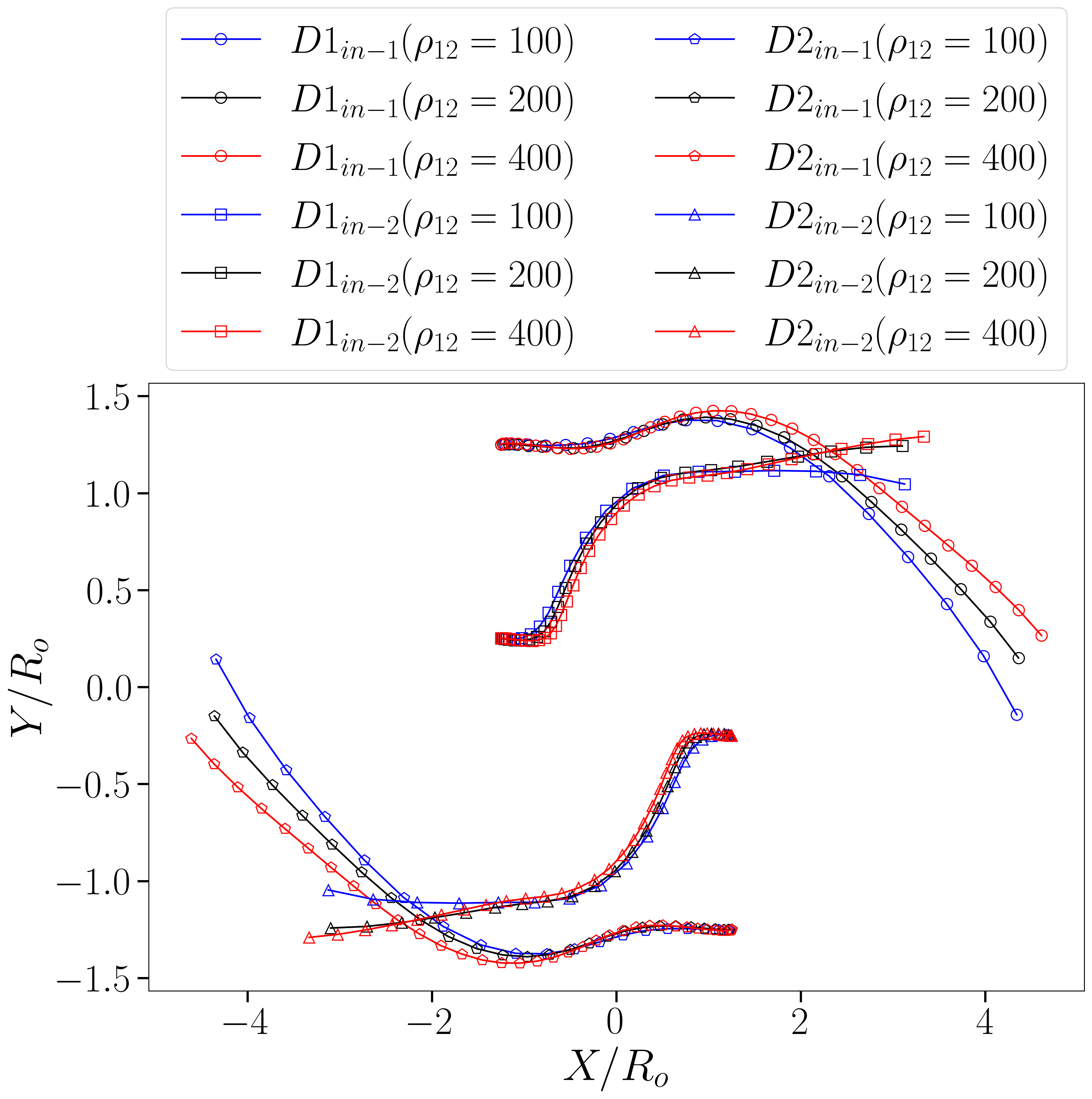} \\
        (a) Shell droplets ($\mu_{12}=24$) &
        (b) Core droplets ($\mu_{12}=24$)  \\ 
    
        \includegraphics[scale = 0.20]{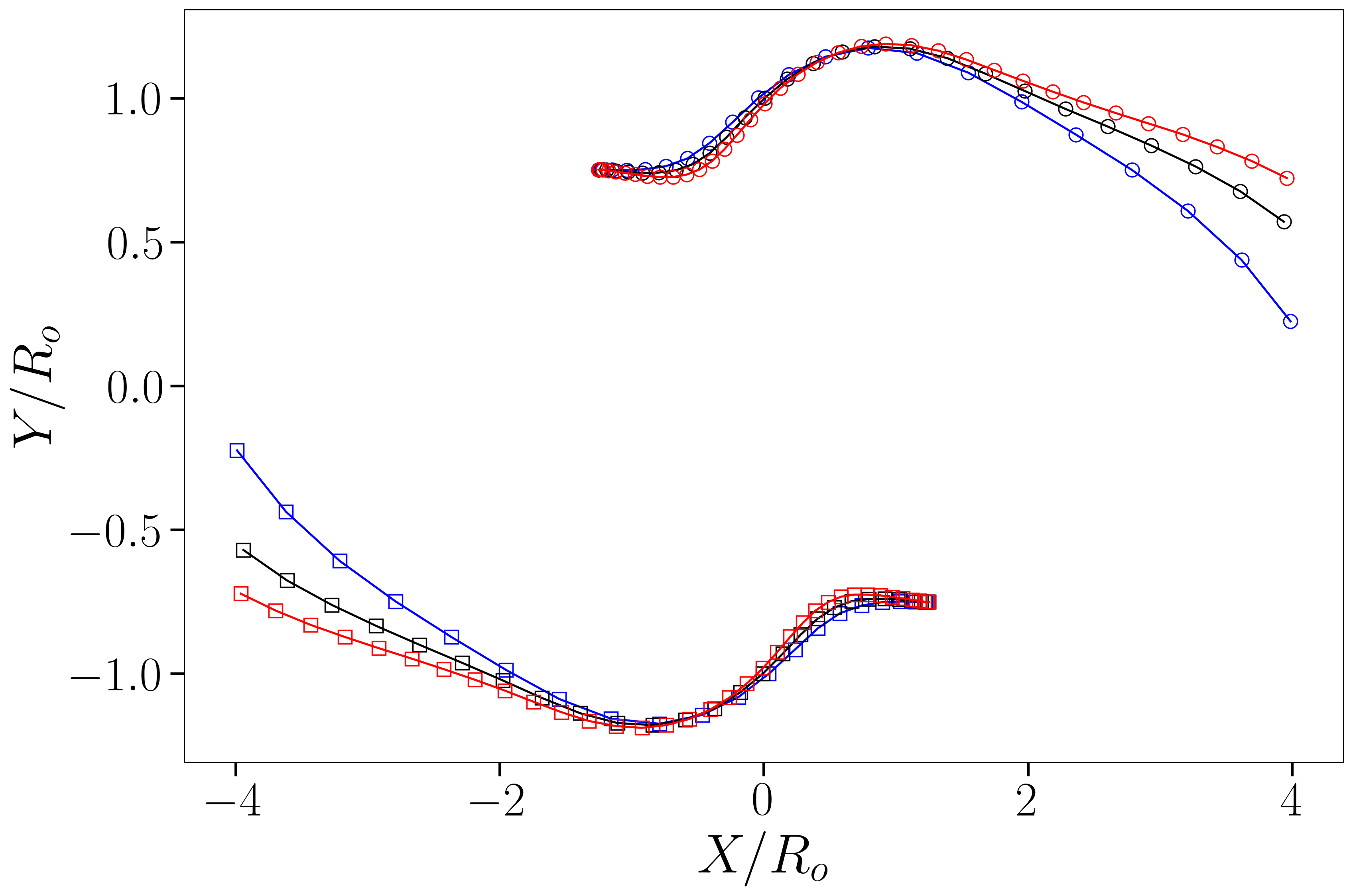} & \includegraphics[scale = 0.20]{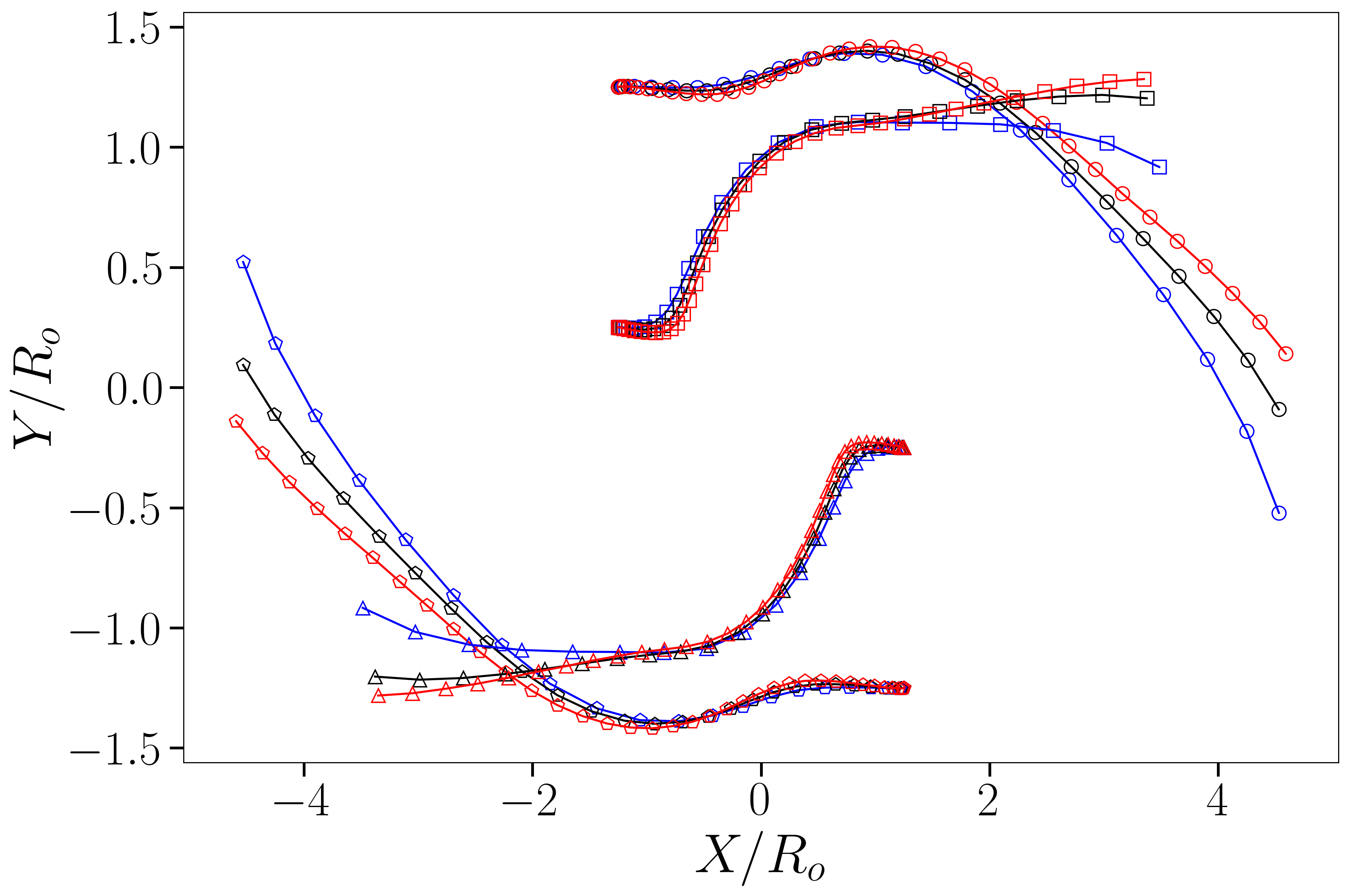} \\
        (c) Shell droplets ($\mu_{12}=42$) &
        (d) Core droplets ($\mu_{12}=42$) \\ 
  
        \includegraphics[scale = 0.20]{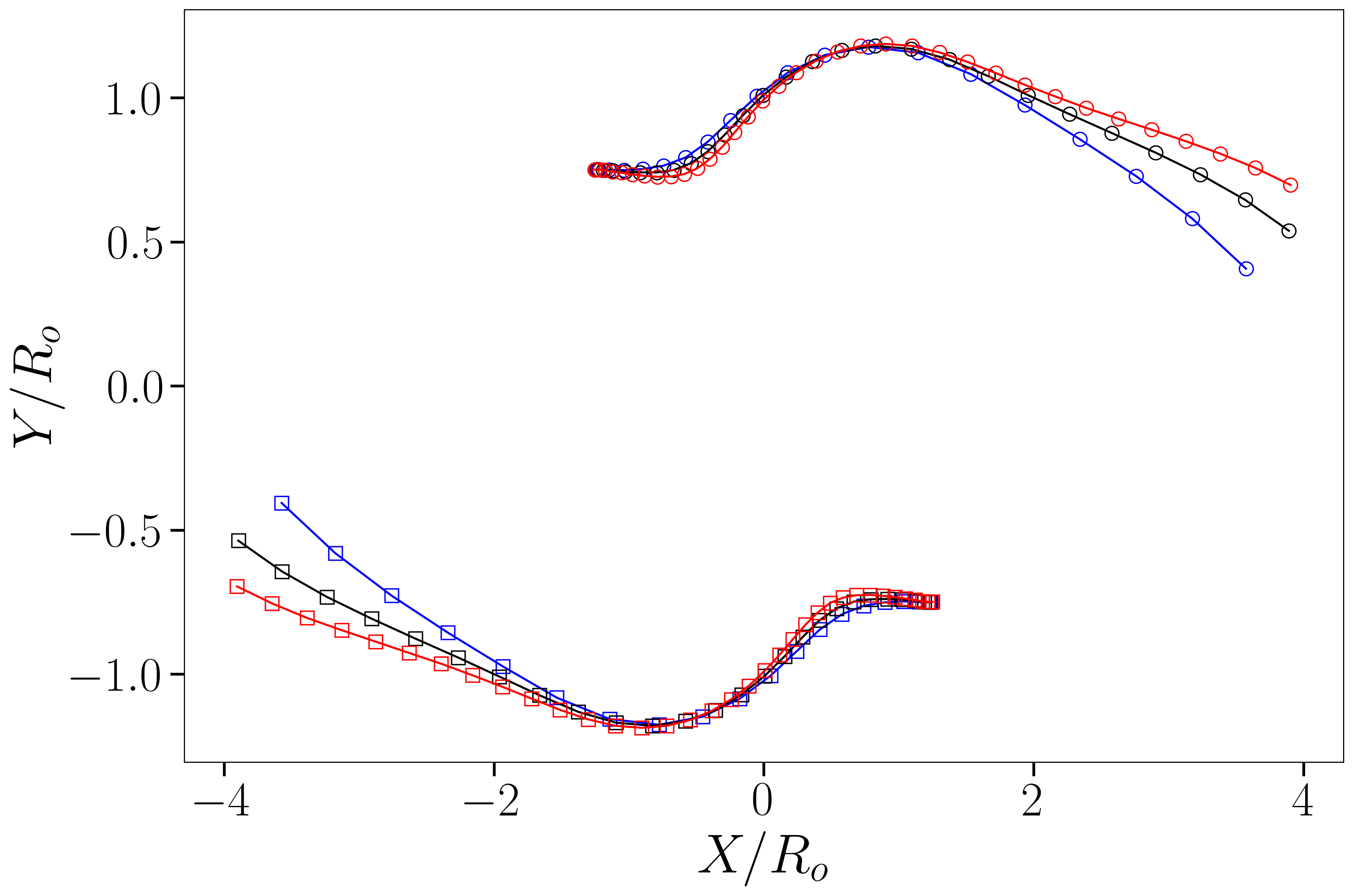} & \includegraphics[scale = 0.20]{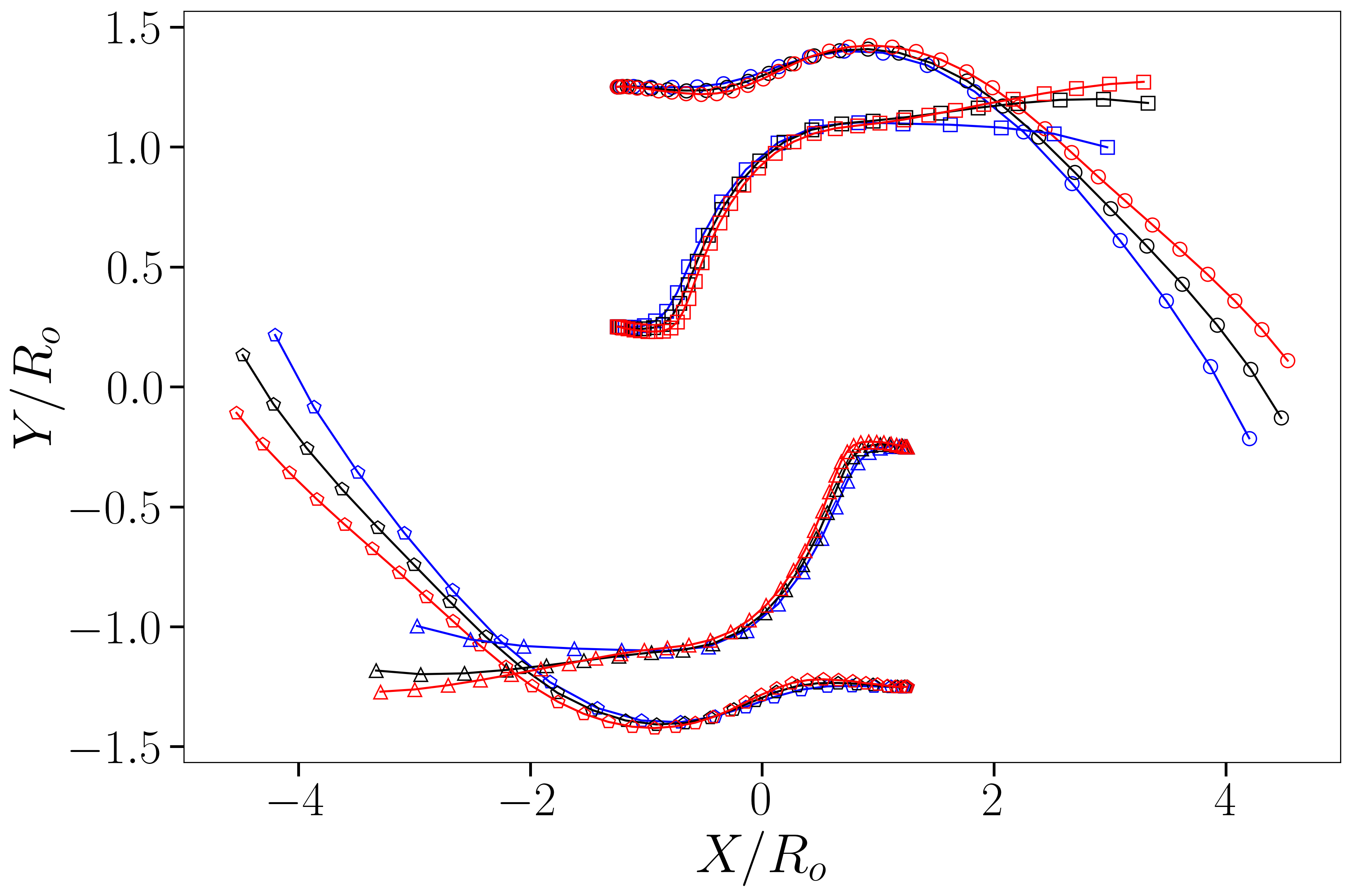} \\
        (e) Shell droplets ($\mu_{12}=60$) &
        (f) Core droplets ($\mu_{12}=60$) \\ 
    \end{tabular}

    \caption{Trajectory of shell and core droplets observed over time for the cases with nine different combination of three distinct density ratios ($\rho_{12}$ = $100$, $200$, $400$) and viscosity ratios ($\mu_{12}$ = $24$, $42$, $60$). The other simulation parameters are $Re = 1.0$, $Ca=0.1$, $\frac{R_o}{H}=0.40$, $\frac{R_i}{R_o} = 0.375$, $\frac{\Delta X}{R_o}=2.50$, and $\frac{\Delta Y}{R_o}=1.50$.}
    \label{fig:two-core-compound-trajectory-dy-1.5}
\end{figure}

In our investigation for the higher value of vertical initial offset, we examined nine different cases, each representing a unique combination of three distinct density ratios ($\rho_{12}$ = $100$, $200$, and $400$) and viscosity ratios ($\mu_{12}$ = $24$, $42$, and $60$). Across all scenarios, the eventual outcome was a pass-over, which was anticipated due to the higher initial vertical offset. This consistent result provided a foundation for further analysis on how the inclusion of multiple core droplets influences the motion of core-shell droplets under different density and viscosity conditions. Figure~\ref{fig:two-core-compound-trajectory-dy-1.5} shows the overall motion trajectories of both shell and core droplets. This Figure is organized into two columns: the left column (Figure~\ref{fig:two-core-compound-trajectory-dy-1.5}(a), (c), and (e)) illustrates the trajectories of the shell droplets across three different viscosity ratio scenarios, each containing trajectory curves for three distinct density ratios. The right column (Figure~\ref{fig:two-core-compound-trajectory-dy-1.5}(b), (d), and (f)) presents the trajectories of the core droplets under the same conditions. From the plots in the left column (Figure~\ref{fig:two-core-compound-trajectory-dy-1.5}), it becomes evident that changes in the viscosity ratio have minimal impact on the trajectories of the shell droplets. This observation suggests that the higher viscosity ratio between the droplet and the continuous fluid does not significantly alter their movement direction or behavior. The shell droplets' paths remain relatively consistent across different viscosity ratios, indicating that viscosity plays a limited role in influencing the overall motion of the droplets in these scenarios. In contrast, the density ratio appears to have a more pronounced effect on the droplet trajectories. Although the general movement trends of both shell and core droplets remain similar across different density ratios, subtle variations can be observed in their courses. Specifically, droplets with higher density ratios tend to move further apart after passing each other. This behavior can be attributed to their increased momentum, underscoring the critical role of density ratios in determining the movement behavior of compound droplets. Collectively, while the viscosity ratio does not significantly impact the droplet trajectories, the density ratio does influence the extent of separation and the overall path of the droplets. 

The increased momentum associated with higher density ratios results in a greater displacement, highlighting the importance of considering density effects in simulations and practical applications. Additionally, it is important to highlight that during the pass-over phenomenon, the core droplets inside the shell exhibit continuous rotational motion around each other, constantly altering their positions. This internal rotation likely contributes to an induced rotation during forward movement of the outer shell, enhancing the overall pass-over dynamics of the shell droplets. The interaction between the inner core droplets generates complex internal forces that influence the outer shell’s behavior, facilitating a more pronounced and efficient pass-over process. This intricate interplay between the core and shell droplets emphasizes the significance of internal dynamics in shaping the overall movement patterns observed in these simulations. 

\begin{figure}[H]
    \centering
    \begin{tabular}{cc}
        \includegraphics[scale = 0.21]{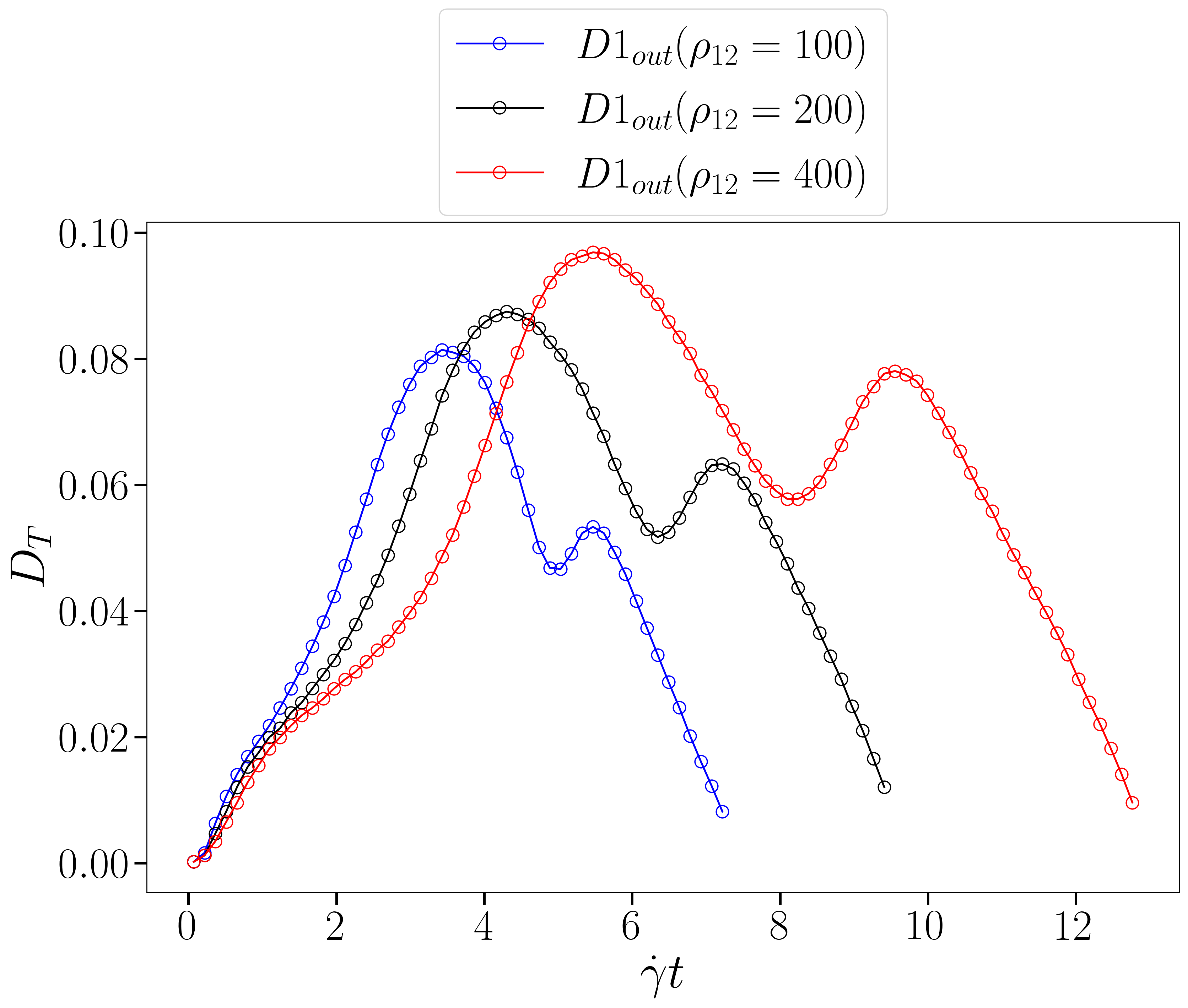} & \includegraphics[scale = 0.21]{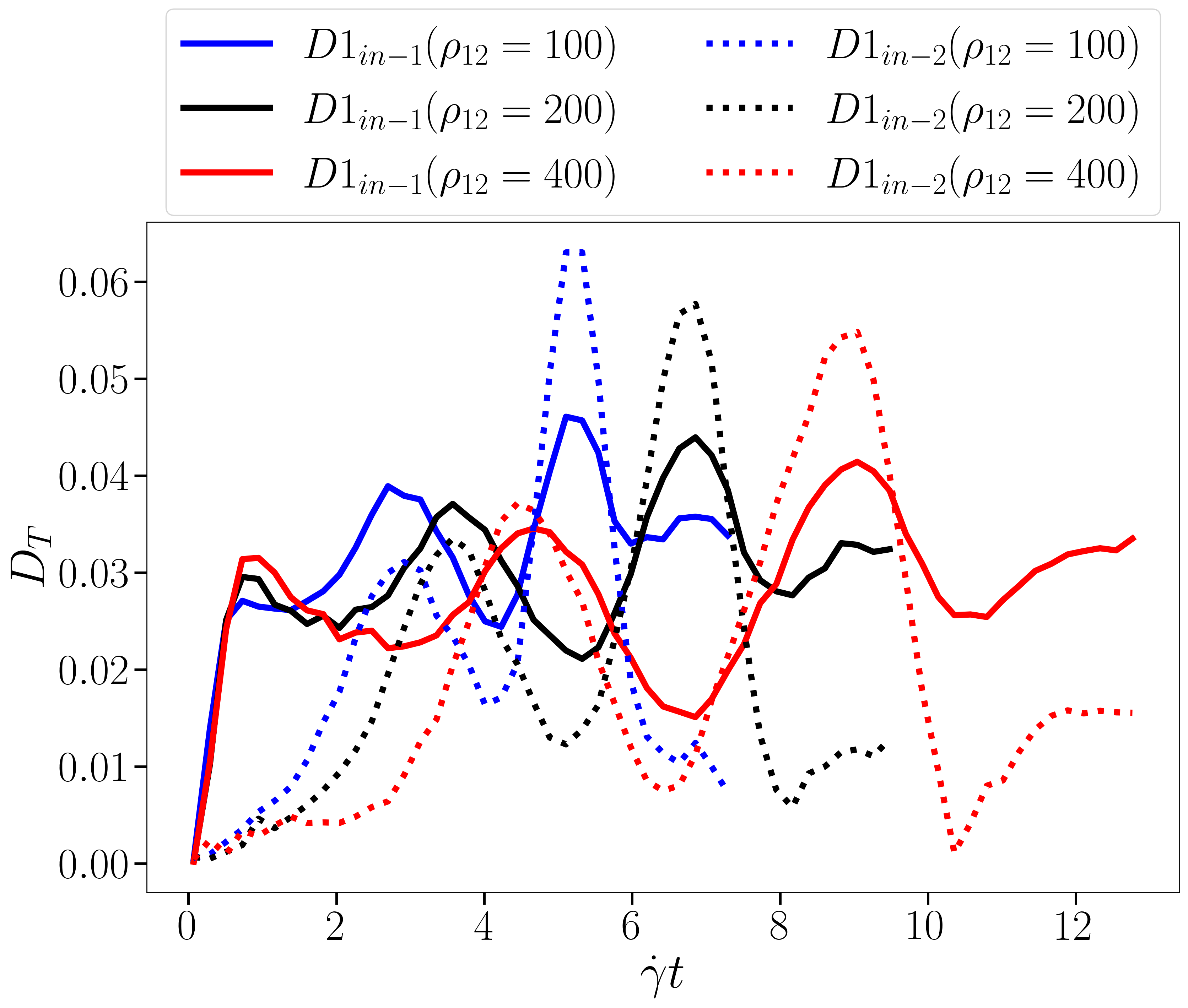} \\
        (a) Shell droplets ($\mu_{12}=24$) &
        (b) Core droplets ($\mu_{12}=24$)  \\ 
    
        \includegraphics[scale = 0.21]{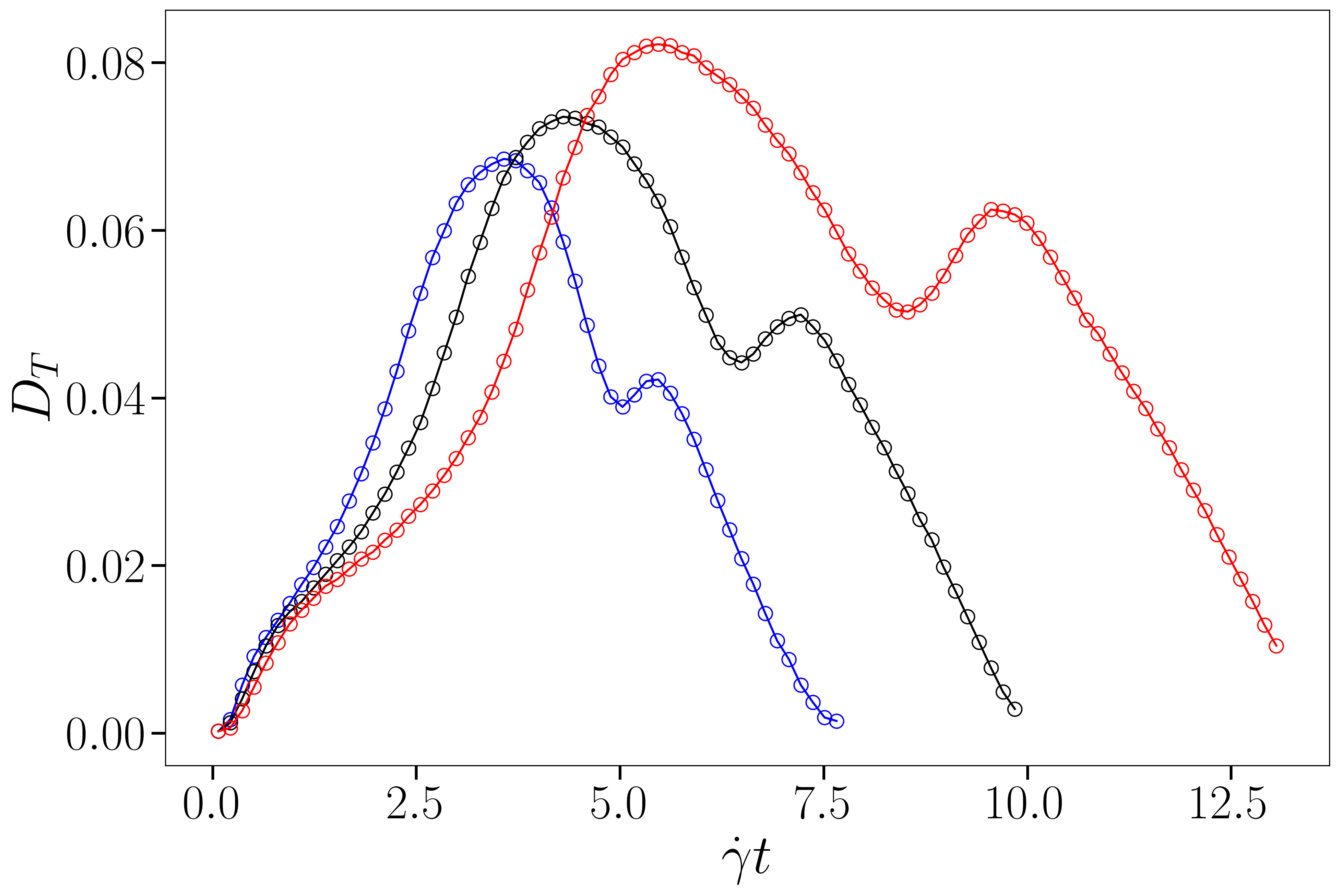} & \includegraphics[scale = 0.21]{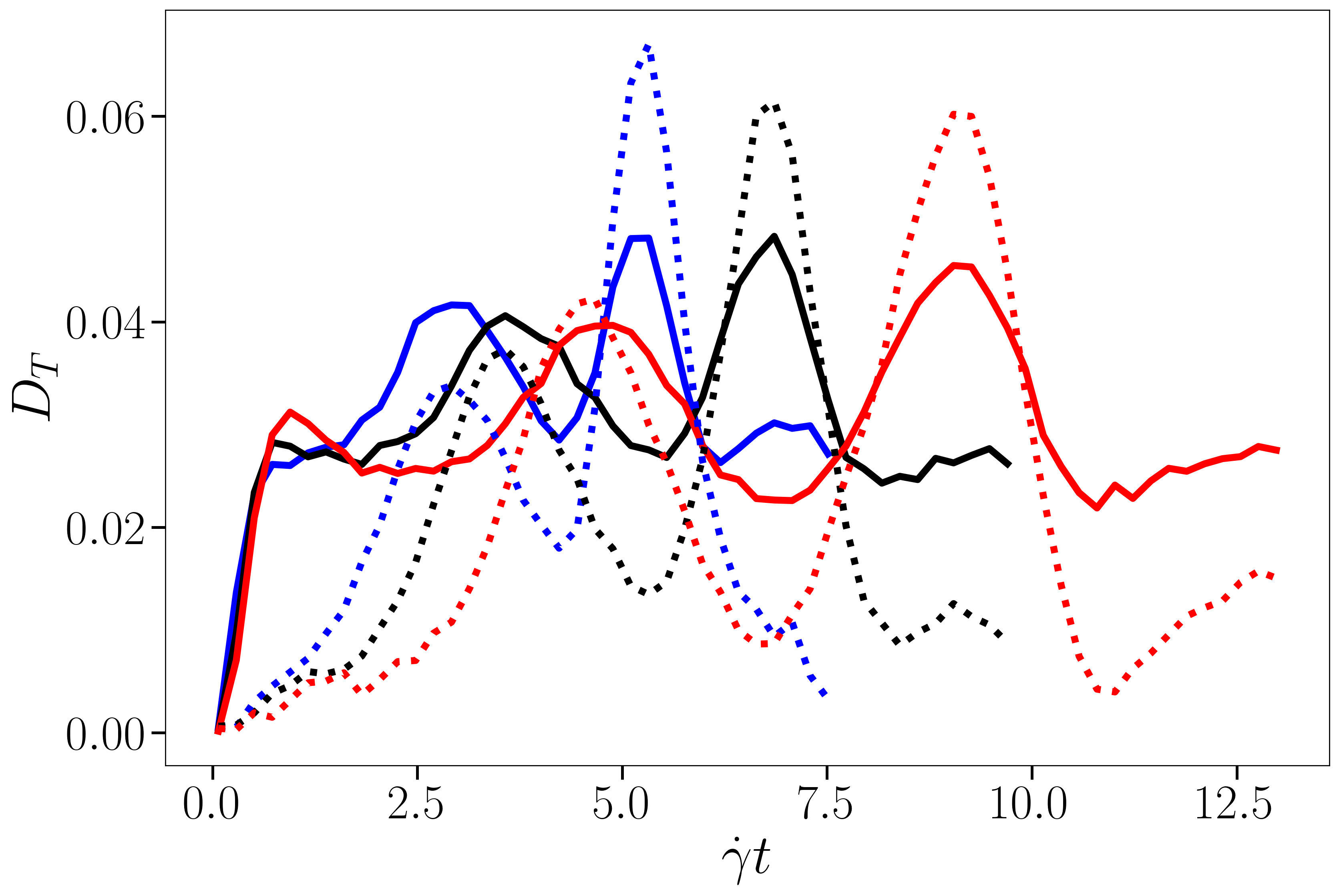} \\
        (c) Shell droplets ($\mu_{12}=42$) &
        (d) Core droplets ($\mu_{12}=42$) \\ 
  
        \includegraphics[scale = 0.21]{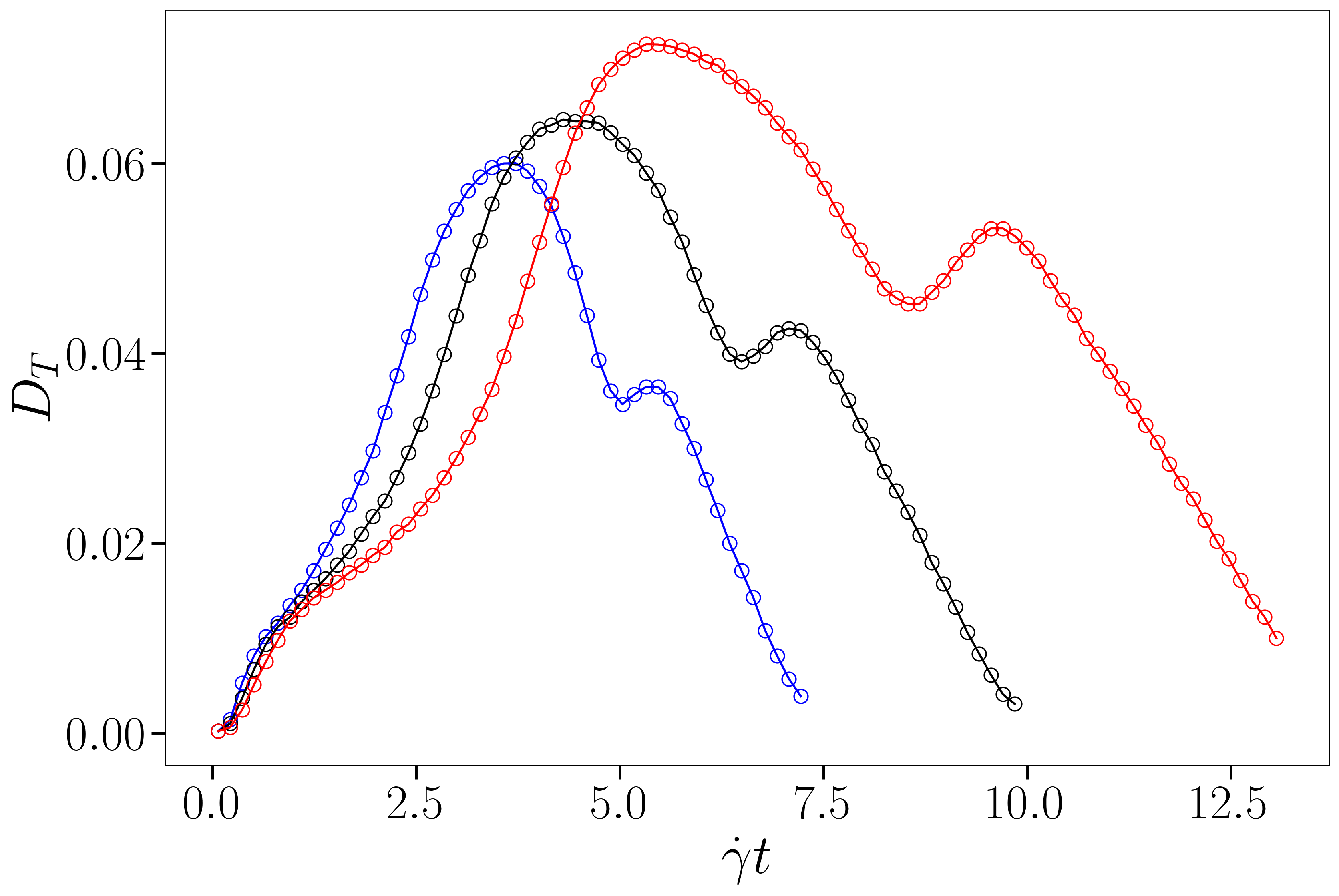} & \includegraphics[scale = 0.21]{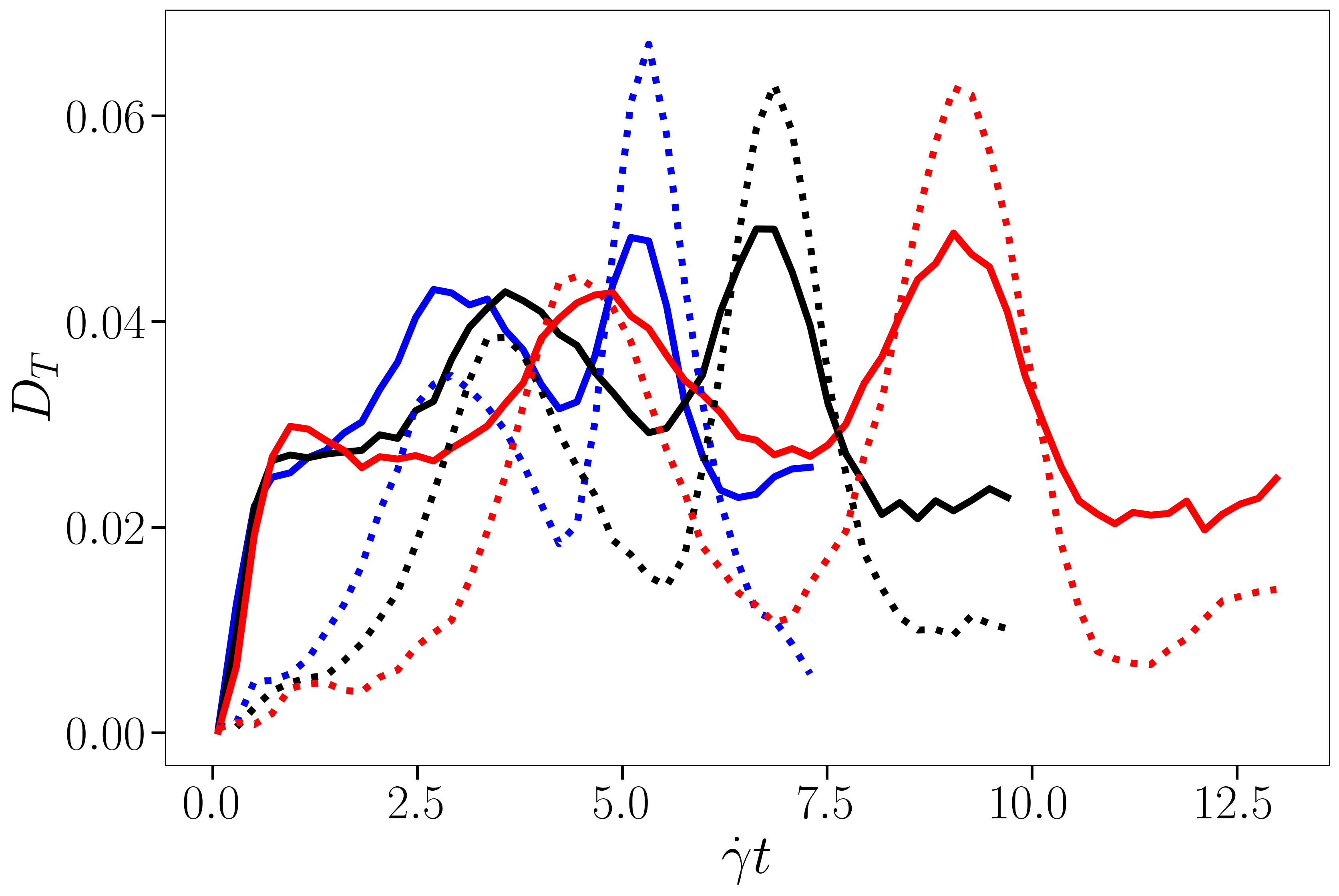} \\
        (e) Shell droplets ($\mu_{12}=60$) &
        (f) Core droplets ($\mu_{12}=60$) \\ 
    \end{tabular}

    \caption{Deformation of shell and core droplets observed over time for the cases with nine different combination of three distinct density ratios ($\rho_{12}$ = $100$, $200$, $400$) and viscosity ratios ($\mu_{12}$ = $24$, $42$, $60$). The other simulation parameters are $Re = 1.0$, $Ca=0.1$, $\frac{R_o}{H}=0.40$, $\frac{R_i}{R_o} = 0.375$, $\frac{\Delta X}{R_o}=2.50$, and $\frac{\Delta Y}{R_o}=1.50$.}
    \label{fig:two-core-compound-deform-dy-1.5}
\end{figure}

Another crucial factor influencing the dynamics of droplet collision is their deformation, which reflects the stress environment they experience. To investigate the deformation of the droplets throughout their evolution, we plotted the deformation curves for both the shell and core droplets in Figure~\ref{fig:two-core-compound-deform-dy-1.5}. Due to the symmetrical nature of the shear effect, we have presented the deformation of only the left-side shell droplet along with its inner droplets. Figure~\ref{fig:two-core-compound-deform-dy-1.5} is organized to illustrate the effects of density and viscosity ratios on the deformation of the shell and core droplets, with the shell droplet's deformation curves displayed in the left column sub-figures and the core droplet's deformation curves in the right column. We observe a general trend where the shell droplets reach a peak deformation during the collision, particularly when the left droplet rises over the top of the right droplet. After this event, their deformation begins to stabilize, showing another small fluctuation caused by another deformation turn during subsequent separation as they continues to pass over each other completely. 

Although viscosity ratio did not significantly impact the trajectory of the droplets, as observed in Figure~\ref{fig:two-core-compound-trajectory-dy-1.5}, Figure~\ref{fig:two-core-compound-deform-dy-1.5} reveals that increased viscosity ratios helps droplets retain their shape, resulting in lower magnitudes of peak deformation throughout the collision process. This indicates that higher viscosity acts as a stabilizing factor, reducing the extent of deformation. Conversely, increased density ratios makes the droplets heavier, slowing their movement and extending the time required to complete the pass-over motion. Despite this, the deformation curves for the shell droplets exhibit similar trends across all three density ratios. This suggests that while density influences the kinetics of the collision process, the overall pattern of deformation remains consistent. The core droplets, having the same viscosity as the surrounding continuous fluid, do not show significant differences in the peak magnitudes of deformation attained during the collision process. However, both inner droplets exhibit continuous fluctuations in deformation due to their internal planetary-like motion and the varying stress environment inside the shell.
Notably, the highest deformation of the inner droplets across all viscosity ratio cases occurs at approximately $\dot{\gamma }t \approx 5.0, 7.0,$ and $10.0$ for $\rho_{12} = 100, 200,$ and $400$, respectively, coinciding with the second peak deformation of the shell droplet. This peak is associated with a low-pressure zone between the shell droplets during their subsequent separation, driven by the rapid vortex flow in the surrounding fluid. At this stage, the additional deformation of the shell droplet exerts a significant influence on the inner droplets, leading to their maximum deformation.

In summary, our findings demonstrate that viscosity and density ratios significantly impact the deformation behavior of compound droplet pairs during collision. Higher viscosity contributes to shape retention and reduced deformation, while increased density affects the duration of the collision process without altering the general deformation pattern. The internal dynamics of the core droplets further highlight the complex interplay of forces within the compound droplet system, emphasizing the intricate nature of droplet interactions under shear flow.

\subsubsection{Moderate vertical initial offset: $\frac{\Delta Y}{R_o}=0.75$}

\begin{figure}[H]
    \centering
    \begin{tabular}{cc}
        \includegraphics[scale = 0.21]{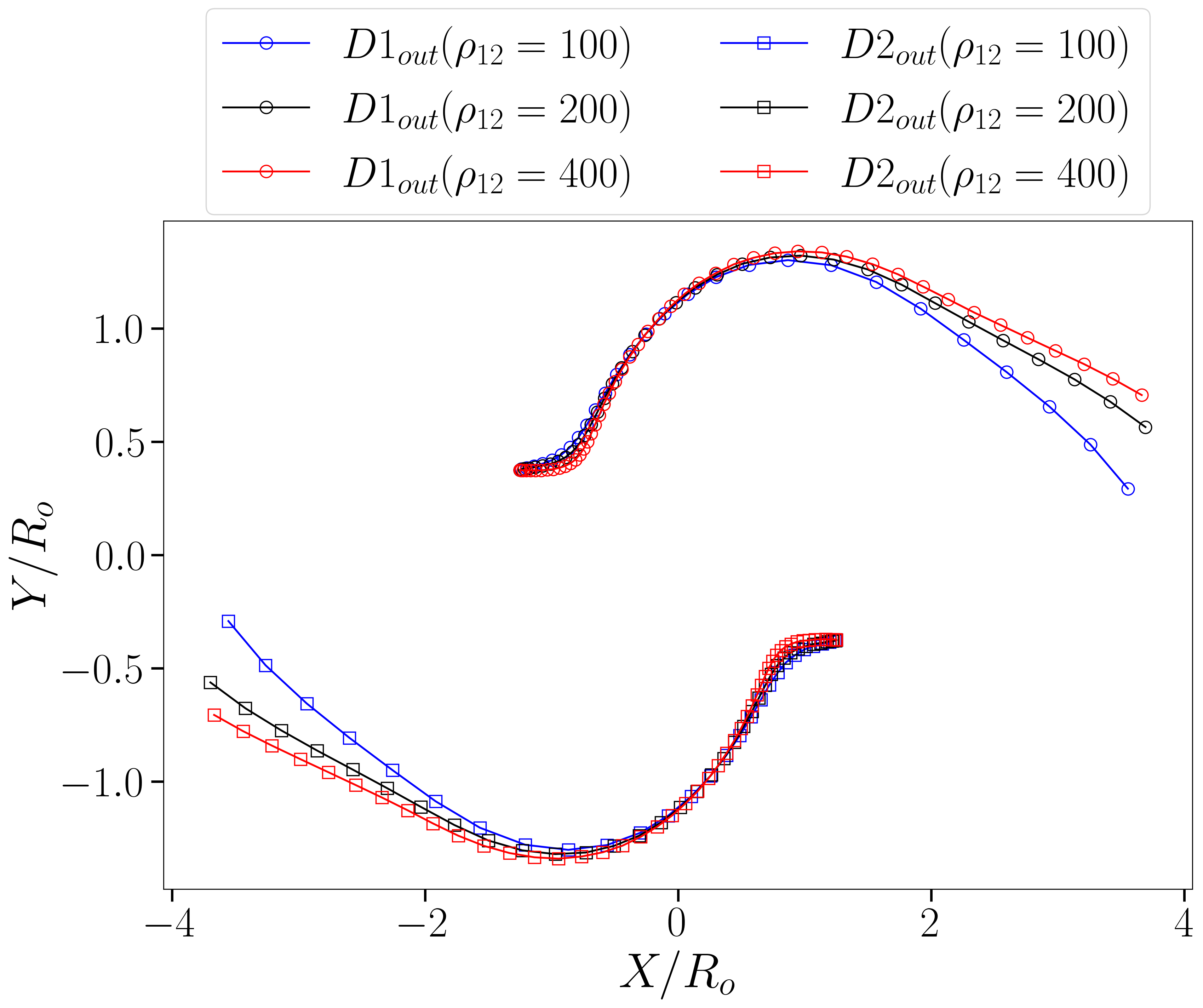} & \includegraphics[scale = 0.21]{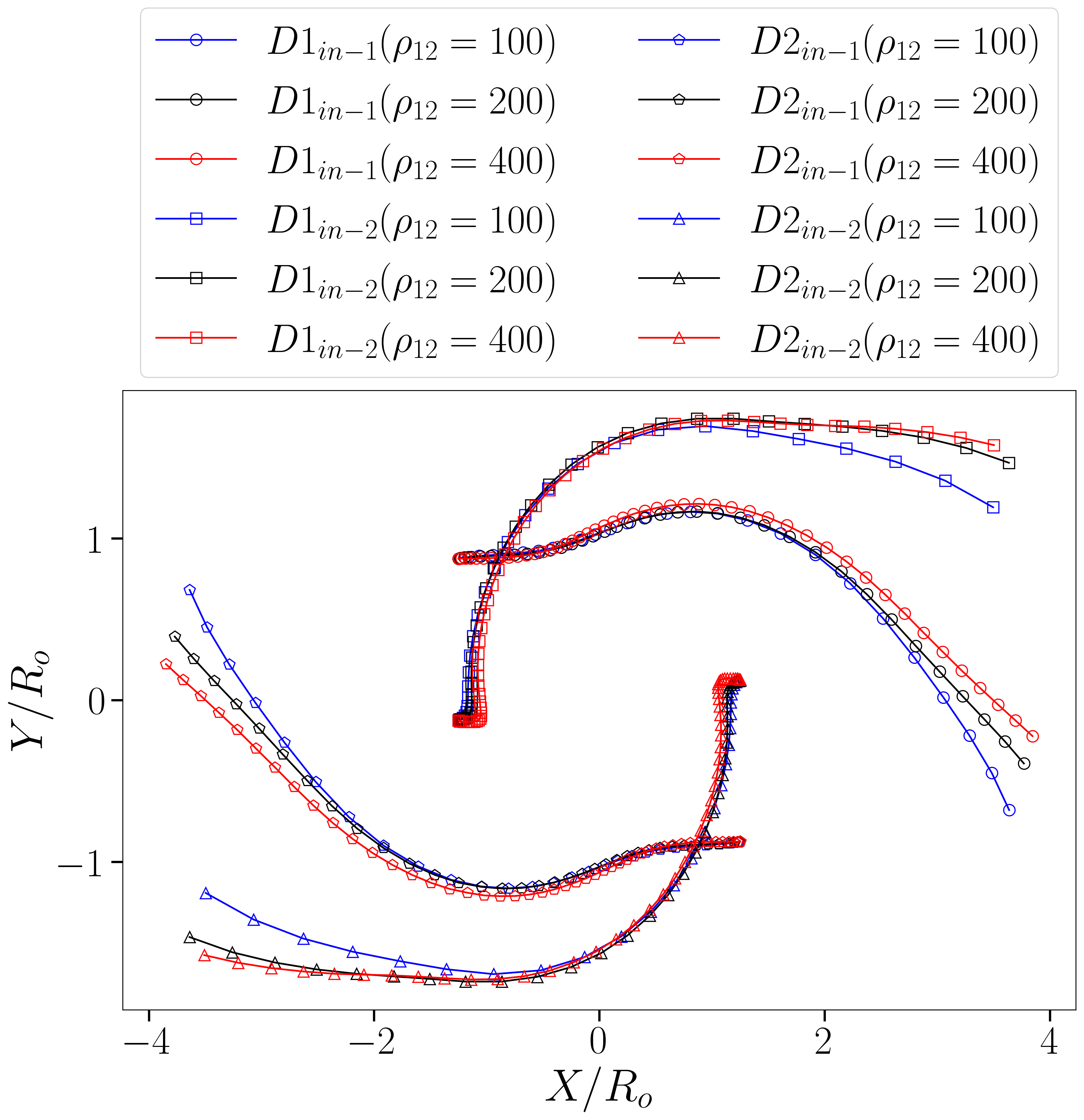} \\
        (a) Shell droplets ($\mu_{12}=24$) &
        (b) Core droplets ($\mu_{12}=24$)  \\ 
    
        \includegraphics[scale = 0.21]{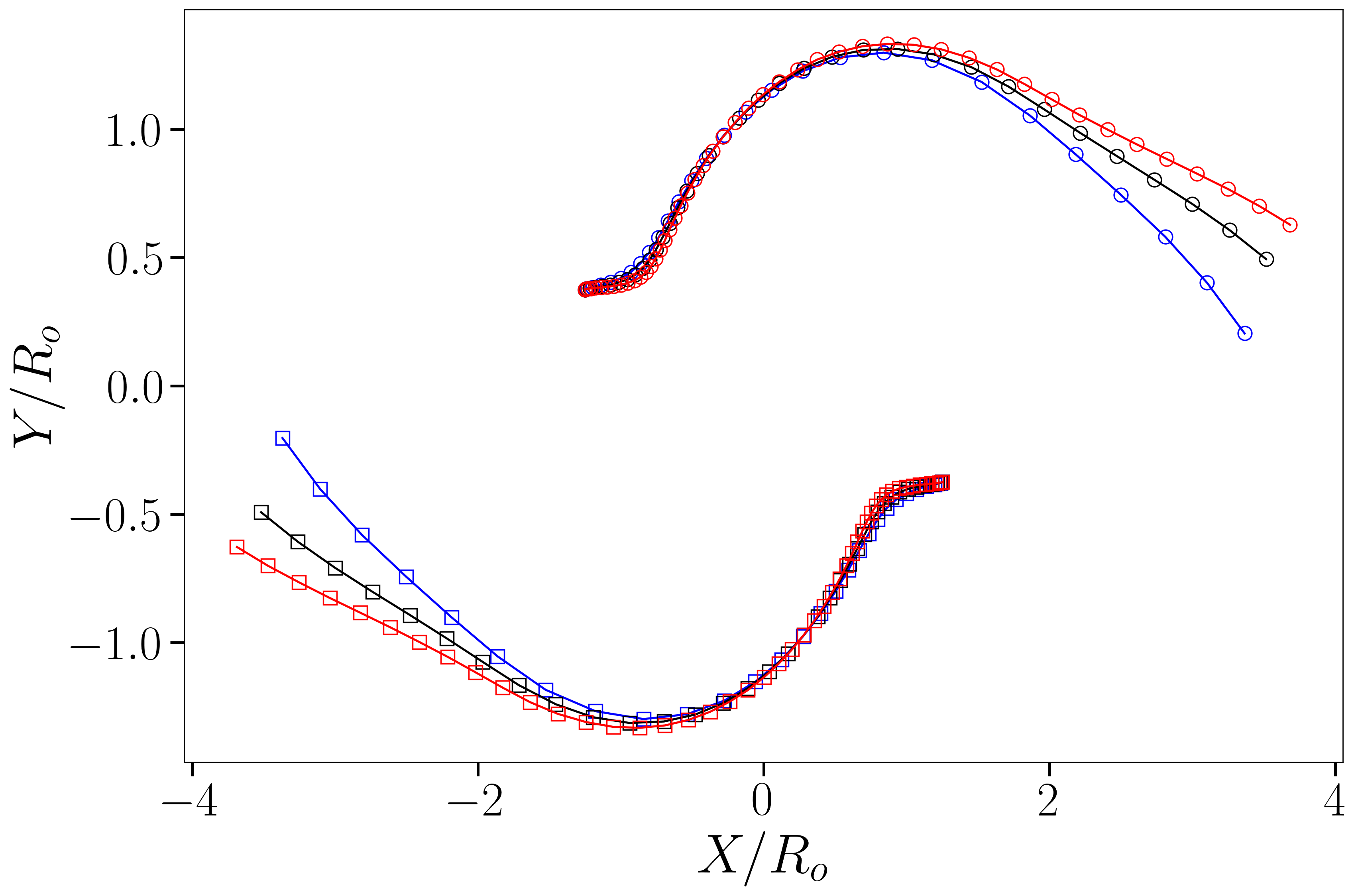} & \includegraphics[scale = 0.21]{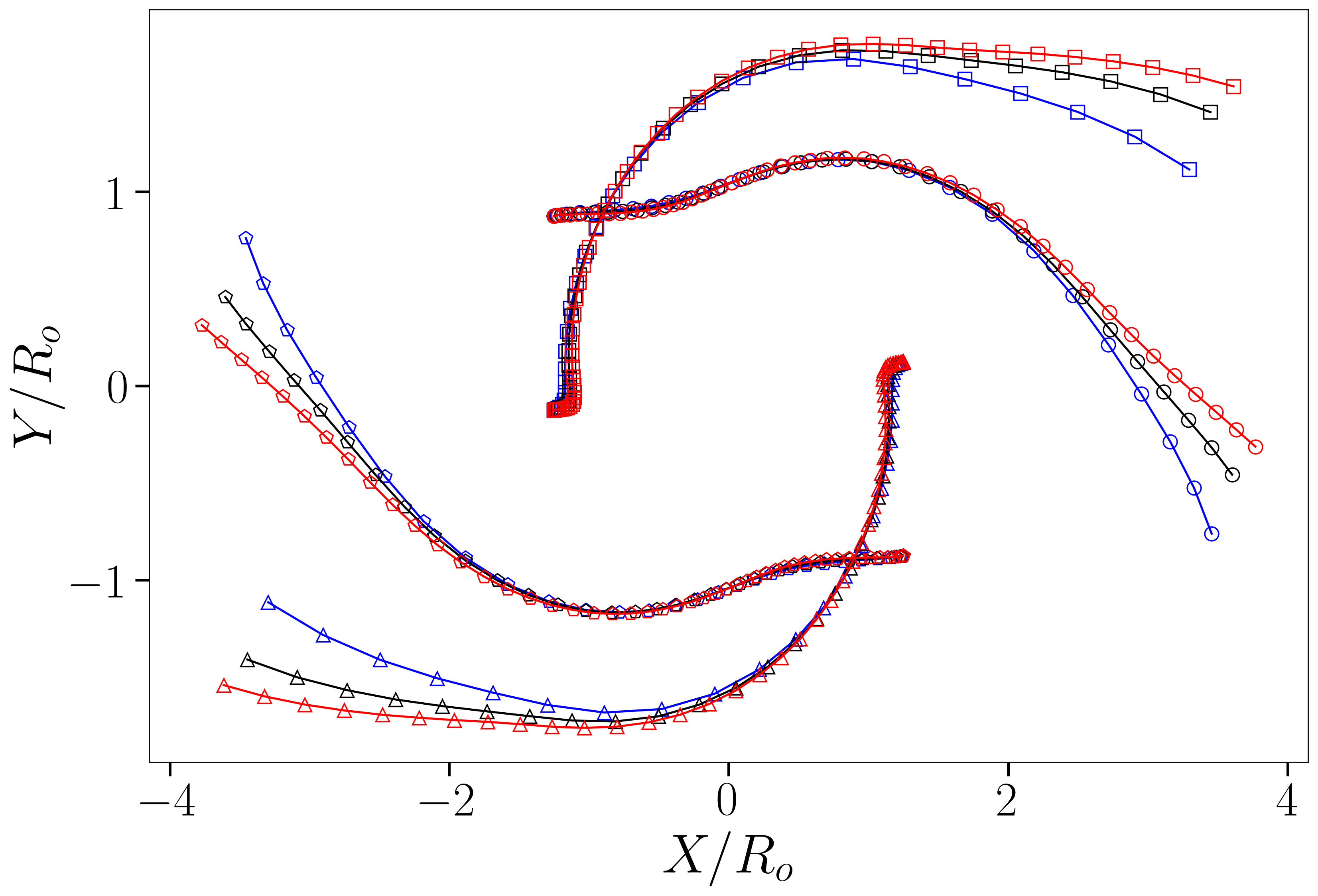} \\
        (c) Shell droplets ($\mu_{12}=42$) &
        (d) Core droplets ($\mu_{12}=42$) \\ 
  
        \includegraphics[scale = 0.21]{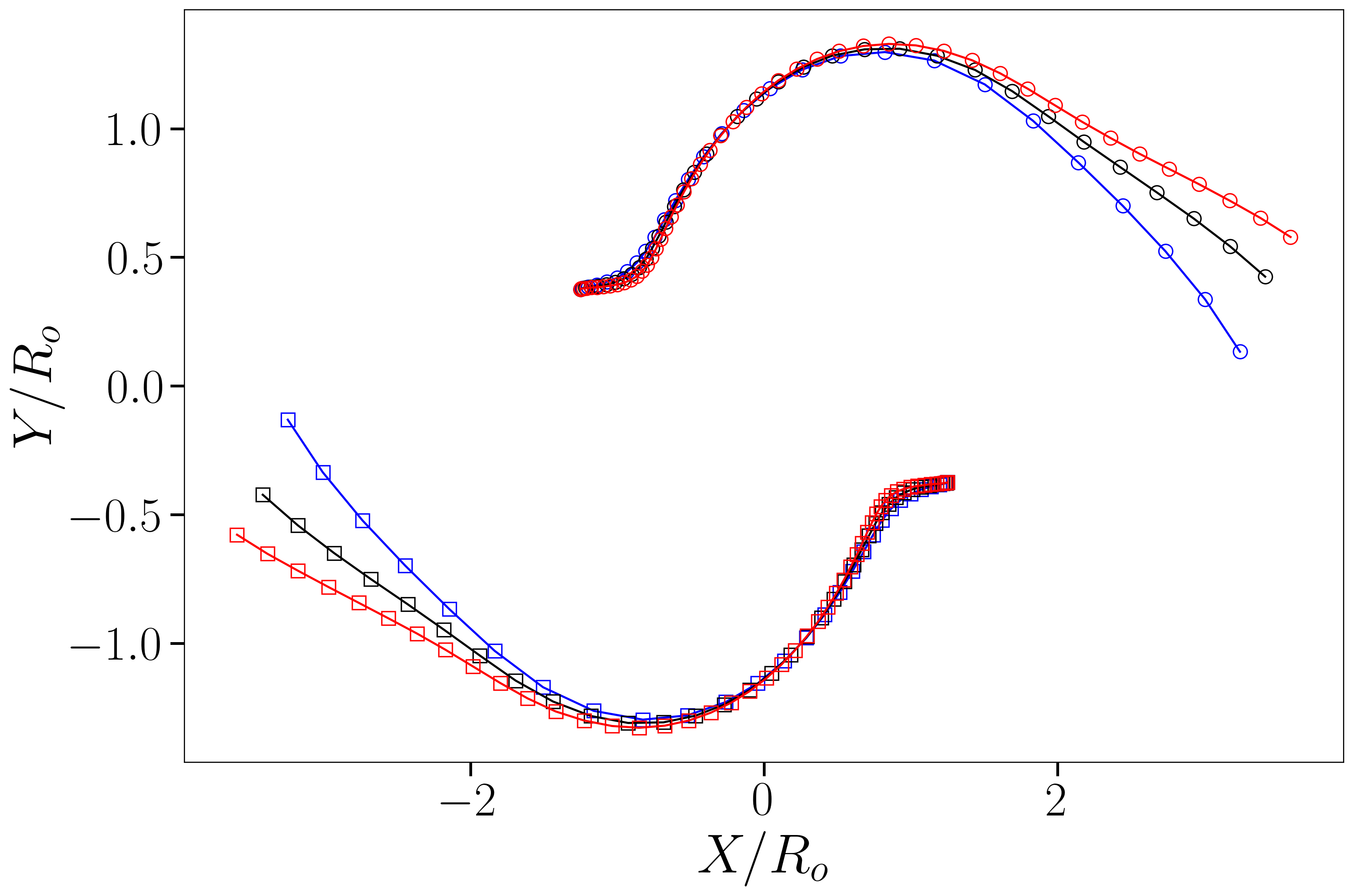} & \includegraphics[scale = 0.21]{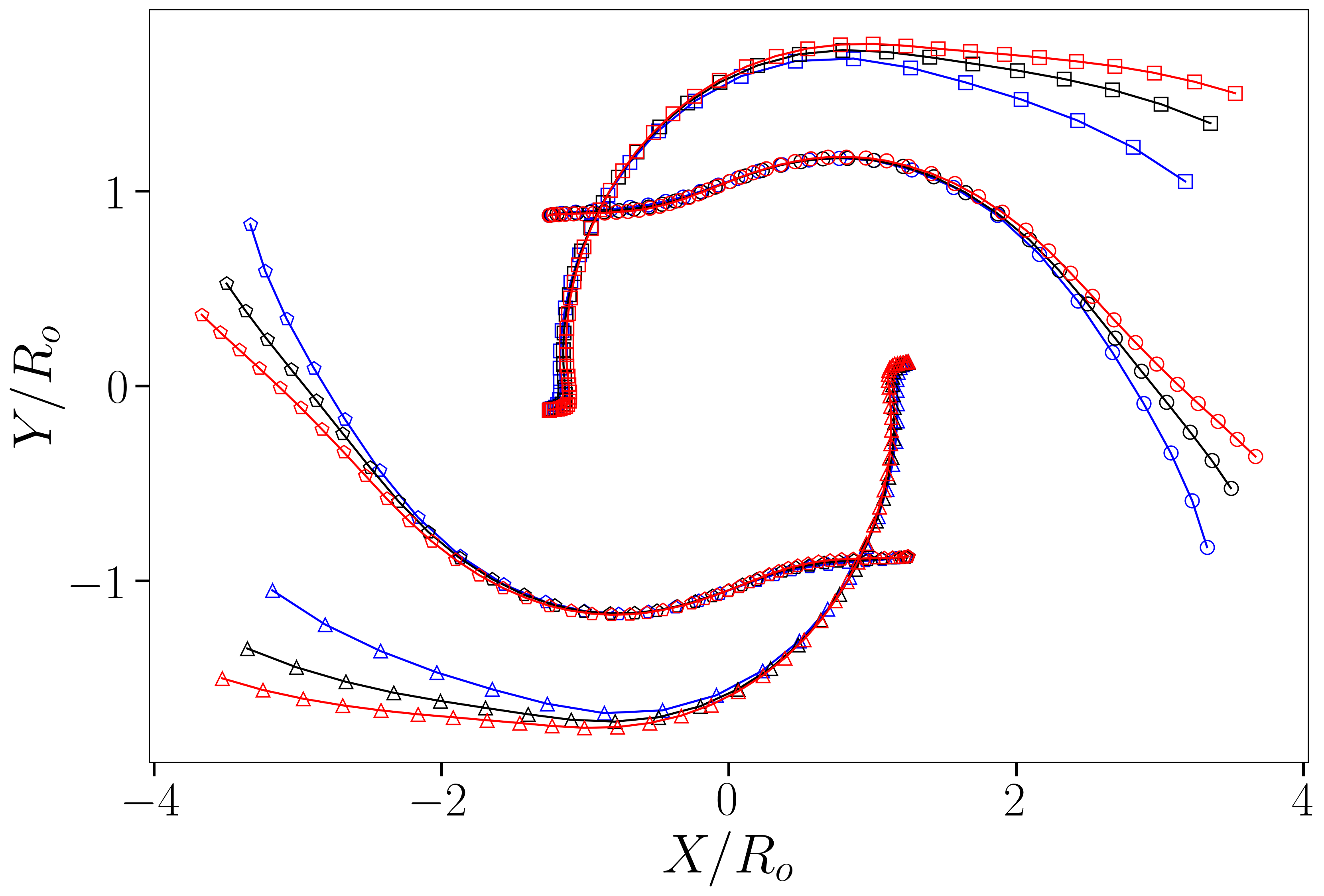} \\
        (e) Shell droplets ($\mu_{12}=60$) &
        (f) Core droplets ($\mu_{12}=60$) \\ 
    \end{tabular}

    \caption{Trajectory of shell and core droplets observed over time for the cases with nine different combination of three distinct density ratios ($\rho_{12}$ = $100$, $200$, $400$) and viscosity ratios ($\mu_{12}$ = $24$, $42$, $60$). The other simulation parameters are $Re = 1.0$, $Ca=0.1$, $\frac{R_o}{H}=0.40$, $\frac{R_i}{R_o} = 0.375$, $\frac{\Delta X}{R_o}=2.50$, and $\frac{\Delta Y}{R_o}=0.75$.}
    \label{fig:two-core-compound-trajectory-dy-0.75}
\end{figure}

Our studies in Section~\ref{sec_5.3}(1) have demonstrated that multicore compound droplet collisions with a high initial vertical offset exhibit different behaviors in terms of deformation and trajectory under varying density and viscosity ratios, although the eventual collision outcome remains a pass-over for all cases. This observation, led us to further investigate the behavior of multi-core compound droplets under different density and viscosity conditions with a reduced initial offset. Recognizing that the initial offset is a critical parameter in altering droplet behavior, we present in this section the results of similar investigations with an initial vertical offset of $\frac{\Delta Y}{R_o}=0.75$, while keeping all other parameters fixed.
Even with this moderate initial vertical offset, all nine combinations of density and viscosity ratios resulted in pass-over events. This consistent outcome suggests that the inclusion of two core droplets and their rotational movement inside the shell droplets plays a significant role in leading to pass-over events. Notably, in our earlier studies, cases with a single core droplet exhibited coalescence at similar initial vertical offsets~\cite{al2024hydrodynamic}, highlighting the unique dynamics introduced by the presence of multiple core droplets. Figure~\ref{fig:two-core-compound-trajectory-dy-0.75}  presents the trajectory plots for both the shell and core droplets at a initial vertical offset of $\frac{\Delta Y}{R_o}=0.75$. 

The analysis reveals that, even with a moderate initial vertical offset, both the shell and core droplets exhibit similar movement trends. 
No significant effect is observed due to varying viscosity. However, droplets with lower density pass over more quickly and direct their motion towards the central regions of the domain, while higher density shell droplets follow a similar path but with slower movement, as indicated by the offset in their trajectory curves in Figure~\ref{fig:two-core-compound-trajectory-dy-0.75}(a), (c), and (e). Additionally, similar to the case with an initial offset of $\frac{\Delta Y}{R_o}=1.5$, the core droplets in this scenario also flip their vertical positions (Figure~\ref{fig:two-core-compound-trajectory-dy-0.75}(b), (d), and (f)) through rotation inside the shell droplet during the pass-over process.

\begin{figure}[H]
    \centering
    \begin{tabular}{cc}
        \includegraphics[scale = 0.21]{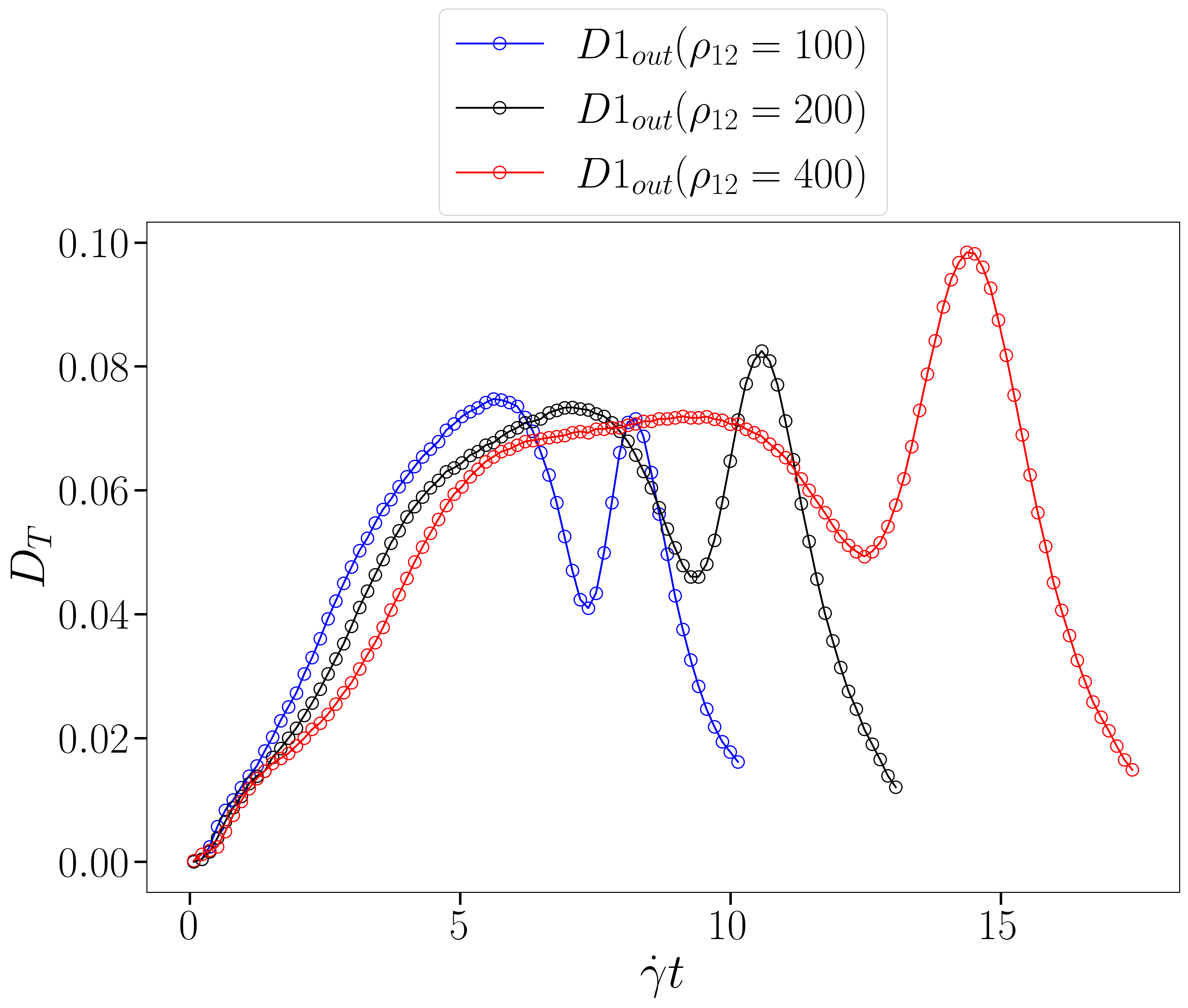} & \includegraphics[scale = 0.21]{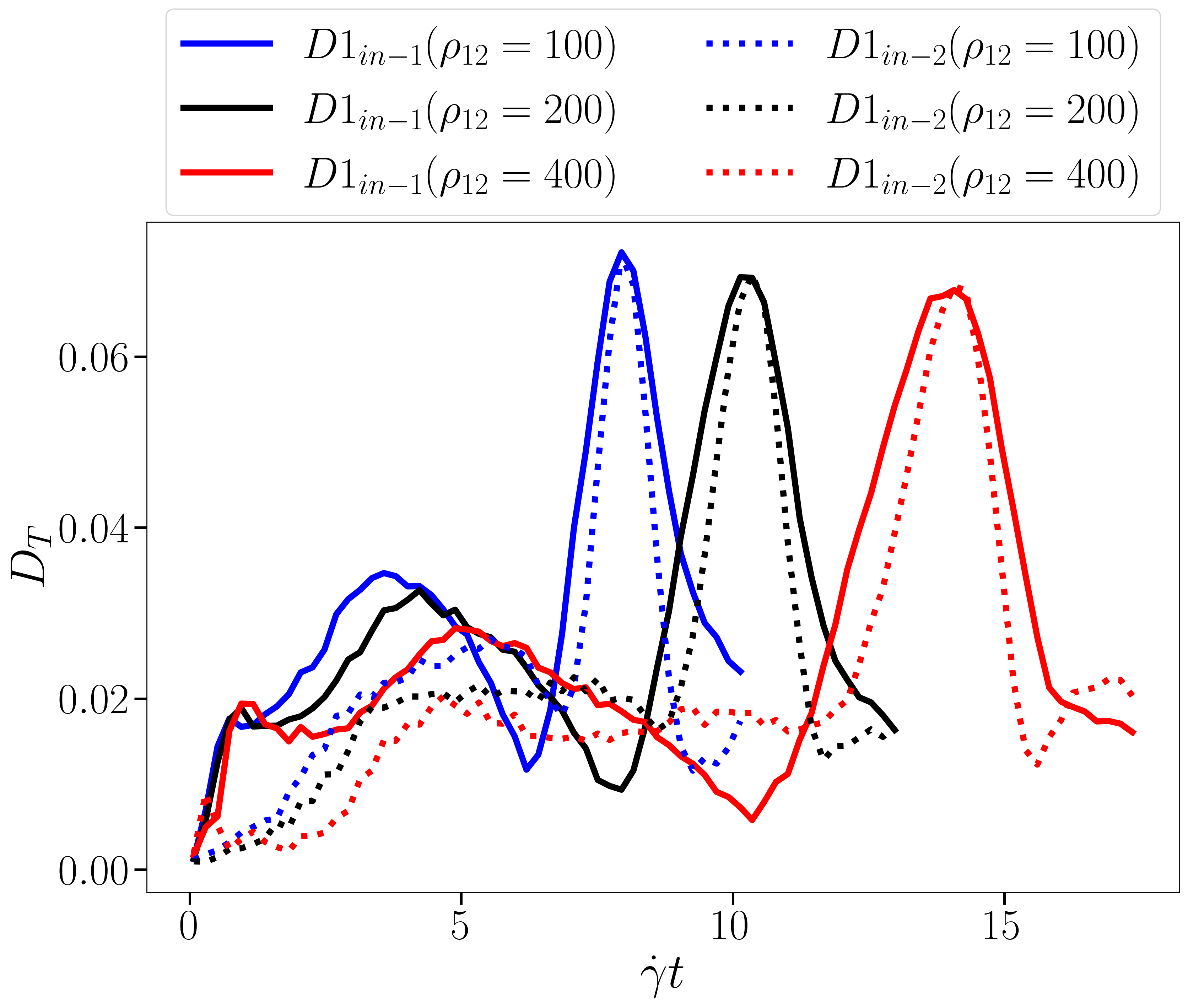} \\
        (a) Shell droplets ($\mu_{12}=24$) &
        (b) Core droplets ($\mu_{12}=24$)  \\ 
    
        \includegraphics[scale = 0.21]{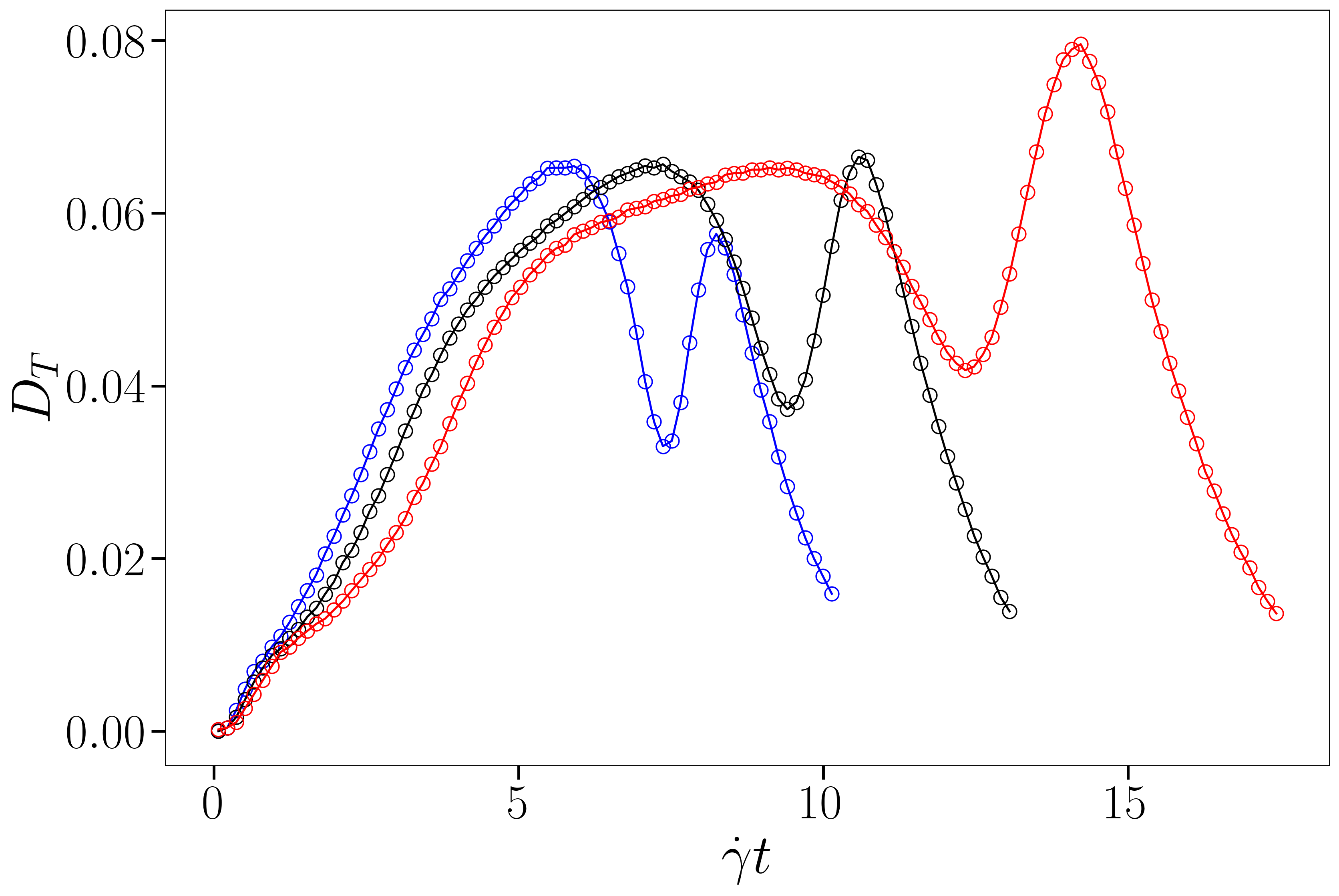} & \includegraphics[scale = 0.21]{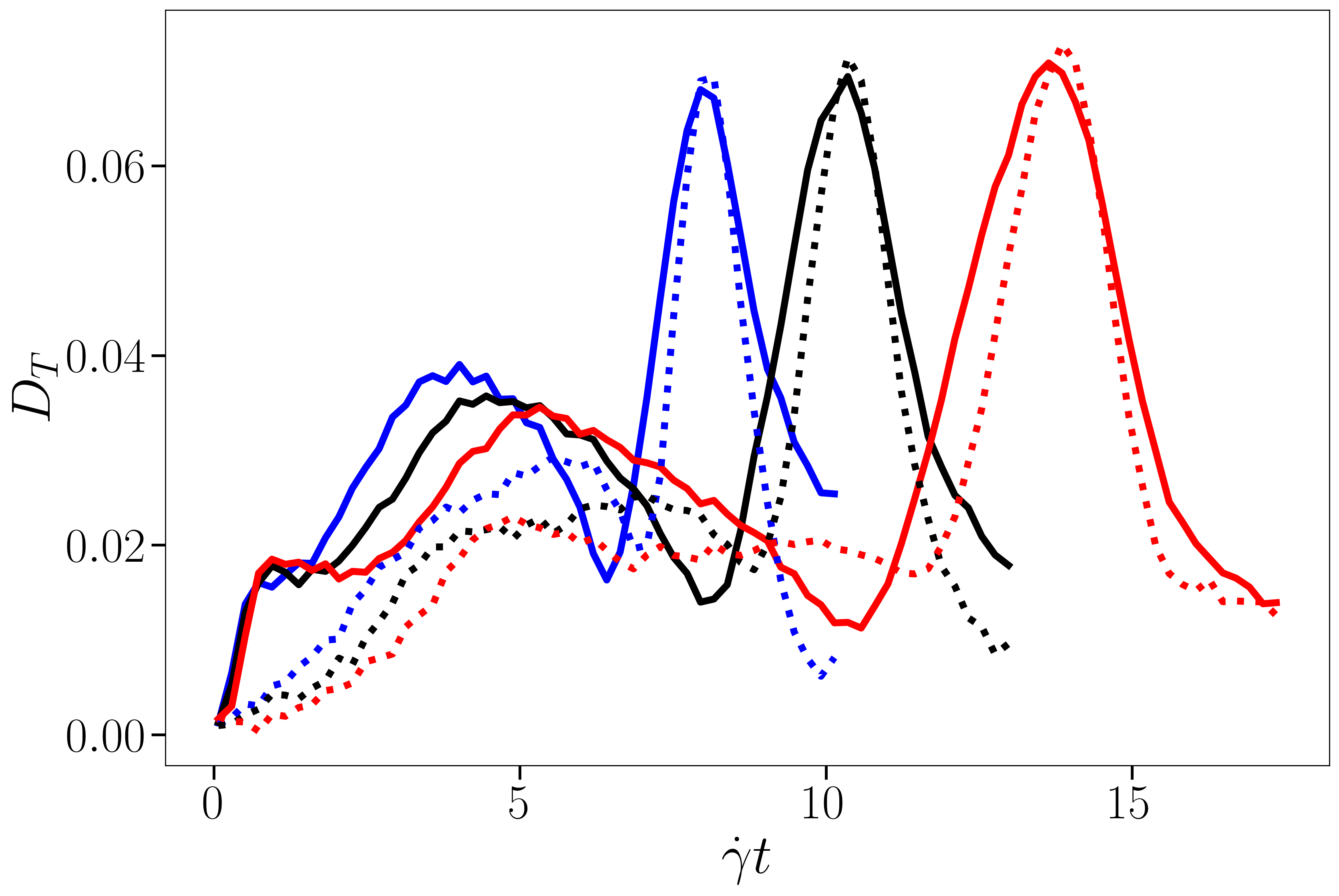} \\
        (c) Shell droplets ($\mu_{12}=42$) &
        (d) Core droplets ($\mu_{12}=42$) \\ 
  
        \includegraphics[scale = 0.21]{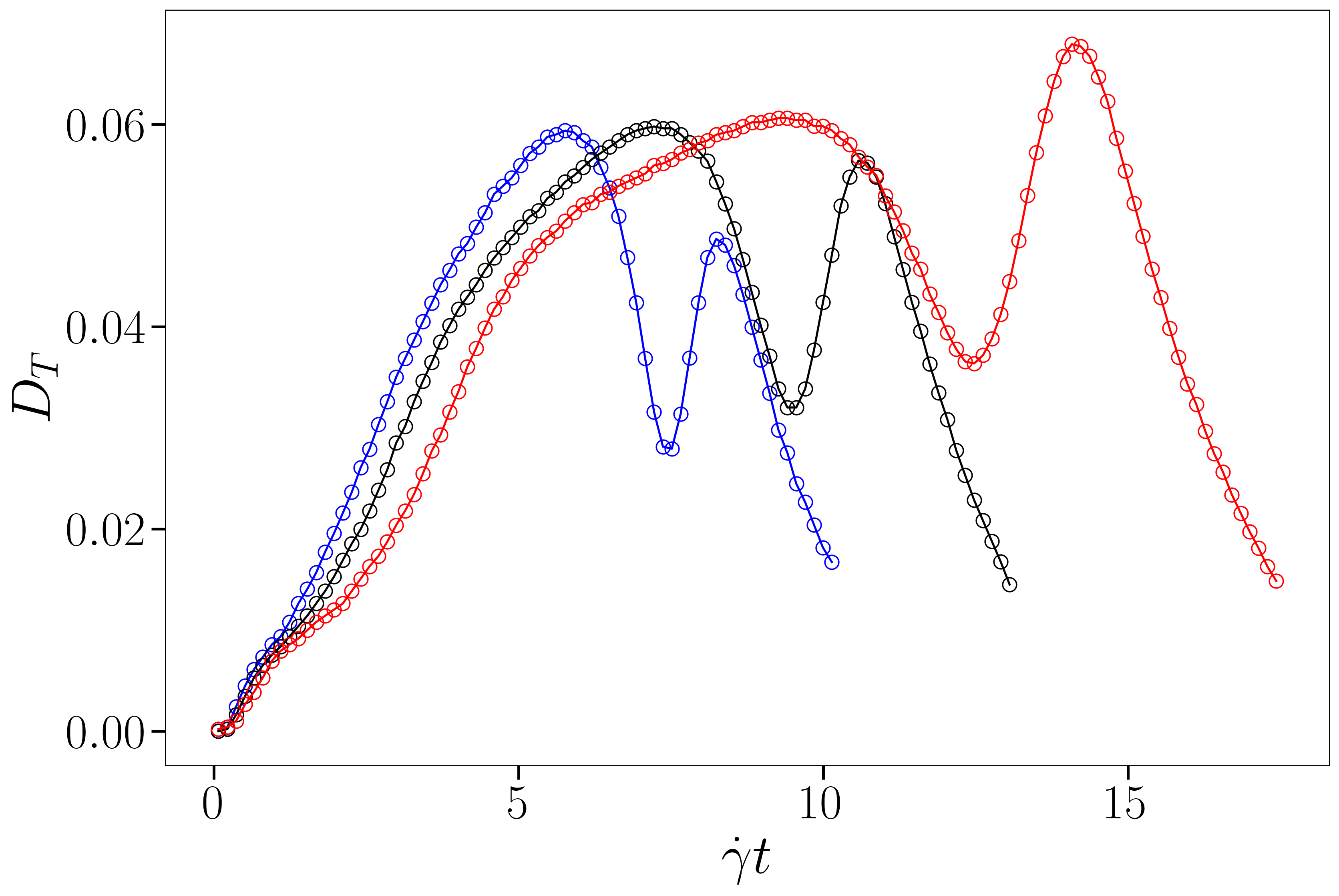} & \includegraphics[scale = 0.21]{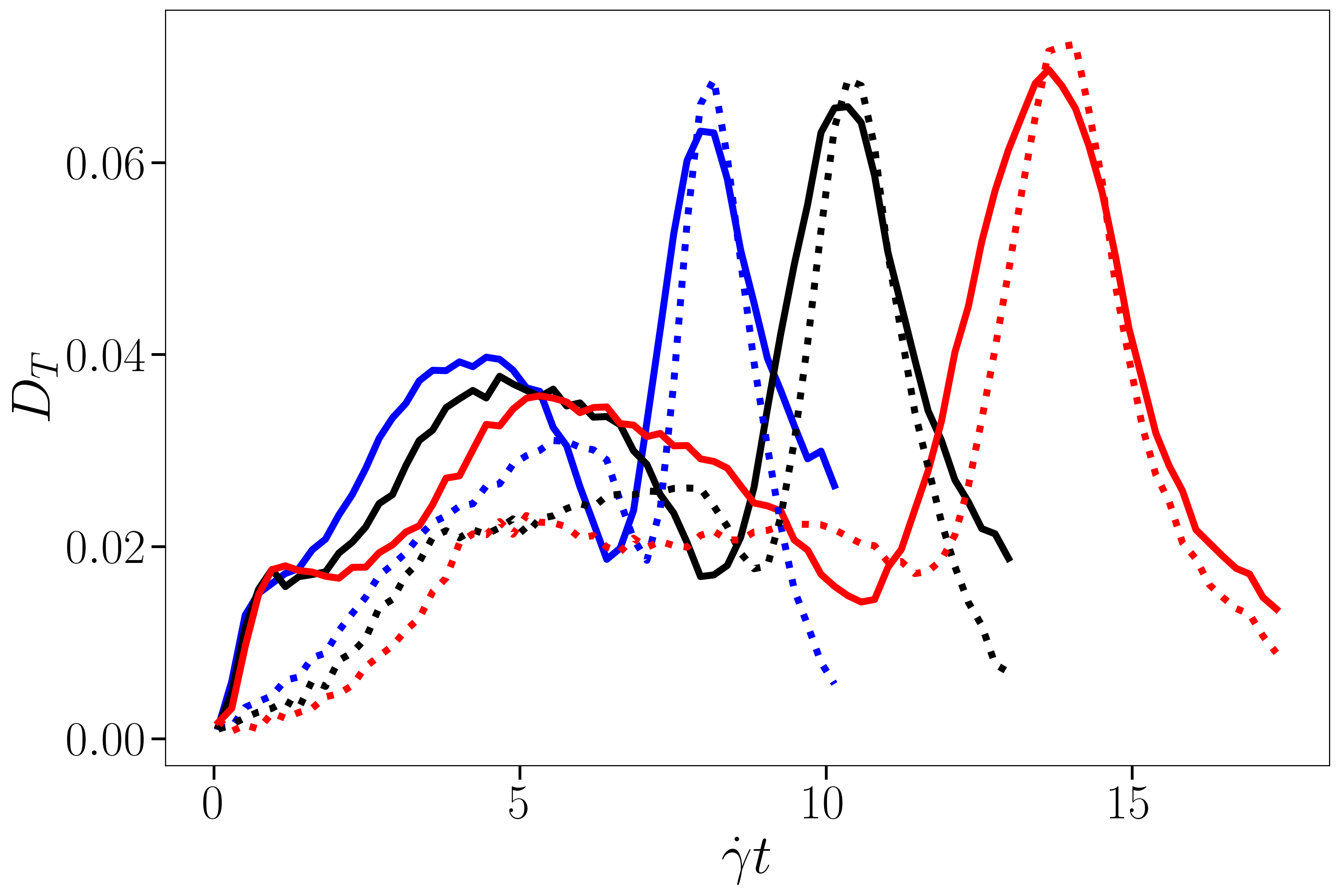} \\
        (e) Shell droplets ($\mu_{12}=60$) &
        (f) Core droplets ($\mu_{12}=60$) \\ 
    \end{tabular}

    \caption{Deformation of shell and core droplets observed over time for the cases with nine different combination of three distinct density ratios ($\rho_{12}$ = $100$, $200$, $400$) and viscosity ratios ($\mu_{12}$ = $24$, $42$, $60$). The other simulation parameters are $Re = 1.0$, $Ca=0.1$, $\frac{R_o}{H}=0.40$, $\frac{R_i}{R_o} = 0.375$, $\frac{\Delta X}{R_o}=2.50$, and $\frac{\Delta Y}{R_o}=0.75$.}
    \label{fig:two-core-compound-deform-dy-0.75}
\end{figure}

The deformation curves, shown in Figure~\ref{fig:two-core-compound-deform-dy-0.75} for both shell and core droplets (with only the left side shell droplet and its inner core droplet presented due to shear symmetry), further elucidate the dynamics of the collision. The shell droplet exhibits two peaks in deformation: the first peak occurs during the initial collision when the left droplet begins to rise over the right droplet, and the second peak appears during the separation phase. As the droplets complete the pass-over, their deformation stabilizes.
Comparing the deformation curves for an initial vertical offset of $\frac{\Delta Y}{R_o}=0.75$ with those for an initial offset of $\frac{\Delta Y}{R_o}=1.5$, it is observed that the first peak is higher in magnitude for $\frac{\Delta Y}{R_o}=1.5$, while it is lower for $\frac{\Delta Y}{R_o}=0.75$. A higher initial offset provides more space for the pass-over, allowing the droplets to deform more readily due to reduced continuous fluid film thickness between them. Conversely, with $\frac{\Delta Y}{R_o}=0.75$, the droplets encounter greater resistance from the bulk continuous fluid film between them, resulting in less deformation during the collision. This effect is more pronounced in cases with higher density ratios, where the first peak remains relatively flat for an extended period, while the second peak exhibits higher magnitudes compared to the other two density ratio cases. This indicates that increased density ratios impede deformation during the initial collision, especially when the initial offset is reduced, highlighting the influence of initial vertical alignment of the shell droplets on their deformation.

In terms of viscosity, the effects observed are consistent with those noted in Section~\ref{sec_5.3}(1), where increased viscosity creates greater resistance to deformation. For the core droplets, the deformation curves show continuous fluctuations. In this scenario, the lower initial vertical offset between the shell droplets causes the inner droplets to experience a more intense stress environment, leading the two core droplets to reach similar magnitudes of peak deformation while exhibiting continuous fluctuations due to their rotational motion.

\subsubsection{Low vertical initial offset: $\frac{\Delta Y}{R_o}=0.15$}

\begin{figure}[H]
    \centering
    \begin{tabular}{cc}
        \includegraphics[scale = 0.21]{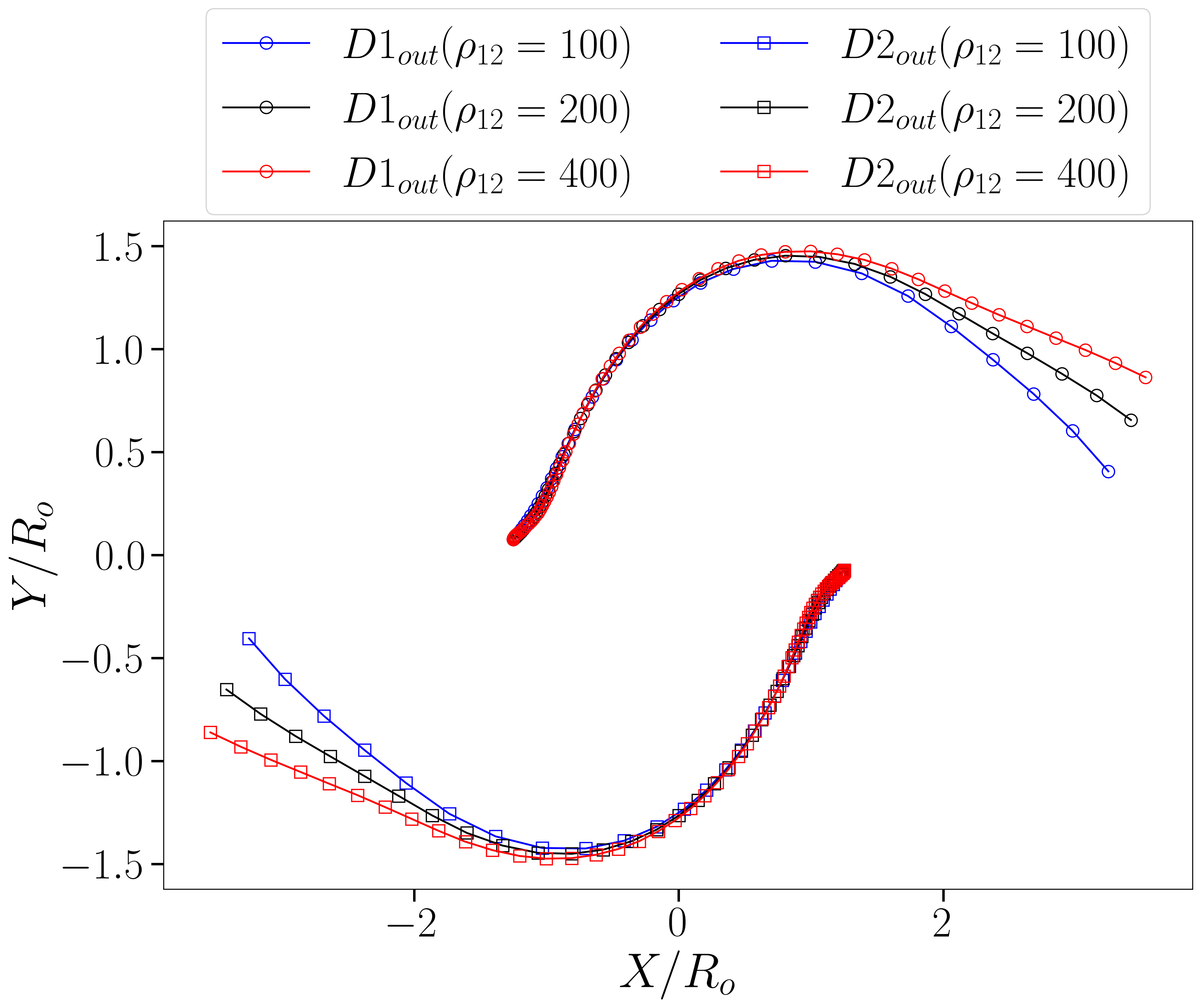} & \includegraphics[scale = 0.21]{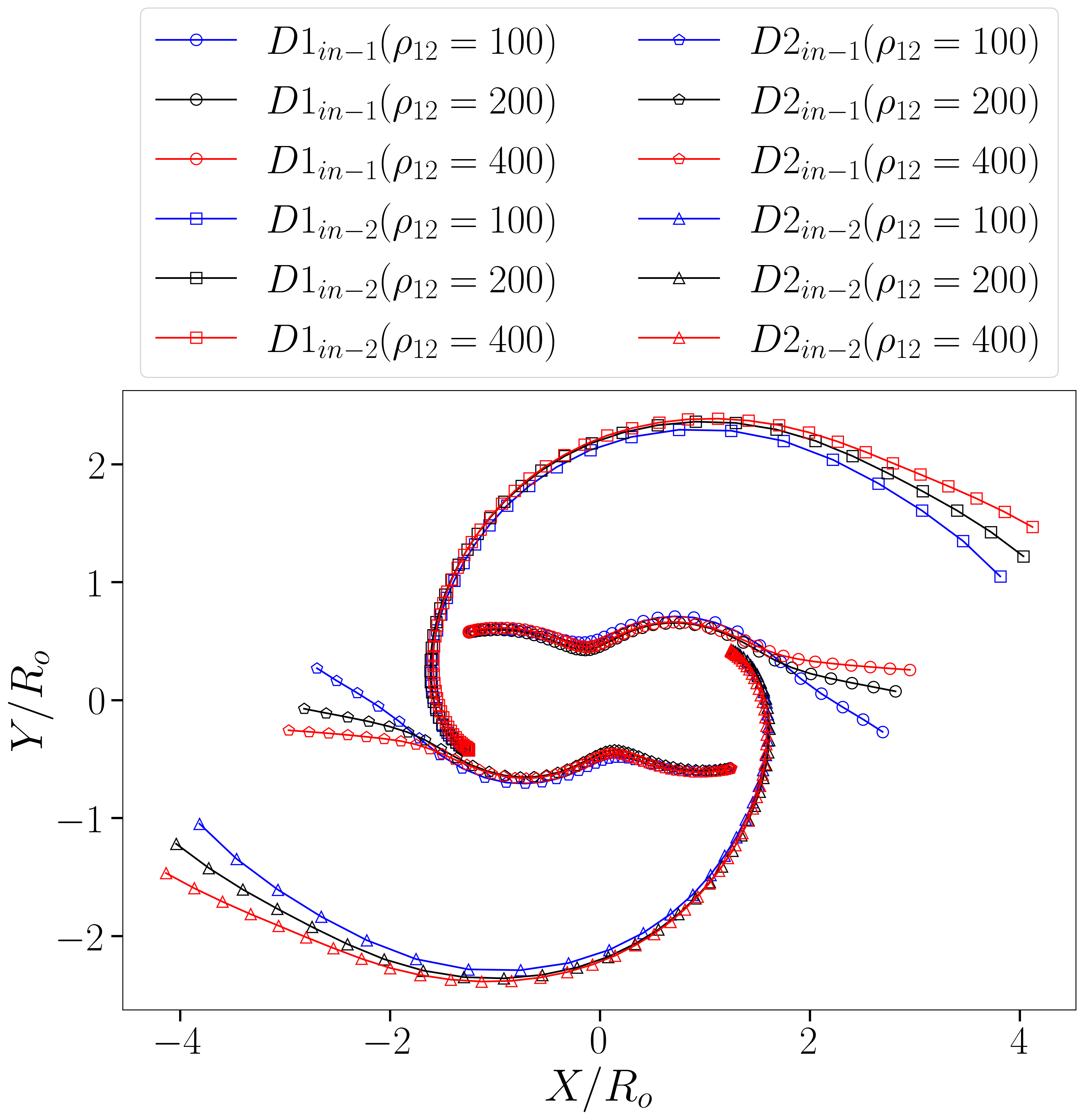} \\
        (a) Shell droplets ($\mu_{12}=24$) &
        (b) Core droplets ($\mu_{12}=24$)  \\ 
    
        \includegraphics[scale = 0.21]{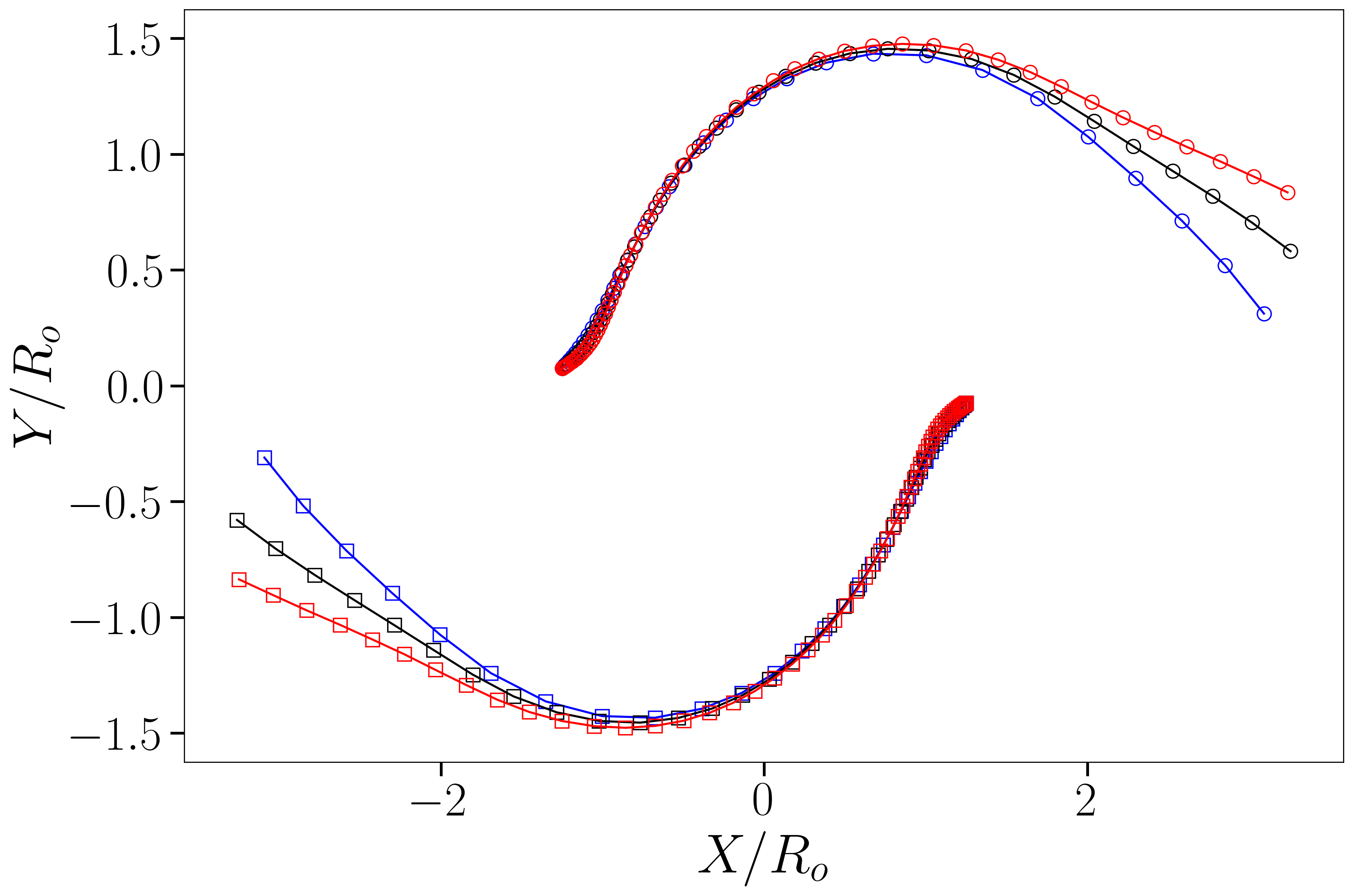} & \includegraphics[scale = 0.21]{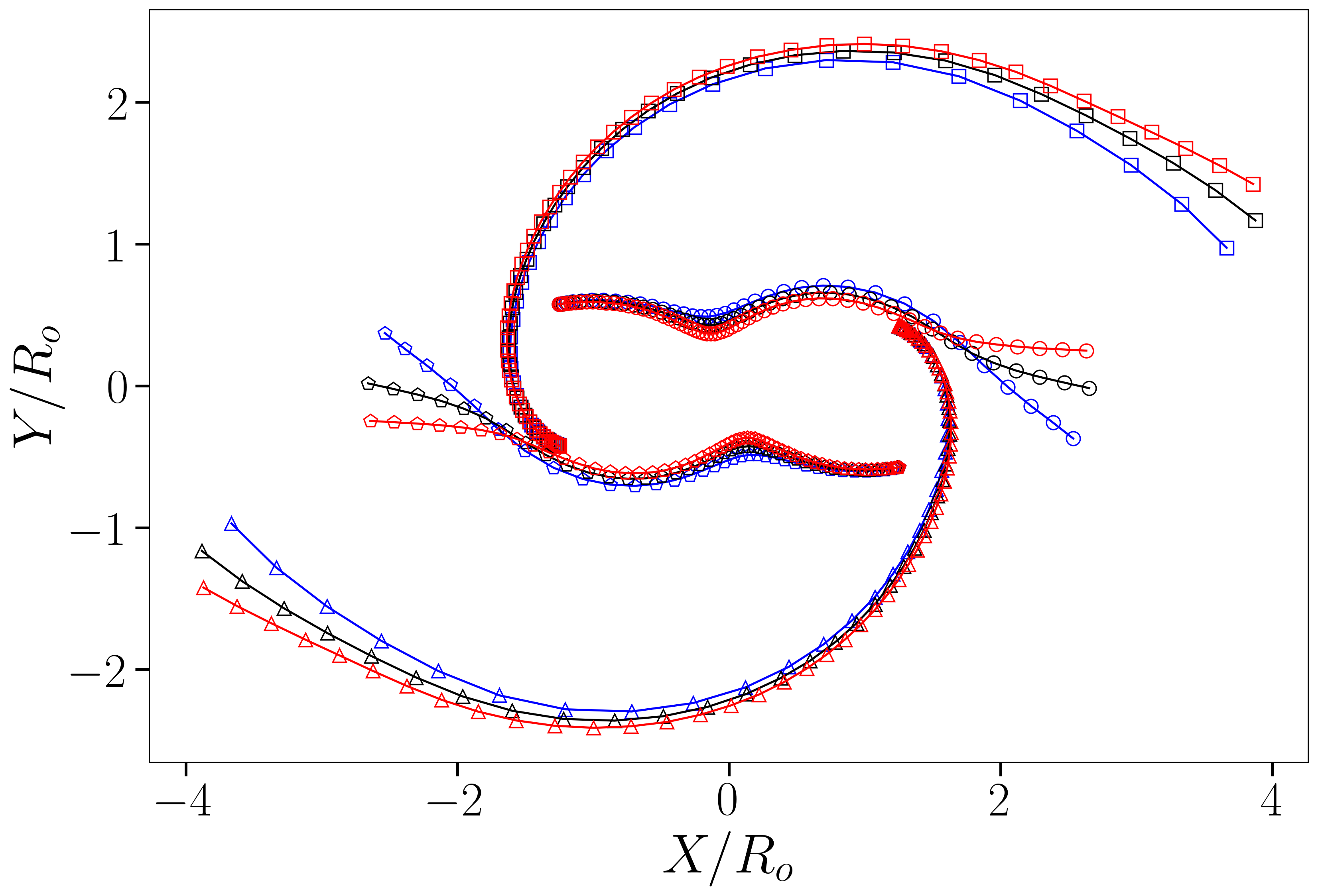} \\
        (c) Shell droplets ($\mu_{12}=42$) &
        (d) Core droplets ($\mu_{12}=42$) \\ 
  
        \includegraphics[scale = 0.21]{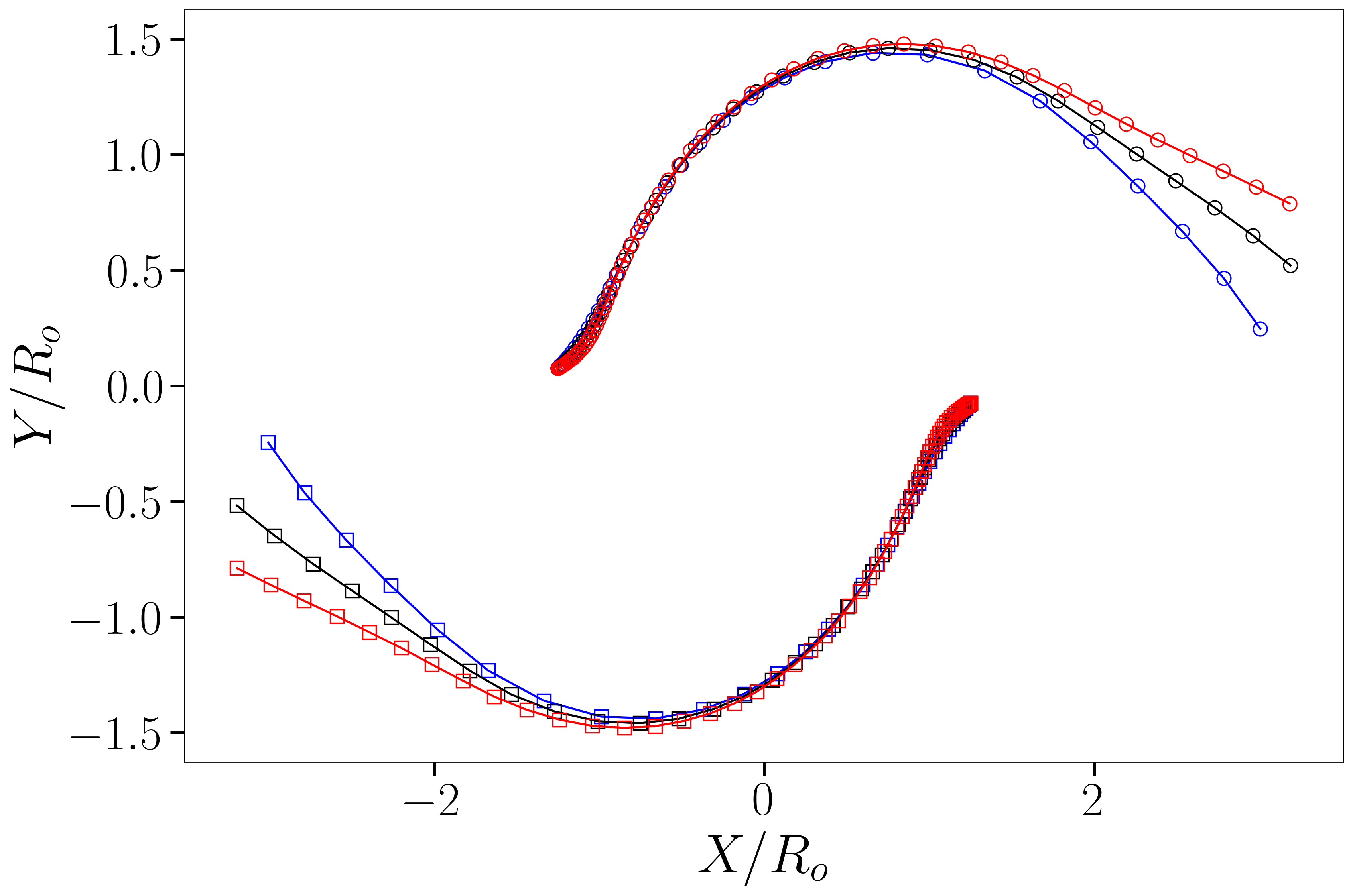} & \includegraphics[scale = 0.21]{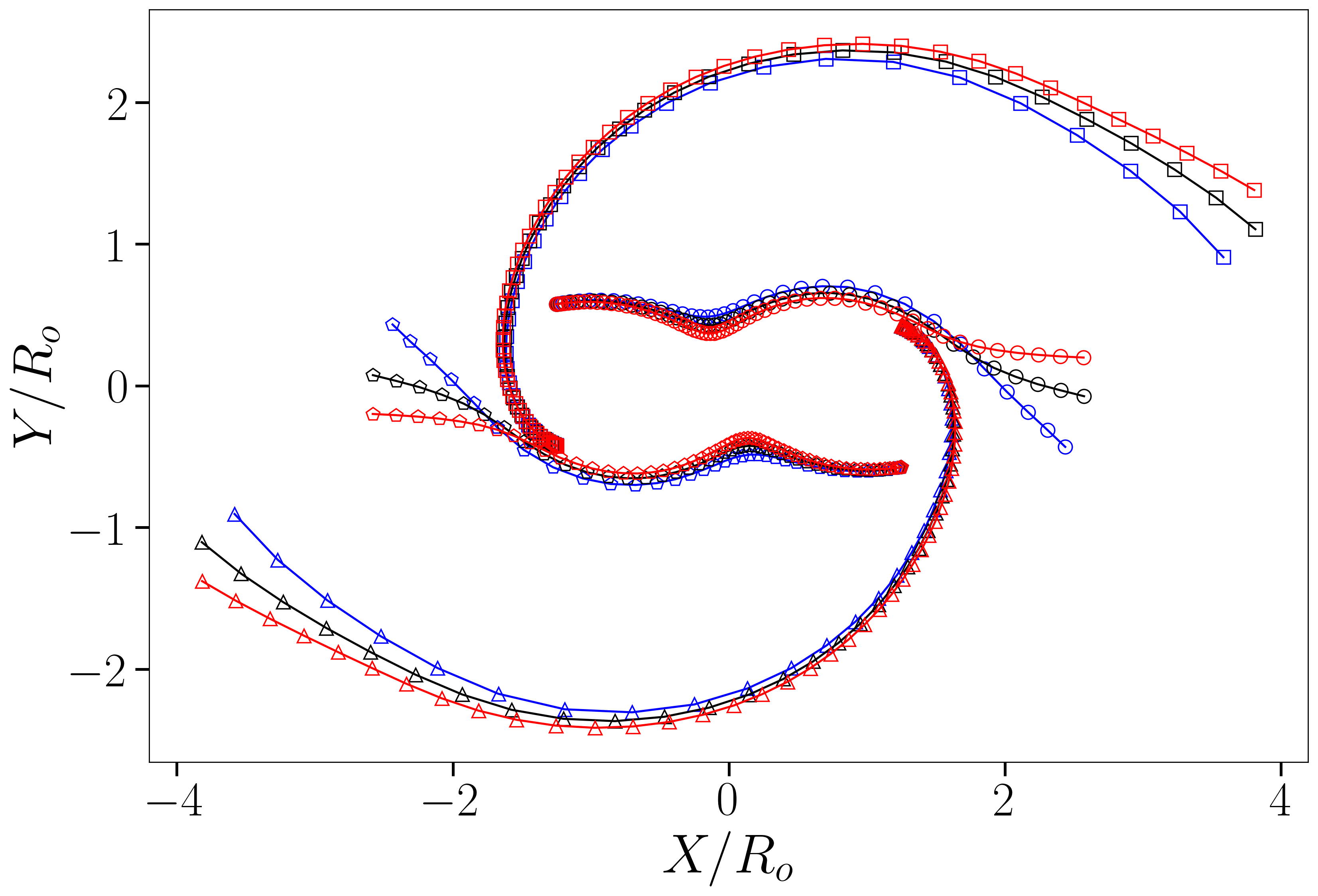} \\
        (e) Shell droplets ($\mu_{12}=60$) &
        (f) Core droplets ($\mu_{12}=60$) \\ 
    \end{tabular}

    \caption{Trajectory of shell and core droplets observed over time for the cases with nine different combination of three distinct density ratios ($\rho_{12}$ = $100$, $200$, $400$) and viscosity ratios ($\mu_{12}$ = $24$, $42$, $60$). The other simulation parameters are $Re = 1.0$, $Ca=0.1$, $\frac{R_o}{H}=0.40$, $\frac{R_i}{R_o} = 0.375$, $\frac{\Delta X}{R_o}=2.50$, and $\frac{\Delta Y}{R_o}=0.15$.}
    \label{fig:two-core-compound-trajectory-dy-0.15}
\end{figure}

Sections~\ref{sec_5.3}(1-2) highlight the significant influence of the interplay between initial vertical offset, density ratio, and viscosity ratio on the collision dynamics of multicore compound droplets, although the ultimate outcome in all scenarios remains a pass-over motion. To further probe the combined effects of these parameters on the multi-core compound droplet collision dynamics, we conducted simulations with an even lower initial vertical offset of $\frac{\Delta Y}{R_o}=0.15$. This specific offset value was previously observed to result in a reverse-back outcome in our studies on single-core compound droplet collisions~\cite{al2024hydrodynamic}. By keeping the vertical offset minimal, the compound droplets remain in close proximity, thereby enhancing the interaction effects. 
The trajectory plots for both shell and core droplets, as shown in Figure~\ref{fig:two-core-compound-trajectory-dy-0.15}, reveal that the shell droplets exhibit similar path trends throughout their evolution, with minor deviations at the end for the three different density ratios. 

This consistency aligns with the findings from our analyses involving higher initial vertical offsets. For the core droplets, a similar behavioral pattern is observed; however, with the minimal vertical offset ($\frac{\Delta Y}{R_o}=0.15$), there are notable differences in their rotational dynamics. The shell droplets, due to their close proximity resulting from the minimal vertical offset, nearly collide head-on and cannot progress much further forward. Instead, the shear flow prompts the shell droplets to begin rotating. This rotational movement significantly influences the behavior of the core droplets. Specifically, the top core droplet of the left shell droplet ($D1_{in-1}$) and the bottom core droplet of the right shell droplet ($D2_{in-2}$) do not change their vertical positions significantly. In contrast, the other two core droplets ($D1_{in-2}$ and $D2_{in-1}$) undergo nearly a full rotation, flipping their initial vertical positions. This rotational behavior can be attributed to the enhanced interaction effects resulting from the minimal vertical offset. The minimal vertical distance between the shell droplets restricts their forward progress, causing them to rotate under the influence of shear flow. 


\begin{figure}[H]
    \centering
    \begin{tabular}{cc}
        \includegraphics[scale = 0.21]{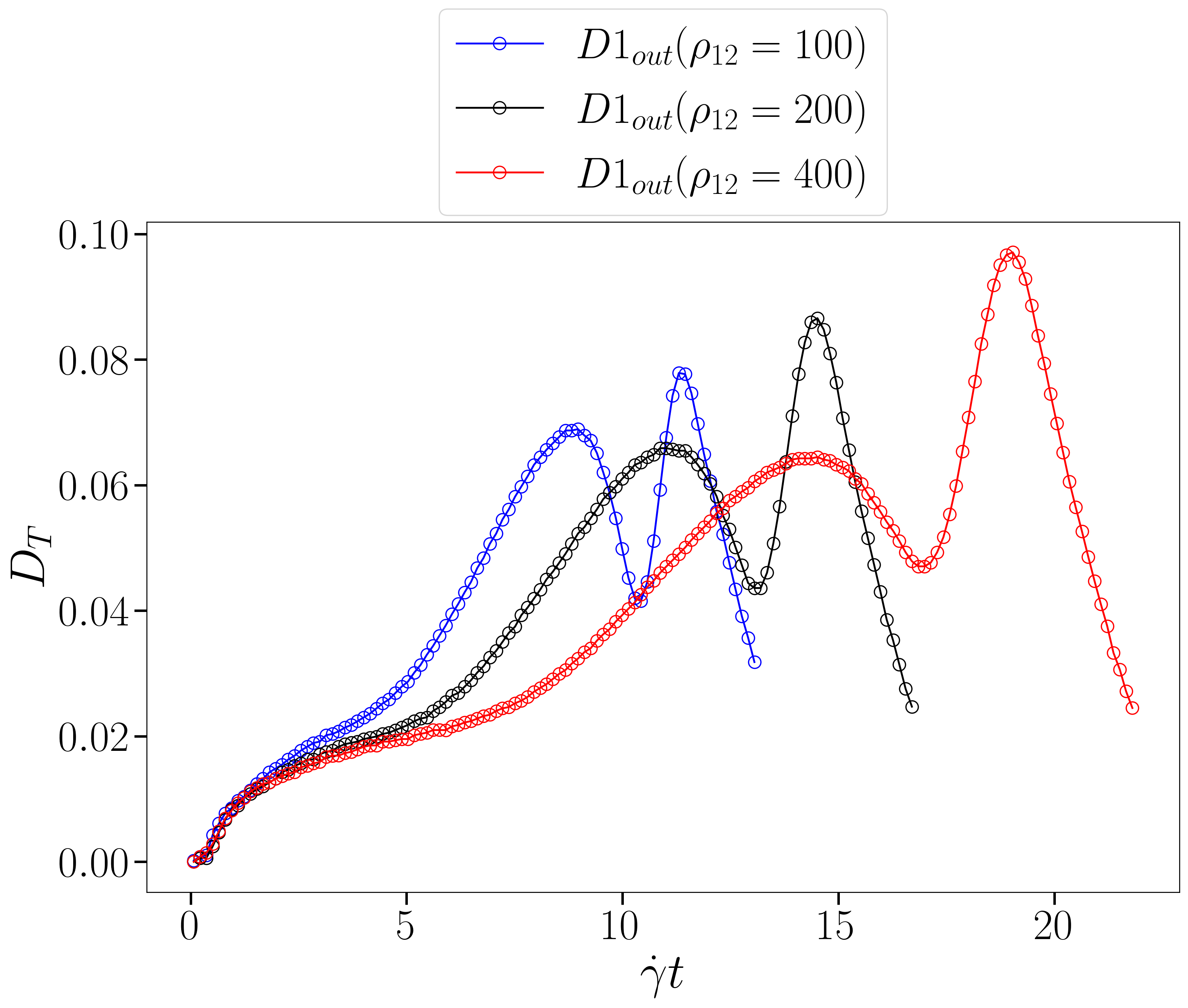} & \includegraphics[scale = 0.21]{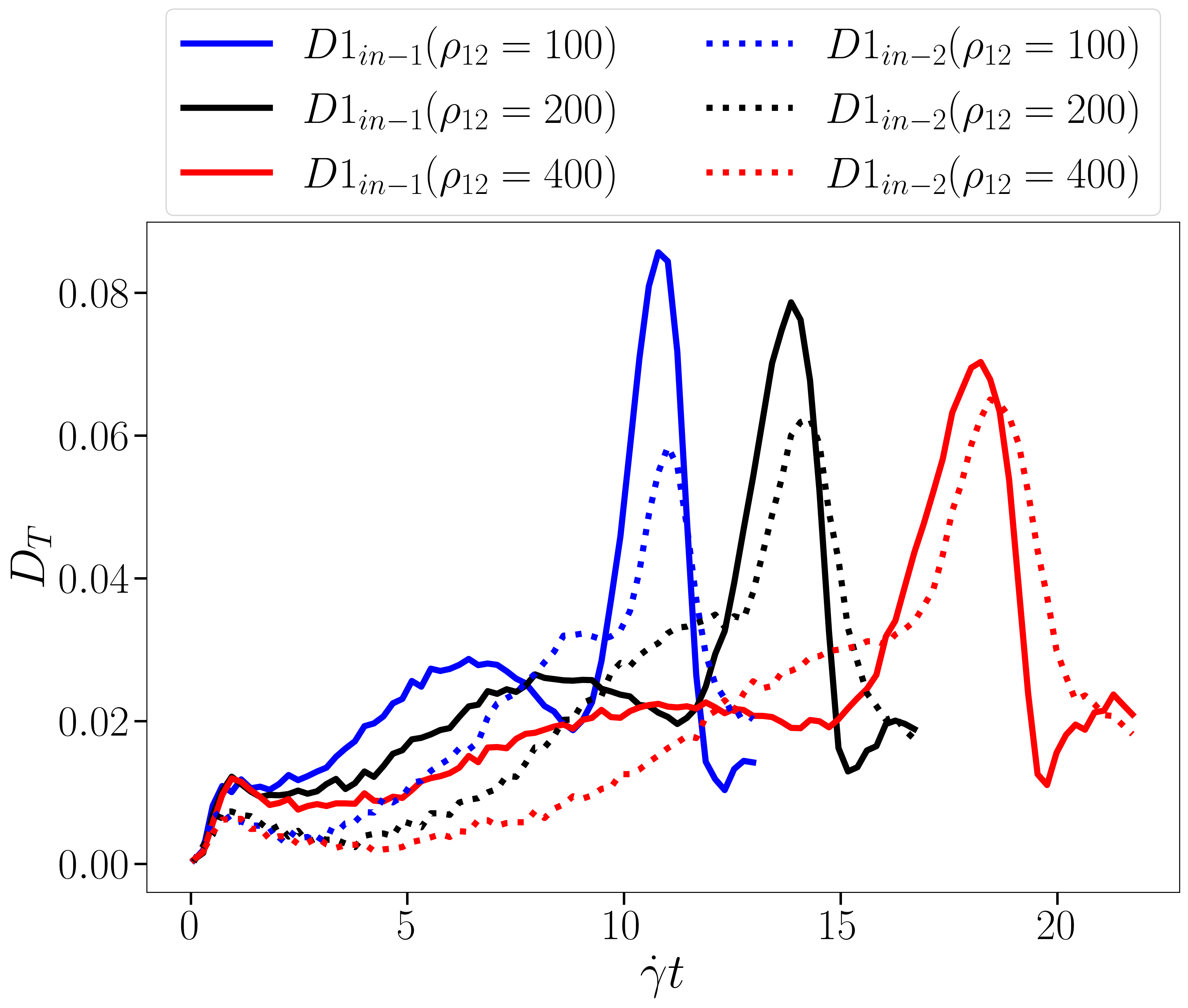} \\
        (a) Shell droplets ($\mu_{12}=24$) &
        (b) Core droplets ($\mu_{12}=24$)  \\ 
    
        \includegraphics[scale = 0.21]{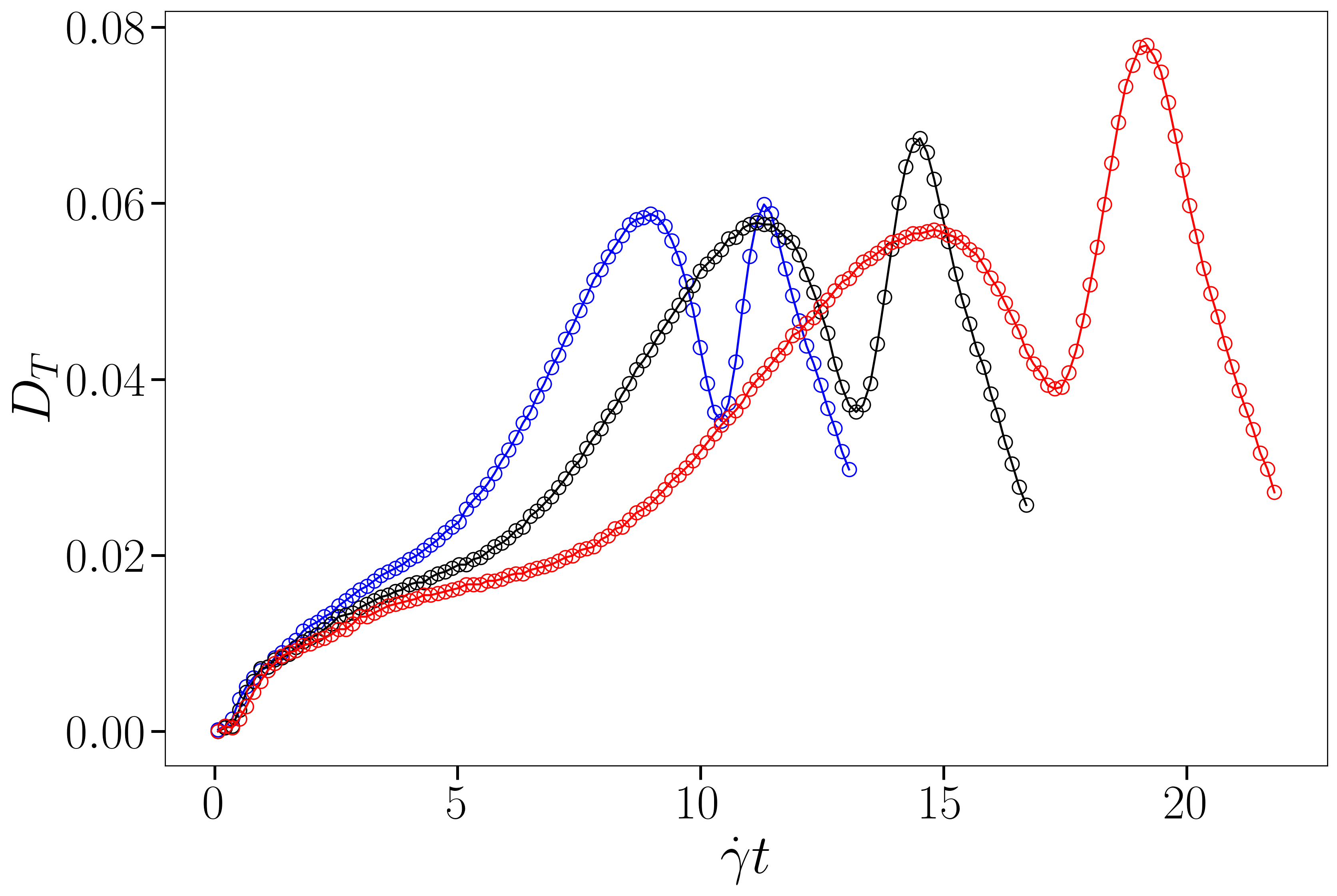} & \includegraphics[scale = 0.21]{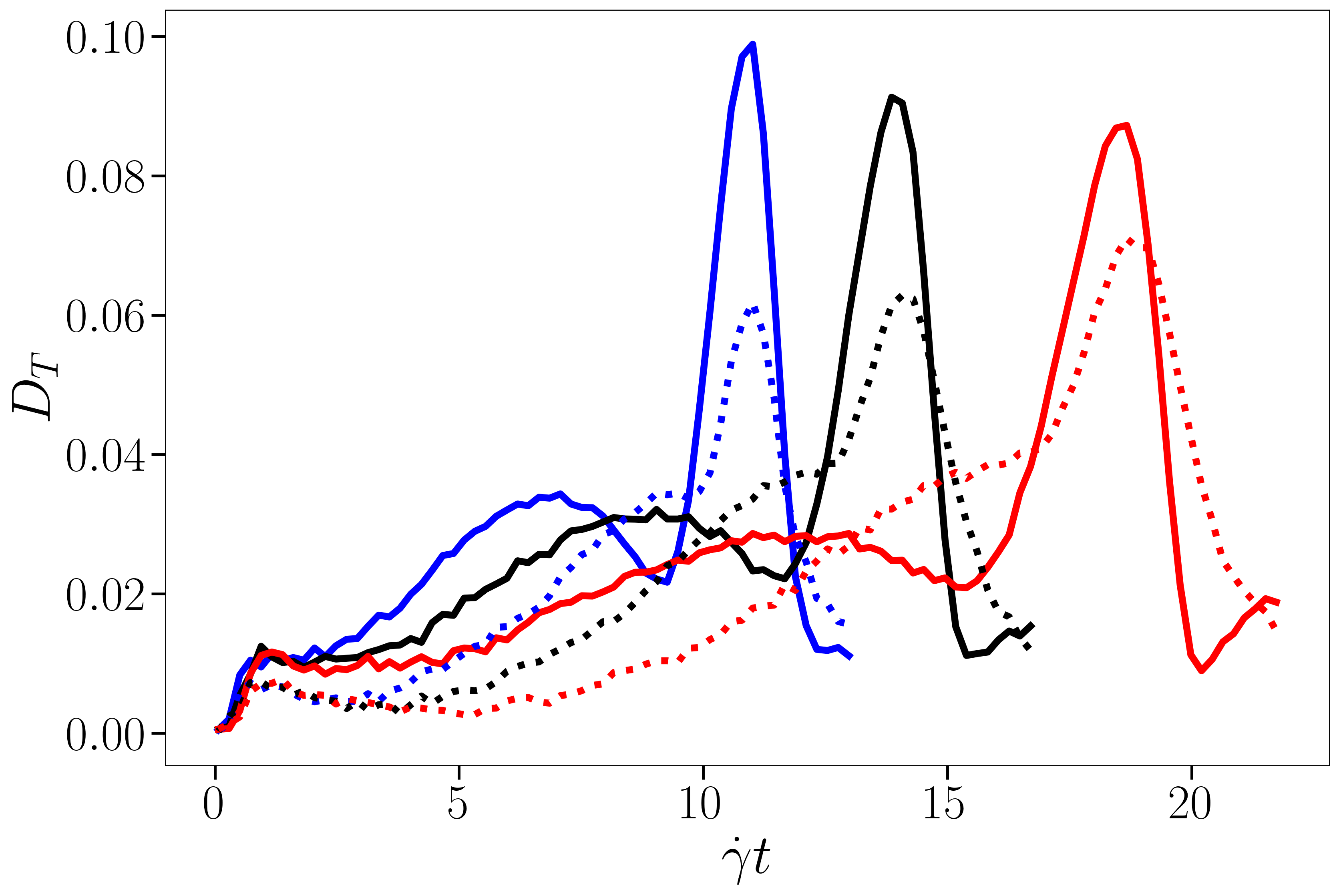} \\
        (c) Shell droplets ($\mu_{12}=42$) &
        (d) Core droplets ($\mu_{12}=42$) \\ 
  
        \includegraphics[scale = 0.21]{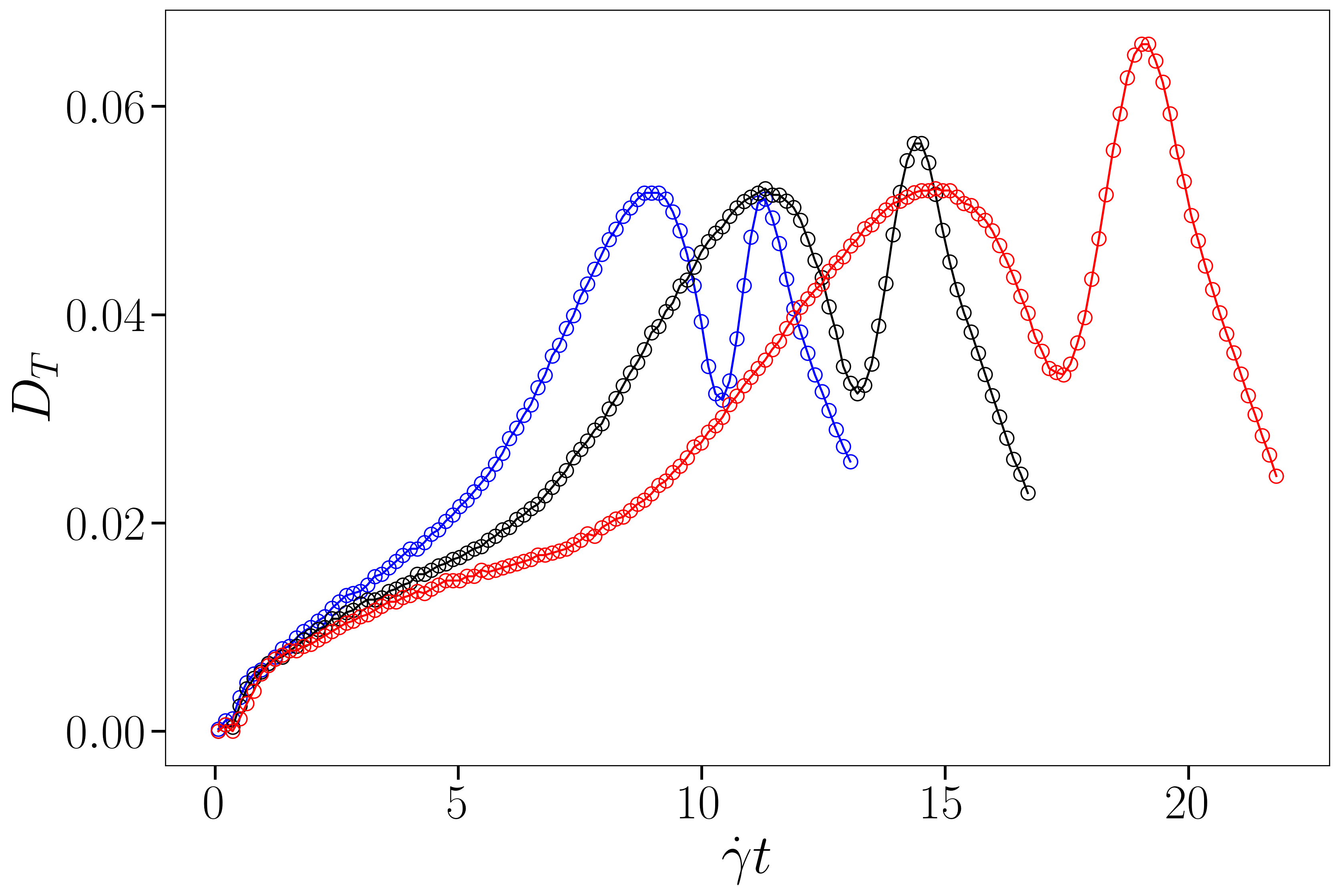} & \includegraphics[scale = 0.21]{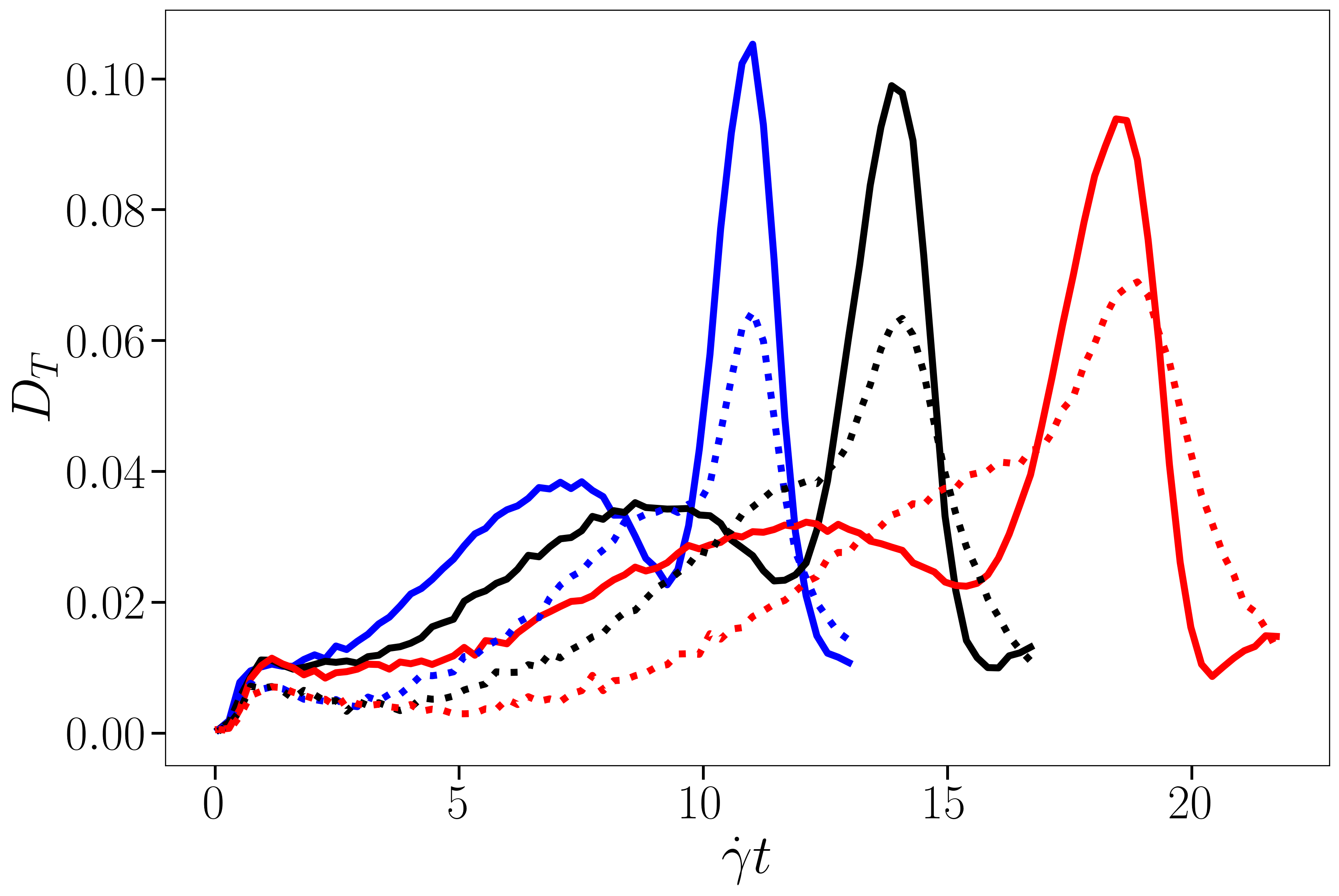} \\
        (e) Shell droplets ($\mu_{12}=60$) &
        (f) Core droplets ($\mu_{12}=60$) \\ 
    \end{tabular}

    \caption{Deformation of shell and core droplets observed over time for the cases with nine different combination of three distinct density ratios ($\rho_{12}$ = $100$, $200$, $400$) and viscosity ratios ($\mu_{12}$ = $24$, $42$, $60$). The other simulation parameters are $Re = 1.0$, $Ca=0.1$, $\frac{R_o}{H}=0.40$, $\frac{R_i}{R_o} = 0.375$, $\frac{\Delta X}{R_o}=2.50$, and $\frac{\Delta Y}{R_o}=0.15$.}
    \label{fig:two-core-compound-deform-dy-0.15}
\end{figure}

The rotation of the shell droplets, in turn, initiates the rotational movement of the core droplets ($D1_{in-2}$ and $D2_{in-1}$), as mentioned earlier. The top core droplet of the left shell ($D1_{in-1}$) and the bottom core droplet of the right shell ($D2_{in-2}$) start with a similar rotational movement but soon stabilize in their vertical positions as the other two core droplets complete their rotations more quickly, flipping their positions. The differential rotational dynamics of the core droplets can be explained by the variations in the local stress environment induced by the shear flow and the proximity of the shell droplets. The core droplets experiencing faster rotations are likely subject to greater shear forces and interactions with the surrounding fluid and other droplets, leading to their positional flips. In contrast, the core droplets maintaining their vertical positions are less affected by these forces, likely due to a relatively stable local environment or initial positioning within the shell droplet. The trajectory analysis with a minimal initial vertical offset reveals that the shell droplets follow consistent path trends with minor deviations, while the core droplets exhibit distinct rotational behaviors influenced by the close proximity and resulting interactions. The shear flow-induced rotation of the shell droplets significantly affects the core droplets, causing varied rotational dynamics that underscore the complex interplay between density, viscosity, and initial positioning in determining the overall behavior of multicore compound droplet collisions.

In examining the deformation behavior of multicore compound droplets with a minimal initial vertical offset ($\frac{\Delta Y}{R_o}=0.15$), we have generated and analyzed deformation plots for both shell and core droplets, as depicted in Figure~\ref{fig:two-core-compound-deform-dy-0.15}. The deformation trends observed in the shell droplets display a characteristic pattern with two peaks during their evolution, although distinct differences are noted compared to cases with a higher initial offset ($\frac{\Delta Y}{R_o}=1.50$). Specifically, the first peak in deformation magnitude is lower than the second peak, indicating different interaction dynamics under minimal offset conditions.
The minimal vertical offset between the shell droplets causes a nearly frontal collision, resulting in significant resistance due to the bulk continuous fluid between them. This resistance limits the extent of deformation during the initial collision phase. Instead, the shear effect from the continuous fluid induces a rotational movement, causing the left shell droplet ($D1$) to rise over the right one ($D2$), eventually leading to a pass-over motion. This contrasts with the higher initial offset scenario where droplets had more convenience to deform initially. The second peak in deformation occurs during the subsequent separation of the shell droplets, a consistent observation across all initial vertical offset cases examined. When considering the effect of varying density ratios, a familiar trend emerges. Increased density results in a longer time required for the pass-over motion, as the droplets, being heavier, move more slowly. Despite this, the deformation behavior of shell droplets remains similar across different density ratios, each exhibiting two peaks. This suggests that while density influences the speed and timing of droplet interactions, the overall deformation pattern is consistent across different density conditions. The influence of viscosity ratio on deformation is also notable. Higher viscosity in the droplets leads to lower peak deformation magnitudes. This is expected, as the continuous fluid, with comparatively lower viscosity, struggles to deform the shape of highly viscous droplets through shear drag. The higher the viscosity of the droplet, the more resistant it is to deformation under the shear forces exerted by the continuous fluid. This behavior underscores the damping effect of viscosity on deformation dynamics. The deformation curves for the core droplets reveal a more complex and fluctuating pattern, characterized by multiple peaks and valleys. This fluctuation is indicative of the dynamic stress environment experienced by the core droplets. From the trajectory plots (Figure~\ref{fig:two-core-compound-trajectory-dy-0.15}), we observed that the top core droplet of the left shell droplet ($D1_{in-1}$) could not rotate much due to the localized stress environment. 

This limited rotation consequently leads to higher deformation based on the localized stress it experiences. Conversely, the bottom core droplet of the left shell droplet ($D1_{in-2}$) rotates more freely, flipping its position and showing lower deformation due to reduced resistance to its movement. Overall, these findings highlight that the minimal vertical offset enhances interaction effects, leading to a near head-on collision that induces rotation rather than extensive initial deformation. The density ratio influences the timing and speed of interactions, while the viscosity ratio affects the magnitude of deformation, with higher viscosity providing greater resistance to shape changes.



\section{\label{sec:level4}Conclusion}

In conclusion, we have numerically investigated the dynamics and deformation behaviors of double-core compound droplet pairs interacting under confined shear flows. The study addresses a gap in existing literature by exploring how physical (e.g., Ca number, viscosity ratio, and density ratio) and geometric (e.g., initial vertical offset ratio) parameters influence collision outcomes, droplet trajectories, and deformation characteristics. 

We first examined how the dual influence of $Ca$ and initial vertical offset dictates both the morphological evolution and overall pairwise interactions of multi-core compound droplets. The simulations span initial vertical offset ratios ranging from $\frac{\Delta Y}{R_o}=0.0$ to $1.5$ and $Ca$ values between $0.05$ to $0.1$, while keeping all other parameters, including density and viscosity ratios fixed. Although the shear rate in our simulation is maintained at low levels to prevent the breakup of compound droplets, notable shape deformations still occur, leading to the emergence of previously unobserved collision dynamics.
Our findings reveal two distinct phenomena at high $Ca=0.1$: (i) shell droplet coalescence, with core droplets remaining separated and rotating in a planetary-like motion, and (ii) pass-over motion, where shell droplets avoid coalescence and core droplets maintain rotational and translational motion. At lower $Ca=0.07$, new collision outcomes emerge, including the coalescence of both shell and core droplets or shell droplet pass-over combined with core droplet coalescence. At the lowest $Ca=0.05$, shell droplet coalescence becomes dominant, with the inner core droplets consistently rotating inside the merged shell. 
Beyond collision outcomes, the interplay between $Ca$ and vertical initial offset also affects deformation and trajectory. At lower $Ca$, increased surface tension enhances droplet stability, leading to reduced deformation and a greater tendency for shell droplets to coalesce. Conversely, higher $Ca$ enables larger deformation, allowing for pass-over events at higher offsets. The trajectories reveal that as initial offset between shell droplets increases, shell droplets exhibit more pronounced pass-over motions, while core droplets maintain distinct rotational paths within their shells.

Extending our study beyond the unit density and viscosity ratios, we then explored the interplay of initial vertical offset ($\frac{\Delta Y}{R_o}=0.0-1.5$), viscosity ratio ($\rho_{12}=100-400$), and density ratio ($\mu_{12}=24-60$) on the hydrodynamic responses and deformation patterns of double-core compound droplets. At high initial vertical offsets of $\frac{\Delta Y}{R_o}=1.5$, the dominant outcome is pass-over motion across all combinations of density and viscosity ratios beyond unity. Higher density ratio results in increased droplet momentum, causing greater post-collision separation of shell droplets, while higher viscosity ratios stabilizes droplet shape and reduces deformation. At moderate ($\frac{\Delta Y}{R_o}=0.75$) and low ($\frac{\Delta Y}{R_o}=0.15$) initial offsets, the interactions become more intense, with shell droplets experiencing higher resistance and deformation due to reduced separation. Increased viscosity ratio dampens deformation, while higher density ratios delays collision processes due to the added inertia. At minimal offsets, the proximity enhances interaction effects, leading to rotational dynamics and unique deformation behaviors. Notably, the inner core droplets exhibit continuous fluctuations in deformation, influenced by their planetary-like motion and the stress environment within the shell.

Our analysis shows that the pairwise dynamics of multi-core compound droplets are substantially more complex compared to those observed in simple droplet pairs or even single-core compound droplet pairs. We aim for our findings to enhance the understanding of the mechanisms governing the dynamic behavior of confined core droplets within shell droplets interacting in microchannels. These insights are particularly relevant for the design and optimization of soft composite materials based on compound droplets, as well as for practical applications such as drug delivery systems, where the release of encapsulated drugs within the cores can be significantly impacted by the interplay of the physical and geometrical properties of both the shell and core droplets.
The deformation, trajectory, and collision outcomes of multi-core compound droplet pairs interacting in confined shear flows have not been addressed in prior research. Thus, the value of this work lies in the discovery of these complex pictures of interfacial evolution. Furthermore, the results offer a robust framework for designing future experiments and simulations to explore the interactions of multi-core compound droplet pairs in diverse flow environments.\\ 
\\
\textbf{AUTHOR DECLARATIONS}\\
\textbf{Conflict of Interest}\\
The authors have no conflicts to disclose.\\ \\
\textbf{DATA AVAILABILITY}\\
The data supporting this study’s findings are available from the corresponding author upon reasonable request.

\providecommand{\noopsort}[1]{}\providecommand{\singleletter}[1]{#1}%

\end{document}